
%
\magnification 1200
\input epsf.tex
%

%
\font\eightrm=cmr8
\font\eighti=cmmi8
\font\eightsy=cmsy8
\font\eightbf=cmbx8
\font\eighttt=cmtt8
\font\eightit=cmti8
\font\eightsl=cmsl8
\font\sixrm=cmr6
\font\sixi=cmmi6
\font\sixsy=cmsy6
\font\sixbf=cmbx6
\catcode`@11
\newskip\ttglue
\font\grrm=cmbx10 scaled 1200

\def\eightpoint{\def\rm{\fam0\eightrm}
\textfont0=\eightrm \scriptfont0=\sixrm \scriptscriptfont0=\fiverm
\textfont1=\eighti \scriptfont1=\sixi \scriptscriptfont1=\fivei
\textfont2=\eightsy \scriptfont2=\sixsy \scriptscriptfont2=\fivesy
\textfont3=\tenex \scriptfont3=\tenex \scriptscriptfont3=\tenex
\textfont\itfam=\eightit \def\it{\fam\itfam\eightit}
\textfont\slfam=\eightsl \def\sl{\fam\slfam\eightsl}
\textfont\ttfam=\eighttt \def\tt{\fam\ttfam\eighttt}
\textfont\bffam=\eightbf
\scriptfont\bffam=\sixbf
\scriptscriptfont\bffam=\fivebf \def\bf{\fam\bffam\eightbf}
\tt \ttglue=.5em plus.25em minus.15em
\normalbaselineskip=6pt
\setbox\strutbox=\hbox{\vrule height7pt width0pt depth2pt}
\let\sc=\sixrm \let\big=\eightbig \normalbaselines\rm}
\newinsert\footins
\def\newfoot#1{\let\@sf\empty
  \ifhmode\edef\@sf{\spacefactor\the\spacefactor}\fi
  #1\@sf\vfootnote{#1}}
\def\vfootnote#1{\insert\footins\bgroup\eightpoint
  \interlinepenalty\interfootnotelinepenalty
  \splittopskip\ht\strutbox 
  \splitmaxdepth\dp\strutbox \floatingpenalty\@MM
  \leftskip\z@skip \rightskip\z@skip
  \textindent{#1}\footstrut\futurelet\next\fo@t}
\def\fo@t{\ifcat\bgroup\noexpand\next \let\next\f@@t
  \else\let\next\f@t\fi \next}
\def\f@@t{\bgroup\aftergroup\@foot\let\next}
\def\f@t#1{#1\@foot}
\def\@foot{\strut\egroup}
\def\footstrut{\vbox to\splittopskip{}}
\skip\footins=\bigskipamount 
\count\footins=1000 
\dimen\footins=8in 

\def\ref#1{$^{#1}$}
\def\flex{\raise 6pt\hbox{$\leftrightarrow $}\! \! \! \! \! \! }
\def\oversome#1{ \raise 8pt\hbox{$\scriptscriptstyle #1$}\! \! \! \! \! \! }
\def\tr{ \mathop{\rm tr}}

\newbox\bigstrutbox
\setbox\bigstrutbox=\hbox{\vrule height10pt depth5pt width0pt}
\def\bigstrut{\relax\ifmmode\copy\bigstrutbox\else\unhcopy\bigstrutbox\fi}
\def\refer[#1/#2]{ \item{#1} {{#2}} }
\def\rev<#1/#2/#3/#4>{{\it #1\/} {\bf#2}, {#3}({#4})}
\def\boxit#1{\vbox{\hrule\hbox{\vrule\kern3pt
\vbox{\kern3pt#1\kern3pt}\kern3pt\vrule}\hrule}}

\def\sqr#1#2{{\vcenter{\hrule height.#2pt
   \hbox{\vrule width.#2pt height#1pt \kern#1pt
    \vrule width.#2pt}
    \hrule height.#2pt}}}
\def\dal{\mathchoice{\sqr{6}{4}}{\sqr{5}{3}}{\sqr{5}3}{\sqr{4}3} \, }


\def\smin{\,\raise 0.06em \hbox{${\scriptstyle \in}$}\,}
\def\smsubset{\,\raise 0.06em \hbox{${\scriptstyle \subset}$}\,}

\def\Natural{\hbox{\hskip 1.5pt\hbox to 0pt{\hskip -2pt I\hss}N}}

\def\Rational{\hbox{\hbox to 0pt{\hskip 2.7pt \vrule height 6.5pt
                                  depth -0.2pt width 0.8pt \hss}Q}}
\def\Real{\hbox{\hskip 1.5pt\hbox to 0pt{\hskip -2pt I\hss}R}}
\def\Complex{\hbox{\hbox to 0pt{\hskip 2.7pt \vrule height 6.5pt
                                  depth -0.2pt width 0.8pt \hss}C}}
\def \E {{{\rm e}}}

\def \tr {{\rm tr}\, }
\def \ln {{\rm ln}\, }

\nopagenumbers
\def \cotg {\rm cotg }
\def \gmu {g^{^{-1}}}
\hfill CERN-TH/95-49

\hfill hep-th/9503002
\vskip 1cm
\centerline {\grrm  Updating QCD$_2$}
\vskip .6cm

\centerline { E. Abdalla\newfoot {${}^*$}{Permanent address: Instituto de
F\'\i sica - USP, C.P. 20516, S. Paulo, Brazil.} and
M.C.B. Abdalla\newfoot
{${}^\dagger $}{Permanent address: Instituto de F\'\i sica
Te\'orica - UNESP, R. Pamplona 145, 01405-000, S. Paulo, Brazil.}}
\vskip .4cm

\centerline{ CERN-TH}

\centerline{ CH-1211 Geneva 23}

\centerline{ Switzerland}
\vskip 1cm
\centerline{\bf Abstract}
\vskip .5cm

\noindent We review two-dimensional QCD. We start with the field theory aspects
since 't Hooft's  $1/N$ expansion, arriving at the non-Abelian bosonization
formula, coset construction and gauge-fixing procedure. Then we consider the
string interpretation, phase structure and the collective coordinate approach.
Adjoint matter is coupled to the theory, and the Landau--Ginzburg
generalization
is analysed. We end with considerations concerning higher algebras,
integrability, constraint structure, and the relation of high-energy scattering
of hadrons with two-dimensional (integrable) field theories.
\vfill
\centerline{To appear in Physics Reports}
\vskip 1.5cm

\noindent CERN-TH/95-49
\vskip .3cm
\noindent March 1995
\eject
\countdef\pageno=0 \pageno=1
\newtoks\footline \footline={\hss\tenrm\folio\hss}
\def\folio{\ifnum\pageno<0 \romannumeral-\pageno \else\number\pageno \fi}
\def\advancepageno{\ifnum\pageno<0 \global\advance\pageno by -1
\else\global\advance\pageno by 1 \fi}

\centerline{\bf CONTENTS}
\vskip 2cm

\noindent 1. Introduction \dotfill   2

\noindent 2. QCD$_2$ as a field theory \dotfill   6

2.1  The $1/N$ expansion and  spectrum \dotfill   10

2.2   Ambiguity in the self-energy of the quark \dotfill  17

2.3   Polyakov--Wiegmann formula and gauge interactions \dotfill   20

2.4   Strong coupling analysis \dotfill   30

2.5   A Lagrangian realization of the coset construction  \dotfill   33

2.6   Chiral interactions  \dotfill   37

\noindent 3. Pure QDC$_2$ and string theory\dotfill 39

3.1 Introduction \dotfill  39

3.2 Wilson loop average and large-$N$ limit  \dotfill   40

3.3 String interpretation    \dotfill   47

3.4 Collective coordinates approach  \dotfill   57

3.5 Phase structure of QCD$_2$  \dotfill   62

\noindent 4. Generalized QCD$_2$ and adjoint-matter coupling \dotfill 64

4.1  Introduction and motivation \dotfill   64

4.2  Scalar and fermionic matter coupling; quantization \dotfill   65

4.3  The Hagedorn transition; supersymmetry \dotfill   69

4.4  Landau--Ginzburg description; spectrum and string theory \dotfill   72

\noindent 5. Algebraic aspects of QCD$_2$ and integrability \dotfill 77

5.1 $W_\infty$ algebras for colourless bilinears \dotfill   77

5.2 Integrability and duality   \dotfill 80

5.3 Constraint structure of the theory \dotfill  87

5.4 Spectrum and comparison with the $1/N$ expansion  \dotfill  92

5.5 Integrability conditions and Calogero-type models \dotfill  95

5.6 QCD at high energies and two-dimensional field theory \dotfill 99

\noindent 6. Conclusions \dotfill   109

\noindent Appendix A \dotfill  110

\noindent Appendix B \dotfill  112

\noindent Appendix C \dotfill  113

\noindent References \dotfill    116
\vfill \eject
\noindent {\bf 1. Introduction}
\vskip .5cm
\noindent
Over several years, gauge theories have proved their efficacy to describe
phenomena in the high energy domain and QCD is a important part of such a
theoretical framework. The high energy behaviour of strong interactions, as
analyzed by means of renormalization group (RG) and Callan--Symanzik (CS)
equations can be investigated by a perturbative expansion, permitting
to confront this facet of the theory with experiment, with excellent
results.\ref{1} Indeed, the momentum-dependent running coupling constant
characterizes the strength of the interaction, in such a way that because of
the negative $\beta$-function of QCD, perturbation theory is legitimate.

But this is only part of the development of gauge theories, after the
long-standing success of quantum electrodynamics.\ref{2} Indeed, classical
solutions (monopoles or instantons) have been obtained, showing the complexity
of the theory and the importance of topology. In this way, more abstract
branches of mathematics came to play an important role in the unravelling of
the structural properties of gauge theories.\ref{3}

For the time being, the use of more sophisticated gauge theories, namely
the use of larger gauge groups, combined with the idea of symmetry breaking and
the Higgs mechanisms, has led to the development of unified theories to
describe the high energy interactions, albeit not including gravity at a first
step. Thus the electroweak $SU(2)\otimes U(1)$ symmetric gauge interaction,
and the strong interaction described by the $SU(3)$ colour gauge group, could
in principle be unified into the framework of higher gauge groups, such as
$SU(5)$, $O(10)$ or even more sophisticated ones.\ref{4}

The inclusion of gravity in such a scheme is provided by the introduction of
supersymmetry, which is, moreover, potentially sufficient to solve the
hierarchy
problem.\ref{5} However, further troubles concerning the introduction of
gravity in such a unifying scheme is only solved in the framework of string
theory.\ref{6,7}

On the other hand, string theory was developed as a model for strong
interactions, as a consequence of ideas related to dual models. Strong
interactions as described by quantum chromodynamics must, in a sense, exist in
two phases, an infrared phase, with confined quarks bound into mesons,
evolving according to a (non-critical) string theory, and a high energy
phase, described by the perturbative expansion as previously mentioned.

In fact, we have to point out that two kinds of string theories exist,
associated to different types of phenomena. Critical strings are appropriate to
implement the unification of all gauge theories, but are not the subject of
concern here. Non-critical strings, on the other hand, should provide the
framework necessary for the description of strong interactions, being
presumably related to gauge theories. These ideas, although appealing, and in
conformity with the general intuition about high energy interactions have not
yet been fully accomplished, in a realistic gauge theory.

In this scenario two-dimensional gauge theories are configurated as a
laboratory\ref{8} where ideas may be tested, and the relation between string
concepts and field theory may be analyzed in detail. Moreover, results such as
the spectrum, less orthodox, perturbative approaches as the one based on the
large-$N$ limit, the algebraic structure of the theory, and the recently
stressed duality properties may be analyzed in detail, and some of them
exactly.

The prototype of two-dimensional gauge theories is QED$_2$, or the Schwinger
model.\ref{9} In this model the gauge field acquires a mass via a dynamical
Higgs mechanism induced by the fermions. There is a spontaneous breakdown of
the chiral $U(1)\times\widetilde U(1)$ symmetry, where the first factor refers
to charge and the second to chirality of the fermion. The ground state
exhibits a double infinite degeneracy labelled by fermion number and
chirality, analogous to four-dimensional quantum chromodynamics.\ref{10} The
so-called ``infrared slavery" of QCD$_4$ is naturally described in QED$_2$ as
a consequence of the linear rise of the Coulomb potential, characteristic of
one space dimension.

The question of confinement is however not settled in this way, since
asymptotic states corresponding to screened quarks might exist as well. By all
means the Fock space is that of a free pseudoscalar bosonic field $\Sigma(x)$
of mass $e/\sqrt\pi$, where $e$ is the electric charge. This is a
manifestation of confinement in the Schwinger model.\ref{10} All states in
the physical Hilbert space ${\cal H}_{phys}$ can be constructed by applying
functionals of the bosonic field $\Sigma(x)$ on the irreducible vacuum state.
However, confinement cannot be fully understood before a flavour quantum number
is assigned to the fermion, since it is otherwise not possible to distinguish
between neutral bosonic states and screened fermionic states.\ref{11}

Bosonization of the Schwinger model\ref{12} provides further insight; the
mass content of the theory can be easily read from the diagonalization of the
quadratic Lagrangian, once we substitute the gauge field for $A_\mu=-{\sqrt\pi
\over e}\widetilde\partial_\mu(\Sigma+\eta)$. In such a case one can also
verify that the massive Schwinger model, which is not soluble, can be written
in terms of a modified sine-Gordon equation, in which the periodic symmetry is
broken by the electromagnetic interaction. The massive theory is far more
complicated than the massless case. The confinement issue can be understood
semi-classically; as it turns out, the screening effects are more violent in
this case, since such computation leads us to a potential whose linear rise is
sacrificed for the advantage of the screening picture. For a long discussion
see chapter 10 of ref. [8]. Moreover one finds that the field $\eta$ obeys a
massive equation. A similar outcome will be true in the non-Abelian case.

Several properties of the Schwinger model provide a realization of analogous
features expected to characterize four-dimensional quantum chromodynamics
(QCD$_4$). Besides the spontaneous breakdown of the chiral symmetry without
Goldstone bosons, infinite degeneracy of the vacuum and confinement, an
enormous amount of physical implications followed. However, the model is
certainly still too simple, and the first missing issue is the generalization
of the mesonic bound state, which in QCD is believed to present itself
according to the Regge behaviour, while in QED$_2$, owing to the simplicity of
the $U(1)$ symmetry, only functionals of the bosonic field $\Sigma (x)$ appear
in the physical spectrum.

Therefore, it is natural to include colour as the next step towards more
realistic models, implementing the non-Abelian character of the fundamental
fields, and consider quantum chromodynamics in two dimensions (QCD$_2$). In
this case, bosonization leads to a simpler theory, but the spectrum is still
rather complex, even if it can be worked out in the large-$N$ limit. The
Hilbert space contains a larger class of bound states, connected with the
Regge behaviour of the mass spectrum, thus leading to a large number of
states. The case of massless fermions is no longer soluble in terms of free
fields. Even pure gauge QCD$_2$ is non-trivial, and cannot be completely
solved in terms of simple fields, see section 3. The situation is even more
difficult in the case of massive fermions, since the exact fermion
determinant, necessary to obtain the bosonized action, is not available, and
the argument leading to bosonization must be based on the principle of form
invariance. However, it is from the latter model that we expect a more
realistic description of the string behaviour, thought to be the most
important character of four-dimensional non-Abelian gauge theories in their
description of the strong interaction. Still the question of confinement must
be analyzed, especially if one considers that it is possible to construct
operators such as the gauge-invariant part of the bound state of two fermions,
by means of the inclusion of strings of the type ${\rm exp}\left[ie\int_x^y
{\rm d}z^\mu\,A_\mu\right]$.

Several authors made efforts in the direction of solving such a difficult
model,\ref{13-27} some of them giving useful results, but an exact solution
is still missing. We mention the $1/N$ expansion introduced by
'tHooft,\ref{13,14} from which one obtains some information about the
spectrum of the theory, and the computation of the exact fermion
determinant\ref{15} in terms of a Wess--Zumino--Witten (WZW)
model,\ref{16,17} by which one arrives at an equivalent bosonic
action.\ref{17-24}

Quantum chromodynamics in two dimensions is a super-renormalizable field
theory with finite field and coupling constant renormalization. As already
mentioned it was first studied by 'tHooft\ref{14} who, working in the
light-cone gauge, $(A_-=0)$, and formulating the problem in terms of light-cone
variables, obtained a non-linear equation for the fermion self-energy, from
which he obtained the approximate spectrum of the theory. This procedure is
however ambiguous, as pointed out by Wu,\ref{25} and implies a tachyon for
small bare fermion masses (therefore also in the massless fermions case), see
also ref. [26]. This situation clearly requires that a non-perturbative and
explicitly gauge invariant approach should be used to obtain information
based on firm grounds. We will also see that Wu's and 'tHooft's results, for
the fermion self-energy, are different and lead to different physical
pictures: while 'tHooft's two-point fermion function corresponds to a simple
free Fermi behaviour\newfoot{$^{1)}$}{It is nevertheless
infrared-cut-off-dependent, due to quark confinement. The confining properties
are made manifest since quark poles are pushed to infinity as the cut-off
disappears. For gauge-independent quantities, such a procedure is equivalent to
interpreting all integrals as principal values.}, Wu's result presents an
anomalous branch cut reflecting the fact that all planar rainbow graphs
contribute, in his scheme, to the self-energy. Differently from the Abelian
case, fermion loops are suppressed in the large-$N$ limit, in a way that at
lowest order in $1/N$ the gluon field remains massless, leading to
speculations that QCD$_2$ might exist in two phases, associated to the weak or
strong coupling regime, where the weak coupling phase -- or 'tHooft's phase --
would be associated with massless gluons, with a Regge trajectory for the
mesons, while in the  strong coupling regime -- or Higgs phase -- the gluons
would be massive and the $SU(N)$ symmetry would be broken to its maximal
Abelian subgroup.

The nature of such strong or weak limits is very delicate. In fact, the theory
is asymptotically free, as it should since it is super-renormalizable. In the
strong coupling, it should be a confining  theory. The introduction of an
explicit infrared cut-off in fact selects the confining properties. In that
case, quarks disappear from the spectrum, which consists of mesons with a
Regge behaviour. As pointed out by Callan, Coote and Gross,\ref{27} for
gauge-invariant quantities one can interpret all integrals as principal
values, and we are led to the solution $\Sigma_{_{SE}}
(p)={e^2N\over \pi}{1\over p_-}$ for the self-energy (SE), with the fermion
two-point function $S_F(p)={\not p+m+{e^2N\over 2\pi}{\gamma_-\over p_-}\over
p^2-m^2+{e^2N\over\pi}}$. Such a procedure is useful to analyze properties
connected to the high-energy scattering amplitudes, displaying properties
connected to what is known as parton-like properties; that is, in the
high-energy limit, the quarks behave like free particles.

This does not mean, however, that confinement does not take place. To see the
confinement mechanism, one has to examine the current two-point function, and
as a result, quark continuum states do not appear, a fact confirming
confinement (for a more precise discussion see refs. [8] and [11]).

Such a procedure has some advantages, namely one can study the high energy
behaviour of the theory, which, because of asymptotic freedom, must exhibit a
free field structure, which is not the case with an infrared cut-off, since the
high-energy limit and the zero infrared cut-off limit do not commute.

In the massless case, there are several non-perturbative results available. In
particular, the external field problem for the effective action has been
analyzed, the computation of gauge current and fermionic Green's
functions can be reduced to the calculation of tree diagrams. There are also
features not covered by 'tHooft's method. As an example, if we take the
pseudo-divergence of the Maxwell equation we arrive at
$$
\left(\nabla^2 + {e^2\over \pi}\right) F_{01} = 0 \quad ,
$$
where $F$ is the gauge field strength; this equation generalizes the analogous
result obtained for the Schwinger model to the non-Abelian case. This suggests
that an intrinsic Higgs mechanism, analogous to the one well-known in QED$_2$,
can also characterize the non-Abelian theory. This is, nevertheless, not
contained in 'tHooft's approach, since the mass arises from a fermion loop,
the same which contributes to the axial anomaly, and it is suppressed in the
$1/N$ expansion.

In spite of difficulties, QCD$_2$ served as a laboratory for gaining insight
into various phenomenological aspects of four-dimensional strong interactions,
such as the Brodsky--Farrar scaling law\ref{28} for hadronic form factors, the
Drell--Yan--West relation or the Bloom--Gilman duality\ref{29} for deep
inelastic lepton scattering.

The next important step towards understanding this theory is its relation to
string theory, or SCD$_2$. It concerns one of the most important applications
of the theory of non-critical strings.\ref{30}

The general problem of strong interactions did not progress substantialy until
recently as far as it concerns low-energy phenomena. Such a problem should be
addressed using non-perturbative methods, since perturbation theory of strong
interactions is only appropriate for the high-energy domain, missing
confinement, bound-state structure and related phenomena. In fact, several
properties concerning hadrons are understandable by means of the concept of
string-like flux tubes, which are consistent with linear confinement and Regge
trajectories, as well as the approximate duality of hadronic scattering
amplitudes, which are the usual concepts of the string idea. In fact, a
similar idea is already present in the construction of the dipole of the
Schwinger model, in which case it is, however, far too simple to be realistic.

The large-$N$ limit of QCD$_2$ is smooth and provides a picture of the string
in
the Feynman diagram space. The large-$N$ limit is expected to provide most of
the qualitative pictures of the low-energy limit of the theory. In certain
low-dimensional systems, the $1/N$ expansion turns out to be {\it the} correct
expansion, for models with problematic infrared behaviour, such as $\Complex
P^{N-1}$ and Gross--Neveu models,\ref{31,32} where properties such as
confinement and spontaneous mass generation are straightforwardly derived in
the large-$N$ approximation, and the $S$-matrices can be explicitly
checked.\ref{33,34}

In short, these ideas support the suggestion  that the understanding of the
theory of strong interactions requires the study of the large-$N$ limit of QCD.
Although several models mimic such a theory in two dimensions, concerning
the confinement aspects, a more thorough comprehension by a simplified
two-dimensional model cannot be complete without the inclusion of QCD$_2$.

For pure QCD$_2$ the $1/N$ expansion of the partition function can be obtained
to arbitrary order, and may be interpreted as a sum over surfaces, thus
describing a string theory. The string action is not exactly known, but it is
described, in the zero area limit, by a topological field theory. Area
corrections are given by the Nambu action and possibly terms in the extrinsic
geometry, forbidding folds. It is clear that the theory without matter is yet
too simple. Even the introduction of fundamental fermions is not sufficient to
describe certain realistic aspects of the higher dimensional theory.

Matter in the adjoint representation of the gauge group provides fields which
mimic the transverse degrees of freedom characteristic of gauge theories in
higher dimensions, and may show more realistic aspects of strong interactions.
The main new point consists in the presence of a phase transition indicating a
deconfining temperature.

There are algebraic structures in QCD$_2$, indicated by canonical methods, and
such algebras point to the integrability of the model. In particular, there are
spectrum-generating algebras of the same type as that appearing in the quantum
Hall effect.\ref{35} Moreover, the relation to Calogero systems and the $c=1$
matrix model confirms such integrability properties, which can finally be
proved by the construction of a Lax pair. This opens the possibility of a
closed solution, at least at the S-matrix level (on-shell physics\ref{8}).

Finally, the high energy description of four-dimensional QCD is also described
by an integrable two-dimensional model, opening a possibility of more realistic
results from such a study. Indeed, at high energies Feynman diagrams simplify
and become effectively two dimensional. The theory may be described in the
impact parameter space, and in the case of QCD$_4$, the Reggeized particles
scatter according to an integrable Hamiltonian.
\vskip 1cm
\penalty-3000
\noindent {\bf 2. QCD$_2$ as a field theory}
\vskip .5cm
\nobreak
\noindent We start with the definition of the theory, from very early
developments, and concentrate on non-perturbative results. Since the theory
confines the fermionic degrees of freedom, it is usefull to integrate over the
fermions. This amounts to the computation of the fermionic determinant. In
terms of the original gauge fields it is given by an infinite series, as given
by eq. (2.15). Later the fermionic determinant will be computed in terms of the
potentials used to define the gauge fields, leading to the
Wess--Zumino--Witten action.

Before entering in the full bosonization of the theory (via the WZW action)
it is possible to investigate the mesonic spectrum through the $1/N$ expansion
of the theory, where the gauge group is $SU(N)$, or $U(N)$. In such a case
there is a simplification in the light-cone gauge, where the gauge fields do
not have self interactions, and the ghosts decouple. The self-energy of the
quarks can be exactly computed. A word of cautious concerning this method has
to be said concerning the infrared cut-off (see section 2.2).

After computing the fermionic determinant and writing the theory in an
appropriate form to be discussed in section 5 in the framework of integrable
models, we discuss the strong coupling limit, where the $1/N$ expansion
possibly breaks down, and later the gauge (first class) constraints imposed to
the theory in the WZW functional language. The case of chiral interactions is
briefly discussed in the last subsection.

The theory is defined by the Lagrange density
$$
{\cal L} = -{1\over 4}\tr F_{\mu\nu}F^{\mu\nu} + \overline \psi_i (i \!\not \!
\partial + e \!\not \!\! A)\psi_i \quad ,\eqno(2.1)
$$
with the notation defined in Appendix A. The fermions $\psi_i$ are in the
fundamental representation of the gauge group. The field equations derived from
this Lagrangian are
$$
\eqalignno{
\nabla_\mu^{ab}F_b^{\mu\nu}+e\overline\psi\gamma^\nu\tau^a\psi= & \, 0\quad,&
(2.2a)\cr
i\gamma^\mu\partial_\mu\psi+e\gamma^\mu\tau^aA_\mu^a\psi= & \, 0\quad. &
(2.2b)\cr}
$$

The current $J_\mu^a = \overline \psi \gamma_\mu \tau ^a \psi$ is covariantly
conserved as a consequence of (2.2$b$), namely
$$
\left(\partial _\mu \delta^{ab} + e f^{acb} A_\mu^c\right)\left(\overline \psi
\gamma^\mu \tau^b \psi \right)= 0 \quad .\eqno(2.3)
$$

For a gauge-invariant regularization this equation holds true in the quantum
theory. We consider in general an external field current
$$
J_\mu^a(x\vert A) = \langle \overline \psi (x) \gamma_\mu
\tau^a\psi(x)\rangle_A\quad,\eqno(2.4)
$$
which depends on the external gauge field $A_\mu$. It is obtained by
differentiating the functional
$$
W[A]=-i\, \ln{\det i\not\!\!D[A]\over\det i\not\!\partial}\quad ,\eqno(2.5)
$$
with respect to $A_a^\mu$, i.e. the current (2.4) is given in terms of (2.5) by
the expression
$$
e J_\mu^a(x\vert A) =  {\delta W\over \delta A_a^\mu(x)}\quad .\eqno(2.6)
$$
The functional $W[A]$ represents an effective action for $A_\mu$.

By the Fujikawa method,\ref{37} making small transformations in $i\!\not\!\!\!
D$, corresponding to classical symmetry transformations, we can analyze the
change in the integration variable (since the action is supposed to be
invariant under symmetry transformation) and search for anomalies. The measure
${\cal D}\overline\psi{\cal D}\psi$ is $U(1)$-invariant, in such a way that
the current $J_\mu^a$ is covariantly conserved. However, it is not invariant
under a chiral transformation. Let us consider the pseudo current
$$
J_{5\mu}^{b}(x\vert A)=\langle \overline \psi (x) \gamma_\mu\gamma_5
\tau^b\psi(x)\rangle_A= \epsilon_{\mu\nu} J^{\nu b}(x\vert A)\quad.\eqno(2.7)
$$
In such a case, use of the Fujikawa method with a gauge-invariant regulator
leads to the  anomaly equation
$$
\nabla_\mu^{ab}{J^{5\mu}}^b=-\widetilde \nabla_\mu^{ab} {J^\mu}^b={e\over 2\pi}
\epsilon_{\mu\nu}{F^{\mu\nu}}^a\quad .\eqno(2.8)
$$

The first consequence of the above anomaly equation is obtained from the
pseudo-divergence of the first Maxwell equation (2.2$a$), which yields (see
notation in Appendix A)
$$
\epsilon_{\nu\rho}{\nabla}^\rho {\nabla}_\mu F^{\mu\nu} + e \widetilde
{\nabla}_\mu J^\mu = - {1\over 2}\left({\nabla}^2 + {e^2\over \pi} \right)
\epsilon_{\mu\nu}F^{\mu\nu}=0\quad ,\eqno(2.9)
$$
showing that one expects, as foreseen in the introduction, a mass generation
for
the gauge field, analogous to the Schwinger model.

Furthermore, it is possible to compute the external field current
$J_\mu^a (x\vert A)$ integrating (2.8). Indeed, introducing the kernel
$K_\mu^{ab} (x,y\vert A)$ by the equations
$$
\matrix{
{\nabla}_\mu^{ab} {K^{\mu}}^{bc} (x,y\vert A)  =  0  & {} &
{\nabla}_-^{ab} K_+^{bc} = -\delta^{ac}\delta (x-y)\quad , \cr
{}& \Longrightarrow &{}\cr
\widetilde{\nabla}_\mu^{ab}{K^{\mu}}^{bc}(x,y\vert A) = -\delta^{ac}
\delta(x-y) &{}& {\nabla}_+^{ab} K_-^{bc} = \delta^{ac}\delta (x-y)\quad ,\cr}
\eqno(2.10)
$$
we have
$$
\eqalignno{
J_\mu ^a & = {e\over 2\pi} \int {\rm d}^2 y \, K_\mu^{ab}(x,y\vert A)
\epsilon_{\rho \sigma}{F^{\rho\sigma}}^b (y)\quad ,\quad {\rm or}\quad
&(2.11a)\cr
J_\pm ^a & = {e\over 2\pi} \int {\rm d}^2y\, K_\pm^{ab}(x,y\vert A) \epsilon_{
\rho \sigma}{F^{\rho\sigma}}^b(y)\cr
& = \mp{e\over 2\pi}\int{\rm d}^2y\,\widetilde K_\pm^{ab}(x,y\vert A)\epsilon_{
\rho\sigma}{F^{\rho\sigma}}^b(y) ,&(2.11b)\cr}
$$
where the kernel $K_\mu$ depends on the external gauge field $A_\mu$. The
kernel $K_\mu$ can be obtained as an expansion in the Lie algebra valued
fields ${\cal A}_\mu^{ab} = f^{acb} A_\mu^c $ with the use of the function
$D_\pm (x-y)$:
$$
D_\pm=\partial_\pm D(x)\quad,\quad D(x)=-{i\over 4\pi}\ln(-x^2+i\epsilon)
\quad , \eqno(2.12)
$$
as\ref{8}
$$
\eqalignno{
\mp K_\pm^{ab}&(x,y\vert A)=\widetilde K_\pm^{ab}(x,y\vert A)=\delta^{ab}D_
\pm(x-y)\cr
&-i\!\!\sum _{n=1}^\infty (-e)^n\!\!\int \!\!{\rm d}^2x_1 \cdots{\rm d}^2x_n \,
D_{_\pm} (x\!-\!x_1)\cdots D_{_\pm} (x\!-\!x_n)\left[ {\cal A}_{_\mp} (x_1)
\cdots {\cal A}_{_\mp}(x_n)\right]^{ab}\!\! . &(2.13a)\cr}
$$

It is sometimes useful to rewrite this expression in the fundamental
representation, where $A_\mu = \sum _c A_\mu ^c \tau^c$, obtaining
$$
\eqalignno{
K_\pm ^{ab}(x,y\vert A) = & \,  \delta^{ab}D_\pm (x-y) - \cr
& - \sum _{n=1}^\infty (-e)^n\int {\rm d}^2x_1 \cdots {\rm d}^2x_n \, D_\pm
(x-x_1)\cdots D_\pm (x-x_n)\cr
& \times \tr \left\{ \tau^a \left[ A_\mp (x_1), [A_\mp (x_2), \cdots [A_\mp
(x_n),\tau^b ]]\cdots \right]\right\} \quad .&(2.13b)\cr}
$$

Equation (2.11) can be linearized by means of  (2.10), since the
field strength $F_{+-}$ may be alternatively written as $F_{+-}=-\partial_-A_+
+\nabla_+A_-$ or $F_{+-}=-\nabla_-A_++\partial_+A_-$, in such a way that after
a partial integration (and use of (2.10)) one has
$$
\eqalignno{
J_\pm^a(x\vert A)= \,& {e\over 2\pi} A_\pm (x) - {e\over 2\pi}\int {\rm d}^2
y\,
K_\pm ^{ab}(x,y\vert A) \partial _\pm A^b_\mp  \cr
= \, & {e\over 2\pi} \left[ A_\pm - \int {\rm d}^2y\, \partial _\pm D_\pm (x-y)
A_\mp^a(y)\right]\cr
& + {i\over 2\pi} \sum _{n=2}^\infty (-ie)^n\int {\rm d}^2x_1 \cdots
{\rm d}^2x_n \, D_\pm (x-x_1)\cdots D_\pm (x-x_n)\cr
& \times \tr \left\{ \tau^a \left[ A_\mp (x_1), [ \cdots  [A_\mp (x_{n-1}),
\partial _\pm A_\mp (x_n)]] \cdots \right]\right\} \quad .&(2.14)\cr}
$$

Notice that we now have a sum of tree diagrams, although this is a one-loop
result. This is a consequence of two-dimensional space-time integration, where
one-loop diagrams can be computed in terms of tree diagrams (see ref. [8]).
In any case, care must be taken about divergences, in order not to lose
anomalous contributions.

Using such expressions, one can compute once more the fermionic determinant as
an expansion in terms of the gauge fields, or following refs. [38, 39] one
finds the effective action functional
$$
\eqalignno{
W[A] = \,& W[0]-{ie^2\over 2\pi}\int {\rm d}^2 x\, \delta^{ab} A_\mu^{a}\left(
g^{\mu\nu} - {\partial ^\mu \partial^\nu\over\partial^2}\right)A_\nu^b (x)\cr
& + {i\over 2}\sum_{n=2}^\infty {(ie)^{n+1}\over n+1} \int {\rm d}^2x\, \tr \,
\left[ A_-(x) T_+^{(n)} (x\vert A) + A_+(x) T^{(n)}_-(x\vert A)\right]\quad ,
&(2.15)\cr}
$$
where
$$
\eqalignno{
T_\pm ^{(n)} (x\vert A) = & - {1\over 2\pi} (-1)^n \int {\rm d}^2x_1 \cdots
{\rm d}^2x_n \, D_\pm (x-x_1)\cdots D_\pm (x-x_n)\cr
& \times\left[A_\mp(x_1),[\cdots [A_\mp (x_{n-1}), \partial_\pm A_\mp (x_n) ]]
\cdots \right] \quad .&(2.16)\cr}
$$

An alternative to such an expression for the effective action is the
Polyakov--Wiegmann method, where a closed expression can be obtained. On the
other hand it depends on potentials used to define the gauge field $A_\mu$,
and even in such a case the result is non-local as a function of those
potentials. Nevertheless the Polyakov--Wiegmann result will prove to be useful
in a wider sense than the above result.
\vskip 1cm
\penalty-1000
\noindent {\bf 2.1 The $1/N$ expansion and spectrum}
\vskip .5cm
\nobreak\noindent
The first successful attempt to obtain an insight into the dynamical structure
of QCD$_2$ was undertaken by 'tHooft, who considered the limit where the
number of colours $N$ is large, i.e. the $1/N$ expansion. In such a case one
considers quarks interacting via an $U(N)$ colour gauge group\newfoot{$^{2)}$}
{Some authors have considered the case of the $SU(N)$ gauge group. In the
large-$N$ limit, for the lowest order, the difference is irrelevant.}. In the
large-$N$ limit, one considers the contributions of graphs with the same
topology.
Therefore, the diagrams with the same number and type of external lines are
classified according to its (non-)planarity, or the number of handles and
holes -- the Euler characteristic. Moreover, in two dimensions it is useful to
work in the light-cone gauge, since the gauge field strength has only one
component; in that gauge the field strength is linear in the fields, since
$$
\epsilon^{\mu\nu}F_{\mu\nu}= F_{+-}=\partial_+A_--\partial_-A_+-ie[A_+, A_-]=-
\partial _-A_+\quad ,\eqno(2.17)
$$
for the gauge $A_-=0$, and all self interactions of the gauge fields
disappear. In such a case the Lagrangian boils down to
$$
{\cal L} = {1\over 8}\tr (\partial_-A_+)^2 + \overline \psi \left( i\gamma^\mu
\partial _\mu + {e\over 2}\gamma_-A_+ - m \right)\psi\quad , \eqno(2.18)
$$
where a mass is allowed for the quarks.

The ghosts decouple in such a gauge. Notice also that using light-cone
quantization one readily sees that the momentum canonically associated to $A_-$
is zero, and that it is not a dynamical field. The Feynman rules are very
simple. We have for the gauge field, and fermion propagator, respectively
$$
\eqalignno{
\left\langle A_+(k)A_+(-k)\right\rangle = &\, {4i\over k_-^2} \quad
,&(2.19a)\cr
\left\langle \psi(k)\overline \psi (-k)\right\rangle = & \,{i\over\not \!k-m}=
i {\not \! k + m \over k^2 - m^2}\quad ,&(2.19b)\cr}
$$
and the only vertex is
$$
\left\langle \psi\overline \psi A_+ \right\rangle ={ie\over 2}\gamma_-\quad .
\eqno(2.19c)
$$

Since $\gamma_-^2=0$, and $\gamma_+\gamma_-\gamma_+=4\gamma_+$, the
$\gamma$-algebra is extremely simple. The fermion propagator simplifies to
${i k_-\over k^2- m^2}$, and the vertex to $ie$.

The limit $N\to\infty$ has to be taken for $\alpha=e^2N$ fixed. In such a case
one generates, as usual, the planar diagrams with no fermion loops in lowest
order in $1/N$, further corrections being classified by the Euler
characteristic of the diagram in Feynman rules space. Therefore we are left
with
ladder diagrams, with self-energy insertions for the fermion lines. It is thus
possible to write the full fermion  propagator in terms of the unknown
function $\Sigma_{SE}(k)$, the self-energy, to be computed later. Adding
contributions as ${ik_-\over k^2-m^2+i\epsilon}\sum_{n=0}^\infty\left[{k_-\over
 k^2-m^2+i\epsilon}\Sigma_{SE}(k)\right]^n$, we find\newfoot{$^{3)}$}{Notice
the absence of the factors $1/2$ from the $\gamma$ matrices. There is a
cancellation of the factor 4 due to $\gamma_+\gamma_-\gamma_+=4\gamma_+$ and
the factors $1/2$ coming from the fermion propagator and the vertex. A final
factor $1/2$ for each outgoing fermion will be implicitly taken into account.}
$$
\left\langle \psi(k)\overline \psi (-k)\right\rangle_{{\rm full}} \equiv S_F(k)
= {ik_-\over k^2 - m^2 - k_-\Sigma_{SE}(k)} \quad .\eqno(2.20)
$$

For the planar approximation it is possible to derive a simple bootstrap
equation for $\Sigma_{SE}(k)$, since following the external line one finds a
vertex, which is connected to the second outgoing fermion line by the
gauge-field propagator (2.19$a$), exact in the large-$N$ limit, and by the full
fermion propagator (2.20); the equation obtained is (see Fig. 1)
$$
-i\Sigma_{SE}(k)=-2e^2N\!\int\!{{\rm d}p_+\,{\rm d}p_-\over (2\pi)^2}\,
{i\over p_-^2} {i (k_-+p_-)\over (k_+\!+\!p_+)(k_-\!+\!p_-)\!-\!\Sigma_{_{SE}}
(k\!+\!p)(k_-\!+\!p_-)\!-\!m^2\!+\!i\epsilon}\quad .\eqno(2.21)
$$
\penalty-2000
$$\epsfxsize=12truecm\epsfbox{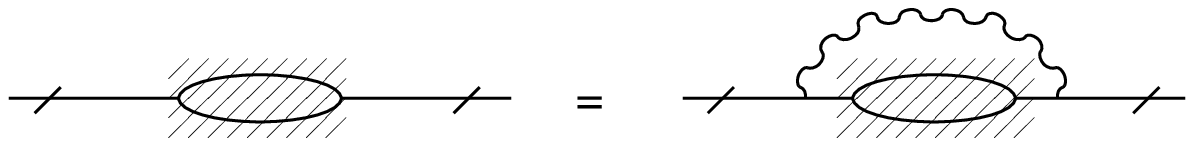}$$
\nobreak
\noindent{\eightpoint Fig. 1: Fermion self-energy equation.}
\vskip.3truecm
\penalty-2000

The right-hand side does not depend on $k_+$, as one readily sees by changing
variables as $p'_+= k_++p_+$. Therefore $\Sigma_{SE}(k)$ depends only on
$k_-$, and as a consequence the ``$+$" integral may be performed independently
of the function $\Sigma_{SE}(k)$ itself, and the equation simplifies to
$$
\Sigma_{SE}(k_-)={ie^2N\over 2\pi^2}\!\int\!{{\rm d}p_-\,(k_-\!+\!p_-)\over p_-
^2}\!\int\!\!{{\rm d}p_+\over p_+(k_-\!+\!p_-) - (k_-\!+\!p_-)\Sigma_{SE}(k_-
\!+\!p_-) - m^2 + i\epsilon}\quad .\eqno(2.22)
$$

There are two kinds of divergences in such an integral. Ultra-violet
divergences are soft, since the theory is super-renormalizable, and in the
present integral it is only logarithmic and disappears using symmetric
integration; for the $p_+$ integration we find, by simple contour integration,
the result ${-\pi i\over\vert k_-+p_-\vert}$, independent of
$\Sigma_{SE}(p)$. Therefore the solution follows straightforwardly after
substituting the result of the $p_+$ integration, that is
$$
\Sigma_{SE}(k_-)={e^2{N}\over 2\pi}\int{{\rm d}p_-\over p_-^2}\,\varepsilon
(k_-+p_-)\quad .\eqno(2.23)
$$

The onus of such a solution is that it is infrared-divergent, as a consequence
of the choice of gauge and of super-renormalizability. There are procedures
to regularize the infrared divergence in $\Sigma_{SE}(k_-)$ above, and obtain
sensible results. The original strategy followed by 'tHooft was to
cut-off a slice in momentum space around $k_-=0$, with width $\lambda$, and
take $\lambda \to 0$ when computing physical (gauge-invariant) quantities. A
second strategy followed by the authors of refs. [26-28] is to define
the light-cone gauge propagator by means of a principal-value prescription as
$$
P \, {1\over q_-^2} = {1\over 2} \left[{1\over (q_- - i \epsilon )^2} +
{1\over (q_- + i \epsilon )^2}\right]\quad .\eqno(2.24)
$$

In the first case, the self-energy is cut-off-dependent, with the result
$$
\Sigma_{SE}(k)={e^2{N}\over\pi}\left[{\varepsilon(k_-)\over\lambda}-{1\over
k_-}\right]\quad ,\eqno(2.25)
$$
while in the latter case one obtains the finite value
$$
\Sigma_{SE}(k) = - {e^2{N}\over \pi}{1\over  k_-}\quad .\eqno(2.26)
$$
In ref. [40] $\lambda$ was interpreted, in the limit $\lambda\to 0$, as a
gauge parameter. In 'tHooft's procedure, confinement was interpreted in
terms of $\lambda$, since quark propagator's poles are removed to infinity,
while in the regular cut-off prescription the fermion propagator is
$$
S_F(k) \sim {ik_-\over k^2 -m^2 + {e^2{N}\over \pi}+ i\epsilon}\quad .
\eqno(2.27)
$$
In this latter case, confinement is obtained from the fact that quark continuum
states do not appear. Coloured operators vanish with the use of the cut-off
procedure in the limit $\lambda \to 0$, since there presumably exist no
finite-energy coloured states. However, as discussed in [27] it is
sometimes useful to consider coloured states in order to understand the
interplay between confinement and the high-energy limit, since the zero
cut-off and high-energy limits do not commute.

Using the singular cut-off we arrive at the dressed propagator
$$
S^\lambda_F(k)={ik_-\over k^2-m^2+{e^2N\over \pi} - {e^2N\over \pi\lambda}\vert
k_-\vert + i\epsilon}\quad ,\eqno(2.28)
$$
which displays the above-mentioned fact that the pole is shifted towards
$k_+\to \infty$ excluding the physical single-quark state. Independently of
keeping or not the $\lambda$-dependence, we can proceed and set a
Bethe--Salpeter equation in order to find information about the bound states.
We consider a blob representing a two quark bound state. In the planar limit,
such a wave functional obeys the equation pictorially depicted in Fig. 2, where
a homogeneous term has been abandoned. The structure of this equation is
extremely simplified in the light-cone gauge, where as mentioned, the gauge
field has no self interaction, and due to planarity only ladder-type diagrams
as in the right-hand side of Fig. 2 survive.
\penalty-2000
$$\epsfysize=2.5truecm\epsfbox{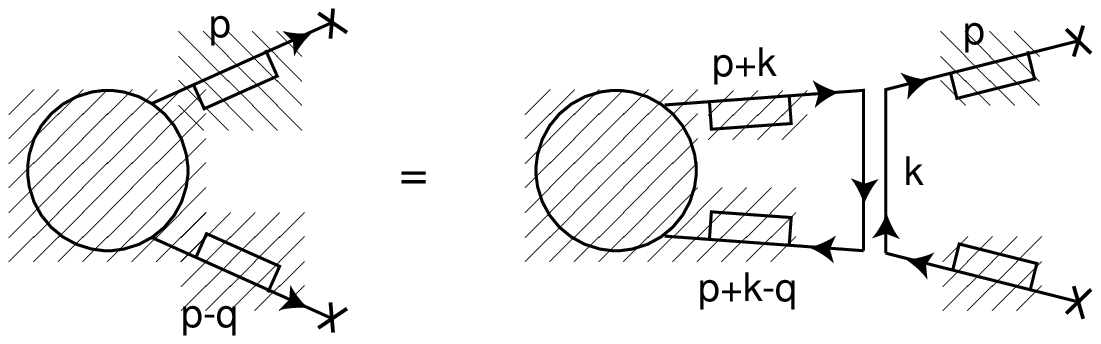}$$
\nobreak\nobreak
\noindent{\eightpoint Fig. 2: Bethe--Salpeter equation.}
\vskip.3truecm
\penalty-2000

This leads to the integral equation for the blob $\varphi(p;q)$
$$
\eqalignno{
\varphi(p;q) =\,&  {e^2N\over (2\pi)^2} {i(p_--q_-) p_-\over \left[(p-q)^2 -
M^2-{e^2N\over \pi\lambda}\vert p_--q_-\vert + i\epsilon\right]}\cr
&\times{1\over\left[p^2-M^2-{e^2N\over\pi\lambda}\vert p_-\vert+i\epsilon
\right]}\int{\rm d}k_+\,{\rm d}k_-\,{\varphi(p+k;q)\over
k_-^2}\quad,&(2.29)\cr}
$$
where $M^2=m^2-e^2N/\pi$. Notice also that $\varphi(p;q)$ is a function of the
momenta, that is $\varphi(p_+,p_-;q)\equiv\varphi(p_+,p_-;q_+,q_-)$. We do not
need the full solution in order to obtain the spectrum; we consider a
simplified equation obeyed by
$$
\varphi(p_-;q) \equiv \varphi(p_-;q_+,q_-)=\int{\rm d}p_+\,\varphi(p_+,p_-;q)
\quad .\eqno(2.30)
$$

The $k_+$ integral in eq. (2.29) corresponds to the definition (2.30), and we
can furthermore integrate over $p_+$, obtaining
$$
\eqalignno{
\varphi(p_-;q)= &\, i {e^2N\over (2\pi)^2}\int {\rm d}p_+ \, \left[ p_+- q_+-
{M^2\over p_--q_-}+ \left( - {e^2N\over \pi\lambda} + i \epsilon \right)
\varepsilon \, (p_--q_-)\right]^{-1}\cr
& \, \times \left[ p_+- {M^2\over p_-} - \left( {e^2N\over\pi\lambda}-i\epsilon
\right)\varepsilon\,\,(p_-)\right]^{-1}\int{\rm d}k_-\,{\varphi(p_-+k_-;q)\over
k_-^2}\quad .&(2.31)\cr}
$$

The $p_+$ integral is zero if $\varepsilon\,(p_--q_-)$ and $\varepsilon\,(p_-)$
are equal, since we have to integrate between the poles, to get a non-zero
result. For $q_->0$, we can satisfy this condition only for $0<p_-<q_-$, in
which case the integral picks up the contribution of one pole, with the result
$$
\varphi (p_-;q) = {e^2N\over 2\pi}\theta (p_-)\theta(q_--p_-)\left[ {M^2\over
p_-}\!+\!{M^2\over q_--p_-}\!+\!{2e^2N\over \pi\lambda}\!+\!q_+\right]^{-1}\!\!
\int\!{\rm d}k_- \, {\varphi(p_-+k_-;q)\over k_-^2}\quad .\eqno(2.32)
$$

Had we used a regular cut-off, such an integral equation would be finite, owing
to the absence of the term ${e^2\over \pi \lambda}$, and to the principal-value
prescription for the distribution $1/k_-^2$. Using the singular cut-off, we
have to separate the divergent piece
$$
\int {\rm d}k_- \, {\varphi(p_-+k_-;q)\over k_-^2}= {2\over \lambda}\varphi
(p_-;q) + \int {\rm d}k_- \, \varphi(p_-+k_-;q){P\over k_-^2}\quad ,\eqno(2.33)
$$
where $P$ is the principal-value prescription for the quadratic singularity
near the origin, equation (2.24) (the first term on the right-hand side arises
from $\varphi(p_-;q)\int_{-\lambda}^\lambda {\rm d}k{P\over k^2}$). One finds,
for $\lambda\to 0$, that the cut-off disappears after inserting (2.33) back
into (2.32). We arrive at the integral equation
$$
-q_+\varphi(p_-;q) = M^2\left( {1\over p_-} + {1\over q_--p_-}\right)\varphi
(p_-;q)- {e^2N\over 2\pi} P \int _{p_-}^{q_--p_-}{\rm d}k_-\, {\varphi
(p_- + k_-;q)\over k_-^2}\quad,\eqno(2.34a)
$$
which in its turn, upon use of
$$
\tau={\pi M^2\over e^2N}={\pi m^2\over e^2N}-1\quad , \quad q^2=e^2N\mu^2
\quad ,\quad {\rm and} \quad  p_-/q_-=x\quad ,\eqno(2.34b)
$$
where $\mu$ is the mass of the two-particle state in units of $e/\sqrt \pi$,
can be rearranged into
$$
\mu^2 \varphi(x) = \tau\left( {1\over x} + {1\over 1-x}\right)\varphi (x) +
P\int _0^1 {\rm d}y \, {\varphi(y)\over (x-y)^2}\quad .\eqno(2.35)
$$

Although it is not possible to solve such an equation, the approximate
spectrum can be obtained. The right-hand side of (2.35) is interpreted as a
Hamiltonian action on the ``wave function" $\varphi(x)$. We suppose that the
eigenstates behave as $x^\beta$ for $x \approx 0$. Using
$$
\int _0^\infty {\rm d}x\, {x^\beta\over (x-1)^2 } = \beta \pi \, \cotg \, \beta
 \pi\quad ,\eqno(2.36)
$$
we verify that such a solution of eq. (2.35) can be found if
$$
\pi \beta \, \cotg \,\pi\beta + \tau=0 \quad ,\eqno(2.37)
$$
for functions that vanish on the boundary, $\varphi(0)=\varphi(1)=0$, the
``Hamiltonian" is Hermitian. However there are still problems when we require
eigentates to be mutually orthogonal. Nevertheless the boundary condition
respects completeness. It is thus natural to consider periodicity. For a
periodic function, the second term in the Hamiltonian is approximated by
$$
\int _0^1 {\rm d}y\, {\E^{iwy}\over (y-x)^2} \approx \int _{-\infty}^\infty
{\rm d}y\, {\E^{iwy}\over (y-x)^2} = -\pi \vert w\vert \E^{iwx}\quad ,
\eqno(2.38)
$$
and for the above-discussed boundary conditions the eigenfunctions are $\varphi
_k=\sin k\pi x$ for $\tau\approx 0$, with eigenvalues $\mu_k^2=\pi^2k$,
leading to a Regge trajectory, without continuum part in the spectrum. This is
a good approximation for large values of $k$.

It is important to know whether the $1/N$ expansion gives trustworthy results.
As remarked in ref. [27], it does. The next-to-leading corrections are
simplified by the fact that quarks are confined, and we need to take the
$\lambda\to 0$ limit. The gauge-field propagator does not get corrections in
such a case. The quark--antiquark scattering amplitude was also studied in
detail, and computed in terms of the eigenfunctions $\varphi_k(x)$.

The consideration of the equation obeyed by the quark--antiquark scattering
amplitude is a straightforward generalization of the previous results. Consider
the quark--antiquark scattering amplitude, which we denote by $T_{\alpha\beta,
\gamma\delta}(p,p';r)$, as given in the left-hand side of Fig. 3. The quark
lines with momenta
$p$ and $p'$ are connected by products of $\gamma_-$ and the quark propagator.
Since $\gamma_-^2 =0$, only a simple $\gamma$-type factor survives, and we have
$$
T_{\alpha\beta,\gamma\delta}(p,p';q)=\gamma_{-\alpha\gamma}\gamma_{-\beta\delta
}\, T(p,p';q)\quad.\eqno(2.39a)
$$
In the large-$N$ limit it obeys the equation graphically displayed in Fig. 3,
which translates into
$$
\eqalignno{
T_{\alpha\beta,\gamma\delta}&(p,p';q) = {ie^2\over (p_--p'_-)^2} (\gamma_-)_
{\alpha\gamma} (\gamma_-)_{\beta \delta}  \cr
& + ie^2N \int {{\rm d}^2k\over (2\pi)^2} {(\gamma_-)_{\alpha\epsilon}
(\gamma_-)_{\beta\lambda}\over (k_--p_-)^2} S(k)_{\epsilon \mu} S(k-q)_
{\lambda\nu} T_{\mu\nu,\gamma\delta} (k,p';q)\quad. &(2.39b)\cr}
$$
\penalty-2000
$$\epsfysize=3.5truecm\epsfbox{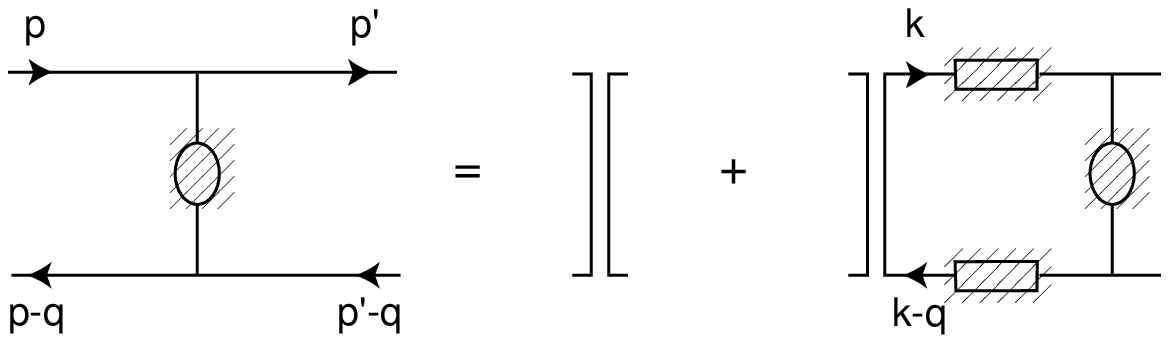}$$
\nobreak
\noindent{\eightpoint Fig. 3: Bethe--Salpeter equation for the
quark--antiquark scattering.}
\vskip.3truecm
\penalty-2000

Using eq. (2.39$a$), and substituting (2.28) in (2.39$b$), after the
$\gamma$-matrix algebra, one gets rid of all $\gamma_-$ factors. We now use
the kind of trick introduced in (2.30), since the $(+)$ variables are
essentially spectators, a fact derived from the instantaneous interaction in
one light-cone variable. We define the functional
$$
\varphi(p_-,p'_-; q) = \int {\rm d}p_+ \,S_F(p)S_F(p-q)T(p,p';q)\quad ,
\eqno(2.40)
$$
from which eq. (2.39$b$) is solved for $T(p,p';q)$ in terms of $\varphi(k_-,
p'; q)$ as
$$
T(p,p';q) = {2ie^2\over (p_--p'_-)^2} + {2ie^2N\over \pi^2}\int {\rm d}k_- \,
{\varphi(k_-,p';q)\over (k_- - p_-)^2} \quad .\eqno(2.41)
$$

We can suppose that the masses of quarks and antiquarks in the two propagators
in (2.40) are different, but we take a simplified point of view where they are
the same. The integral
$$
\eqalignno{
\int &{\rm d}p_+\,  S_F(p) S_F(p-q) \cr
&\!=\!\int \!{\rm d}p_+\,{i\over p_+-A(p_-)}{i\over p_+-q_+-A(p_- -q_-)}\cr
&\!=\!\int \!{\rm d}p_+\, {1\over A(p_-)\!-\!A(p_-\!-\!q_-)\!-\!q_+} \!\left[
{1\over p_+\!-\!A(p_-)} - {1\over p_+\!-\!q_+\!-\!A(p_-\!-q_-)}\right]\,
&(2.42a)\cr}
$$
is non-zero if it is performed between the poles. With $A(p_-)$ given by
$$
A(p_-)={e^2N\over \pi}\left[{1\over p_-}-{\varepsilon(p_-)\over\lambda}\right]+
{m^2-i\epsilon\over p_-}\quad ,
$$
we find the result
$$
\int {\rm d}p_+\,S_F(p)S_F(p-q)={1\over {M^2\over p_-} + {M^2\over q_--p_-} +
\left( {e^2N\over \pi\lambda}-i\epsilon \right) \varepsilon (p_-)} \theta(p_-)
\theta(q_- -p_-)\quad .\eqno(2.42b)
$$

We use (2.39$a$) in (2.39$b$), multiply it by $S_F(p)S_F(p-q)$, integrate over
$p_+$, and substitute definitions (2.34$b$), in the final equation, arriving at
$$
\mu^2\varphi (x,x';q) = {\pi^2\over Nq_-(x-x')^2} + \tau\left({1\over x} -
{1\over 1-x}\right)\varphi (x,x';q)+\int_0^1{\rm d}y\,{\varphi(x,x';q)-\varphi
(y,x';q)\over (x-y)^2}\, .\eqno(2.43a)
$$

The homogeneous equation is of the Schr\"odinger type, that is
$$
H\varphi_k=\mu^2_k\varphi_k=\tau\left({1\over x}+{1\over 1-x}\right)\varphi_k(x
)+\int_0^1{\rm d}y\,{\varphi_k(x)-\varphi_k(y)\over (x-y)^2}\quad,\eqno(2.43b)
$$
and the eigenfunctions $\varphi_k(x)$ can be found as before. In terms of
$\varphi_k(x)$, the authors of ref. [27] constructed
$$
\varphi(x,x';q) = - \sum {\pi e^2\over q^2-q_k^2} {1\over q_-} \int _0^1
{\rm d}y\, {\varphi_k(x)\varphi^*_k(y)\over (x'-y)^2} \quad ,\eqno(2.44)
$$
and subsequently the quark--antiquark scattering amplitude
$$
\eqalignno{
T(x',x;q)&={ie^2\over q_-^2(x'-x)^2}-{ie^2(e^2N)\over\pi q_-^2}\sum_k{1\over
(q^2-q_k^2)}\int_0^1{\rm d}y\,\int_0^1{\rm d}y'\,{\varphi^*_k(y')\varphi_k(y)
\over (y'-x')^2(y-x)^2}\cr
& =  {2ie^2\over q_-^2(x'-x)^2}\cr
& -\! \sum_k {2i\over (q^2-q_k^2)}\! \left\{\varphi^*_k(x') {2e\over \lambda}
\left(\!{e^2N\over \pi}\!\right)^{^{1\over 2}}\!\!\left[ \theta(x'(1\!-\!
x'))\!+\!{\lambda\over 2\vert q_-\vert }\!\left({\tau\over x'}+{\tau\over 1-x'}
-\mu_k^2 \right)\!\right] \!\right\} \cr
&\times \left\{ \varphi_k(x){2e\over\lambda}\left({e^2N\over\pi}\right)^{^
{1\over 2}}\left[ \theta(x(1-x)) + {\lambda\over 2\vert q_-\vert}\left({\tau
\over x} + {\tau\over 1-x}-\mu_k^2\right)\right]\right\}\, ,&(2.45)
\cr}
$$
which shows no continuum states, but again only bound-state poles at $q^2=q_k^
2=\pi ke^2N$, making once more the confinement features explicit. The
normalized
bound state has also been computed, and it can be shown to be of order $1/
\lambda$ as $\lambda\to 0$, compensating the fact that the quark propagator
vanishes in the same limit, which finally leads to finite-bound-state
amplitudes.

The results of 'tHooft represented a profound breakthrough, since they were
precursors of more recent attempts to write down differential equations to be
obeyed by bound states or collective excitations. The results about the Regge
behaviour of the physical spectrum lead to a strong link to string theory
developments, which had to wait almost two decades to be followed further.
\vskip 1cm
\penalty-1000
\noindent {\bf 2.2 Ambiguity in the self-energy of the quark}
\vskip .5cm
\nobreak
\noindent The basis of 'tHooft's result about the $1/N$ expansion is an
infrared cut-off procedure, which consists in drilling a hole in momentum
space around
the infrared region $(k\sim 0)$, the size of the hole $(\lambda)$ being the
cut-off. For $\lambda \to 0$, when calculating some gauge-invariant objects,
one observes that $\lambda$-dependent constants cancel. Such a procedure is
equivalent to defining the light-cone propagator with a principal-value
prescription, that is, following ref. [40]
$$
\eqalignno{
\partial^{-1}_- = \,& {1\over 4}\int {\rm d}^2y \,\varepsilon (x_+ - y_+)\delta
(x_- - y_-) \quad ,& (2.46a) \cr
\partial^{-2}_- = \,& {1\over 8}\int {\rm d}^2y \, \vert x_+ - y_+\vert \delta
(x_- - y_-) \quad ,& (2.46b)  \cr}
$$
or, in momentum space
$$
\eqalignno{
P\, {1\over k_-} = \,& {1\over 2} \left( {1\over k_- + i\epsilon } +
{1\over k_- - i\epsilon } \right) \quad , & (2.47)\cr
P\, {1\over k_-^2} = \,& {1\over 2} \left( {1\over (k_- + i\epsilon)^2 } +
{1\over (k_- - i\epsilon)^2 } \right) \quad . & (2.48)\cr}
$$

The difference between 'tHooft's procedure, and the principal-value
prescription for $\lambda \to 0$, is not dificult to obtain. Using the cut--off
procedure and Fourier transforming we obtain the expression
$$
\partial^{-2}_- =  {1\over 8}\left( \vert x_+ \vert - {2\over \pi\lambda}
\right) \delta (x_-) \quad ,\eqno(2.49)
$$
which differs from eq. (2.46$b$) by the extra term $-{1\over 4\pi\lambda}
\delta(x)\equiv -{1\over 4\pi\lambda}\delta (x^+)$ corresponding to a gauge
ambiguity in the Coulomb equation
$$
\partial_-^2A_+ = -J_-\quad .\eqno(2.50)
$$

The cut--off dependence may be gauged away. The procedure is by all means
ambiguous. Indeed, using the principal-value prescription, the momentum
integrals in a Feynman diagram do not commute, but rather obey the
Poincar\'e--Bertrand formula\ref{25}
$$
\int {{\rm d}k'\over k-k'} \int {{\rm d}k''\over k'-k''} f (k',k'') -
\int {\rm d}k''\int {\rm d}k'{1\over (k-k')(k'-k'')}f(k',k'')= -
\pi^2f(k,k)\quad.\eqno(2.51)
$$

The choice made by Wu\ref{25} was to Wick-rotate the Feynman integral,
going to Euclidian space, in order to compute the fermion self-energy. In such
a case, the fermion propagator is
$$
S_F(k)={i(k_1-ik_2)\over k^2 +m^2-(k_1-ik_2)\Sigma_{SE}(k)}\quad ,\eqno(2.52)
$$
where the self-energy $\Sigma_{SE}(k)$ satisfies the integral equation
$$
\Sigma_{SE}(p)=-{e^2N\over \pi^2} \int_{-\infty}^\infty {{\rm d}k_1 {\rm d}k_2
\over (k_1 - ik_2)^2} {k_1 + p_1 - i(k_2 + p_2)\over (k+p)^2 + m^2 - [k_1 + p_1
- i (k_2+p_2)]\Sigma_{SE}(k+p)} \quad .\eqno(2.53)
$$

With the ansatz
$$
\Sigma_{SE}(k) = (k_1+ik_2)f(k^2)\quad ,\eqno(2.54)
$$
one can integrate over the angular variables and obtain the algebraic equation
$$
f(k^2) = {e^2N\over \pi^2} {1\over k^2[1-f(k^2)] + m^2}\quad ,\eqno(2.55)
$$
with the solution
$$
f(k^2) = {1\over 2k^2} \left\{ k^2 + m^2 - \left[ (k^2 + m^2)^2 - {4\over \pi}
e^2N k^2 \right]^{1/2} \right\} \quad .\eqno(2.56)
$$
Therefore, $\Sigma_{SE}(p)$ (and hence the fermion propagator) has a
square-root-type singularity. The propagator pole has not appeared. The
principal-value result is obtained in the large momentum limit.

This result is significantly different from the one obtained by 'tHooft. In
fact, there are further inconsistencies, as pointed out in ref. [26], in a
detailed analysis of the infrared cut-off procedures defined by either the
cut-off $\lambda$ around the origin in momentum space, or by the
principal-value prescription. Those authors analyzed general gauges of the
type $n^\mu A_\mu =0$, where $n^\mu$ is a fixed vector. In the case $n^2=-1$,
the most singular terms in the gauge-field propagator are of the form
$1/(n^\mu A_\mu)^2$, with integrals of the form
$$
\int {{\rm d}(n \cdot k)\over (n\cdot k)^2 }\,\,f\left(n\cdot (k-p)\right)\quad
.\eqno(2.57)
$$
Notice that 'tHooft's analyses were done in the light-cone gauge, where
$n^2=0$.

Although, as discussed in ref. [27], the use of either 'tHooft's
regularization or the principal-value prescription leads to the same bound
spectrum, the interpretation in terms of fermion propagators is yet unclear.
The authors have proved that there is no solution for the self-energy equation
if the bare mass of the quark is small and the principal-value prescription
is used. On the other hand, the Euclidian symmetric integration was applied to
the Schwinger model, and the results obtained agree with the exact solution
of the theory.\ref{42}

Since the quark is after all not an observable, the question seems to be one of
interpretation of the results. In ref. [43] it is argued that the tachyonic
character of the $1/N$ correction computed in the Minkowski light-cone gauge
can be seen as a reinterpretation of Nambu's pion in terms of spontaneously
broken chiral symmetry. Wu's solution, in the zero quark-mass limit,
transforms into 'tHooft's result in the infinite momentum frame. By all means,
the theory as presented in Minkowskian version has an appealing interpretation
when one analyzes the confined quarks as compared to the high-energy behaviour,
 providing a model of hadrons with the expected properties of confinement as
displayed by the spectrum, and description by a parton model in the
high-energy limit.

A QCD$_2$ physical interpretation of the results obtained from the
large-$N$ limit, using also the experience with the Schwinger model, can
already be drawn. Such a question was studied in detail in
[27], where the authors presented QCD$_2$ as a good model for confined quarks
in spite of the huge simplification due to the reduced number of dimensions.

The fundamental property of quarks, which evaded solution in terms of a
realistic theory such as four-dimensional QCD, is that they are confined, in
the sense that only colour singlets appear in the Hilbert space; nevertheless,
high-energy scattering is described in terms of the
parton model, where quarks are essentially free -- although forming a bound
state. Therefore one needs apparently contradictory ideas, the infrared slavery
leading to confinement, and asymptotic freedom describing the parton model. The
string idea seems correct to describe the mesonic spectrum or to explain the
Regge behaviour; but the ultraviolet scattering of strings is too soft, since
amplitudes fall off too quickly with energy.\ref{7} On the other hand,
Yang--Mills theory has been proved to be asymptotically free, in excellent
agreement with experiment at high energy, but its description of bound states
is far from being realized.

It is very exciting that some of these gaps have been covered over the years in
two-dimensional QCD. In fact, in two dimensions, the theory is ``infrared
enslaving" due to the properties of the Coulomb law. If the experience with the
(soluble) Schwinger model can be taken into account, several phenomenologically
appealing properties of long-distance physics of hadrons may be described.
Moreover we will see that the best description of QCD$_2$ with fermions is
achieved by the non-Abelian bosonization, where the bosons are bound states of
the fermions, providing a natural description of the mesons. Besides that,
two-dimensional Yang--Mills theory is far better than the Schwinger model,
since it is highly non-trivial, and can provide a more realistic model of
interacting particles, while the ``meson" of the Schwinger model is free.
Moreover, although in two dimensions it is not possible to analyze important
questions such as large-angle scattering, hadronic scattering amplitudes and
further short-distance properties of high-energy, scattering of bound states
are satisfactorily described by QCD$_2$, since the coupling constant has
positive mass dimension, leading to obvious asymptotic freedom.

In ref. [27] it is argued that the confinement properties found using the
infrared cut-off in the $1/N$ expansion, by means of which the quark pole is
shifted towards infinity in the vanishing cut-off limit, hold true in the
principal-value prescription. Therefore, in spite of the possible criticisms
presented in the previous section, the interpretation of the theory in terms
of confined quarks is solid, since every approach leads to a rather well
defined bound-state structure, although the quark propagator itself is an
ill-defined quantity, depending on the cut-off procedure. The  second question
is whether such properties hold true at higher orders in $1/N$, and the answer
is positive.

The large-$N$ limit of QCD is suitable to describe the dual resonance model, as
is also clear from the topological structure seen in the Feynman diagrams in
such a limit. This issue will be studied in detail in section 3 for the pure
gauge theory. The effectiveness of such an expansion depends on the relative
size of the higher corrections. The correction to the gluon propagator for
large $N$ is given by the fermion loop. If the full fermion propagator is used
with the cut-off procedure, such a correction is seen to vanish, since the
quark propagator itself vanishes in the limit $\lambda\to
0$. A more careful analysis shows that one has to take into account terms of
order $\lambda$ in the gluon propagator, but one can prove that they do only
change the infrared-divergent part of the propagator. Likewise, an analysis of
the quark--antiquark gluon vertex as well as of the quark self-energy shows, by
arguments based on rescaling of the momentum variables, that the relevant
properties obtained at lowest order in $1/N$ remain unchanged as $\lambda\to
0$.

If it is obvious that quarks are confined in the cut-off procedure due to the
shift of the quark propagator pole towards infinity in the $\lambda \to 0$
limit, this is less transparent in the principal-value prescription. On the
other hand, the free-field structure at high energies turns out to be clearer
in the principal-value prescription. We shall now see how this works.

Using the regular cut-off (principal-value prescription) the final
Bethe--Salpeter equation (2.43$a$) remains unchanged since the $\lambda$-terms
disappear in favour of the principal-value terms. Quark continuum states do
not appear in the solution, signalling confinement. On the other hand, one can
consider also coloured operators and the corresponding expectation values, such
as the two-point expectation $\langle 0\vert T\overline\psi_i\psi_j(x)\overline
\psi_j\psi_i(y)\vert 0\rangle$, which vanishes using the singular cut-off
procedure; for the regular cut-off case it gives a finite result, compatible
with free-field perturbation theory. The results are correct, and are not
contradictory, since the high-energy and zero cut-off limits do not commute, as
exemplified for the explicit case of the integral
$$
q^2 \int {{\rm d}x\over q^2 x (1-x) - e^2N/\pi\lambda}\quad , \eqno(2.58)
$$
which appears in the coloured two-point function. The integral (2.58)
approaches a constant for large $q^2$, but vanishes if the small cut-off limit
is taken first.

Therefore, the principal-value procedure (regular cut-off) is appropriate to
describe the high-energy behaviour of the theory, i.e. reproduces parton model
results. Green functions containing currents and computed using such a
procedure show a short-distance behaviour compatible with the free-quark model,
showing the interplay between asymptotic freedom and confinement.

Form factors have been discussed by Einhorn.\ref{40} A ``parton model" has
been constructed there, and hadronic form factors have been shown to be
power-behaved, with a power determined by the coupling constant. For meson
scattering
amplitudes see ref. [41].
\vskip 1cm
\penalty-1000
\noindent {\bf 2.3 Polyakov--Wiegmann formula and gauge interactions}
\vskip .5cm
\nobreak
\noindent In the case of $U(1)$, namely the Schwinger model, only the first
term in eq. (2.13) survives, and one just obtains the gauge-field mass
generation, well known in the model. In the general non-Abelian case, the
computation of the fermion determinant is accomplished (see eqs. (2.5--16)),
but is given in terms of an infinite series. A detailed account of the method
is described in chapter 11 of ref. [8].

By all means, the most interesting and clear way of computing the fermion
determinant is the Polyakov--Wiegmann method, starting from the equations
obeyed by the current, and ending up with the WZW functional, which represents
a summation of the series. As a bonus, we obtain the bosonized version of the
fermionic action. The implementation of the bosonization techniques of a
non-Abelian symmetry is well known\ref{15-19} and we shall review the
Polyakov--Wiegmann identities and the consequent form of the action using
functional methods.

In two dimensions we can locally\newfoot{$^{4)}$}{In general we are considering
the fields as maps from Minkowski space to the $SU(N)$ algebra and such a
decomposition holds globally. In section 3, where we find a more general
situation, one has to be cautions about this choice.} write the gauge field in
terms of two matrix-valued fields $U$ and $V$ as
$$
A_+ = {i\over e}U^{-1}\partial _+U \quad , \quad
A_- = {i\over e}V\partial _-V^{-1} \quad , \eqno(2.59a)
$$
or
$$
e A_\mu = {i\over 2}(g_{\mu\nu} + \epsilon_{\mu\nu} )V\partial ^\nu V^{-1} +
{i\over 2}(g_{\mu\nu} - \epsilon_{\mu\nu} )U^{-1}\partial ^\nu U\quad .
\eqno(2.59b)
$$

Since the fermionic determinant must be gauge-invariant, there is no loss in
choosing $V=1\,(A_-=0)$. We start out of the current conservation and anomaly
equation
$$
\eqalignno{
\partial _\mu J^\mu - ie [A_\mu , J^\mu] & =  0 \quad , &(2.60a)\cr
\widetilde \partial _\mu J^\mu - ie [\widetilde A_\mu , J^\mu] & = - {e\over
2\pi} \epsilon_{\mu\nu}F^{\mu\nu}\quad , &(2.60b)\cr}
$$
for which, using the gauge potentials (2.59), we find the solution
$$
J_\pm = \pm {i\over 2\pi}U^{-1}\partial_\pm U \quad .\eqno(2.61)
$$
Notice that eq. (2.60$a$) was given by (2.3), while (2.60$b$), resp. (2.8), is
a one-loop effect, but maintained to all orders.

The effective action $W[A]$ is obtained by noticing that it is exactly its
variation with respect to the gauge field that leads to the current, that is
$$
J_-= {2\over e}{\delta W\over \delta A_+} \quad ,\eqno(2.62)
$$
where the factor 2 comes from the definitions in Appendix A. Therefore
$$
\delta W = {e\over 2}\int {\rm d}^2x\, J_-\delta A_+ \quad .\eqno(2.63)
$$

In terms of the $U$-variation, we have
$$
e\delta A_+ = i U^{-1}\partial _+\delta U - i U^{-1} \delta U U^{-1} \partial
_+U = i {\nabla}_+(U^{-1} \delta U)\quad ,\eqno(2.64)
$$
where the operator ${\nabla}_+$ acts as
$$
{\nabla}_+ f = \partial _+ f + [U^{-1} \partial _+U, f]\quad. \eqno(2.65)
$$

Therefore we find for the variation of the effective action, after integrating
${\nabla}_+$ by parts:
$$
\delta W = - {1\over 4\pi} \tr \, \int {\rm d}^2 x \, U^{-1}\delta U
{\nabla}_+(U^{-1}\partial _-U)\quad .\eqno(2.66)
$$
We can use the simple identity ${\nabla}^\mu(U^{^{-1}}\partial_\mu U)=
\partial^\mu (U^{^{-1}}\partial_\mu U)$, as well as the relation $\partial_+
(U^{^{-1}}\partial_-U)={\nabla}_-(U^{^{-1}}\partial _+U)$, to rewrite
$\delta W$ as
$$
\delta W = -{1\over 4\pi} \tr \, \int {\rm d}^2x\, U^{-1}\delta U \partial
_-(U^{-1} \partial _+U)\quad .\eqno(2.67)
$$

Such an equation may be integrated in terms of the WZW action. Consider the
action of the principal $\sigma$-model $S_{P\sigma M}$:
$$
S_{P\sigma M} = {1\over 8\pi}\int {\rm d}^2x\, \tr \, \partial ^\mu U^{-1}
\partial _\mu U\quad, \eqno(2.68)
$$
and the Wess--Zumino term $S_{WZ}$
$$
S_{WZ}= {1\over 4\pi} \tr \int_0^1 {\rm d}r\, \int {\rm d}^2x\,
\epsilon^{\mu\nu} \widetilde U^{-1}\dot {\widetilde U} \widetilde
U^{-1}\partial_\mu \widetilde U \widetilde U^{-1} \partial _\nu \widetilde U
\quad ,\eqno(2.69)
$$
where $\widetilde U(r,x)$ is the extension of $U(x)$ to a space having the
Euclidian two-dimensional space as a boundary, such that $\widetilde U(0,x)=1$
and $\widetilde U(1,x)=U(x)$. The variation of $S_{P\sigma M}$ is trivially
performed
$$
\delta S_{P\sigma M} = - {1\over 4\pi}\tr \int {\rm d}^2 x\, U^{-1}\delta U
\partial _\mu (U^{-1}\partial ^\mu U)\quad,\eqno(2.70)
$$
while the variation of $S_{WZ}$ is a total derivative in $r$, such that we can
integrate over that auxiliary parameter, which leads to
$$
\eqalignno{
\delta S_{WZ} = & -{1\over 4\pi} \int_0^1 {\rm d}r\, \int {\rm d}^2x\,
{\partial \over \partial r} \epsilon^{\mu\nu} \tr \widetilde U^{-1}
\delta {\widetilde U} \partial_\mu (\widetilde U^{-1} \partial _\nu \widetilde
U)\quad ,\cr
= & - {1\over 4\pi} \epsilon^{\mu\nu} \int {\rm d}^2 x\, \tr U^{-1} \delta U
\partial _\mu (U^{-1} \partial _\nu U)\quad . &(2.71)\cr}
$$

Adding $S_{P\sigma M}$ and $S_{WZ}$ we obtain the corresponding variation which
matches (2.67); therefore we find the solution to the effective action
$$
\eqalignno{
W[A]= & - {1\over 8\pi}\int {\rm d}^2x\, \partial ^\mu U^{-1}\partial _\mu U -
{1\over 4\pi}\int_0^1{\rm d}r \int {\rm d}^2x\, \epsilon^{\mu\nu} \widetilde
U^{-1}\dot {\widetilde U} \widetilde U^{-1}\partial_\mu \widetilde U
\widetilde U^{-1} \partial _\nu \widetilde U \quad ,\cr
= & - \Gamma [U] \quad , &(2.72)\cr}
$$
defining the Wess--Zumino--Witten (WZW) functional.\ref{16,17,44}
The WZW theory has a right- and a left-moving component of Noether current,
which generate an affine Lie algebra. Witten proved that the minimal theory
(i.e. with unit coefficient) is equivalent to free fermions, and the
above-mentioned currents turn into the fermionic currents. The elementary
fields of the theory build a representation of the affine algebra. The
equation of motion can be interpreted, in the quantum theory, as equations
defining the realization of the current algebra -- or the so-called null
states of conformal field theory, for whose detailed definition we refer to
the original literature.\ref{45} In the case of quantum chromodynamics, the
gauge-field self interaction will correspond to an off-critical perturbation
of such a model.

{}From the gauge invariance, we recover the full $U\, ,\,V$ dependence,
replacing
$U$ by the gauge-invariant $\Sigma=UV$. Notice, at this point, that the
determinant does not factorize in terms of chiral and anti-chiral
components\newfoot{$^{5)}$}{This is expected on grounds of vector gauge
invariance, since it forces us into a definite type of regularization. If one
defines the determinant as factorizing into definite chiral components, vector
gauge invariance is not respected. The difference is a contact term, as a
result of eq. (2.78). See refs. [18,19] for a detailed discussion.}.

Such a problem motivates us to consider the relation between the WZW functional
computed as a function of the product $\Sigma=UV$, namely $\Gamma[UV]$, and
$\Gamma[U]\,,\,\Gamma[V]$. Each part of the WZW functional can be considered
as follows. For the principal sigma model piece, we have, using the cyclicity
of the trace:
$$
\eqalignno{
S_{P\sigma M}[UV] = \,& {\tr \over 8\pi} \int {\rm d}^2x\, \partial ^\mu
(V^{-1}
U^{-1})\partial_\mu (UV)\cr
= \,& {\tr \over 8\pi}\! \int {\rm d}^2x\, \partial ^\mu U^{-1} \partial _\mu
U + {\tr \over 8\pi} \int {\rm d}^2x\, \partial ^\mu V^{-1} \partial _\mu V +
{\tr\over 4\pi} \int{\rm d}^2x\, U^{-1} \partial_\mu U V\partial^\mu V^{-1}\cr
= \,& S_{P\sigma M}[U] + S_{P\sigma M}[V] + {\tr \over 4\pi} \int {\rm d}^2x\,
U^{-1} \partial_\mu U V\partial^\mu V^{-1} \quad . &(2.73)\cr}
$$

For the WZ term (2.69) we use
$$
(UV)^{-1} \partial_\mu (UV) = V^{-1}\left[ U^{-1} \partial _\mu U + \partial
_\mu VV^{-1}\right] V\quad ,\eqno(2.74)
$$
and after some calculation we find
$$
S_{WZ}[UV] = S_{WZ}[U] + S_{WZ}[V] + {1\over 4\pi}\, \tr \int_0^1{\rm d}r\,\int
{\rm d}^2 x\, \epsilon_{\mu\nu}W^{\mu\nu}\quad,\eqno(2.75)
$$
where
$$
W_{\mu\nu} =  {{\rm d}\over {\rm d}r}\, \widetilde U^{-1}\partial _\mu
\widetilde U \widetilde V \partial _\nu \widetilde V^{-1} - \partial_\mu\left[
\widetilde V \partial _\nu \widetilde V^{-1}\widetilde U^{-1} \dot {\widetilde
U}\right] - \partial _\nu \left[ \widetilde U^{-1} \partial _\mu \widetilde U
\widetilde V \dot {\widetilde U}^{-1}\right]\quad . \eqno(2.76)
$$

Since the last two terms are total derivatives, they drop out, while the first
one turns out to be local, not depending on the extensions $\widetilde U\, ,\,
\widetilde V$. Therefore
$$
S_{WZ}[UV] = S_{WZ}[U] + S_{WZ}[V] + {1\over 4\pi}\tr \,\int_0^1{\rm d}r\, \int
{\rm d}^2 x\, \epsilon^{\mu\nu} U^{-1} \partial _\mu U V \partial _\nu V^{-1}
\quad .\eqno(2.77)
$$

Adding eqs. (2.73) and (2.77) we find
$$
\eqalignno{
\Gamma[UV]=\Gamma[U] &+\Gamma[V]+{1\over 4\pi}\tr\,\int{\rm d}^2x\,(g^{\mu\nu}+
\epsilon^{\mu\nu})U^{-1} \partial_\mu U V \partial_\nu V^{-1}\quad ,&(2.78a)\cr
 = \Gamma[U] & +\Gamma[V]+{1\over 4\pi}\tr\,\int{\rm d}^2x\, U^{-1}\partial_+UV
\partial_- V^{-1} \quad ,&(2.78b)\cr}
$$
which we call the Polyakov--Wiegmann formula from now on. The last term is an
obstacle for the factorizability of chiral left and right parts of the
fermionic determinant.

In order to implement the change of variables (2.59), in the quantum theory, we
still have to compute its Jacobian, that is
$$
{\cal D}A_+ {\cal D}A_- = J \, {\cal D}U{\cal D}V\quad ,\eqno(2.79)
$$
where
$$
J=\det{\delta A_+\over\delta U}{\delta A_-\over\delta V}=\det {\nabla}\quad .
\eqno(2.80)
$$

Notice that ${\nabla}$ is the covariant derivative in the adjoint
representation. Therefore it corresponds to the previously computed
determinant in fundamental representation, raised to the power given by the
quadratic Casimir ($c_V$), leading to the result
$$
J= \E^{-i c_V \Gamma[UV]}\quad .\eqno(2.81)
$$

It is well known that the invariances of the fermionic part of the Lagrangian
(2.1) are local gauge transformations $SU(N)$, as well as $SU(N)_L\times
SU(N)_R$, for both right $(R)$ and left $(L)$ components, namely
$$
\eqalignno{
\psi _{\scriptstyle{R\atop L}} & \to w_{\scriptstyle{R\atop L}}\psi
_{\scriptstyle{R\atop L}} \quad , &(2.82)\cr
A_\pm&\to w_{\scriptstyle{R\atop L}}\left(A_\pm+{i\over e}\partial_\pm\right)
w^{-1}_{\scriptstyle{R\atop L}}\quad ,&(2.83)\cr}
$$
corresponding to pure vector gauge transformation when $w_R=w_L=w$, while for
$w_R=w^{-1}_L=w$  it corresponds to a pure axial vector transformation. If we
use the change of variables $(L)$, the transformations
$$
\eqalignno{
\psi &\to \E^{i\gamma _5\theta }\psi \quad ,&(2.84)\cr
A_+&\to wA_+w^{-1}+{i\over e}w(\partial _+w^{-1})\quad ,&(2.85)\cr
A_-&\to w^{-1}A_-w+{i\over e}w^{-1}(\partial _-w)\quad ,&(2.86)\cr }
$$
reduce to $U\to U w^{-1}$ and $V\to w^{-1}V$.

The above transformations are not symmetries of the effective action $W[A]$ due
to the axial anomaly. This non-invariance may be understood in terms of a new
bosonic action $S_F[A,w]$ for  the fermions in a background field $A_\mu$.
Indeed
$$
S_F[A,g]\equiv\Gamma[UgV^{-1}]-\Gamma[UV]\quad,\eqno(2.87)
$$
in such a way that using the invariance of the Haar measure, we find
$$
\det i\not\!\!D\equiv\E^{iW[A]}=\int{\cal D}g\,\E^{iS_F[A,g]}\quad.\eqno(2.88)
$$

In fact $S_F(A,g)$ plays the role of an equivalent bosonic action and its
explicit form may be obtained by repeated use of the Polyakov--Wiegmann formula
(2.78)
$$
S_F[A,g]=\Gamma[g]+{1\over 4\pi}\!\int\!{\rm d}^2x\,{\rm tr}\,\left[e^2A^\mu A_
\mu-e^2A_+gA_-\gmu -eiA_+g\partial_-\gmu -eiA_-\gmu \partial _+g\right]\quad ,
\eqno(2.89)
$$
representing $\det i\!\not\!\!D$ in terms of bosonic degrees of freedom.
Therefore we recover a formulation in terms of $A_\mu$ and of the independent
field $g$.

The advantage of such a result over (2.16) is plural, and largely compensates
the problems posed by the fact that a local formulation does not exist for the
WZW fields $g$. As we have seen, (2.89) represents a bosonized version of
QCD$_2$, and the inclusion of gauge interactions was natural. Moreover,
algebraic properties of the Polyakov--Wiegmann functional will later permit to
obtain drastic simplifications of the theory, with a thorough separation of
some gauge fields, whose appearance will be effective only by means of the BRST
constraints. Therefore such results point to an extraordinary parallel to the
line of development of the Schwinger model, as presented for instance in [8].

Observe that we may reobtain Witten's non-Abelian bosonization formulae
$$
\eqalignno{
j_+&=-{i\over 2\pi }g^{-1}\partial _+g\quad ,&(2.90a)\cr
j_-&=-{i\over 2\pi }g\partial _-g^{-1}\quad ,&(2.90b)\cr }
$$
from eq. (2.88), by first writing down the vacuum expectation value of products
of the currents $j^\mu$ of the free fermion theory and then functionally
differentiating it with respect to $A_\mu$ and setting $A_\mu=0$. For the
correlators of $j_\pm$ we have
$$
\displaylines{
\langle j_+^{i_1j_1}(x_1)\cdots j_+^{i_nj_n}(x_n)\rangle_F\hfill \cr
\hfill =\left({-1\over 2\pi }\right)^n \left\langle [g^{-1}(x_1)i\partial_+g
(x_1)]^{i_1j_1}\cdots [g^{-1}(x_n)i\partial _+g(x_n)]^{i_nj_n}\right\rangle_B
\quad , \quad (2.91a)\cr
\langle j_-^{i_1j_1}(x_1)\cdots j_-^{i_nj_n}(x_n)\rangle _F\hfill \cr
\hfill =\left( {-1\over 2\pi }\right)^n \left\langle [g(x_1)i\partial _-g^{-1}
(x_1)]^{i_1j_1}\cdots [g(x_n)i\partial _-g^{-1}(x_n)]^{i_nj_n}\right\rangle_B
\quad .\quad (2.91b)\cr }
$$

In the $U(1)$ case, where the Wess--Zumino term in (2.72) vanishes, only the
principal $\sigma$-model is left, namely
$$
W[A]=-{1\over 8\pi }\int {\rm d}^2x \, (\partial _\mu \Sigma^{-1})(\partial_\mu
\Sigma)\quad. \eqno(2.92)
$$
Recall that $\Sigma  \equiv UV$. Now if we substitute $U=\E^{i(\varphi-\phi)}$
and $V=\E^{-i(\varphi+\phi)}$, in eqs. (2.59) one obtains for $A_\mu$ and
$F_{\mu\nu}$ respectively
$$
\eqalignno{
eA_\mu=\,&(\partial_\mu\varphi+\widetilde \partial _\mu \phi)\quad ,&(2.93)\cr
eF_{\mu\nu}=\,&(\partial_\mu\widetilde\partial_\nu-\partial_\nu\widetilde
\partial _\mu )\phi =-\epsilon _{\mu \nu }\dal \phi \quad . & (2.94)\cr}
$$

In the above set of equations, (2.94) can be solved for $\phi$ in terms of
$F_{\mu\nu}$, and we get
$$
\phi={e\over 2}\int{\rm d}^2y\,D(x-y)\epsilon^{\mu\nu}F_{\mu\nu}(y)\quad,
\eqno(2.95)
$$
where $D(x-y)$ is the massless propagator. If we make the identification
$\Sigma=\E^{2\phi}$, and use that $\dal D(x-y)=\delta^2(x-y) $ we obtain
$$
W[A]={1\over 2\pi }\int {\rm d}^2x\, (\partial _\mu \phi )^2
={1\over 8\pi }\int {\rm d}^2x\, \epsilon^{\mu \nu }F_{\mu \nu }(x)
D(x-y)\epsilon^{\lambda \rho}F_{\lambda \rho}(y)\quad .\eqno(2.96)
$$

If now we return to the equivalent bosonic action $S_F[A,g]$, eq. (2.89),
in the general non-Abelian case, we rewrite the integrand of the second term as
$$
e^2A^\mu A_\mu-e^2A^\mu g\left(g_{\mu\nu}+\epsilon_{\mu\nu}\right)A^\nu\gmu-ie
A^\mu\left(g_{\mu\nu}+\epsilon_{\mu\nu}\right)g\partial^\nu\gmu-ieA^\mu\left(
g_{\mu\nu}-\epsilon_{\mu\nu}\right)\gmu\partial^\nu g\, ,
$$
take the variational derivative with respect to $A_\mu$, and we obtain the
current
$$
J_\mu={e\over 2\pi}A_\mu-{ei\over
4\pi}\left\{\left(g_{\mu\nu}+\epsilon_{\mu\nu}
\right)gD^\nu g^{-1}+\left(g_{\mu\nu}-\epsilon_{\mu\nu}\right)g^{-1}D^\nu g
\right\}\quad .\eqno(2.97)
$$
Using eq. (2.59), this current may also be written in the form
$$
\eqalignno{
J_+ &= {i\over 2\pi}\left\{ U^{-1}\partial_+U-\left(Ug\right) ^{-1}\partial_+
\left( U g\right)\right\}\quad , & (2.98)\cr
J_-&={i\over 2\pi}\left\{ V^{-1}\partial_-V-\left( V g ^{-1}\right)^{-1}
\partial_-\left( V g^{-1}\right)\right\}\quad .&(2.99)\cr}
$$
Under local gauge transformations in the extended bosonic space, $U\to U\omega$
, $V\to V\omega$ and $g\to\omega^{-1}g\omega$, the above currents (2.98) and
(2.99) transform covariantly: $J_\pm\to\omega^{-1}J_\pm\omega$, and the
effective action realizes the local symmetry, $S_F[A,\omega^{-1}g\omega]=S_F
[A,g]$.

Only in the Abelian case, the components (2.98) and (2.99) reproduce the form
of the Witten current (2.90). In the non-Abelian case, however, it
involves the gauge field itself. This leads to an effective action $S_F[A,g]$,
which in terms of the current (2.97) reads
$$
S_F[A,g]=\Gamma[g]+\int{\rm d}^2x\,\tr\left\{ eJ_\mu A_\mu-{1\over 4\pi }\left(
A_+A_--g^{-1}A_+gA_-\right)\right\}\quad .\eqno(2.100)
$$
The second term in the integrand cancels only in the Abelian case, where $S_F[
A,\omega]$ reduces to the conventional form.

We now turn to the partition function related to the QCD$_2$ Lagrangian,
given by eq. (2.1). It reads
$$
{\cal Z} \left[ \overline \eta , \eta , i_\mu \right] = \int {\cal D}\psi
{\cal D}\overline \psi {\cal D}A_\mu \, \E ^{i\int {\rm d}^2 z\,{\cal L}\,+ \,
i\int{\rm d}^2 z \,(\overline\eta\psi+\overline \psi \eta  + i_\mu A_\mu ) }
\quad ,\eqno(2.101)
$$
where $\eta,\overline\eta $ are the external sources for the fermions
$\overline \psi, \psi$, and $i_\mu$ is the external source for the gauge field
$A_\mu$.

The bosonized version of the theory was obtained by rewriting the fermionic
determinant $\det i\!\!\not\!\!\! D$ as a bosonic functional integral as eq.
(2.88), where now we identify $W[A]\equiv -\Gamma[UV]$. The external sources
have been used to redefine the fermionic field as
$$
\overline\psi i\not\!\!D\psi+\overline\eta\psi+\overline\psi\eta=\Big[\overline
\psi+\overline\eta(i\not\!\!D)^{-1}\Big]i\not\!\!D\Big[\psi+(i\not\!\!D)^{-1}
\eta\Big]-\overline\eta(i\not\!\!D)^{-1}\eta\quad.\eqno(2.102)
$$

The non-linearity in the gauge-field interaction can also be disentangled by
means of the identity
$$
\E^{-{i\over 4}\int{\rm d}^2z\,\tr F_{\mu\nu}F^{\mu\nu}}=\int{\cal D}E\,
\E^{-i\int{\rm d}^2z\,\left[{1\over 2}\tr E^2+{1\over 2}\tr EF_{+-}\right]}
\quad,\eqno(2.103)
$$
where $E$ is a matrix-valued field. Taking into account the previous set of
information we arrive at
$$
\eqalign{
&{\cal Z}\left[ \overline \eta , \eta , i_\mu\right]= \int {\cal D}E{\cal D}
U{\cal D}V{\cal D}g\cr
& \times \E^{i\Gamma[UgV] -i(c_V+1) \Gamma[UV] -i \int {\rm d}^2 z \,\tr
[{1\over 2} E^2 + {1\over 2}E F_{+-}] +i \int {\rm d}^2z\, i_\mu A_\mu
-i \int {\rm d}^2z {\rm d}^2w \, \overline \eta  (z) (i \not D)^{-1}(z,w)
\eta (w) }\quad .\cr}\eqno(2.104a)
$$

We should also include a term $\, m\,\tr (g\!+\!g^{^{-1}})$ in the effective
action,\ref{23} if we were considering massive fermions, but we shall avoid
such a complication and consider only the massless case. We should mention,
repeating the introduction, that already in the case of the Schwinger model the
inclusion of mass for the fermion couples the previously free and massless
``$\eta$" excitations to the theory, rendering results technically more
complicated to be obtained.

Gauge fixing is another ingredient and, in fact, the process of introducing
ghosts here is standard. We perform the procedure implicitly, until it is
necessary to explicitly take into account the ghost degrees of freedom. Up to
considerations concerning the spectrum, our manipulations do not explicitly
depend on the gauge fixing/ghost system, and we keep it at the back of our
minds and formulae. We return to this problem in section 2.6.

It is not dificult to see that the field $\widetilde g$ decouples (up to the
BRST condition) after defining a new gauge-invariant field $\widetilde g=UgV$.
Using the invariance of the Haar measure, ${\cal D}g={\cal D}\widetilde g$,
the partition function turns into
$$
\eqalign{
&{\cal Z}\left[\overline\eta,\eta,i_\mu\right]=\int{\cal D}\widetilde g\,\E^{i
\Gamma[\tilde g]}\int{\cal D}E{\cal D}U{\cal D}V{\cal D}({\rm ghosts})\cr
& \times \E^{-i(c_V+1)\Gamma[UV]-i\int{\rm d}^2z\,\tr[{1\over 2}E^2+{1\over 2}E
F_{+-}]+iS_{\rm ghosts}+i\int{\rm d}^2z\,i_\mu A_\mu -i\int{\rm d}^2z {\rm d}^2
w\, \overline \eta  (z) (i \not D)^{-1}(z,w) \eta  (w) }\, ,\cr}\eqno(2.104b)
$$
where the $A_\mu$ variables are given in terms of the two-matrix-valued fields
$U$ and $V$, as in eq. (2.59).

In the way the gauge-field strength $F_{+-}$ is presented, it hinders
further developments; however if we write it in terms of the $U$ and $V$
potentials, we arrive at the identity                    aqui
$$
\tr EF_{+-}={i\over e}\tr UEU^{-1}\partial_+(\Sigma\partial_-\Sigma^{-1})
\quad .\eqno(2.105)
$$
We have used the variable $\Sigma=UV$ and this will imply a further
factorization of the partition function. In fact, $\Sigma$ is a more natural
candidate for representing the physical degrees of freedom, since $U$ and $V$
are not separately gauge-invariant. We redefine $E$, taking once more
advantage of the invariance of the Haar measure, in such a way that the
effective action depends only on the combination $\Sigma$. The variables
$U$ and $V$ will then appear separately only in the source terms, which are
gauge-dependent, as they should; there the gauge fields may be described as
$$
A_+= {i\over e} U^{-1}\partial_+ U\quad ,\quad A_-={i\over e}(U^{-1}
\Sigma)\partial_-(\Sigma^{-1} U)\quad .\eqno(2.106)
$$
If we eventually choose the light-cone gauge, we will have e.g. $U=1,\,A_-={i
\over e}\Sigma\partial_-\Sigma^{-1}$ and $A_+=0$. From the structure of
(2.105), it is natural to redefine variables as $\widetilde E'=UEU^{-1},{\cal
D} E={\cal D}\widetilde E'$. Notice that already at this point the $E$
redefinition implies, in terms of the gauge potential, an infinite gauge tail,
which captures the possible gauge transformations. It is also convenient to
make the rescaling $\widetilde E'=2e(c_V+1)\widetilde E$, with a constant
Jacobian. In terms of the field $\widetilde E$, consider the change of
variables
$$
\partial_+\widetilde E = {i\over 4\pi} \beta^{-1}\partial_+\beta \quad ,\quad
{\cal D}\widetilde E = \E^{-ic_V\Gamma[\beta]}{\cal D} \beta\quad ,\eqno(2.107)
$$
introducing $\beta$, the analogous of a Wilson-loop variable. Now we use the
identity (2.78) to transform the $\beta \Sigma$ interaction into terms that
can be handled in a more appropriate fashion. Writing both steps separately, we
have
$$
\eqalign{
&{\cal Z}\left[\overline\eta,\eta,i_\mu\right]=\int{\cal D}\widetilde g\,\E^{i
\Gamma[\tilde g]}{\cal D}U{\cal D}({\rm ghosts})\,\E^{iS_{{\rm ghosts}}}
\int {\cal D} \Sigma{\cal D}\widetilde E\E^{-i(c_V+1)\Gamma[\Sigma]}\cr
& \!\times \!\E^{-(c_V+1)\tr\!\int\!{\rm d}^2z\partial_+\widetilde E\Sigma
\partial_-\Sigma^{^{-1}}\!-\!2ie^2 (c_V+1)^2 \!\int \!{\rm d}^2z\tr
\widetilde E^2 +i \!\int\!{\rm d}^2z\,i_\mu A_\mu-i\int{\rm d}^2z {\rm d}^2w
\,\overline \eta (z)(i\not D)^{^{-1}}(z,w)\eta (w)}\cr}\eqno(2.108a)
$$
in such a way that after replacement of (2.107) in (2.108$a$) and using
(2.78) for $\Gamma [\beta \Sigma]$, we arrive at
$$
\eqalign{
&{\cal Z}\left[\overline\eta,\eta,i_\mu\right]=\int{\cal D}\widetilde g\,\E^{i
\Gamma[\tilde g]}{\cal D}U{\cal D}({\rm ghosts})\,\E^{iS_{{\rm ghosts}}}\int
{\cal D}\Sigma{\cal D}\beta\cr
&\!\times \!\E^{-i(c_V+1)\Gamma[\beta\Sigma]+i\Gamma[\beta]+ {2ie^2(c_V+1)^2
\over (4\pi)^2}\tr\!\int\!{\rm d}^{^2}\!z[\partial_+^{^{-1}}\!(\beta^{^{-1}}\!
\partial_+\beta)]^2+i\!\int\!{\rm d}^{^2}\!z i_\mu A_\mu-i\!\int\!{\rm
d}^{^2}\!
z{\rm d}^{^2}\!w\overline\eta(z)(i\not D)^{^{-1}}\!(z,w)\eta(w)}\cr}
\eqno(2.108b)
$$

Defining the massive parameter $\mu = (c_V+1)e/2\pi$ and the field
$\widetilde \Sigma = \beta\Sigma$, the partition function reads
$$
\eqalign{
&{\cal Z}\left[ \overline \eta , \eta , i_\mu\right]=\int{\cal D}\widetilde g
\,\E^{i\Gamma[\tilde g]}{\cal D}U{\cal D}({\rm ghosts})\,\E^{iS_{{\rm ghosts}}}
\int {\cal D}\widetilde\Sigma\,\E^{-i(c_V+1)\Gamma[\widetilde\Sigma]}\cr
&\times \int {\cal D} \beta\, \E^{i\Gamma[\beta] + {\mu^2i\over 2}\tr \int
{\rm d}^2 z\,[\partial_+^{-1}(\beta^{-1}\partial_+\beta)]^2}\,\E^{i\int{\rm
d}^2
z\,i_\mu A_\mu -i\int{\rm d}^2z\,{\rm d}^2w\overline\eta(z)(i \not D)^{-1}(z,w)
\eta (w) }\quad ,\cr}\eqno(2.108c)
$$
where now $A_+={i\over e}U^{-1}\partial_+U\,,\,A_-={i\over e}(U^{-1}\beta^{-1}
\widetilde \Sigma )\partial_-(\widetilde\Sigma^{-1}\beta U)$, and we used the
Haar invariance of the $\Sigma$ measure.

Up to BRST constraints and  source terms, the above generating functional
factorizes in terms of a conformal theory for $\widetilde g$, which represents
a gauge-invariant bound state of the fermions, of a second conformal field
theory for $\widetilde \Sigma$, which represents some gauge condensate, and of
an off-critically perturbed conformal field theory for the $\beta$ field,
which also describes a gauge-field condensate, which we interpret as an
analogue of the Wilson-loop variable in view of the change of variables
(2.107).  The conformal field theory representing $\widetilde\Sigma$ has an
action with a negative sign (see (2.108$c$)). Therefore we have to carefully
take into account the BRST constraints in order to arrive at a positive metric
Hilbert space. This is reminiscent of the commonly encountered negative metric
states of gauge theories, and appeared already in the Schwinger model.\ref{8}
In that case the requirement that the longitudinal current containing the
negative metric field vanishes, implies the decoupling of the unwanted fields
from the physical spectrum. The only trace of such massless fields is the
degeneracy of the vacuum. In that case, the chiral densities commute with the
longitudinal part of the current, and it is possible to build operators
carrying
non-vanishing fermion number and chirality. They are, however, constant
operators commuting with the Hamiltonian, and the ground state turns out to be
infinitely degenerate. There are definite vacua superpositions where the above
states are just phases -- the so-called $\theta$-vacua.

Notice that in eq. (2.108$a$) the only place where the charge shows itself
is in the $\widetilde E^2$ term. The limit $e\to 0$ corresponds to a
topological
theory, since $\widetilde E$ integration would restrain the $F_{\mu\nu}$ field
to be a pure gauge. This will be used later in connection with the string
formulation. Moreover, we might also generalize such a term to $f(\widetilde
E)$, for an arbitrary function $f$ not quadratic in $\widetilde E$. In such a
case, possible in two dimensions, we arrive at a Landau--Ginzburg
generalization (see section 4.4).
\vskip .5cm
\penalty-1000
\noindent {\it Reobtaining the $U(1)$ case}
\vskip .5cm
\nobreak
\noindent If we write the $\beta$ field as an exponential, $\beta = \E^{2i\sqrt
\pi \vartheta}$, we find for the $\beta$ Lagrangian
$$
L= {1\over 2}\left(\partial _\mu \vartheta\right)^2 - {1\over 2}{e^2\over \pi}
\vartheta^2 \quad ,\eqno(2.109)
$$
which describes a bosonic excitation of mass
$$
m_\vartheta = {e\over \sqrt \pi}\quad ,\eqno(2.110)
$$
which is well known from the Schwinger-model analysis. The remaining fields are
massless excitations, and the full Lagrangian reads
$$
L = {1\over 2}\left(\partial_\mu \vartheta\right)^2 - {1\over 2}{e^2 \over \pi}
\vartheta^2 - {1\over 2}\left(\partial _\mu \eta\right)^2 + {1\over 2}
\left(\partial _\mu \varphi\right)^2 \quad ,\eqno(2.111)
$$
which describes the Schwinger model.
\vskip 1cm
\penalty-1000
\noindent {\bf 2.4 Strong coupling analysis}
\vskip .5cm
\nobreak
\noindent The 'tHooft analysis is well suited for the weak coupling limit of
the theory, namely for very heavy quarks. For strong coupling there is a
problem signalled by the presence of the tachyonic pole in the quark
propagator for light quarks. Since the quarks should not appear asymptotically,
the issue of the strong coupling remains. Indeed, the question of whether we
have the screening/confinement picture characteristic of the Schwinger model
is not yet clear.

Some authors tried to give an answer to such a question.\ref{20,22} The only
available method to study strongly coupled fields is bosonization. Bosonization
 of fermions in a representation of non-Abelian symmetry groups is trivial
and leads to complicated $\sigma$-model interactions. Borrowing methods
used in the Abelian case, one is led to non-local terms, and the symmetry is
not
preserved. However, some calculations may still be performed, and it was used
in this case to arrive at a generalized sine-Gordon interaction rendering some
non-trivial results. Abelian bosonization methods have been used for the first
time in non-Abelian gauge theories in ref. [48].

The analysis of gauge theories is rather complex by itself. First there is the
question of gauge fixing. In fact there are procedures which simplify our work
enormously. In ref. [49] it has been proved that it is possible to
choose a gauge where the electric field is diagonal and traceless, and where
the diagonal gauge fields are in the Coulomb gauge. In such a case the ghosts
decouple.\ref{49} We shall use such a procedure.

The Lagrangian
$$
{\cal L} = - {1\over 2} F^{\mu\nu}\left( \partial _\mu A_\nu - \partial _\nu
A_\mu - ie [A_\mu ,A_\nu] \right) + {1\over 4} F_{\mu\nu} F^{\mu\nu} +
\overline \psi (i \not \! \partial + e \not \! A)\psi - m \overline \psi \psi
\eqno(2.112)
$$
leads to canonical momenta given by the expressions
$$
\Pi_0=0\quad,\quad \Pi_1=F^{01}\quad,\quad {\rm and}\quad \Pi_\psi=-i\psi^+
\quad .\eqno(2.113)
$$

The Hamiltonian is easily computable, and one finds
$$
H = - {1\over 2} \Pi_1^2 + i \overline \psi \gamma_1\partial _1\psi + m
\overline \psi \psi - e A^\mu \overline \psi \gamma_\mu \psi - \Pi_1 (\partial
_1 A_0 + ie [A_0, A_1])\quad .\eqno(2.114)
$$

The Gauss law, which is a consequence of the constraint ($\Pi_0=0$), is written
$$
\partial _1\Pi_1 - ie [A_1,\Pi_1] - e \overline \psi \gamma_0\psi = 0 \quad .
\eqno(2.115)
$$

We define new fields $e_i$ in terms of the canonically conjugated momenta,
which
are diagonal in the group index
$$
\sqrt \pi \Pi_1^{ii} = e_i - {1\over N}\sum_{i=1}^N e_i\quad ,\eqno(2.116a)
$$
from which it follows that
$$
\sum_i\left( \Pi_1^{ii}\right)^2 = {N\over \pi} \sum _{i,j} (e_i-e_j)^2 \quad .
\eqno(2.116b)
$$
This is very useful due to the gauge fixing discussed before eq. (2.112):
non-diagonal momenta $\Pi_1^{ij}$, for $i\ne j$, can be chosen as zero. With
such a choice, the Gauss law (2.115) is very simple for off-diagonal terms,
and determines $A_1$. One finds
$$
- {ie\over \sqrt \pi}A_1^{ik}(e_k - e_i) - e \overline \psi_i \gamma_0\psi_k =
0 \quad , \quad i\ne k \quad .\eqno(2.117)
$$

Therefore $A_1^{ik}={\sqrt\pi i\over e_k-e_i}\overline\psi_i\gamma_0\psi_k$;
upon its insertion into the Hamiltonian (2.114), using (2.116$b$) and the
Gauss law explicitly for the last term, we obtain
$$
H = - {N \over 2\pi} \sum_{i,j}(e_i - e_j)^2 + i \overline \psi
\gamma_1\partial _1\psi + m \overline \psi \psi - ie\sqrt \pi \,\overline
\psi_i
\gamma_0\psi_k {1\over e_k - e_i}\overline \psi _k \gamma_1\psi_i\quad .
\eqno(2.118)
$$

The last term may be Fierz-transformed in order to obtain only the diagonal
terms of the current, by means of the formula
$$
\eqalign{
\overline \psi_i\gamma_0\psi_j\overline \psi_j\gamma_1\psi_i & = -{1\over 2}
\left[ \overline\psi_i\gamma_0\gamma^\mu\gamma_1\psi_i \overline \psi_j
\gamma_\mu \psi_j + \overline\psi_i\gamma_0\gamma_5\gamma_1\psi_i \overline
\psi_j \gamma_5 \psi_j + \overline\psi_i\gamma_0\gamma_1\psi_i \overline\psi_j
\psi_j \right] \cr
& = - {1\over 2}\left[ \overline \psi_i \gamma_1\psi_i \overline \psi_j
\gamma_0\psi_j + \overline \psi_i \gamma_0\psi_i \overline \psi_j \gamma_1
\psi_j + \overline \psi_i \psi_i \overline \psi_j \gamma_5\psi_j -
\overline \psi_i \gamma_5\psi_i \overline \psi_j \psi_j \right]\quad .\cr}
\eqno(2.119)
$$

The first two terms of the above equation vanish, since they are symmetric for
$i \leftrightarrow j$, while we sum over $i,j$, multiplying by ${1\over e_i -
e_j}$, which is antisymmetric; using further such symmetry properties we
obtain
$$
ie\sqrt\pi\sum\limits_{i,j}M_+^iM_-^j{1\over e_i - e_j}\quad ,\eqno(2.120)
$$
where $M_\pm^i= \overline\psi_i (1\pm \gamma_5)\psi_i$.

At this point one uses the Abelian bosonization procedure, by means of which
the diagonal part of the current is written as
$$
j_\mu^i = {1\over \sqrt \pi} \epsilon _{\mu\nu}\partial ^\nu \phi^i \quad .
\eqno(2.121)
$$

Using the fact that $\Pi_1$ is diagonal, in the diagonal part of the Gauss
law, and substituting (2.120) therein for the current, one identifies $\phi^i$
and $e_i$, that is $e_i = \phi^i$.

The last term in eq. (2.118) depends on the regularization procedure employed
to define the severe divergences appearing in the product of fields as given,
but it is nevertheless clear that such a term depends on the difference $\phi_i
-\phi_j$. Baluni\ref{20} was able to obtain an integral representation for
such a term, but we just write the bosonic version of the Hamiltonian as
$$
H = \sum _i \left[{1\over 2} \Pi_{\phi_i}^2 + {1\over 2}(\phi'_i)^2 + m'
{\rm cos}\, \phi_i\right] - {N\over 2\pi}\sum _{i,j}(\phi_i -\phi_j)^2 +
\sum_{i,j} f(\phi_i - \phi_j) \quad ,\eqno(2.122)
$$
where $\Pi_{\phi_i}$ is the momentum canonically conjugated to the bosonized
field $\phi_i$, $f(\phi)$ is a function of the differences and contains
regularization-dependent constants. Notice that the terms $\Pi_{\phi_i}^2$ and
${\phi'_i}^2$ correspond, in the Hamiltonian procedure, to the bosonized
version of the kinetic term. Recently,\ref{142} it has been argued that the
bosonized field $\phi_i$ fulfills an equation of motion, in the presence of
(background) gauge fields, which corresponds to the Bethe--Salpeter equation
(2.35). In such a case, the integral term arises from the Vandermonde
determinant. We wish here to obtain also information about the bosonic spectrum
in the strong coupling limit, using the Hamiltonian (2.122). It is useful to
define a new basis of fields as
$$
\varphi = {1\over \sqrt N}\sum \phi_i\, \, {\rm and }\quad \chi_l=\sum_{i=1}^N
t^l_{ii}\phi_i\quad , \quad {\rm with \, \, inverse}\quad \phi_{N-i} =
{\varphi\over \sqrt N}+ \sum M_{ij}\chi_{N-j}\quad ,\eqno(2.123)
$$
where $l = 1,\cdots ,N$ and the above constants are
$$
t_{ii}^l = \cases{0 & $i> l+1\quad ,$\cr
-\sqrt{l(l-1)} & $i=l\quad ,$\cr
1/\sqrt{l(l-1)} & $i<l\quad ,$\cr} \eqno(2.124a)
$$
and
$$
M_{ij}= \cases{0 & $i<j-1$\quad ,\cr
- \sqrt{{N-j\over N-j+1}} & $i=j-1$\quad ,\cr
{1\over \sqrt{(N-j)(N-j+1)}}& $i>j-1$\quad .\cr}\eqno(2.124b)
$$
In terms of the above fields one has
$$
\sum_{i,j} (\phi_i-\phi_j)^2 = N \sum \chi_l^2\quad ,\eqno(2.125)
$$
and the last term in the Hamiltonian does not depend on the ``centre-of-mass"
coordinate $\varphi$. In the strong coupling limit $\varphi$ has a very small
mass compared to the $(N-1)$-plet described by the field $\chi_l$. For the
cosine term one has
$$
{\rm cos}\, \left[ 2\sqrt\pi\left( {\varphi\over N} + \sum M_{ij}\chi_j \right)
\right] = {\rm cos}\, \left[ 2\sqrt{{\pi\over N}}\varphi \right] \prod_j
{\rm cos}\, (2\sqrt \pi M_{ij}\chi_j)+ {\rm sin \,terms}\quad .\eqno(2.126)
$$

For very massive $\chi$ fields (strong coupling limit) we shall put $\chi\sim
0$, and the terms in sine disappear. According to the usual renormalization
procedure for the cosine terms, it must be renormalized according to the mass
of the fields (see refs. [8] and [12]), which leads to
$$
m\Lambda \left({e\sqrt \pi\over \Lambda}\right)^{N-1\over N} {\rm cos}\,\left[
2\sqrt{{\pi\over N}}\varphi \right]\quad ,\eqno(2.127)
$$
where $\Lambda$ is an arbitrary renormalization parameter, providing $\varphi$
with an interaction of the form defined by the Hamiltonian
$$
H=N_{m'}\left[{1\over 2}\Pi_\varphi^2+\left(\partial_1\varphi\right)^2-{m'}^2
{\rm cos}\,\left( 2\sqrt{{\pi\over N}}\varphi\right)\right]\quad ,\eqno(2.128)
$$
where
$$
{m'}^2 = \left\{ Nm \left( {e\over \sqrt\pi}\right)^{N-1\over N}\right\}^
{2N\over 2N-1}\quad ,\eqno(2.129)
$$
and $N_{m'}$ is a normal product with respect to the renormalized mass $m'$.

The baryon, interpreted as a soliton with quark number $N$, thus has a mass
$$
M_B= {2\over \pi} 2 \sqrt{{\pi\over N}} m' (2N-1)\quad  {\mathop {\sim}\limits_
{N\to \infty}} \quad {8\over \sqrt\pi}\sqrt{{me\over \sqrt\pi}}N\quad .
\eqno(2.130)
$$

Therefore it vanishes for small quark mass, in accordance with the idea that
its mass approaches zero in the chiral limit, and such baryons are reminiscent
of Goldstone states. Baryonic masses may also be interpreted as
soliton--antisoliton bound states of the sine-Gordon field. Further properties
of the strong coupling limit can be found in ref. [50].
\vskip 1cm
\penalty-1000
\noindent {\bf 2.5 A Lagrangian realization of the coset construction}
\vskip .5cm
\nobreak
\noindent As we have seen, fermionic gauge theories are naturally written in
terms of gauged WZW theories. These in turn provide a Lagrangian realization
of the coset construction.\ref{51,52}

For a moment we delete the ``mass term" $\mu$ in (2.108$c$). We are left with a
gauged WZW theory as  explained in section 2.3. The pure WZW functional is
invariant under a $G\times G$ symmetry transformation given by
$$
g(x^+,x^-)\to G(x^-)g(x^+,x^-)G_+(x^+)\quad .\eqno(2.131)
$$
In general, the anomaly-free vector subgroup $H\subset G\times G$ (in the
QCD$_2$ case, $H$ corresponds to $G$) can be gauged by adding the term
$$
{1\over 4\pi}\tr \int {\rm d}^2x \left[ e^2 A_+A_- - e^2 A_+gA_-g^{-1} + ie
A_-g^{-1} \partial _+ g + ie A_+ g \partial _- g^{-1}\right] \quad
.\eqno(2.132)
$$

Such a gauging procedure introduces constraints in the theory.\ref{53} In
order to understand this point in more detail, we have to consider the effect
of the ghost sector. In general, ghosts are implemented by considering a
gauge-fixing function ${\cal F}(A)$ and introducing a factor
$$
\det \left( {\partial {\cal F}\over \partial A_\mu } {\partial A_\mu
\over \partial\epsilon }\right)\delta \left( {\cal F}(A)\right)\eqno(2.133)
$$
in the partition function, where $\epsilon $ is the gauge parameter. However,
if we are to render explicit the conformal content of the theory, it is more
useful here to represent all possible chiral determinants in terms of ghost
integrals. The reparametrization invariance is thus explicit and one can
verify that the gauge-fixing procedure, as outlined above, and which is more
frequently used in the gauge-field literature, is trivial in the sense that
one is led to a unit Faddeev--Popov determinant.

Therefore we assume that ghosts are introduced by writing determinants in
terms of ghost systems decoupled from the gauge fields by a chiral rotation, a
procedure which is possible in two-dimensional space-time. This is equivalent
to writing all determinants as
$$
\det {\nabla}_+ = \E^{-ic_V\Gamma[U]} (\det \partial_+ )^{c_V}\quad ,\quad
\det {\nabla}_-=\E^{-ic_V\Gamma[V]} (\det \partial_-)^{c_V}\quad ,\eqno(2.134)
$$
and substituting the free Dirac determinant in terms of ghosts as
$$
\eqalignno{
(\det\partial_+)^{c_V}= & \int {\cal D}b_{--}{\cal D}c_+\,\E^{i\tr\int {\rm d}
^2 x\,b_{--}\partial_+c_+}\quad ,&(2.135)\cr
(\det\partial_-)^{c_V}= & \int {\cal D}b_{++}{\cal D}c_-\,\E^{i\tr\int {\rm d}
^2x\,b_{++}\partial_-c_-}\quad .&(2.136)\cr}
$$

In fact the determinant of the Dirac operator does not factorize as in eq.
(2.134) because of the regularization ambiguity. At every step, one has to
ensure vector current conservation. Such determinants cancel out by changing
some of the variables (as in eq. (2.107)) but do not cancel in (2.108$c$), from
which we are led to the contribution
$$
\int{\cal D}b_{--}{\cal D}b_{++}{\cal D}c_+{\cal D}c_-\,\E^{i\tr\int{\rm d}^2x
\,(b_{++}\partial_-c_-+b_{--}\partial_+c_+)}\quad .\eqno(2.137)
$$

Although decoupled at the Lagrangian level, such terms are essential due to
constraints arising in the zero total conformal charge sector, and lead to BRST
constraints on physical states. The constraints are obtained in a system of
interacting conformally-invariant sectors $(g,\Sigma,b_{++},b_{--},c_+,c_-)$
described by the partition function
$$
{\cal Z}=\int{\cal D}g{\cal D}\Sigma{\cal D}b_{++}{\cal D}b_{--}{\cal D}c_-
{\cal D}c_+\,\E^{ik \Gamma[g]-i(c_V+k)\Gamma[\Sigma]+i\tr\int{\rm d}^2x\,(b_
{++}\partial_-c_-+b_{--}\partial_+c_+)}\quad .\eqno(2.138)
$$

Such a construction is a particular one out of a general
equivalence\ref{51,53} between the algebraic construction of $G/H$
coset models, and an $H$-gauged WZW theory on a group manifold $G$.

Indeed, starting out of the WZW functional $\Gamma[G]$ one can gauge the
anomaly-free vector subgroup $H$ by means of the gauged WZW functional
$$
\Gamma[g,A]=\Gamma[g]+{1\over 4\pi}\tr\int{\rm d}^2x\,\left\{-iA_+\partial_-g
g^{-1}+iA_-g^{-1}\partial_+g - A_+gA_-g^{-1}+A_+A_-\right\}\quad,\eqno(2.139)
$$
where $A_\mu$ belongs to the adjoint representation of $H$; we rewrite the
gauge field in terms of the potentials as in (2.59) and use the
Polyakov--Wiegmann identity (2.78), regainning (2.87). Taking into account the
ghost system and the Jacobian, and moreover using the gauge $V=1$, we arrive at
$$
Z= \int {\cal D}g{\cal D}h{\cal D}b_{++}{\cal D}b_{--}{\cal D}c_-{\cal D}c_+
\,\E^{ik\Gamma[g]-i(k+c_H)\Gamma[U]+i\tr\int{\rm d}^2x(b_{++}\partial_-c_- +
b_{--}\partial _+c_+)}\quad ,\eqno(2.140)
$$
where $c_H$ is the quadratic Casimir for the subgroup $H$ and arises from the
Jacobian induced by (2.59). As before, the partition function factorizes
into non-interacting sectors at the Lagrangian level. However, the BRST
condition couples them. The reason is the existence of constraints in the
theory. They can be derived by coupling all the Lagrangian fields to an
external gauge
field by means of the minimal coupling such as exemplified by (2.139), with
$A_+^{^{\rm ext}}=\partial_+U^{^{\rm ext}}{U^{^{\rm ext}}}^{-1}\,,\,A_-^{^
{\rm ext}}=\partial_- V^{^{\rm ext}}{V^{^{\rm ext}}}^{-1}$, which again, by
means of eq. (2.78), leads to a
partition function independent of $U^{^{\rm ext}}$ and $V^{^{\rm ext}}$ due to
the
vanishing of the total central charge. By computing the derivative of the
partition function with respect to $A_+$ and $A_-$ we obtain currents which
must vanish for consistency.

Let us prove this assertion. The interaction of the fields from the WZW theory
with such external gauge fields is equivalently obtained from (2.87), that is
$$
\eqalignno{
ik \Gamma[g,A] & = ik\Gamma[U_{_{\rm ext}} gV_{_{\rm ext}}] - i k \Gamma[U_{_
{\rm ext}} V_{_{\rm ext}}] \quad ,&(2.141)\cr
-i(c_H + k)\Gamma[\Sigma,A]&=-i(c_H\!+\!k)\Gamma[U_{_{\rm ext}}\Sigma V_{_
{\rm ext}}]\!+i (c_V\!+\!k)\Gamma[U_{_{\rm ext}}V_{_{\rm ext}}]\, ,&(2.142)\cr}
$$
and
$$
i\tr\!\!\int\!\!{\rm d}^2x\,[b_{++}D_-^{^{\rm ext}}\!\!c_-+b_{--}D_+^{^
{\rm ext}}\!\!c_+] =i\tr\!\!\int\!\!{\rm d}^2x\,[b_{++}V_{_{\rm ext}}
\partial_-(V_{_{\rm ext}}^{-1}c_-)+b_{--}U_{_{\rm ext}}^{-1}\partial_+(U_{_
{\rm ext}}c_+)],\eqno(2.143)
$$
where $k$ is the central charge. We recall that in section 2.3 we had $k=1$.
In the first two cases, namely eqs. (2.141) and (2.142), the invariance of
the Haar measure permits a change of variables as $\widetilde g=U_{_{\rm ext}}g
V_{_{\rm ext}}\,,\,\left({\cal D}\widetilde g={\cal D}g\right)$ and $\widetilde
\Sigma=U_{_{\rm ext}}\Sigma V_{_{\rm ext}}\,\,\left({\cal D}\Sigma={\cal D}
\widetilde
\Sigma\right)$, while in the latter case (2.143) a chiral rotation can be done,
leaving back the free ghost system and a WZW term, $c_H\Gamma[U_{_{\rm
ext}}V_{_
{\rm ext}}]$. Therefore, the $\Gamma[U_{_{\rm ext}}V_{_{\rm ext}}]$ term
cancels, owing to
the balance of central charges, and the partition function does not depend on
the external gauge fields. This implies, in particular, that the functional
derivative of the partition function with respect to the external gauge fields
vanishes, and therefore
$$
{\delta{\cal Z}(A_+^{^{\rm ext}},A_-^{^{\rm ext}})\over\delta A_+^{^{\rm ext}}}
\Big\vert_{A_+^{^{\rm ext}},A_-^{^{\rm ext}}=0}=0={\delta{\cal Z}(A_+^{^
{\rm ext}},A_-^{^{\rm ext}})\over \delta A_-^{^{\rm ext}}}\Big\vert_{A_+^{^
{\rm ext}},A_-^{^{\rm ext}}=0}\quad,\eqno(2.144)
$$
which are equivalent, because of the minimal coupling (see eq. (2.89)), to the
set of constraints
$$
\left\langle kg^{-1}\partial_+ g-(c_H+k)\Sigma^{-1}\partial_+\Sigma-4\pi[
b_{++},c_-]\right \rangle=0=\left\langle J_{+g}+J_{+\Sigma}+J_{+{\rm ghost}}
\right\rangle\quad \eqno(2.145)
$$
and
$$
\left\langle k\partial_- gg^{-1}-(c_H+k)\partial_-\Sigma\Sigma^{-1}-4\pi[
b_{--}, c_+]\right\rangle=0=\left\langle J_{-g}+ J_{-\Sigma} + J_{-{\rm ghost}}
\right\rangle\quad .\eqno(2.146)
$$

Each of the above currents satisfies a current algebra with a central
charge. One can build up a BRST charge $Q$ as
$$
Q^{(+)} = \sum \colon {c_-^i}_{-n} \left( {J_{+g}^i}_n+ {J_{+\Sigma}^i}_n
\right) \colon -{i\over 2}f^{ijk}\sum\colon {c_-^i}_{-n}{b_{++}^j}_{-m}
{c_-^k}_{n+m}\colon\quad,\eqno(2.147)
$$
where the indices $i, j, k$ refer to the adjoint representation of the
symmetry group, $f^{ijk}$ are the structure constants, and the mode expansion
of
the fields reads
$$
\eqalignno{
c_-^i & =  \sum {c_-^i}_n {x^+}^{-n} \quad ,&(2.148a)\cr
b^i_{++} & = \sum {b_{++}^i}_n {x^+}^{-n-1}\quad ,&(2.148b)\cr
J_{+g,\Sigma}^i&=\sum\left(J_{+g,\Sigma}^i\right)_n{x^+}^{-n-1}\quad.&(2.148c)
\cr}
$$
As it should, the charge (2.147) is nilpotent: $\,{Q^{(+)}}^2={1\over 2}\{Q^{
(+)},Q^{(+)}\}=0$. This implies that the above system is a set of first-class
constraints (a similar set of constraints $Q^{(-)}$ is obtained for $J_-,
b_{--}$, and $c_+$).

The stress tensor can be computed in terms of such currents, and we have three
contributions, namely $T^{\rm tot}(z)=T_g(z)+T_\Sigma(z)+T^{\rm ghost}(z)$,
that is
$$
T^{\rm tot}(x^+)={1\over k+c_G}\colon\left(J_+^g\right)^2\colon-{1\over k+c_H}
\colon\left(J_+^\Sigma\right)^2\colon-\colon b_{++}^i\partial_+c_-^i=T^G+T^H+
T^{\rm gh} \quad .\eqno(2.149)
$$

The central charges corresponding to the right-hand side of eq. (2.149) can be
computed, being respectively $c(G,k)$, $c(H,-k-c_H)$ and $c_{\rm gh}=-2d_H$;
adding them we obtain the total central charge which is
$$
\eqalignno{
c^{\rm tot}=\,&{2kd_G\over 2k+c_G}+{2(-k-c_H)d_H\over 2(-k-c_H)+c_H} - 2d_H \cr
=\, & {2kd_G\over 2k+c_G} - {2kd_H\over 2k+c_H}\quad .&(2.150)\cr}
$$

Therefore the total central charge coincides with the one found using the coset
construction. The energy momentum tensor, when written in the form
$$
T^{\rm tot} = T^G - T^H + T' = T^{\rm coset} + T'\quad ,\eqno(2.151)
$$
is such that one can prove that the central charge $c'$ corresponding to $T'$
vanishes, while $T'$ itself commutes with $T^{\rm coset}$. The unitary
representations of $T'$ are thus trivial,\ref{54} and there is a strong
equivalence with the coset construction, once the unitarity of the physical
spectrum is established. This has been done in ref. [53].

The gauged WZW model is equivalent to the coset construction of $G/H$ conformal
field theories. The physical subspace is generated by a product of matter and
ghost sectors, obeying the equation $Q\vert{\rm phys}\rangle=0$. This also
solves the problem of the sector with negative central charge, which should not
be considered separately, being coupled through the BRST condition. Had we no
such condition we would expect problems concerning negative metric states.
Therefore we cannot consider each sector separately.

In the case of the inclusion of QCD$_2$ in such a scheme, we shall see that
there are further constraints. Although the new constraints seem to be of the
first-class type when considered alone, there is a combination that is
second-class due to the cancellation of the ghost contribution. Therefore, in
the case of QCD$_2$ we have to deal with a Dirac quantization procedure of
second-class constraints!\ref{55}

However, we shall see (in section 5) that several interesting properties,
characteristic of the model, as well as part of the conformal structural
relations, still hold, and the QCD$_2$ problem can be understood as an
integrable perturbation of a coset construction of conformal field
theory.
\vskip 1cm
\penalty-1000
\noindent {\bf 2.6 Chiral interactions}
\vskip .5cm
\nobreak
\noindent Fermionic gauge theories with chiral coupling of the fermions to the
gauge field exhibit an anomaly in the covariant divergence of the external
field gauge current, $J_{\mu,ch}^a(x\vert A)$, which is referred to as the
non-Abelian anomaly.\ref{56} Such an anomaly implies, at first sight, an
inconsistency with the gauge field equations of motion, and a breakdown of
gauge invariance. The requirement that the physical particles belong to safe
representations, where the anomaly has a vanishing group theoretic factor
leads to the prediction of a certain number of quarks (balancing the leptons).
Although such predictions seem to be successful, the study of anomalous
theories reveals a consistent field theoretic structure.\ref{8}

The issue can be better understood from the fact that because of the
non-invariance of the fermionic measure under chiral transformations, all
dynamical variables are observable, and gauge fixing is neither required nor
allowed. However, one can follow the line drawn by refs. [57, 58], introducing
the unity in the ``Faddeev--Popov form" (see chapter 13 of ref. [8] for
details), namely
$$
\Delta_{\cal F}[A]\int {\cal D}g \,\delta[{\cal F}(A^g)] =1 \quad ,\eqno(2.152)
$$
where ${\cal F}$ is an arbitrary gauge fixing function. We are lead to a
gauge-invariant formulation in terms of a larger set of fields, where the
partition function is
$$
Z[J,\eta,\overline\eta]=\int{\cal D}g{\rm d}\mu[A]{\cal D}\psi{\cal D}\overline
\psi\,\E^{iI[A,g,\psi,\overline \psi]+i\int [J\,{}^g\!\!A+\overline\eta^g\psi+
{}^g\!\overline\psi \eta] }\quad ,\eqno(2.153)
$$
and the gauge measure is now given by the usual Faddeev--Popov procedure,
$$
{\rm d}\mu[A] ={\cal D}A_\mu\Delta_{\cal F}[A]\delta({\cal F}(A))\quad ,
\eqno(2.154a)
$$
and the notation is defined as
$$
\eqalignno{
{}^g\!\!A_\mu & \equiv {A_\mu^g}^{^{-1}}=g\left(A_\mu + {i\over e}\partial_\mu
\right)g^{-1}\quad ,&(2.154b)\cr
I[A,g,\psi,\overline \psi] &= S[A,\psi,\overline \psi] +
\alpha_1(A,g^{-1})\quad ; &(2.154c)\cr}
$$
$\alpha_1(A,g^{-1})$ is the 1-cocycle defined by the Wess--Zumino consistency
condition. The usual ``unitary gauge" $g=1$, leads to the ``ordinary" (in the
sense of na\"{\i}ve) discussion of the theory. Such a gauge defines the
so-called gauge-non-invariant formulation.\ref{57--59} The ``classical"
equation of motion acquires a modification due to the 1-cocycle above and reads
$$
\nabla_\mu F^{\mu\nu}+J^\nu+{\delta\alpha_1\over\delta A_\nu}=0\quad,
\eqno(2.155)
$$
displaying no inconsistency. Now the theory displays a quantum extended gauge
symmetry. Both representations, namely the full gauge theory, and the theory at
unitary gauge, are equivalent only after integration over the gauge  field
$A_\mu$. In such a case, we can see that the previously mentioned
inconsistency of the equations of motion disappears.

The integration of the fermionic sector may be performed, leading to a
bosonized
action, obtained integrating the fermions. The effective action reads
$$
L_{\rm eff}= -{1\over 4} \tr F_{\mu\nu}F^{\mu\nu} + \Gamma^{(R)}[A]\quad ,
\eqno(2.156)
$$
where
$$
i\Gamma^{(R)}[A] = \ln \,\det \,\left(i\! \not \! \partial + e\!\not \!\! A
{1-\gamma_5\over 2}\right)\quad,\eqno(2.157)
$$
which is equal to the usual WZW action in the gauge $A_+=0$. However, we
cannot apply the vector gauge symmetry as before to select a particular
combination of both potentials, and we must allow for a regularization
arbitrariness which, as shown by Jackiw and Rajaraman,\ref{60} has the form of
the square of the gauge field with an arbitrary coefficient $(a)$. One
therefore obtains a partition function given by
$$
Z = \int {\cal D}A_\mu {\cal D}g \, \E^{iS_{\rm eff}[A,g]}\quad ,\eqno(2.158)
$$
with
$$
S_{\rm eff}[A,g]= \int{\rm d}^2x\, \left[ -{1\over 4} \tr F_{\mu\nu}F^{\mu\nu}
+
{ae^2\over 8\pi} A^\mu A_\mu \right] - \Gamma[g] - {ie\over 4\pi}\int{\rm d}^2x
\,\tr \left[ g^{-1}\partial _+gA_-\right]\quad .\eqno(2.159)
$$

The Euler--Lagrange equations derived from above read
$$
\eqalignno{
\left(g^{\mu\nu}-\epsilon^{\mu\nu}\right) \partial _\mu (g^{-1}\partial_\nu
g)+ie\left(g^{\mu\nu}+\epsilon^{\mu\nu}\right){\nabla}_\mu A_\nu & = 0\quad ,
&(2.160a)\cr
{\nabla}_\mu F^{\mu\nu}+{ae^2\over 4\pi}A^\nu-{ie\over 4\pi}\left(g^{\nu\mu}-
\epsilon^{\nu\mu}\right)\tr (g^{-1}\partial _\mu g\,t)&=0\quad,&(2.160b)\cr}
$$
where ${\nabla}_\mu\chi=\partial_\mu\chi+[g^{-1}\partial_\mu g,\chi]$ is the
adjoint covariant derivative. The canonical quantization may be performed
using the Dirac method. In the Abelian case the theory simplifies. A full
account of such developments is beyond the scope of the present review; it is
presented in chapters 13 and 14 of ref. [8], to which we refer, as well as in
the references presented therein.
\vskip 1cm
\penalty-5000
\noindent {\bf 3. Pure QCD$_2$ and string theory}
\nobreak
\vskip .5cm
\penalty-1000
\noindent {\bf 3.1 Introduction }
\vskip .5cm
\nobreak
\noindent Quantum electrodynamics in four dimensions hit enourmous successes
after the establishment of renormalized perturbation theory. Due to the
smallness of the fine structure constant, perturbation theory led to results
that could be tested experimentally, and the relative errors were ten orders
of magnitude smaller than unity. The renormalization prescription, although
very awkward, was later precisely defined in the mathematical sense,
in such a way that all predictions were reliable. Important lessons were drawn
for general quantum field theories off the perturbative scheme, such as the LSZ
formalism\ref{61} or the axiomatic approach.\ref{62} However, dynamical
calculations were restricted to
perturbation theory, and results concerning the strong interactions remained
unreliable. In particular, information about the spectrum of the theory was
only
accessible via approximative, often non-unitary schemes, as the Bethe--Salpeter
equation in the ladder approximation. Weak interactions, although well
described by perturbation theory, was known to be, in the case of the Fermi
theory, ill-behaved in the high-energy domain. Therefore, quantum field theory
fell into stagnation due to the difficulty in going beyond QED.

This motivated the works on the S-matrix theory, which subsequently played a
dominant role.\ref{63} It was thought that the bootstrap idea might
substitute the dynamical principles and provide a more fundamental
formulation, implying a very radical position towards conventional
developments. This led to the concept of duality.\ref{64} The explicit
realization of such ideas was implemented by the Veneziano formula,\ref{65}
leading to full development of dual models. However, its predictive power
was very low, due to the lack of an underlying dynamical principle, since the
idea of having a Lagrangian was abandoned, or at least avoided. A number of
incorrect results concerning the description of strong interactions led
physicists to discard the dual model formulation. In particular, the
high-energy behaviour of strong interactions is extremely well described by
perturbative QCD if use is  made of the RG and CS equations to impose
perturbation theory.\ref{1} On the other hand, string theory
was reinterpreted as a theory of unified interactions.

Nonetheless the problem of strong interactions could still not advance for the
understanding of low-energy phenomena, which should only be addressed using a
non-per\-tur\-ba\-ti\-ve method. In fact, several properties concerning
hadrons are understandable by means of the concept of string-like flux tubes,
consistent with linear confinement and linear Regge trajectories, properties
derived also for the large-$N$ limit of QCD$_2$ with fermions. As we have
already observed, the $1/N$ expansion classifies the Feynman diagrams
according to their topology. For a fixed topology, the sum of diagrams is like
a sum of triangulated surfaces, with a structure very similar to string
theory. But it is only recently that more concrete results were obtained with
a direct relation between the large-$N$ expansion of two-dimensional QCD
without fermions and the string expansion. It should be clear that the string
dynamics is not that of critical strings, or even Liouville strings, and that
terms depending on the extrinsic geometry must be present.
\vskip 1cm
\penalty-3000
\noindent {\bf 3.2 Wilson loop average and large-$N$ limit }
\vskip .5cm
\nobreak
\noindent The biggest difficulty in the analysis of strong interactions is the
question of the large-distance behaviour, which cannot be understood by
perturbation theory. In this way, important phenomena in the description of
strong interactions, such as confinement, $\theta$-vacua structure, as well as
the bound-state spectrum, are poorly understood. On the other hand one knows
that the high-energy theory is well described by perturbation theory, which is
especially revigorated by the use of asymptotic freedom in connection with the
Callan--Symanzik and regularization-group equations. In such a description
quarks are called partons, and are essentially free, contrasting with the
confinement picture. Such different behaviours point to different pictures.

There are some attempts to deal with the strong limit by means of the
discretization of space-time, where one can obtain strong coupling expansion,
treating the system by methods borrowed from statistical mechanics.

The fact that phases are described, in general, in terms of such local-order
parameters facilitates the understanding of statistical systems displaying a
complex phase behaviour. However, Elitzur\ref{66} proved that every
non-gauge-invariant local quantity has a vanishing expectation value at all
temperatures. But a phase transition should be described by a parameter that
would indicate spontaneous symmetry breaking and, due to the above results,
such a local quantity does not exist, and local observables cannot be used to
indicate the possibility of different phases.

There are however, in gauge theories, observables that are not local. Indeed,
the Aharanov--Bohm effect shows that the exponential of the path-ordered
integration of the gauge field is a meaningful physical quantity, and contains
non-trivial information. Actually Feynman had already used phases as meaningful
objects in quantum mechanics, in order to describe amplitudes. Even in the
absence of the (``physical") electric and magnetic field, the effect of the
phase $\E^{ie\int {\rm d}x^\mu\, A_\mu}$ can be measured in an electron wave
function, which gives such a phase a physical meaning by itself.

The Wilson loop\ref{67}
$$
W[C] = \tr \, P \, \E^{ie\oint_C {\rm d}x^\mu\, A_\mu }\quad \eqno(3.1)
$$
may be defined as a function of the loop $C$ for any gauge theory, where $P$
means that we have to order the group indices according to the loop location.
It is colourless and will have a definite r\^ole in the description of
confinement, as we shall see. The Wilson line attached to a fermion $\E^{ie
\int_y^xA_\mu{\rm }dx^\mu}\psi(y)$, has an amplitude interpretation; such a
phase may describe the fermionic interaction.

It is not a very easy task to obtain information concerning Wilson-loop
expectation values in four-dimensional gauge theories. A strong coupling
expansion\ref{68} is available in lattice gauge theories, but general results
are very hard to obtain. The situation in two-dimensional Yang--Mills theories
in the absence of dynamical fermions is drastically simplified, and by a clever
choice of gauge the Wilson-loop expectation value can be exactly computed in
terms of the gauge group parameters. Let us consider the Wilson-loop
expectation value  for the pure gauge theory
$$
W[C]={\cal N}^{-1}\tr\,P\int{\cal D}A_\mu\,\E^{ie\oint_C {\rm d}x^\mu A_\mu}
\E^{-{i\over 4} \tr \int {\rm d}^2x\, F_{\mu\nu}F^{\mu\nu}}\quad .\eqno(3.2)
$$

If we consider the Coulomb gauge $A_0=0$, there is no ghost contribution; it
partially cancels with the multiplicative normalization of the functional
integral ${\cal N}^{-1}$, and we are left with
$$
W[C]={{\cal N}'}^{-1}\tr\,P\int{\cal D}A_1\E^{ie\oint_C{\rm d}x^1\,\tau^a_{x_1}
A^a_1}\E^{{i\over 2}\tr\int{\rm d}^2x\,(\partial_0A_1)^2}\quad.\eqno(3.3)
$$
Here we have written explicitly the generators of the group $\tau^a$.
Introducing now the Green function $G(x,y)$ satisfying
$$
{\partial_{x_0}}^2 G(x,y) = \delta^{(2)} (x-y)\quad,\eqno(3.4)
$$
with solution
$$
G(x,y)= {1\over 2} \vert x^0 - y^0\vert \, \delta(x^1-y^1)\quad;\eqno(3.5)
$$
completing the square in the functional integral we find\ref{69}
$$
W[C]= \tr \, P \, \E^{{i\over 2}e^2\oint \oint {\rm d}x^1{\rm d}y^1\,
\tau^a_{x_1}G(x,y) \, \tau^a_{y_1}}\quad .\eqno(3.6)
$$

Since the Green function is local in $(x^1-y^1)$, the indices carried by $\tau^
a_{x_1},\tau^a_{y_1}$ are unimportant, and we obtain their square, which is
the quadratic Casimir, namely $C_2(R)=\tau^a\tau^a$. Therefore the $P$ symbol
can be immediately deleted, and the trace operation leads to the dimension of
the representation of the gauge group under consideration. In formulas we write
$$
W_R[C] = ({\rm dim \, } R) \,\, \E^{{i\over 2} e^2C_2(R)A(C)}\quad ,\eqno(3.7)
$$
where $R$ is the representation, and $A(C)$ the area enclosed by the loop $C$,
obtained in
$$
\oint{\rm d}x_1\,\oint{\rm d}y_1\,G(x,y)=2\times{1\over 2}\vert x_0-y_0\vert
\oint{\rm d}x_1=\vert x_0-y_0\vert\vert x_{\rm init}-x_{\rm fin}\vert=A\quad.
\eqno(3.8)
$$

The fact that the result is given by the area is equivalent to saying that,
while computing the Wilson loop, the potential at a point of the loop, which
is due to the quark at other points, is proportional to the distance,
signalling a
confining situation. Were the Wilson-loop exponent is proportional to the
perimeter, as is the case for a typical three-dimensional massless propagator
instead of (3.5), we would have no confinement, as in weakly coupled QED in
four dimensions. Such is the Wilson criterion for confinement,\ref{67} and
such is also the use of the Wilson loop as an order parameter of the theory.

The language used here, as well as the introduction of a lattice, makes it more
natural to use Euclidean rather than Minkowski space. Thus until (and
including) section 3.3 we work on Euclidean space.

As discussed in great detail in ref. [68] there is a natural interpretation
of the theory on the lattice in terms of a string theory, which describes the
flux tubes, and the string tension is defined by
$$
k = \lim _{C\to \infty} -{1\over A(C)}\ln W[C]\quad .\eqno(3.9)
$$
The non-confining case thus corresponds to $k=0$, which is the case if the
Wilson loop displays the perimeter behaviour. It follows that the Wilson
loop is a good parameter to verify whether we have a confinement phase or not.

This may be considered as the starting point of extremely important results
relating pure QCD$_2$ and string theory. The large-$N$ limit of
two-dimensional pure Yang--Mills theory can be obtained as a consequence of
some exact results concerning Wilson loops. First we have the fact that the
expectation value of the Wilson loop in two dimensions was exactly computed in
terms of the quadratic Casimir, eq. (3.7). We refer the reader to Appendix C
as well as refs. [70-77] for more details  concerning group theory. Here we
shall repeat only some of the main results, recalling that given the
representation, a $1/N$ expansion of such a result for the expectation of a
Wilson loop when using a gauge group $G=SU(N)$ or $G=U(N)$ can be obtained
from group theoretical values of the dimension and the Casimir of the
representation, which is computed as a function of $N$ and the lengths $n_1>n_2
\cdots >n_r$ of the horizontal lines of the Young tableau $Y(R)$, defining the
representation $R$. Given a representation defined by a Young tableau with
rows $(n_1,\cdots,n_r)$, $\sum n_i = n $, the result\ref{75} for $G=U(N)$ is:
$$
C_2^{U(N)}(R)=Nn+\sum n_i(n_i+1-2i)=Nn+\widetilde
C_2^{U(N)}(R)\quad,\eqno(3.10)
$$
where $\sum \limits_{i=1}^{r}n_i = n $. For $SU(N)$ one substitutes $n_i$ by
$n_i-{n\over N}$, and drop the first term (arising from $N\sum n_i$), obtaining
$$
C_2^{SU(N)}(R)= Nn + \sum n_i (n_i + 1 - 2i) - {n^2\over N} = Nn - \widetilde
C_2^{SU(N)}(R) \quad .\eqno(3.11)
$$

For the sake of completeness we also recall that the dimension of the
representation is
$$
{\rm dim }\,R={\prod\limits_{i\le j\le N}(n_i-i-n_j+j)\over\prod\limits_{i\le
j\le N}(i-j)} \quad . \eqno(3.12)
$$
Notice that when we use (3.12) and (3.11) in the expression for the Wilson loop
(3.7), we end up, after a suitable redefinition of the charge $(e^2=\alpha/N)$,
 with a result that can be analyzed for large values of $N$. Such an analysis
is still premature, and we should first rewrite the theory in a convenient way
for string interpretation.

That step is realized from results arising from the lattice
formulation.\ref{68} The introduction of gauge fields on a lattice may be done
analysing a general pure matter action of the type ${\cal L}\sim\varphi^\dagger
_i\varphi_j$, where $(i,j)$ are two sites on the lattice. In a case where the
interaction is invariant under a symmetry group acting linearly on $\varphi$,
such a symmetry can be raised to a local symmetry by introducing a gauge field
taking values on the link $(i,j)$ and by writing the interaction as ${\cal L}
\sim\varphi_i^\dagger U_{ij}\varphi_j$, so that $U$ transforms as $U\to gUg^
{-1}$ under an element $g$ of the symmetry group $G$, while $\varphi\to g
\varphi$, and $\varphi^\dagger\to\varphi^\dagger g^{-1}$. A gauge field
self-interaction can be taken as a trace of $U$ on an elementary closed loop:
a plaquette, namely ${\cal L}_U\sim{\rm Re\,}(\tr\,U_{ij}U_{jk} U_{kl}U_{li})$.

Thus, in general, one works with group-valued objects defined on an elementary
link $U$, after dividing the space in elementary plaquettes, with such links
as edges. The field $U$ will represent the gauge degrees of freedom, and in the
continuum limit, where the size of the link goes to zero, we have $U=1+ieAa$.

The so-called Wilson action reproduces correctly the Yang--Mills theory in the
continuum limit; it is advantageous for certain lattice computations, and can
be described very easily in terms of group-valued elements. Indeed, if one
considers a lattice, which we suppose here for simplicity to be a regular
square lattice, each scalar field is defined on a site, while the natural
definition of a gauge field is on a link, due to the vectorial character. One
thus takes $U$ as above to describe the gauge field, and the action as given
by the cyclic product of gauge fields as above, that is\ref{68}
$$
S_W=\sum_p{N\over 2\alpha a^2}\tr\,\left(U_p+U^\dagger_p\right)\quad.
\eqno(3.13)
$$

The fact that the Wilson-loop average displays a string-like behaviour in the
large-$N$ limit, has been known for some time, due to the possibility of
relating it to a sum over surfaces with minimal area. Thus representing
Yang--Mills theory in terms of Wilson-loop averages is the corner-stone of its
relation to string theory. But there is a further technical point, which in
practice is a crucial device for performing some exact computations, which is
the introduction of the heat-kernel action. We start discussing this latter
aspect of the problem.

Using the lattice formulation, the Wilson loop is defined as the trace of the
product of group-valued operators over the edges of a plaquette. Consider the
(Euclidian) Wilson action, as given before. Loop averages, such as
$$
W[U]=\int{\rm d}U\,\,\E^{-{N\over 2\alpha a^2}\tr\,(U+U^+)}\tr\,U\quad,
\eqno(3.14)
$$
have been considered in the literature, see refs. [78--80]. In two
dimensions Migdal\ref{69} found a way to systematically integrate the
above quantity over a given edge variable. In other words, in a partition
function representation in terms of such quantities, one integrates out a link
that is common to two plaquettes, in such a way that the action does not
acquire modifications from such a procedure. This means that one has to modify
the lattice action, keeping the continuum limit, arriving at a
renormalization-group-invariant action. As a result of such requirements one
arrives at the
heat kernel action, to be properly defined below. Such an improved action is
thus exact in the sense that it describes either small or large lattices,
implying the important property of being almost independent of the type of
triangulation of the two-dimensional world-sheet. One considers the Boltzman
factor $Z[U,e^2,a^2]$ and first develops it in an expansion on characters of
the group $\chi_R[U]$, which for a single plaquette of area $a^2$ reads
$$
Z[U,e^2,a^2]=\sum_Rf_a(R)\,({\rm dim \,} R)\,\chi_R[U]\quad ,\eqno(3.15)
$$
where the sum is taken over all representations $R$ and $f_a(R)$ are the
coefficients of the expansion. Such a series is the Fourier representation in
terms of the group characters, being therefore very general, and is valid for
any arbitrary area $A$. Therefore we are allowed to write
$$
Z[U, e^2, A] = \sum_R f(R)\, ({\rm dim\, }R) \, \chi _R[U] \quad .\eqno(3.16)
$$

The heat kernel formulation is established once one knows the coefficients
$f(R)$. We compute them by imposing that the product $Z[UL]\,Z[L^\dagger V]$
\newfoot {$^{6)}$}{Here we shortened the notation, but the Boltzman factor
$Z[U,e^2,a^2]$, denoted shortly by Z[U], is still a function of $e^2$ and
$a^2$,
consequently $Z[UV]$ is a function of the charge $e^2$ and the sum of both
areas.}, after integration over the common link, namely after integrating out
the variable $L$, is identical to the Boltzman factor $Z[UV]$. Therefore we
have
$$
Z[UV]= \int {\cal D}L \sum _{R,S} f(R)\, f(S)\,({\rm dim\, R})
\, ({\rm dim\, S})\,\chi_R[U L]\,\chi _R[L^\dagger V]\quad .\eqno(3.17)
$$

Integration over characters is a common procedure in group theory. The rules
are summarized in Appendix C. Using eq. (C.12) we have
$$
Z[UV] = \sum_R (f(R))^2 \, ({\rm dim R})\, \chi[UV]\quad .\eqno(3.18)
$$

Such a procedure can be repeated indefinitely, and one can start out of a
single plaquette of area $a^2$, ending up with a macroscopic area $A$. One
thus obtains as a result
$$
\E^{-S} = Z[U] = \sum _R (f(R))^{A/a^2}\left( {\rm dim \, R}\right)\,
\chi[U]\quad . \eqno(3.19)
$$

The form of $f(R)$ must be such that it goes to unity as $a^2\to 0$, that is
$$
f(R)\, \sim \, 1- a^2 \epsilon_R\quad ;\eqno(3.20)
$$
therefore, for a finite area $f(R)= \E^{-A\epsilon_R}$, and the Boltzman factor
reads
$$
Z[U]= \sum _R \E^{-A\epsilon_R}\, ({\rm dim \, R})\, \chi[U]\quad .\eqno(3.21)
$$

The form of $\epsilon_R$ is obtained upon computation of the expectation value
of the Wilson loop $W[C] = \tr _{_C} \, U$, using $Z[U]$ as a Boltzman factor.
The (Euclidian counterpart of the) result (3.7) should be reproduced. We use
the fact that the Wilson loop, being the trace over the group-valued field,
projects out a character from the expansion (3.19) and one obtains
$$
\left\langle W[C] \right\rangle = \E^{-A\epsilon_R}\, ({\rm dim \, R}) =
({\rm dim \, R})\, \E^{- {\alpha^2\over 2N} C_2(R)A}\quad ,\eqno(3.22)
$$
thus fixing $Z[U]$ as given by
$$
Z[U]=\sum_R\E^{-{\alpha^2\over 2N}C_2(R)A} ({\rm dim \, R})\,\chi_R [U]\quad .
\eqno(3.23)
$$

The heat-kernel action is the corner-stone of all subsequent developments. It
is not difficult to see that the heat-kernel action diagonalizes the
Hamiltonian operator in a given representation. The Hamiltonian for the pure
QCD$_2$ model corresponds to the square of the momentum operator, that is, in
the temporal gauge
$$
H= {e^2\over 2} \int {\rm d}x\, {\delta^2\over\delta
{A^a_1}^2}\quad.\eqno(3.24)
$$
It acts on functionals of the loop operator on a compact space $(0<t<L)$,
$U= P\,\E^{ie\int _0^L {\rm d}x\, A_1}$, as
$$
H={e^2L\over 2}\tr\,\left(U{\partial\over\partial U}\right)^2\quad ,\eqno(3.25)
$$
where we have considered a space of total length $L$. Now $\left(U{\partial
\over \partial U}\right)^a\chi_R(U)=\chi_R(T^aU)$, and the diagonalized
Hamiltonian is just
$$
H= {1\over 2}e^2 LC_2(R)\quad .\eqno(3.26)
$$

Before proceeding, notice that these issues are not inherent in the lattice
formulation, but rather in the fact that gauge theory can be expressed in terms
of loops averages, upon defining the matrix $U_{xy}$ along an arbitrary contour
$C_{xy}$ connecting the points $x$ and $y$ as
$$
U_{xy}=\tr\,P\,\E^{ie\int_{C_{xy}}{\rm d}x_\mu\,A^\mu(x)}\quad,\eqno(3.27)
$$
and for a closed contour $C$, the expectation value of a Wilson loop in the
continuum is
$$
\left\langle W[C]\right\rangle = Z^{-1}\int {\cal D}A_\mu\, \E^{iS}\tr \, P\,
\E^{ie\oint {\rm d}x^\mu\, A_\mu(x)}\quad .\eqno(3.28)
$$

In particular, making small variations of the loop, with respect to its area,
it is possible to obtain a differential equation obeyed by the Wilson loop
described above, the Makeenko--Migdal equation.\ref{81}

Such results may be generalized for the product of two neighbouring Wilson
loops. We consider two neighbouring loops with a common link $l$, to which we
associate the group element $L$. Suppose we have a loop $(lp_1)$, and another
$(p_2l^\dagger)$, where $l^\dagger$ runs in the opposite direction to $l$.
Integration over the link is
$$
\eqalignno{
W[p_1\, p_2] = & \int {\rm d}L \,\, W[l\, p_1]\, W[p_2\, l^\dagger] = \cr
= & \sum _{R_1}\sum _{R_2} ({\rm dim}\, R_1) ({\rm dim}\, R_2) f(R_1)
f(R_2)\int
{\rm d}L\,\chi_{R_1}[U_{p_1}L]\,\chi_{R_2}[L^\dagger U_{p_2}]\, .& (3.29)\cr}
$$

In view of the orthogonality relation of the characters (C.12) we obtain
$$
W[p_1\,p_2]=\sum_R({\rm dim}\,R)\,{f(R)}^2\,\chi_R[U_1 U_2]\quad.\eqno(3.30)
$$

The form of $f(R)$ was obtained before, that is $f(R)=\E^{-{\alpha^2\over 2N}
C_2(R)A}$. Furthermore, it is not difficult to obtain, using such heat-kernel
action, the partition function in the case of a genus $g$ surface. We consider
the particular case of a sphere with $n$ holes, and cut it into parts without
holes. Subsequently we integrate over the link variables used in the cutting
procedure, as in Fig. 4.
\penalty-2000
$$\epsfxsize=3.5truecm\epsfbox{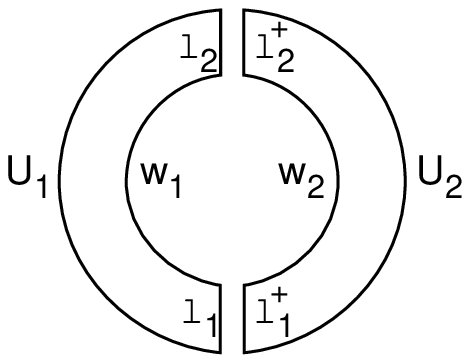}$$
\noindent{\eightpoint Fig. 4: Cutting procedure.}
\vskip.3truecm
\penalty-2000

We find
$$
W_{_{A_1 A_2}} \!=\!\!\sum_{R_1, R_2}\!\!{\rm dim}\,R_1{\rm dim }\,R_2\,
\E^{^{-A_1 C_2(R_1) - A_2 C_2(R_2)}}\!\!\!\int\!\!{\rm d}L_1{\rm d}L_2
\chi_{_{R_1}}[L_{_1}U_{_1} L_{_2}W_{_1}] \chi_{_{R_2}}[U_{_2}L_{_2}^\dagger
W_{_2}L_{_1}^\dagger]\, . \eqno(3.31)
$$

The integration is now performed using again (C.12), and we find
$$
W_{_{A_1 A_2}} = \sum _R \E^{-{\alpha^2\over 2N}(A_{1+2})C_2(R)} \chi_R[U_1
U_2]
\, \chi_R[W_1 W_2]\quad .\eqno(3.32)
$$

Notice the absence of one factor of ${\rm dim}\, R$ in such a integral. We can
now proceed with the gluing procedure as before, attaching with further
elements
to glue over the link variables $U_1 U_2$. We have to take into account that
the presence of a hole lowers by a unit the power of ${\rm dim}\, R$. Moreover
we are left with a factor $\chi[W]$, where $W$ represents the ``Wilson loop" of
the hole. We thus obtain, for genus $g$
$$
W^{(g)}_A = \sum _R ({\rm dim}\, R)^{2-2g}\, \E^{-{\alpha^2\over 2N}C_2(R) A}
\prod_1^h\chi_R[W_i]\quad,\eqno(3.33)
$$
where the handles can be obtained gluing holes and repeating the argument.

\vskip 1cm
\penalty-1000
\noindent {\bf 3.3 String interpretation}
\vskip .5cm
\nobreak
\noindent The partition function thus obtained can be expanded in powers of
$1/N$, as is clear from its form. The idea developed by Gross,\ref{70} to be
reviewed in what follows, is that such an expansion is equivalent to a string
theory expansion, where the string coupling is identified  with $1/N$, and the
string tension is related to the coupling constant ($\alpha= e^2N$).

The general intuition comes from the simplicity of the geometrical properties
of the pure Yang--Mills action. The Euclidian Yang--Mills partition function
over a manifold ${\cal M}$ is
$$
Z_{\cal M}=\int{\cal D}A^\mu\,\E^{-{1\over 4}\tr\int{\rm d}^2x\, \sqrt {\hat g}
F_{\mu\nu}F_{\mu\nu}}\quad .\eqno(3.34)
$$
Here $\hat g$ is the determinant of the induced metric, i.e. $\hat g=\det
g^{\alpha\beta}=\det\left(\!{\partial x^\mu\over\partial\xi^\alpha}{\partial
x^\nu\over\partial\xi^\beta}G_{\mu\nu}\!\right)$.

The dynamics is very simple in the absence of quarks, since gauge-vector fields
in two dimensions have zero degrees of freedom, as a simple counting reveals;
nevertheless the theory is non-trivial if ${\cal M}$ contains non-contractible
loops.
Indeed, if $C$ is such a loop, the Wilson loop $\tr\,P\E^{ie\oint_C{\rm d}x^\mu
\,A_\mu}$ cannot be gauged to unity. Gross\ref{70} argued that this is due
to the fact that in two dimensions the field strength can be defined in terms
of a scalar field $F$ by
$$
F_{\mu\nu} = \epsilon_{\mu\nu}\, F\quad , \eqno(3.35)
$$
in terms of which the action is geometrically very simple:
$$
S = {1\over 2}\tr \int {\rm d}^2 x\, F^2 \sqrt {\hat g}\quad ,\eqno(3.36)
$$
being independent of the metric, except for the volume form. The model is
invariant under area-preserving diffeomorphisms $(W_\infty)$. Therefore the
resulting theory can only depend on the topology of the manifold, its area, and
the parameters $N$ and $e$, that is
$$
Z_{\cal M} = Z_{\cal M} [G, N, e^2A]\quad ,\eqno(3.37)
$$
where $G$ is the genus of the target manifold.

The above partition function could be mapped to the partition function of a
string theory with target space ${\cal M}$. Such a conjecture reads
$$
\eqalignno{
\ln Z_{\cal M}[G,N,e^2A]&=Z_{\cal M}^{\rm string}[g_{st}= 1/N,\alpha=e^2 N]\cr
& = \sum _{{\rm genus}\,\equiv\, g}\left( g_{st} \right)^{2g-2}\int {\cal D}
x^\mu \E^{-\int {\rm d}^2 \xi \sqrt {\hat g}+ \cdots }\quad , &(3.38)\cr}
$$
where the Nambu action is used to define the string theory.  In such a
relation we have to define the theory on the right-hand side and relate the
genera $g$ and $G$, in such a way as to obtain equality. In fact, as we will
see later, the Nambu action turns out to be only part of the full story.
Indeed,
for vanishing area the right-hand side is a topological theory, and the Nambu
action will describe the effect of the area.\ref{73} We do not consider the
topological theory; in any case, the action appearing in the right-hand side
will not be considered dynamically. Thus the dots in the right-hand side of
eq. (3.38) are momentarily not relevant.

The Nambu--Goto action is certainly invariant under area-preserving
diffeomorphism, but it is very difficult to be quantized. The Polyakov action
on the other hand also displays a $W_\infty$ symmetry, but this is realized
non-linearly. Moreover, besides the difficulty of quantizing the Polyakov
action, there are further facts that make the direct use of the Nambu--Goto
theory more appealing. Indeed, the question of the singularity of the maps
defining the embedding of the (two-dimensional) world-sheet into a
two-dimensional target space is
clearly treated in the case of the Nambu--Goto theory, while such singularities
are unseen in the conformal gauge. In fact, the area of the surface described
by the string is only non-trivial due to folds. The Nambu--Goto action for
non-singular maps (non-vanishing Jacobians $\vert\partial x^\mu/\partial\xi^
\alpha\vert$) is a topological number, measuring how many times one covers the
target space. This raises doubts about the equivalence with the usual Liouville
description of non-critical strings for two-dimensional target spaces if one
does not take folds into account, since in that case one uses a conformal
gauge. The singularities would presumably describe the sources of the theory.
Moreover, string theory usually contains also graviton and dilaton fields, a
nuisance for the string theory interpretation of QCD$_2$, since the latter
does not contain those fields. The absence of folds may be a cure of such a
problem, since we have to include terms in the extrinsic geometry forbidding
them; this would presumably also prevent gravitons and dilatons, which is
necessary, since this is a theory of strong interactions without gravity.
This mechanism is also a way to prevent the tachyon, which is a centre-of-mass
degree of freedom; but since there are no propagating particles due to the
absence of maps with zero winding number, tachyons disappear as well.

Now, given the partition function of the pure Yang--Mills theory, performing
the $1/N$ expansion is just a matter of computation of group theoretical
factors expanded in a Taylor series in $1/N$, which is technically feasible.
One thus has to interpret the terms by relating them to a sum over geometric
objects.
\vskip .5cm
\penalty-300
\noindent {\it $1/N$ expansion  of the Yang--Mills partition function}
\vskip .5cm
\nobreak
\noindent In order to keep track of several terms and be able to interpret
them, it is useful to rewrite the quadratic Casimir eigenvalues as\ref{75}
$$
\eqalignno{
{1\over N}C_2^{U(N)}(R) = & \sum _i n_i + {1\over N}\widetilde C(R)\quad ,
& (3.39a)\cr
{1\over N}C_2^{SU(N)}(R) = \,&  {1\over N} C_2^{U(N)}(R) - {1\over N^2} \left(
\sum _i n_i\right)^2 \quad ,& (3.39b)\cr
\widetilde C (R) = & - \widetilde C(\overline R)\quad ,& (3.39c)\cr
\left({\rm dim}_{-N}\, \overline R\right)^2 = & \left( {\rm dim}_N\,R\right)^2
\quad , & (3.39d)\cr}
$$
where $\overline R$ is the representation conjugate to $R$ and we made the $N$
dependence of the dimension of the representation explicit.

The exponent can be expanded as a series in $1/N$, with the result
$$
\E^{-{\alpha A\over N}C_2(R)} = \sum_{i,j} \E^{-\alpha A n }{1\over i!}\left(
- {\alpha A \over N}\widetilde C(R)\right)^i {1\over j!} \left( - {n^2\over
N^2}\right)^j \quad ,\eqno(3.40)
$$
where the last term drops out in the case where the group is $U(N)$, instead of
$SU(N)$. Moreover, summing over the representation and its conjugate one finds
a factor
$$
\sum {1\over i!}\left( - {\alpha A\over N} \widetilde C(R) \right)^i
\left[ \left({\rm dim}_{N}\,R\right)^{2-2G} + (-1)^i
\left({\rm dim}_{-N}\, \overline R\right)^{2-2G}\right] \quad .\eqno(3.41)
$$
Only even powers of $1/N$ survive. This is a correct result for a theory of
closed orientable strings.

Gross uses some usefull relations. Consider the dimension of the representation
(${\rm dim}\,R$), given by eq. (3.12), and the dimension ($d_R$) of the
representation of the symmetric group of $n = \sum n_i $ objects,
$$
d_{[n_1\cdots n_r]} = n! {\prod_{i\le j\le r}(h_i-h_j)\over\prod_{i\le j\le r}
(i-j)}\quad ,\quad h_i = n_i - i + N\quad .\eqno(3.42)
$$
One finds the relation
$$
{\rm dim \, }R = {d_R\over n!} \prod _1^r {(N+n_i-i)!\over (N-i)!} \quad ,
\eqno(3.43)
$$
which is useful to find a $1/N$ expansion since
$$
{(N+n_i-i)!\over (N-i)!} = N^{n_i} \prod_{k=1}^{n_i} \left( 1 + {k - i\over N}
\right) \quad ,\eqno(3.44)
$$
where $k (i)$ runs over the columns (rows) of the Young tableau. The dimension
is thus given by
$$
{\rm dim \, }R = d_R {N^n\over n!} \prod_v \left( 1 + {\Delta_v \over N}\right)
\quad ,\eqno(3.45)
$$
where $\Delta_v$ is, for each cell of the Young tableau, the column index minus
the row index. Finally, as a function of $N$, we have the relation
$$
\vert {\rm dim \, }\overline R \vert = d_R {N^n\over n!} \prod_v \left( 1 -
{\Delta_v \over N}\right) = \vert {\rm dim \, }R \, (-N)\vert \quad
.\eqno(3.46)
$$

The large-$N$ analysis of the pure Yang--Mills results has to be done in three
qualitatively different cases, depending on the genus, due to the term $({\rm
dim\,}R)^{2-2G}\sim N^{-2n(G-1)}$. Indeed, for $G>1$, there is a
simplification, and the leading term in the $1/N$ expansion of the Yang--Mills
partition function is
$$
Z_G = \sum_{n=0}^\infty N^{-2n(G-1)}\E^{-n\alpha A}\sum_{{\rm rep}\, S_n}
\left( {n!\over f_r}\right)^{2(G-1)}\quad ,\eqno(3.47)
$$
where ${\rm rep}\, S_n$ are the representations of the symmetric group $S_n$.
There are corrections from the dimension of the group as well as from the
Casimir eigenvalue. For genus $G=1$, the torus, the result can be computed in
closed form in the large-$N$ limit. One has
$$
\eqalignno{
Z_{G=1} = & \sum_R \E^{-{\alpha A\over N} C_2(R)} =  \sum_{l_1\ge l_2 \ge
\cdots \ge l_n\ge 0} \E^{-\alpha A \sum l_i}\cr
= & \sum _{k_i =  l_{i+1} - l_i \ge 0} \E^{- \alpha A \sum n k_n } =
\prod_{l=1}^N {1\over 1- \E^{-n\alpha A}} = \eta \left( \E^{-\alpha
A}\right)\quad ,&(3.48)\cr}
$$
that is, one obtains the Dedekind function.

If we consider the logarithm of the above result, which should, as
conjectured, be interpreted as the string partition function, we find
$$
Z_{str} = \ln \eta \left( \E^{-\alpha A}\right)  = \sum _n \E^{- n\alpha
A}\sum _{ab = n} (a+b)\quad .\eqno(3.49)
$$
The coefficient of $\E^{-n\alpha A}$ counts the number of different maps
of a torus onto a torus $n$ times, where the two above cycles are winding $a$
and $b$ times around the two cycles of the target space torus.

The specific case of the sphere $(G=0)$ is more delicate due to the positive
power of the group dimension, and we refer for the specific treatment of
this case to the original publication, since it involves the technique of
discrete orthogonal polynomials.

The case of physical interest is the torus, where we have a flat target
manifold. There, as shown above, and according to the string interpretation
to be given below, every term can be simply understood. For higher genus there
are corrections due to the dimension of the representation of higher order in
$1/N$, which are not given a natural interpretation, and one needs corrections
to the Nambu--Goto action.

Before delving further into these points, it is natural to consider, with some
further detail, the $1/N$ expansion of pure Yang--Mills theory. We have, with
the considerations already stated about the dimension of the representation,
the expansion
$$
\eqalign{
& Z(G, \alpha A, N) =  \sum _{n=0}^\infty \sum _{R \in Y_n}({\rm dim \,}R)
^{2-2G} \E^{-{\alpha A\over 2N} C_2(R)}\cr
& = \! \sum_n \!\!\sum _{R\in Y_n}\!\! \left( {n!\over d_R}\right)^{2G-2}\!\!
\E^{-{n\alpha A\over 2}}\! \sum_{i=0}^\infty\!\!\left[{1\over 2^i i!}\! \left(
 \! -\alpha A\widetilde C(R)\! \right)^i \! N^{n(2-2G)-i} \! +\! {\cal O}
\!\left( \!N^{n(2-2G) - i -1}\!\right)\!\right]\, .\cr}\eqno(3.50)
$$

Such an expansion corresponds to considering the partition function of the
genus $G$ surface with no holes
$$
Z(G, \alpha A, N) = \sum _R ({\rm dim \,} R)^{2-2G} \E^{-{\alpha A\over 2}n}
\E^{-{\alpha A\over N}\widetilde C(R)} \E^{{\alpha A \over 2N^2}n^2}\quad ,
\eqno(3.51)
$$
and expanding the second exponent in powers of $1/N$ (the third in a feature
of $SU(N)$ theory). Notice that the first exponential corresponds to the
exponential of the area of the string, which winds $n$ times around the target
space area $A$. We will interpret it together with the next exponential as the
branched covering. The last term will be understood in terms of tubes and
collapsed handles.

However, one has to be cautious about counting the large-$N$  contributions at
this point. Expression (3.51) is correct. When one expands, and takes the sum
over Young tableaux with $n$ boxes, one loses, however, several important
representations - in fact, it has been argued that this only contains half of
the theory, the so-called chiral perturbation. The order in $1/N$ depends on
the factor $({\rm dim \,}R)^{-2(G-1)}$, which is of order $N^{-2n(G-1)}$,
for Young tableaux with $n$ boxes.

In general, representations are obtained from symmetrizing and antisymmetrizing
tensor products of the fundamental representation. For large $N$, we have to
take the leading-order contributions for quadratic Casimir and dimension, as we
shall do. However, there are also representations obtained from products of a
smaller representation $R$ and the conjugated of another representation,
$\overline S$. Such ``composite" representations are defined by the Young
tableau as in Fig. 5,
\penalty-2000
$$\epsfysize=3truecm\epsfbox{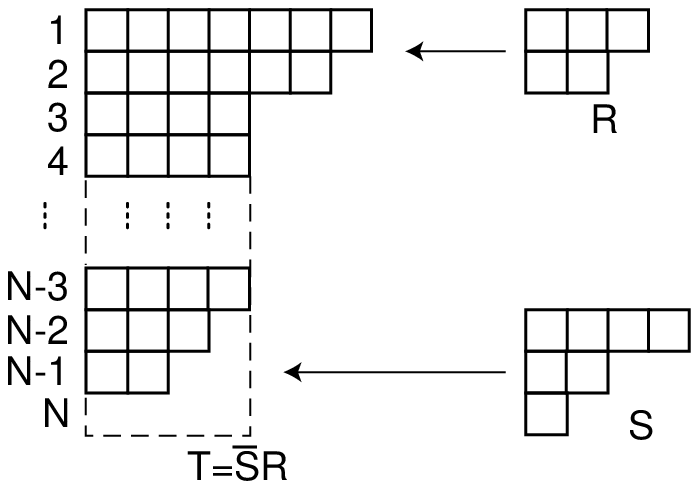}$$
\noindent{\eightpoint Fig. 5: A composite representation.}
\vskip.3truecm
\penalty-2000
\noindent where $S,R$ are two given representations, $\overline S$ is the
adjoint of $S$. As it turns out, the quadratic Casimir (almost) factorizes;
consider the product representation $T=\overline S\,R$, for which the
quadratic Casimir is
$$
C_2(T) = C_2 (R) + C_2(S) + {2n_Rn_S\over N}\quad .\eqno(3.52)
$$
For the dimension (at large $N$), we have
$$
{\rm dim \, }T =({\rm dim \, }R) \, ({\rm dim \, }S)\,(1+ {\cal O}(N^{-1}))
\quad .\eqno(3.53)
$$

Therefore, at large $N$, the total partition function factorizes into two
chiral contributions, as defined by (3.50), with a coupling term $\E^{-{\alpha
A\over N^2}n_R n_S}$. In fact, the problem of taking into account all
representations is very delicate. (For the sphere, Douglas and
Kazakov\ref{82} showed that for $\alpha A < \pi^2$ there must be further
contributions due to the phase transition at $\alpha A=\pi^2$.) From such
counting of ``composite" representations, we obtain
$$
Z(G, \alpha A, N) = \sum _{n_R}\sum _{n_S}\sum _{R \in Y_{n_R}}\sum _{S\in
Y_{n_S}} ({\rm dim \, }\overline S R)^{-2(G-1)} \E^{-{\alpha A\over N}
\left[ C_2(R) + C_2(S) + 2 {n_R n_S\over N}\right]}\quad ,\eqno(3.54)
$$
allowing the interpretation of the above in terms of two coupled chiral
sectors, one being orientation-preserving and the other orientation-reversing.
We will interpret first the simplest case of a single chiral sector, and then
couple the two sectors. We write the chiral sector as
$$
Z(G, \alpha A, N) = \sum _{g=-\infty}^\infty \sum _n \sum _i \zeta_{g,G}^{n,i}
\, \E^{-{n\alpha A\over 2}}(\alpha A)^i N^{-2(g-1)} \quad,\eqno(3.55)
$$
where
$$
2(g-1) = 2n(G-1) + i \quad \eqno(3.56)
$$
is the Kneser formula (see Appendix B), and
$$
\zeta _{g,G}^{n,i} = \sum _R \left( {n!\over d_R}\right)^{-2(G-1)} {1\over i!}
\left( {\widetilde C(R)\over 2}\right)^i \quad .\eqno(3.57)
$$

Still we are looking at the $U(N)$ theory, since the $SU(N)$ term contains the
contribution $\E^{{\alpha A\over 2}{n^2\over N}}$. Notice that the sum over
the base space genus $g$ goes from $-\infty $ to $\infty$, whose meaning is the
inclusion of
disconnected diagrams. The relation (3.56) is crucial and we present a
pedestrian proof in Appendix B.

The interpretation of $\zeta_{g,G}^{n,i}$ in terms of maps becomes clear when
considering maps with a given winding number $n$, singular at a finite set of
points. These are the branch points, such as those appearing in the maps $w=z^n
$. Actually, this is the most general case, as one finds from the following
result.\ref{83} Consider a non-constant holomorphic mapping between Riemann
surfaces, $f\,\colon {\cal M}\to {\cal N}$. Let $P\in {\cal M}$, and choose
local coordinates $\widetilde z\in {\cal M}$ vanishing at $P$ and $w\in
{\cal N}$ vanishing at $f(P)$. Thus we have $w=f(\widetilde z)=\sum_{k\ge n}
a_k\widetilde z^k\,,\,n>0$, and $w=[\widetilde zh(\widetilde z)]^n$, where
$h(\widetilde z)$ is holomorphic and
$h(0)\ne 0$, which is equivalent to $w=z^n$. The value of $n$ is the
ramification number, or equivalently $(n-1)$ is the branch number of $f$ at
$P$. The theorem of Riemann--Hurwitz states that\ref{83}
$$
2(g-1) = 2n(G-1) + B \quad ,\eqno(3.58)
$$
where $B = \sum _i (n_i -1)$ is the total branch number. Gross and Taylor
proved that for $2(g-1) = 2n(G-1) + i$, the coefficient $(\zeta_{g,G}^{n,i})$
is given in terms of the set $\sum(G,n,i)$ of $n$-fold covers of ${\cal M}_G$
with $i$ branch points, and $\nu\in\sum(G,n,i)$, $\nu\,\colon {\cal M}_g\to
{\cal M}_G$. To every cover there is a symmetry factor $S_\nu$, which is the
number of distinct homomorphisms from ${\cal M}_g$ to itself leaving $\nu$
invariant. The above-mentioned relation is
$$
i! \zeta _{g,G}^{n,i} = \sum _{\nu \in \sum (G, n, i)}{1\over \vert S_\nu\vert}
\quad .\eqno(3.59)
$$

It is important to note that there is a very natural interpretation of the
numerical coefficient in (3.55) in terms of maps. In the case of the torus
this is the number of partitions of $(n)$, and we are led to the result
$$
Z = \sum_n p(n) \E^{-{n\alpha A\over 2}} = \eta \left( \E^{-{\alpha A\over 2}}
\right)\quad .\eqno(3.60)
$$
In that case the geometrical interpretation is quite direct.

The last piece deserving interpretation concerns the remaining part of the
Casimir, and the coupling of the two chiral sectors. It is given by (the
exponential of)
$$
{\alpha A \over 2 N^2} n^2 = {\alpha A\over 2 N^2} n + {\alpha A\over 2 N^2}
n (n-1)\quad .\eqno(3.61)
$$

Such a partition is useful, because each term has a definite interpretation;
the first one in terms of handles in ${\cal M}_g$ mapped entirely onto a
single point
in target space, and which is in one of the $n$ sheets of the cover, whose
position has to be integrated, giving a factor of the area. Moreover each
handle in string perturbation theory comes with a factor $1/N^2$ since the
genus is increased by 1. Since the handles are infinitesimal their positions
are the only moduli. Finally, the factor $1/2$ accounts for the
indistinguishability of the ends. Shrinking the length of the handle, the
points coalesce; nevertheless the factor $1/2$ remains. This contribution to
the free energy accounts for the first factor above, $\E^{{\alpha A\over 2N^2}
n}$. Moreover, if we have $n_h$ handles, there is a symmetry factor ${1\over
n_h!}$ leading to exponentiation of such a term. For the second term in eq.
(3.61), we have the interpretation of pinched tubes. Since the tubes now
connect two different sheets, we have a factor ${1\over 2}n(n-1)$ arising from
counting the number of them.

The whole previous interpretation in terms of maps preserves orientation and
is consistent with the chiral component partition. There remains to understand
the coupling between the two chiral components, namely $\E^{-{\alpha A\over
N^2}
n_R n_S}$. A coupling between the chiral components of the form $\left({\alpha
 A\over N^2} n\widetilde n\right)$ can be interpreted as a gluing of the two
surfaces by removing two disks and connecting them by an orientation-preserving
cylinder. An example of such a cylinder is\ref{71} $C=S^1\times [0,1]=\{(z,x),
\,\vert z\vert=1,\,a<x\le 1\}$ such that the map
$$
\nu(z,x)= \cases{z(1-2x) & ${\rm for}\, x\le 1/2\quad $\cr \overline z(2x-1)&
${\rm for}\,x\ge 1/2\quad $\cr}\eqno(3.62)
$$
is orientation-preserving, while
$$
\nu(z,x) = z(1-2x)\quad \eqno(3.63)
$$
is orientation-reversing.

The factor of the area accounts for the arbitrariness in the location of the
tube. As before, the genus increases by a unit, and there is a factor $-1$ for
each. The symmetry factors lead to exponentiation.

One may now ask which string theory is being described by such maps. Although
it is not possible to get a full account of the result, some conclusion may be
drawn. As anticipated, it must contain a $W_\infty$ symmetry. The free energy
is given as an expansion in $\E^{-{\alpha\over 2}A}$, or more precisely by
terms $\E^{-{n\alpha\over 2}A}$; we can thus interpret this as an expansion
in terms of the exponential of an action proportional to the area. It is not
quite the Nambu action. Indeed, the result $nA$ in the exponent
signals maps wrapping $n$ times over the target space, but unlike the Nambu
action there are no folds: the area of the maps $\xi\to x$ is $\int{\rm d}^2x\,
\det{\partial x^\mu\over\partial\xi^\alpha}$, while the Nambu action is $\int{
\rm d}^2x\,\left\vert\det\,{\partial x^\mu\over\partial\xi^\alpha}\right\vert$.

Therefore, if we write an ansatz beginning with the Nambu action, we are
obliged to have terms suppressing folds. This forbids, in particular, maps with
zero winding number. This is a fact in accordance with pure QCD$_2$, which
contains no particle, while terms with zero winding would describe particles.
However, at the moment there is not much more to be said about such a
formulation.

There are ways of rewriting the expansion of the partition function in terms
of known group theoretic factors, which will be convenient to arrive at a
further interpretation and at the possible Lagrangian formulation. First, the
Frobenius formula relates the observable
$$
\Upsilon_\sigma[U]=\prod_{j=1}^s \tr U^{n_j}\quad ,\eqno(3.64)
$$
where $\sigma$ is an element of the permutation group with cycles $n_1\cdots n_
s$, to the characters, since the above functions also build a complete set. We
have
$$
\eqalignno{
\chi_R[U]&=\sum_{\sigma\in S_n}{\chi_R[\sigma]\over n!}\Upsilon_\sigma [U]\quad
,&(3.65a)\cr
\Upsilon_\sigma[U]&=\sum_{R\in Y_n}{\chi_R[\sigma]}\chi_R[U]\quad ,&(3.65b)\cr}
$$
where $S_n$ is the permutation group of $n$ elements and $Y_n$ is the Young
tableau of dimension $n$. In particular, for $U=1$
$$
{\rm dim}\, R = \sum _{\sigma \in S_n} N^s \chi_R[\sigma]\quad ,\eqno(3.66)
$$
holds true, where $s$ is the number of cycles.

The partition function of the chiral contribution may be expanded as
$$
\eqalignno{
Z[N,\alpha A,G]= & \sum_{n=0}^\infty\sum_{R\in Y_n}\sum_{i,t,h}\E^{-n\alpha A}
{(\alpha A)^{i+t+h}(-)^i n^h[n(n-1)]^{t+i}\over i!t!h!(n!)^{2-2G} 2^{i+t+h}
d_R^i}\cr
&\times \{\chi_R[T_2]\}^i N^{n(2-2G)-i-2(t+h)}\left(\sum_{\sigma \in S_n}
{\chi_R(\sigma)\over N^{n-s}}\right)^{2-2G} \quad ,&(3.67)\cr}
$$
where $i$ is the number of branch points, $t$ is the number of
orientation-preserving tubes, and $h$ is the number of handles mapped to
points. Moreover the Casimir has been computed in terms of the character
$\chi_R[T_2]$, of the element $T_2$ containing a single cycle of length 2 and
$(n-2)$ cycles of length 1 as
$$
\widetilde C[R] = {n(n-1)\chi_R[T_2]\over d_R}\quad .\eqno(3.68)
$$

The single cycle of length $2$ is a building block and will be used in the
collective field interpretation of the theory. Some rearrangement is still
required to achieve the final appropriate formulation. Frobenius relation for
unit matrix gives the expression for the dimension, leading to
$$
{\rm dim \,}R={1\over n!}\sum_{\sigma\in S_n}N^{S_\sigma}\chi_R[\sigma]={N^n
\over n!}\chi_R\left(\sum_{\sigma\in S_n}N^{S_\sigma -n}\sigma\right)\quad,
\eqno(3.69)
$$
where $S_\sigma$ is the number of cycles in the permutation $\sigma$. The
leading term for large $N$ is simply ${N^n\over n!}d_R$, where $d_R$ is the
dimension of the representation of the permutation group. We define the group
element
$$
\Omega_n = \sum_{\sigma \in S_n}N^{S_\sigma -n}\sigma\quad ,\eqno(3.70)
$$
which will correspond to extra twists on the covering space. From the
combination formula for characters, eq. (C.13), we get
$$
(\chi_R[\Omega_n])^{-1} = d_R^{-2}\chi_R [\omega_n]\quad ,\eqno(3.71)
$$
from which, for any $l$ (positive or negative) one finds
$$
({\rm dim}\, R)^l = \left({N^nd_R\over n!}\right)^l {\chi_R[\Omega_n^l]\over
d_R}\quad .\eqno(3.72)
$$

Using again eqs. (C.13) and (C.12), we derive
$$
\sum_{\sigma,\varphi\in S_n}d_R^{-1}\chi(\sigma\varphi\sigma^{-1}\varphi^{-1})
={n!\over d_R^2}\sum_\varphi \chi(\varphi)\chi(\varphi^{-1})= \left({n!\over
d_R}\right)^2\quad .\eqno(3.73)
$$

Therefore, the dimension may be written in terms of factors of $N$ and $d_R$
times a character associated with the special operator $\Omega_n$,
$$
({\rm dim}\,R)^{2-2G} = \left({N^nd_R\over n!}\right)^{2-2G}\chi_R(\Omega^{2-
2G}_n) \quad ;\eqno(3.74)
$$
moreover, the counting factor associated with the given Young tableau is
$$
\left({n!\over d_R}\right)^{2G} = \sum{1\over d_R}\prod_1^G\chi_R(\sigma_j
\varphi_j \sigma^{-1}_j\varphi_j^{-1}) \quad .\eqno(3.75)
$$

All characters in the chiral partition function can be combined using eq.
(C.13)
$(G+i+1)$ times, and one arrives at
$$
\eqalignno{
Z^+(G,\alpha A,N)&=\sum_{n,i,t,h}\E^{-n\alpha A/2}{(\alpha A)^{i+t+h}\over
i!t!h!} N^{n(2-2G)-i-2(t+h)}{(-1)^i n^h(n^2-n)^t \over 2^{t+h}} \cr
& \times \!\!\!\!\sum_{p_1,\cdots,p_i\in T_2}\sum_{s_1,t_1,\cdots,s_G,t_G\in S_
n}\!\left[{1\over n!}\delta(p_1\cdots p_i\Omega_n^{2-2G}\prod_{j=1}^Gs_jt_js_j^
{-1}t_j^{-1})\right] , &(3.76)\cr}
$$
where $\delta (\sigma) = {1\over n!}\sum_R d_R \chi_R(\sigma)$.

The terms $\Omega^{2-2G}_n$ in the delta function show extra twists
(permutations of sheets) in the covering at $(2-2G)$ points. For $2-2G >0$,
this
is easier to understand.

Such a result can be generalized along the same lines of reasoning to the full
theory, that is for the coupled chiral and antichiral composite partition
function. Details concerning such a derivation are too long to be reported
here, and we refer to [72] and [73]. The final result reads
$$
\eqalignno{
Z&(G,\alpha A,N) \cr
& \sim \sum_{n^\pm, i^\pm=0}^\infty \, \, \sum_{p_1^\pm ,\cdots ,p^\pm_{i^\pm}
\in T_2\i S_{n^\pm}} \,\, \sum_{s_1^\pm,t_1^\pm, \cdots , s_G^\pm, t_G^\pm \in
S_{n^\pm}} \left({1\over N}\right)^{(n^++n^-)(2G-2)+(i^++i^-)}\cr
& \times {(-)^{(i^++i^-)}\over i^+!i^-!n^+!n^-!}(\alpha A)^{(i^++i^-)}\E^{-{1
\over 2}((n^+)^2 + (n^-)^2 - 2n^+n^-)\alpha A/N^2}\cr
&\times \delta_{S_{n^+}\times S_{n^-}}\left( p_1^+ \cdots p^+_{i^+}p_1^- \cdots
p_{i^-}^- \Omega_{n^+,n^-}^{2-2G} \prod_{j=1}^G[s_j^+,t_j^+] \prod_{k=1}^G
[s_k^-,t_k^-]\right) \quad ,&(3.77)\cr}
$$
where $[s,t] = sts^{-1}t^{-1}$. Here $\delta$ is the delta function on the
group algebra of the product of symmetric groups $S_{n^+}\times S_{n^-}$, $T_2$
is the class of elements of $S_{n^\pm}$ consisting of transpositions, and
$\Omega^{-1}_{n^+,n^-}$ are certain elements of the group algebra of the
symmetric group $S_{n^+}\times S_{n^-}$ with coefficients in I$\!\!\!$R$(1/N)
$. Let us remain with the chiral component, eq. (3.76). The theory defined
thereafter is related to a topological field theory, as shown in ref. [73]. In
order to see this, one first considers the limit of the partition function
(3.76) for vanishing area, which is still non-trivial. One considers the
homotopy group of a punctured surface ($L$ punctures), and the homomorphism of
such a group into $S_n$. One also defines the Hurwitz space of branched
coverings: consider $H(n,B,G,L)$, the set of equivalence classes of the
manifold $\Sigma_t$\newfoot{$^{7)}$}{Recall that one considers here the maps
from world sheet to the target space: $\Sigma_{ws}\to\sigma_t$.} with degree
$n$, branching number $B$; $S$ is a set of points on $\Sigma_t$; $H(n,B,G,L)$
is the equivalence class of branched coverings of $\Sigma_t$.

The chiral amplitude at zero area
$$
Z(G,0,N)=\sum_{n=0}^\infty N^{n(2-2G)}\sum_{s_1\varphi_1\cdots s_G\varphi_G\in
S_n}{1\over n!}\delta\left(\Omega_n^{2-2G}\prod_1^G s_j\varphi_js_j^{-1}\varphi
_j^{-1}\right)\quad \eqno(3.78)
$$
has been shown\ref{73} to be given by the simpler expression
$$
Z(G,0,N) = \sum_{n=0}^\infty\sum_{B=0}^\infty \sum_{L=0}^B \left({1\over N}
\right)^{2G-2}d(2-2G,L)\sum_{f\in H(n,B,G,s)}{1\over \vert {\rm Aut}\,f\vert}
\quad ,\eqno(3.79)
$$
where $\vert{\rm Aut}\, f\vert$ is the order of the automorphism group of the
branched covering map $f$, and
$$
d(2-2G=\chi_G,L) = {\chi_G!\over (\chi_G-L)!L!} \quad \eqno(3.80)
$$
is an Euler character (see ref. [73] for details); moreover one finds for the
last factor in (3.79) the result
$$
d(2-2G,L)\sum{1\over\vert{\rm Aut}\,f\vert}=\chi(H(n,B,G,L))\quad.\eqno(3.81)
$$

This leads to a topological theory described by (3.76). To fully demonstrate
such results a rather heavy mathematical instrumentation is needed, which
goes far beyond the scope of the present review. It is useful to quickly state
some results. The topological theory describing the above partition function
is the topological gravity, with a topological sigma model, and a further
topological term, the so-called co-$\sigma$ sector. The area is restored by
perturbing the topological action with the Nambu action, which does reappear in
this context as a perturbation of the topological theory. The full theory can
also be discussed along these lines.

The area in the above formulation is always multiplied by the charge, for
dimensional reasons. The limit $e\to 0$ corresponds to a topological theory
(see
discussion after eq. (2.108$c$)).
\vskip 1cm
\penalty-1000
\noindent {\bf 3.4 Collective coordinates approach}
\vskip .5cm
\nobreak
\noindent Further information and insight in two-dimensional string theory can
be obtained by the method of collective coordinates, which are useful to
characterize the properties of string theory as a whole. Indeed, non-critical
strings have very rich and detailed descriptions by means of either the
two-dimensional Liouville theory or matrix models (see ref. [30] and
references therein). In the Liouville approach, the $d$-dimensional string
corresponds effectively to a $(d+1)$-dimensional theory, since the Liouville
field itself plays the role of the extra coordinate. Therefore a
two-dimensional string is described by a $c=1$ model. In the matrix-model
approach, this is equivalent to considering the dynamics of a Hermitian matrix
$M(t)$, depending on a single coordinate, with Lagrange density
$$
{\cal L} = \tr {1\over 2} \dot M ^2 - \tr \, V(M)\quad ,\eqno(3.82)
$$
where the kinetic term is actually a simplification of the exponential
propagator,\ref{84} and the potential $V(M)$ is an arbitrary function of the
matrix $M(t)$. The current
$$
J = i [M, \dot M]\quad \eqno(3.83)
$$
is conserved. Diagonalizing the matrix $ M = {\rm diag\, } \lambda (t)$, the
eigenvalues describe a system of free fermions. The analogue of the
Wilson-loop average, the trace of the exponential of $M$, is computable as a
simple sum:
$$
\widetilde\phi_k(t)=\tr\,\E^{ikM}=\sum_{j=1}^N\E^{ik\lambda_j(t)}\quad ,
\eqno(3.84)
$$
whose Fourier transform $\phi (x,t)$ can be interpreted as a density of
fermions. We can study the theory in terms of the two-dimensional scalar field
$\phi(x,t)$. Therefore, the model defined by eq. (3.82) will be replaced by an
effective two-dimensional field theory. Such a description has several
advantages; in particular it provides a global description of the string.

As it stands, the problem of computing wave functionals is very complicated.
Jevicki and Sakita\ref{85} handled a similar problem by introducing the change
of variables (3.84), in such a way that a Schr\"odinger wave equation in terms
of the $\phi(x,t)$ variables is simpler. After the transformation (3.84) one
finds that the Laplacian operator appearing in the Hamiltonian corresponding
to eq. (3.82) is given by
$$
{\partial^2 \over \partial M^2} = {\partial^2 \phi \over \partial M^2}
{\partial \over \partial \phi} + \left( {\partial \phi\over \partial M}
\right)^2  {\partial^2 \over \partial \phi^2} = -k^2 \phi_k {\partial \over
\partial \phi_k} - kk'\phi_{k+k'}{\partial \over \partial \phi_k} {\partial
\over \partial \phi_{k'}}\, .\eqno(3.85)
$$

Therefore, it is possible to find a Hamiltonian in terms of the one-component
field $\phi(x,t)$ and its conjugate $\Pi(x,t)\sim -i{\partial \over \partial
\phi(x,t)}$; it reads
$$
H = \int {\rm d}x\, \left\{ {1\over 2} \partial _x \Pi \phi \partial _x \Pi +
{1\over 6} \Pi^2 \phi^3 + V(x)\phi \right\}\quad .\eqno(3.86)
$$

The Hamiltonian shows a cubic interaction and a tadpole term, indicating a
string-type interaction and annihilation into the vacuum. Notice the fact that
$\phi$ is a two-dimensional field; therefore one is describing strings in a
two-dimensional target space, as already observed in the description of the
relation between Liouville and matrix models. One can also introduce chiral
components
$$
\alpha_\pm (x,t) = \partial_x \Pi \pm \pi \phi (x,t)\quad ,\eqno(3.87)
$$
with Poisson brackets
$$
\{ \alpha_\pm (x), \alpha_\pm (y) \} = \pm 2\pi \delta'(x-y)\quad ,\eqno(3.88)
$$
in terms of which the Hamiltonian (3.86) turns into
$$
H_{coll} = \int {{\rm d}k\over 2\pi} \left\{ {1\over 6} (\alpha_+^3-\alpha_-^3)
 + \left( V(x)+{1\over 2}\mu\right)(\alpha_+-\alpha_-)\right\}\quad
,\eqno(3.89)
$$
where $\mu$ is a constant indicating the energy level. The simplifying case of
a harmonic oscilator potential $V(x) = - {1\over 2} x^2 $ has been largely
studied.\ref{84} An infinite number of conservation laws is found for such a
model.

It is not difficult to derive a Das--Jevicki-type\ref{86}  Hamiltonian
describing the string interaction directly from the $SU(N)$ representation in
terms of a sum of representations in the case of QCD$_2$. Consider the
observables defined by
$$
\Upsilon_\sigma (U) = \prod_{j=1}^s\tr U^{n_j}\quad ,\eqno(3.90)
$$
in terms of the group variable $U$, where we considered as usual the
partition $n=\sum_{i=1}^s n_i$.

Such objects satisfy the basic assumption that as one goes around a Wilson
loop one induces a permutation of the sheets covering the loop. Therefore, to
any Wilson loop can be associated an element $\sigma$ of the permutation group
$S_n$, where $n$ is the number of sheets of the world sheet covering the
Wilson loop. Consider a manifold with special points, and loops that possibly
go around such points; as one encircles any (homotopically non-trivial) loop,
the $n$ sheets on the basis space loop interchange, defining an element in the
permutation group.

Therefore it is natural to assign an element in $S_n$ to each string state.
Moreover, the states are orthogonal if the elements characterizing the two
given
states are not in the same conjugacy class. The normalization of the
observables (3.90) is given in terms of the characters as
$$
\eqalignno{
\int {\rm d}U\, \Upsilon_\sigma (U)\Upsilon_\tau(U^\dagger) = & \int {\rm d}U\,
\sum_{R,R'}\chi_R(\sigma)\chi_{R'}(\tau) \chi_R(U)\chi_{R'}(U^\dagger)\cr
= & \sum_R \chi_R(\sigma)\chi_{R}(\tau) = \delta_{T_\sigma,T_\tau}{n!\over
C_\sigma}\quad ,&(3.91)\cr}
$$
where $C_\sigma$ is the number of elements in the conjugacy class $T_\sigma$;
we can define the scalar product
$$
\langle s' \vert s\rangle = \sum_{t \in S_n} \delta_{s',tst'}\, \,=\,\cases {
{n!\over C_\sigma}&${\rm if\,} s\sim s'\quad ,$ \cr 0 & ${\rm otherwise}\quad
, $\cr }\eqno(3.92)
$$
where $s\sim s'$ means that the elements are in the same conjugacy class, and
the number of elements in such a conjugacy class is $C_s$. As argued in ref.
[87] a state in $S_n$ is typically
$$
\vert S\rangle=\prod_{l=1}^n(a_l^\dagger)^{n_l}\vert 0\rangle\quad ,\eqno(3.93)
$$
where $\vert 0 \rangle$ is the vacuum, and $a_l^\dagger$ is the creation of a
string with wind $l$, satisfying
$$
[a_l, a_m^\dagger] = \vert l\vert \delta_{l,m}\quad ,\eqno(3.94)
$$
which leads to eq. (3.92) for the scalar product of two elements of the type
(3.93). Since the ``energy" is ${1\over 2}e^2Ln$, the free part of the
Hamiltonian  is
$$
H_0 = {1\over 2}e^2L\sum a_l^+a_l \quad .\eqno(3.95)
$$

The first interaction describes the joining of two strings; it should be
represented by a sum over conjugacy classes, weighted by the factor ${e^2L
\over 2N}$; it must be described by the expectation of an operator belonging to
the conjugacy class of $S_n$ with one two-cycle, namely $T_2$, and the rest
are one-cycles, as
$$
H_{int}^{(3)} = {e^2 L\over 2N} \sum _{p \in S_{\tilde n}} \langle s' \vert p
\vert s \rangle = {e^2 L\over 2N} \sum_{ {\mathop {t\in S_n} \limits_ { p\in
S_{\tilde n}} } } \delta_{s'p,tst^{-1}}\quad , \eqno(3.96)
$$
where $S_{\tilde n}\subset S_n$ is in the conjugacy class with one two-cycle,
namely the elementary permutation; the remaining elementary strands are
one-cycles.  Therefore, $s$ has cycles $n_1$ and $n_2$, while $s'$ contains a
cycle
$n_1 + n_2$, and $s'p=tst^{-1}$. A simple counting reveals that conjugation
leads to  $n_1 n_2$ elements, while one has $n_1+n_2$ distinct conjugates,
which is correctly described by the operator
$$
H^{(3)} = {e^2L\over 2N} \sum _{{\mathop {n,n'>0}\limits_{n,n'<0}}}
(a_{n+n'}^\dagger a_n a_{n'} + {\rm c.c.})\quad . \eqno(3.97)
$$

The remaining interaction (quartic)  is such that either two strings of the
same chirality disappear and are subsequently generated or the opposite
chirality strings interact, but the sign is opposite with respect to the
previous (equal chirality) possibility. This is described by the interacting
Hamiltonian
$$
H^{(4)} = {e^2L\over 2N^2}\left[ \sum_{n>0} (a^\dagger_n a_n - a_{-n}^\dagger
a_{-n}) \right]^2 \quad ,\eqno(3.98)
$$
completing the total Hamiltonian. One can also show that relating $a_k$ with
the field $\varphi (x)$ and its momentum canonically conjugated $\pi(x)$ by
$$
a_k = {1\over 2} \int {\rm d}x\, \E^{-ikx}\left[ \varphi (x) + {1\over \pi}
\epsilon (k) \partial \Pi(x)\right]\quad,\eqno(3.99)
$$
one finds a slightly modified Das--Jevicki\ref{86} Hamiltonian
$$
H = {4\over e^2LN}\int {\rm d}x\, \left\{ {1\over 2} \partial \Pi \varphi
\partial \Pi + {1\over 6} \pi^2 \varphi^3 - \left( {e^2LN\over 4}\right)^2
\varphi \right\} + \Delta H\quad ,\eqno(3.100)
$$
with the constraint
$$
\int {\rm d}x\, \varphi (x) = N\quad ,\eqno(3.101)
$$
which takes the zero-mode problem into account, and $\Delta H$ is the quantum
correction to the free energy. Since one arrives at a $c=1$ matrix model, it
is natural to ask what is the relation with the fermion picture of the latter.
As a matter of fact, one can show that there is a simple description of pure
QCD$_2$ on a circle of radius $L$ in terms of free fermions.

In the gauge $A_0=0$, the pure QCD$_2$ Hamiltonian is just the square of the
electric field
$$
H= {1\over 2} \int _0^L {\rm d}x\, \tr F_{01}^2  = {1\over 2} \int _0^L
{\rm d}x\, \tr \dot A_1^2 \quad .\eqno(3.102)
$$

One can define the gauge-invariant quantity\ref{88}
$$
V(x) = W[0,x] \dot A_1 (x)W[x,L]\quad ,\eqno(3.103)
$$
with the Wilson line given by
$$
W[a,b] = P \E^{ie\int _a^b {\rm d}x\, A_1}\quad ,\eqno(3.104)
$$
in such a way that the Gauss law
$$
\nabla_1F_{10} = \partial _1\dot A_1 + ie [A_1, \dot A_1] = 0\quad \eqno(3.105)
$$
boils down to
$$
\partial _1 V = 0\quad .\eqno(3.106)
$$

Therefore we find $V(0)=V(L)$, implying, (from $V(x) = W[0,x] \dot A_1 W[x,L]$)
that $\dot A_1$ commutes with $W=W[0,L]$. The time derivative of $W$ may also
be
computed by means of the well-known formula
$$
\dot W = ie \int _0^L {\rm d}x\, W[0,x] \dot A_1 (x)W[x,L]=ie \int {\rm d}x\,
V(x)\quad ,\eqno(3.107)
$$
from which we can prove that $W^{^{-1}}$ commutes with $\dot W$; from the
constancy of $V(x)$, we can also compute $\dot A_1(x)\!=\!
W[x,0]\dot A_1(0) W[0,x]$, finding $\dot A_1(x)\!=\! {1\over ieL} W[0,x]\dot W
W^{^{-1}}W[x,0]$, which permits us to rewrite the Hamiltonian as
$$
H={1\over 2}\int_0^L{\rm d}x\,\tr\left(W[x,0]\dot A_1(0)W[0,x]\right)^2=-{1
\over 2e^2L}\tr\left(W^{-1}\dot W\right)^2\quad.\eqno(3.108)
$$

This Hamiltonian describes a one-dimensional unitary matrix model. Since $W$
and $\dot W$ commute, the problem is reduced to the consideration of
eigenvalues of $W$.

Therefore we arrive at a system of $N$ fermions on a circle with
Hamiltonian\ref{88,89}
$$
H = -\left( {e^2L\over 2}\right) \sum {\partial ^2\over \partial \theta_i^2}
\quad .\eqno(3.109)
$$
The reduction of QCD$_2$ with matter to a simple dynamical system will be
considered later. Moreover the infinite symmetry of the pure QCD$_2$ theory
mirrors itself here in the fact that the relativistic fermions display also a
$W_\infty$ symmetry, corresponding to area-preserving diffeomorphisms of the
Fermi sea.\ref{35, 36}
\vskip 1cm
\penalty-1000
\noindent {\bf 3.5 Phase structure of QCD$_2$}
\vskip .5cm
\nobreak
\noindent We have seen that QCD$_2$ may accommodate different phases, although
we have not seen such a structure yet. The phase structure of the model has
not been studied in the case where matter fields are present. However, in the
large-$N$  limit it is possible to prove that depending on the value of the
``fine structure" constant, $\alpha = e^2 N$, the theory shows a different
behaviour.

The first observation came long ago, when Gross and Witten\ref{78} obtained a
possible third-order phase transition for the large-$N$  limit of
lattice QCD$_2$. The argument relies on the large-$N$  limit as employed by
Brezin, Itzykson, Parisi and Zuber.\ref{90} The leading term can be computed.
One starts with the lattice formulation, where the Wilson action reads
$$
S[U] = \sum _P {1\over 2e^2a^2}\tr\left(\prod_PU + {\rm h. \, c.}\right)\quad ,
\eqno(3.110)
$$
and a single plaquette action is the product
$$
\prod_P U = U_{\vec n, \hat x_0} U_{\vec n + \hat x_0, \hat x_1} U_{\vec n +
\hat x_0 + \hat x_1, \hat x_0} U_{\vec n + \hat x_1, -\hat x_1}\quad
,\eqno(3.111)
$$
that is one starts out of the point $\vec n$, in the direction $\hat x_0$,
returning back at the end of each round. Gross and Witten considered such a
problem, used the invariance of the Haar measure, and gauge invariance,
changing
$U_{\vec n, \hat x} \to V_{\vec n}U_{\vec n,\hat x}V^\dagger_{\vec n+\hat x}$,
choosing the gauge $U_{\vec n,\hat x_0} =1$ for all $\vec n$, corresponding
to $A_0=0$, after which the  action reads
$$
S[U] = {1\over 2e^2a^2}\sum \tr \left( U_{\vec n,\hat x_1}
U^\dagger _{\vec n + \hat x_0,\hat x_1} + {\rm h. \, c.}\right)\quad
.\eqno(3.112)
$$

After such a procedure, one changes the variables to $W_{\vec n}$, defined as
$$
U_{\vec n + \hat x_0,\hat x_1}=W_{\vec n} U_{\vec n, \hat x_1}\quad ,
\eqno(3.113)
$$
which leads to a partition function that is the product of partition functions
for each site, i.e.
$$
Z = \int \prod_n {\rm d}W_n\,\E^{\sum_n {1\over 2e^2a^2} \tr (W_{\vec n} + W^
\dagger_{\vec n})} = z^{V/a^2}\quad ,\eqno(3.114)
$$
where $z$ is the one-site partition function and $V$ the total volume. Each
integral is computable using the results of the appendix of ref. [91] and one
obtains
$$
\eqalignno{
Z(e^2,N) = \, &  \det M\quad , & (3.115)\cr
M_{i,j}  = \, & I_{i-j} (1/e^2a^2)\quad , & (3.116)\cr}
$$
where $I_i(x)$ is the Bessel function of order $i$. In terms of the
eigenvalues of $W$, one can also write $W=TDT^\dagger$, where the diagonal
matrix $D_{ij}=\delta_{ij}\E^{i\theta_j}$; since the angular piece is
directly integrated, one is left with
$$
{\rm d}W \sim \prod_1^N \Delta^2(\theta_i)\, {\rm d}\theta_i\quad ,\eqno(3.117)
$$
where the Jacobian is a Vandermonde determinant
$$
\Delta^2(\theta_i) = \prod {\rm sin }^2 {\theta_i-\theta_j\over 2} =
4^{-N}\vert \det \Delta \vert ^2 \quad ,\eqno(3.118)
$$
where $\Delta_{j,k} = \E^{ij\theta_k}$. The (one-site) partition function is
$$
Z(e^2, N) = \int _0^{2\pi} \prod {\rm d}\theta_i \Delta^2(\theta_i)\, \E^{
{1\over e^2a^2}\sum_i{\rm cos}\, \theta_i}\quad ,\eqno(3.119)
$$
and the energy $E=-{1\over N^2}\ln Z(e^2,N)$ can be computed in the limit $N\to
\infty\,(\alpha=e^2N)$ by the steepest-descent method, where the eigenvalues
are given by the stationary condition
$$
{2\over\alpha a^2}\,{\rm sin\,}\theta_i=\sum _{j\ne i}{\rm cot\,}\left\vert{
\theta_i-\theta_j\over 2} \right\vert \quad ;\eqno(3.120)
$$
we define a function $\theta(x)$ such that $\theta_i=\theta{i\over N}$ since,
for large $N$, $x_i\sim i/N$ can be seen as a continuous number, and find:
$$
\eqalignno{
E(g) = & - \lim_{N\to \infty}\left\{ {1\over N\alpha a^2} \sum_1^N {\rm cos\, }
\theta_i + {1\over N^2} \sum_{i\ne j} \ln \left\vert {\rm sin\, }{\theta_i
-\theta_j\over 2}\right\vert\right\}\cr
&={1\over\alpha a^2}\!\!\int_0^1\!\!{\rm d}x\,{\rm cos\,}\theta(x)+P\!\int_0^1
\!\!{\rm d}x\!\int_0^1\!\!{\rm d}y\,\ln\left\vert{\rm sin\,}{\theta(x)-\theta
(y)\over 2}\right\vert + {\rm constant}\, ,&(3.121)\cr}
$$
while
$$
{1\over\alpha a^2}{\rm sin\,}\theta(x)=P\int_0^1{\rm d}y\,{\rm cot\,}{\theta(x)
- \theta(y)\over 2}\quad .\eqno(3.122)
$$

At this point we introduce the density of eigenvalues
$$
\rho(\theta) = {{\rm d}x\over {\rm d}\theta}\ge 0\quad .\eqno(3.123)
$$

There are two regions, depending now on the value of $\alpha a^2$. For strong
coupling one expects an eigenvalue distribution over the whole circle $[-\pi,
\pi]$, and eq. (3.122) is solved using
$$
\eqalignno{
\int_0^1{\rm d}y\, {\rm cos\, }{\theta(x)- \theta(y)\over 2} = \, & \int_{-\pi}
^\pi {\rm d}\beta \,\rho(\beta)\, {\rm cot\, }{\theta-\beta\over 2}\cr
= \, & 2 \sum_1^\infty \int {\rm d}\beta\, \rho(\beta)\,({\rm sin\,}n\theta \,
{\rm cos\,}n\beta - {\rm cos\,}n\theta\, {\rm sin\,}n\beta)\quad ,&(3.124)\cr}
$$
which, together with the normalization condition $\int_{-\pi}^\pi{\rm d}\theta
\, \rho(\theta)=1$, fixes the density of eigenvalues
$$
\rho(\theta)= {1\over 2\pi}\left[ 1+ {1\over \alpha a^2}\,{\rm cos\, }\theta
\right]\quad ,\eqno(3.125)
$$
which is positive for $\alpha a^2\ge 2$. For $\alpha a^2\le 2$, the
integration over the ``angle" $\beta$ must be performed in a region
$[-\theta_c,\theta_c]$. In ref. [78] the function
$$
F(\zeta) = \int_{-\theta_c}^{\theta_c}{\rm d}\beta\, \rho(\beta)\, {\rm cot\,}
{\zeta - \beta\over 2}\quad ,\eqno(3.126)
$$
has been computed, and the eigenvalues distribution is found to be
$$
\rho(\theta)={1\over\pi\alpha a^2}{\rm cos\,}{\theta\over 2}\left( {\alpha a^2
\over 2} - {\rm sin}^2{\theta\over 2}\right)^{1/2}\quad,\eqno(3.127)
$$
showing that there is a phase transition at $\alpha a^2=2$, a point where the
expressions (3.125) and (3.127) coincide.

The origin of the phase transition is the fact that the functional integral,
in the strong coupling, has most contributions from the Vandermonde
determinant,
which shows a non-relativistic fermion character, and the fields interact
repulsively due to Pauli's principle. Therefore the density distribution is
almost constant, $\rho(\theta)\sim {1\over 2\pi}$. In the weak coupling limit,
on the other hand, the Wilson action becomes important, and the interaction is
attractive, therefore $\rho(\theta)$ is given by a semi-circle law:
$$
\rho(\theta)={1\over\pi}\sqrt{{1\over\alpha a^2}}\left(1-{\theta^2\over 4\alpha
a^2}\right)^{1/2}\quad ,\quad \vert \theta\vert ^2 \le (2g)^2 \quad
.\eqno(3.128)
$$
\vskip 1cm
\penalty-3000
\noindent {\bf 4. Generalized QCD$_2$ and adjoint-matter coupling}
\vskip .5cm
\penalty-1000
\noindent {\bf 4.1  Introduction and motivation }
\vskip .5cm
\nobreak
\noindent The study of matrix models relevant for four-dimensional QCD is
spoiled by the existence of the ``barrier" at the value of the conformal
central charge $c=1$, as is clear from the expression of the dressed
conformal dimension (see ref. [30] and references therein). The case $c=1$
describes effectively two-dimensional string theory, since the time dimension
is described by the Liouville field. Thus $c>1$ describes, in an analogous way,
$D>2$ string theory. Such theories are in fact much richer than the
two-dimensional counterpart due to the role played by the transverse
oscillators. As it turns out, the most interesting case is also the
most difficult sometimes: nature seems to hide itself in folds unreachable to
perscrutation by the available means.

A matrix model in two dimensions $(c=2)$ can be defined by means of the
Lagrangian
$$
L = \tr \left[ {1\over 2} \partial ^\mu M\partial _\mu M + {1\over 2} \mu M^2 -
{\lambda \over 3!\sqrt N}M^3\right]\quad ,\eqno(4.1)
$$
where $M$ is an $N \times N$ matrix field, and $\lambda, \mu$ arbitrary
parameters. One should look for the singular limit in terms of the parameter
$\lambda/\mu$. Such a critical behaviour is difficult to obtain. As it stands,
the model cannot correctly describe string effects. This assertion derives from
the fact that the Fock space constructed out of the momentum space components
of the $M$ field, namely\ref{92}
$$
{\cal F} = \sum a^\dagger_{i_1j_1}(k_1)\cdots a^\dagger_{i_nj_n}(k_n)\vert
0\rangle \quad,\eqno(4.2)
$$
where
$$
M_{ij} (x) = {1\over \sqrt{2\pi}}\int{{\rm d}k^+\over\sqrt{2k^+}}\left[ a_{ij}
(k)\E^{-ik^+x^-} +a^\dagger_{ji}(k)\E^{ik^+x^-}\right]\quad,\eqno(4.3)
$$
contains all multiplets, while closed string states should be singlets under
global $SU(N)$ symmetry, whose action on $M$ is defined as
$$
M \to \Omega^\dagger M \Omega \quad ;\eqno(4.4)
$$
therefore the Hilbert space should be spanned by states of the form
$$
\tr a^\dagger (k_1) \cdots a^\dagger (k_n)\vert 0\rangle \quad .\eqno(4.5)
$$

Moreover, numerical results indicate also that one expects tachyons for purely
bosonic models as above.

One possible cure of these problems would be a gauging of the symmetry,
confining  the non-singlet states, so that the singlets (4.5) span the physical
Hilbert space of the theory. In fact, such a gauging procedure has further
motivations, from the point of view of two-dimensional QCD, which is still
very simple since the gauge field has no degree of freedom, by
na\"{\i}ve counting. One might consider instead the more realistic case of
three-dimensional QCD, and dimensionally reduce the $(2+1)$ dimensions to
$(1+1)$, compactifying one of the spatial dimensions to a vanishingly small
box. In such a case the third component of the gauge field becomes, in $(1+1)$
dimensions, bosonic matter in the adjoint representation.
\vskip 1cm
\penalty-1000
\noindent {\bf 4.2  Scalar and fermionic matter coupling; quantization}
\vskip .5cm
\nobreak
\noindent Light-cone quantization of $(1+1)$-dimensional QCD with adjoint
matter
fields in the light-cone gauge has been considered both for fermionic as well
as
bosonic matter.\ref{92,93} We will restrain ourselves here to the fermionic
case.

The procedure will be based on a choice of the light-cone gauge, in such a way
that the action is quadratic in the remaining component of the gauge field,
thus making its integration possible. Proceeding with the light-cone
quantization, we see that one of the fermionic components obeys a constraint
equation (no light-cone time derivative). Such a fact can be used in
performing the
light-cone quantization, choosing one light-cone variable as the ``time" (or
``light-cone time"). It is possible, in the Hamiltonian formalism, to fix the
time, and the theory turns out to be described in terms of simple oscillators.
The resulting Schr\"odinger equation leads to the bound-state structure. Thus
we consider the action
$$
S = \tr \int {\rm d}^2x\, \left[ i\overline \Psi \not \!\! D \Psi + m \overline
\Psi \Psi - {1\over 4}F^{\mu\nu}F_{\mu\nu}\right] \quad ,\eqno(4.6)
$$
and separate the chiral components of the Fermi field by means of the
decomposition
$$
\Psi=\left(\matrix{\psi\cr i\chi\cr}\right)\quad,\quad\chi^\dagger=\chi\quad
,\quad \psi^\dagger = \psi \quad ,\eqno(4.7)
$$
with the covariant derivative given as usual in the adjoint representation
by the expression
$$
D_\mu \psi = \partial _\mu \psi - ie [A_\mu ,\psi]\quad .\eqno(4.8)
$$
One can also choose the light-cone gauge $A_-=0$, in which the ghosts
decouple. We find for the action the expression
$$
S_f= \tr \int {\rm d}x^+\, {\rm d}x^-\, \left[ i\psi \partial _+ \psi + i \chi
\partial _- \chi -2im\chi\psi + {1\over 8}(\partial _-A_+)^2 + J^+A_+\right]
\quad ,\eqno(4.9)
$$
where $J^+_{ij}=\tr \psi_{ik}\psi_{kj}$. If we choose $x^+$ as the time
variable as described above, it is clear from (4.9) that $\chi$ does not obey
any equation of motion in the Hamiltonian sense, but rather an equation of
constraint, namely
$$
\partial _-\chi - 2 m\psi =0\quad ,\eqno(4.10)
$$
since no light-cone time derivative is involved. The Gauss constraint,
equivalent to the equation of motion of $A_-$ is
$$
{1\over 4}\partial _+\partial _- A_+ + J_+ =0\quad ,\eqno(4.11)
$$
and is equivalent to the equation of motion for $A_+$, namely
$$
{1\over 4}\partial^2_-A_+ - J_- =0\quad ,\eqno(4.12)
$$
once one uses current conservation. Moreover the gauge field $A_+$ is quadratic
after gauge fixing, and one may perform the corresponding Gausssian
integration, equivalent to the substitution of its equation of motion back into
the action to obtain the light-cone components of the energy--momentum tensor
$$
\eqalignno{
P^+ = \, & \tr \int {\rm d}x^-\, i\psi\partial _-\psi \quad ,&(4.13a)\cr
P^- = \, & \tr \int {\rm d}x^-\, \left[ - 4 m^2\psi {1\over\partial _-} \psi +
2 J^+{1\over \partial ^2_-}J^+\right]\quad .&(4.13b)\cr}
$$

In the Hamiltonian formalism, working at fixed time, as we also mentioned, it
is convenient to consider the mode expansion of the fermion field as given by
$$
\psi_{ij}(x) = \int {{\rm d}k^+\over 2\sqrt{2\pi}} \, b_{ij}\E^{-ik^+x^-}
\quad .\eqno(4.14)
$$

The canonical procedure implies anticommutation relations for the fermions as
given by
$$
\eqalignno{
\{\psi_{ij}(x^-), \psi_{kl}({x'}^-)\} & =  {1\over 4}\delta(x^--{x'}^-)
\delta_{il}\delta_{jk}\quad,&(4.15a)\cr
\{b_{ij}(k^+), b_{kl}({k'}^+)\} & =  \delta(k^++{k'}^+)\delta_{il}
\delta_{jk}\quad .&(4.15b)\cr}
$$
Therefore $b(k)$ are creation operators for $k\le 0$, while for $k\ge 0$ they
are annihilation operators, and $b(k)_{k\ge 0}\vert 0\rangle=0$. Computation
of the light-cone components of the energy--momentum tensor follows from (4.9)
by the usual procedure, leading to (4.13). One substitutes (4.14), obtaining
first the expressions for the current
$$
\widetilde J _{ij}(k)= \int {\rm d}p\, b_{ik}(p)b_{kj}(k-p)\quad .\eqno(4.16a)
$$
We can compute now the energy momentum tensor components in terms of the
creation and annihilation operators $b_{ij}(k)$ as
$$
\eqalignno{
&P^+ =  \int_0^\infty {\rm d}k\, k b_{ij}(-k) b_{ij}(k)\quad,&(4.16b)\cr
&P^- =  m^2\int_0^\infty {{\rm d}k\over k}\, b_{ij}(-k) b_{ji}(k) - 2e^2N
\int_0^\infty {{\rm d}k\over k}\, b_{ij}(-k) b_{ji}(k)\cr
& + {e^2\over 2\pi}\!\!\int_0^\infty \!\!\!{\rm d}k_1\,{\rm d}k_2\,{\rm d}k_3
\,{\rm d}k_4\,\Big\{\!A \delta(k_1\!+\!k_2\!-\!k_3\!-\!k_4)b_{kj}(-k_3)b_{ji}
(-k_4) b_{il}(k_1)b_{lk}(k_2) &(4.16c)\cr
& +\!B\delta(\!k_1\!+\!k_2\!+\!k_3\!-\!k_4\!)[b_{kj}(\!-k_4\!)b_{ji}(\!k_3\!)
b_{il}(\!k_2\!)b_{lk}(\!k_1\!)\!+\!b_{kj}(\!-k_1\!)b_{ji}(\!-k_2\!)b_{il}
(\!-k_3\!)b_{kl}(\!k_4\!)]\Big\}\, ,\cr}
$$
where the coefficients are given, after eq. (4.13), by
$$
\eqalignno{
A=\, & {-1\over (k_1 - k_4)^2}+{1\over (k_1+k_2)^2}\quad ,&(4.17a)\cr
B=\, & {-1\over (k_2 + k_3)^2}+{1\over (k_1+k_2)^2}\quad .&(4.17b)\cr}
$$

A discretized version of the model was discussed by Dalley and
Klebanov,\ref{92}
who found that for any value of $e/m$ the spectrum is real, without any phase
transition, contrary to a previous analysis of the pure-matrix-model case where
they found numerical evidence of a tachyon. They also found that in the
strong coupling limit $(e\to \infty)$ the masses are pushed to infinity, and
the theory becomes trivial. The conclusion is based upon the fact that for
zero bare-fermion mass, there is still a non-vanishing mass gap in the
spectrum. See also ref. [94] for numerical results.

It is quite remarkable that a wave functional obeying a Schr\"odinger equation
with the Hamiltonian operator given by (4.16$b$) can give calculable
eigenvalues for the high-energy part of the spectrum\newfoot{$^{8)}$}{In what
follows one actually neglects pair creation and annihilation effects. The
results can only be valid for the highly excited part of the
spectrum.\ref{93}}. Such an eigenvalue problem, namely
$$
P^- \psi_B = \lambda \psi_B \quad ,\eqno(4.18)
$$
is formidable, mainly due to fermion-number-changing terms in $P^-$; indeed,
consider a general bosonic wave functional
$$
\psi_B = \sum _{n=1}^\infty \psi_{2n} \quad ,\eqno(4.19)
$$
where
$$
\psi_n = \int_0^1\cdots \int_0^1 {\rm d}x_1\, \cdots {\rm d}x_n \delta \left(
\sum x_i - 1\right)\phi_n(x_1\cdots x_n)\tr \psi(-x_1) \cdots \psi(-x_n)\vert
0\rangle \quad ,\eqno(4.20)
$$
and the total momentum $P^+$ has been chosen to be unity.

The fermion-changing part of $P^-$ couples different wave functions $\psi_n$ in
(4.20), preventing a closed solution. In 'tHooft's solution one has also to
make
the symplifying assumption that one was computing the high-energy part of the
spectrum, in which case the integral equation (2.35) simplifies, and in fact it
has contribution for the left-hand side only from the singularity in (4.16$c$).
The singular term is simpler, and one considers only such a contribution here,
in which case the problem simplifies to the diagonalization of the operator:
$$
P=-e^2\int_0^\infty{\rm d}k_1\,\cdots{\rm d}k_4\,{1\over (k_1-k_4)^2}\delta(k_1
+k_2-k_3-k_4)\tr\left[b(-k_3)b(-k_4)b(k_1)b(k_2)\right]\quad,\eqno(4.21)
$$
which is effectively the result of the two-dimensional Coulomb force, and acts
diagonally on the wave functionals $\psi_n$ in (4.20). The eigenvalue equation
$P\psi_B = \widetilde \lambda\psi_B$, can be written in terms of the wave
functions $\phi_n$, noticing that two annihilation operators in $P$ produce
anticommutator terms when acting on the wave functional, and the remaining
terms are of the same form as the original function. It is enough then to
change variable as $x_1 \to y_1\, , \, x_2 \to x_1 + x_2 - y_1$ as well as
make cyclic permutations to arrive at the integral equation
$$
{\lambda\over e^2N}\phi_n(x_1\cdots x_n)=-\!\!\int_{-\infty}^\infty\!\!{{\rm d}
y_1\over (x_1-y_1)^2}\phi_n(y_1, x_1+x_2-y_1, x_3,\cdots ,x_n)\,\pm\,{\rm
cyclic
\, permutations} ,\eqno(4.22)
$$
where the sign on the right-hand side depends on how many times each fermion
has jumped a fermion creation operator from the wave functional.

The general analysis of such an equation has been performed by
Kutasov.\ref{93} He found solutions for the bound states of two adjoint
quarks of the type found already by 'tHooft, that is
$$
\phi_2(x) = {\rm sin \, }\pi n_1x \quad ,\eqno(4.23)
$$
corresponding to the eigenvalue
$$
\lambda _1 = 2e^2 N \pi^2 n\quad ,\eqno(4.24)
$$
for $n_1$ even and large. But he found also higher bound states, whose wave
function is a product of several symmetrized sine functions (see ref. [95] for
further details), leading to the spectrum
$$
M_{n_1 \cdots n_k}^2 = 4 e^2 N \pi^2 \sum _{i=1}^k n_i \quad ,\eqno(4.25)
$$
for $n_i$ even, and the sum large.

There is thus an exponentially growing density of states, and a consequent
Hagedorn transition. However, we have to stress that not all states have been
uncovered, as already mentioned.\ref{93}

\vskip 1cm
\penalty-1000
\noindent {\bf 4.3  The Hagedorn transition; supersymmetry }
\vskip .5cm
\nobreak
\noindent The fact that the spectrum of the model points, once more, to the
Regge behaviour, leads us to consider again the relation of QCD$_2$ to string
theory. Moreover since we have the transverse degrees of freedom, whose role
is played by the adjoint matter, we can consider a more realistic scenario.
Effective interactions\ref{96} foresee string type descriptions of the
theory for the transverse oscillators. If such a string description is correct,
one should expect a phase transition to occur, since the high-temperature
theory is in a plasma phase, while the low-temperature physics is described
by confinement. The subject is motivated by Polchinski's remark,\ref{97}
relating the statistical mechanics of string theory and large-$N$ gauge
theory, where in spherical topology the free energy is temperature
independent, and one expects a transition at some critical temperature $1/
\beta_c$, beyond which the leading order free energy is temperature-dependent.

The natural order parameters of the theory are Wilson loops, which at finite
temperature $1/\beta$ wrap around the compactified ``time" dimension used to
describe temperature; therefore one considers the Wilson loop wrapped $k$
times around the time, i.e.
$$
W_k (x) = {1\over N}\tr P \, \E^{i e\int_0^{k\beta} {\rm d}\tau\, A^2(\tau,x)}
\quad .\eqno(4.26)
$$

One can consider the two-point function of the Wilson line. At low temperature
it falls exponentially with distance, as we saw. Therefore we can write that
$$
W_k(x) W_{-k}(0)\, \to \, \E^{-M_k(\beta)\vert x\vert}\quad \eqno(4.27)
$$
as $\vert x\vert \to \infty$. The Wilson line for $k=1$ corresponds to an
external quark, while for general $k$ one has sources at higher
representations. For low temperatures the theory confines, and one expects the
area law to hold, and we have
$$
M_k^2 (\beta) \simeq (k\beta)^2 \quad ,\eqno(4.28)
$$
as expected. If a phase transition to a plasma occurs, $M^2(\beta)$ will
decrease, and at some critical temperature some winding modes will become
tachyonic.\newfoot{$^{9)}$}{In fact, a transition can occur even before the
critical value is reached.}

The study of gauge theory at finite temperature can be studied in the Coulomb
gauge, which however cannot be reached, as usual $(A_0=0)$, due to the
boundary conditions;\newfoot{$^{10)}$}{Since the time variable is compact, in
the gauge $A_0=0$ one would have $\oint {\rm d}t\,A_0=0$, thus a trivial value
for the Wilson loop.} we are forced to generalize it to
$$
A_{ab}^0 (\tau, x) = {1\over e\beta}\,\theta_a(x)\,\delta_{ab}\quad
.\eqno(4.29)
$$

As before, the other gauge-field component can be integrated out, and besides
the fermionic self-interation, one is also left with a $\theta$-field
interaction. Such an integration leads to a quadratic term in the $J_1$ current
as in the zero-temperature case, and we arrive at the effective Lagrangian
$$
L_{\rm eff}= {1\over e^2\beta^2} (\theta'_a)^2 + \overline \psi \gamma^\mu
D_\mu^\theta\psi+m\overline\psi\psi-J_1(D_0^\theta)^{-2} J_1\quad,\eqno(4.30)
$$
where
$$
D_0^\theta = \delta_{ab}\partial_0 - {i\over \beta}(\theta_a - \theta_b)\quad ,
\quad D_1 = \partial_1\quad .\eqno(4.31)
$$

It is not difficult to deal with such an effective Lagrangian in the
high-temperature phase, where the charge and the mass are small, and one can
sum the one-loop fermion diagrams. One has to sum over trajectories winding
$n$ times around the compact time, which leads to the expression\ref{95}
$$
S_{\rm eff}=-{1\over 2}\sum_{-\infty}^\infty\tr\int {{\rm d}\tau\over \tau}\int
_{{\rm periodic \, } x(t)}\!\!\!{\cal D}[x(t)]\,\E^{-\int_0^\tau {\rm d}\tau'\,
\left[ {1\over 4}\dot x^2 + {n^2\beta^2\over 4\tau^2} - ieA_0\left( \dot x +
{n\beta\over \tau}\right) - m^2\tau\right]}\, .\eqno(4.32)
$$

Since $A_0$ is $\tau$-independent, the term containing it can be integrated,
leading, for periodic $x(t)$, to the result  $\tr\,(-ien\beta A_0)$; taking the
trace over the adjoint representation of $U(N)$, which is equivalent to the
modulus squared of the trace in the fundamental representation, we obtain a
factor
$$
\sum_{a,b=1}^N \E^{in(\theta_a-\theta_b)} \, \simeq \, 2\sum_{a>b}{\rm cos\,}
n\theta \quad .\eqno(4.33)
$$

Now the integration over $x$ is a usual quantum mechanical procedure, and
leads to the potential
$$
V(\theta) = {\beta L\over 2\pi} \sum _{n=1}^\infty (-1)^n \int _0^\infty
{{\rm d}\tau\over \tau^2}\,\E^{-{n^2\beta^2\over 4\tau} - \tau m^2}{\rm cos\,}
n\theta\quad ,\eqno(4.34)
$$
with the effective Lagrangian
$$
L_{\rm eff} = {1\over e^2 \beta}{\theta'}^2 + V(\theta)\quad .\eqno(4.35)
$$

We scale the integration variable $\tau \to {1\over 4}n^2\beta^2 \tau$ in eq.
(4.34), finding the relevant term in the high-temperature limit, contributing
to the effective Lagrangian, which is given by
$$
L_{\rm eff} = {1\over e^2\beta} (\theta'_a)^2 + {2\over \pi} \sum_{a,b=1}^N
\sum_{n=1}^\infty {(-1)^n\over n^2\beta}\, {\rm cos \,}(n\theta_{ab})\quad .
\eqno(4.36)
$$

We could also follow the line of ref. [90] and introduce, at large $N$, the
density of eigenvalues
$$
\rho(\theta, x) = {1\over N}\sum_a \delta(\theta - \theta_a (x))\quad ,
\eqno(4.37)
$$
in terms of which the action becomes
$$
S={N\over e^2\beta}\int{\rm d}x\,{\rm d}\theta\,\rho^{-1}(\partial_\theta^{-1}
\rho')^2 + {N^2\over L}\int {\rm d}x\, {\rm d}\theta_1\, {\rm d}\theta_2 \,
\rho(\theta_1) \rho(\theta_2)V(\theta_{12})\quad ,\eqno(4.38)
$$
where one uses $\partial _\theta^{-1}\rho' = {1\over N}\sum_a {\partial
\theta_a\over \partial x}\delta(\theta -\theta_a(x))$.

At large $N$ both terms must be kept since $e^2\sim 1/N$. The minimum of the
potential implies that a possible classical configuration is where the
eigenvalues are equal, that is $\rho \sim \delta(\theta -\theta_0)$, which
breaks the $U(1)$ symmetry $\theta \to \theta + \epsilon$, while a
symmetric phase would favour a constant distribution $\rho=1/2\pi$. In order to
study the stability of the  latter, we expand $\rho$ around it
$$
\rho(\theta, x) = {1\over 2\pi}\left( 1+ \sum_{n\ne 0}\rho_n(x)\,\E^{-in\theta}
\right)\quad ,\eqno(4.39)
$$
which leads to the action
$$
S = N^2\sum_{n\ne 0}\int {\rm d}x\, \left[ {1\over \beta e^2 n^2 N}(\rho'_n)^2
+ (-1)^n {2\over \pi n^2 \beta}\rho_n^2\right]\quad .\eqno(4.40)
$$

The mass of the winding states can be read from above, and we obtain
$$
M_n^2(\beta \to 0) = {2e^2N\over \pi} (-1)^n \quad .\eqno(4.41)
$$

For $n$ odd the winding states are tachyonic. In four-dimensional gauge theory
this computation has been performed by Polchinski, who found for
$V(\theta)\simeq {1\over 24\pi^2\beta^3}\sum \theta^2_{ab}(2\pi - \theta_{ab})
^2$, leading to a mass of the form
$$
M_k^2(\beta \to 0) = -{2e^2N\over \pi^2\beta^2k^2}\quad .\eqno(4.42)
$$

These results imply deconfinement, since the area law is no longer attainable.
A phase transition is related to the appearance of tachyons, since as argued in
ref. [98] a divergence of the free energy in a given (Hagedorn) temperature
takes place when a tachyon starts playing a role, as formed in a new mode,
above
that temperature.

The situation for the case of fermions in a fundamental representation seems to
be different. Kutasov\ref{93} computed the effective action for the
distribution function $\rho$, and found no instability for any temperature, so
that confinement seems to be settled for that case. For adjoint bosonic matter
he found
$$
M_k^2 (\beta \to 0) = - {22\over \pi} e^2 N\quad ,\eqno(4.43)
$$
thus also exhibiting deconfinement.

Such a transition in adjoint matter indicates a rising density of single
particle
states at high energy, while for the fermion in the fundamental representation
there is no such growth of the density of states with energy.
\vskip .5cm
\noindent {\it Supersymmetry}
\vskip .5cm
\nobreak
\noindent The QCD$_2$ action with fermions in the adjoint representation may be
supersymmetric.\ref{93} In fact, in the zero coupling limit, and with massless
fermions, this can be immediately seen, the supersymmetry generator being
given by
$$
G = {\tr \over 3} \int {\rm d}x^-\, \psi\psi\psi\quad .\eqno(4.44)
$$

However, such a charge of fermionic character commutes with $P^-$ as given by
eq. (4.16$b$), when the theory is interactive as long as the fermions have a
mass $m^2=e^2N$. Moreover
$$
G^2 = NP^+\quad .\eqno(4.45)
$$

In string theories with space-time fermions, infrared stability is achieved by
a (rather fine) cancellation of bosonic and fermionic degrees of freedom, a
fact
requiring asymptotic supersymmetry (at high energy). Due to the enormous
content of mesonic states in QCD$_2$ with adjoint matter the question is
important as well.\ref{93}
\vskip 1cm
\penalty-1000
\noindent {\bf 4.4  Landau--Ginzburg description; spectrum and string theory }
\vskip .5cm
\nobreak\noindent
One of the goals in the study of QCD$_2$ is to understand some technical
details also existing in QCD$_4$, but which in two dimensions are rendered
understandable or perhaps calculable, albeit not trivial. Several models serve
as laboratories, as the two-dimensional non-linear $\sigma$-models, discussed
in detail in [8], or QCD$_2$, where one finds a mass gap from dimensional
transmutation, and confinement in some cases. However, as a two-dimensional
counterpart of the theory of strong interactions, the pure Yang--Mills action
$-{1\over 4}\tr F^{\mu\nu}F_{\mu\nu}$ is not unique. It is possible to
generalize such an interaction without losing several properties whose
maintenance have been important up to now. Such a clever generalization,
obtained in [99], maintains Migdal's heat-kernel formulation, where the
partition function
$$
Z[U]=\sum_R({\rm dim\,}R)\,\E^{-{ie^2\over N}a^2C(R)}\chi_R[U]\quad\eqno(4.46)
$$
has the self-reproducing property (3.18), thus being the best approximation of
the continuum. Such a property is valid for an arbitrary function $C(R)$, and
one can write a continuum action as
$$
S = \tr \int {\rm d}^2x\, [EF - f(E)] \quad ,\eqno(4.47)
$$
generalizing the pure Yang--Mills action where
$$
f_{YM}(E)= 2E^2\quad .\eqno(4.48)
$$

According to the two-dimensional power counting, any arbitrary function $f(E)$
can be allowed instead of the quadratic term. Thus, in general we suppose
that $f$ is expandable as $f(E)= \sum_n f_n E^n$. In ref. [99], the theory of
generalized QCD$_2$ with fermions in the fundamental representation was
studied in the large-$N$ limit. The pure gauge model was discussed in
ref. [100], where they found a $1/N$ expansion similar to the one discussed in
section 3, and a string interpretation.\ref{101} Such a case is however
still very simple, due to the absence of local degrees of freedom.
Including matter fields, the theory is non-trivial, but still tractable since
the coupling has dimension of mass, implying power-counting renormalizability.

In the $U(1)$ case one has a generalized Schwinger model, including a
Landau--Ginzburg potential, and the Lagrangian density reads
$$
L = E \epsilon_{\mu\nu}F^{\mu\nu} - f(E) + \overline \psi i \not \!\! D \psi -
m \overline \psi \psi \quad .\eqno(4.49)
$$

Upon bosonization of the fermion and using the well known formulae\ref{8}
$$
\eqalignno{
\overline \psi \gamma^\mu \psi \simeq \, & {1\over \pi} \epsilon ^{\mu\nu}
\partial _\nu \phi \quad , & (4.50) \cr
\overline \psi i \not \! \! \partial \psi \simeq \, & {1\over 2\pi}\partial
^\mu
\phi \partial _\mu \phi \quad , & (4.51) \cr
\overline \psi \psi = \, & m \gamma {\rm cos\,}(2\phi)\quad , &(4.52)\cr}
$$
one finds the equivalent bosonic action, that is
$$
L = {1\over \pi}\epsilon ^{\mu\nu}\partial _\mu A_\nu (E - e\phi)- f(E) +
{1\over 2\pi}\partial ^\mu \phi \partial _\mu \phi - m^2\gamma {\rm cos\,}
(2\phi)\quad .\eqno(4.53)
$$

One can fix the light cone requiring that $A_1=0$. The $A_0$ equation of motion
is a constraint equation (Gauss law), which demands
$$
E = e (\phi + \theta/2)\quad ,\eqno(4.54)
$$
where $\theta$ is a constant, interpreted as the label of the vacuum of the
theory. One can substitute it back into the action, redefining $ \phi \to \phi
- {\theta\over 2}$ to obtain the Lagrangian
$$
L = {1\over 2\pi}\partial ^\mu \phi \partial _\mu \phi - f(e\phi) - m^2\gamma
{\rm cos\,}(2\phi - \theta)\quad .\eqno(4.55)
$$

In the massless case we have a meson interacting via a Landau--Ginzburg
potential $f(e\phi)$. Therefore, although there are still mesons $(\phi)$ that
can be described as bound states of fermions, they now have complicated
interactions dictated by the Landau--Ginzburg potential $f(e\phi)$.

The vacuum is described now by the potential
$$
V(\phi)= f(e\phi) + m^2 \gamma {\rm cos\,}(2\phi - \theta)\quad ;\eqno(4.56)
$$
notice the possibility of phase transitions at some critical points $e= e_c$.

Concerning the non-Abelian case, we have to consider the Lagrangian
$$
L = {N\over 8\pi}\tr E \epsilon_{\mu\nu}F^{\mu\nu} - {N\over 4\pi} e^2
\sum_{n=2}^\infty f_n \tr \left({E\over e}\right) + \overline \psi (i\not\!\!D
- m)\psi\quad .\eqno(4.57)
$$

The light-cone gauge can be used as in section 2.1. The large-$N$ limit is
obtained as before once a generalized vertex $E^n$ is included, the lines must
inevitably finish in a quark line, where the $f_2E^2$ vertex is viewed as an
interaction in the weak-coupling limit. The effect of a vertex $f_nE^n$ as
given in Fig. 6 is
$$
\eqalignno{
I_n(p,p')=\int{{\rm d}^2k_1\over (2\pi)^2}\cdots{{\rm d}^2k_{n-1}\over
(2\pi)^2}&{1\over p_- - {k_1}_-}S(k_1){1\over {k_1}_- - {k_1}_-}S(k_2)\cdots\cr
&\cdots {1\over {k_{n-2}}_- - {k_{n-1}}_-} S(k_{n-1}){1\over {k_{n-1}}_--p'_-}
\quad .&(4.58)\cr}
$$
\penalty-2000
$$\epsfysize=3truecm\epsfbox{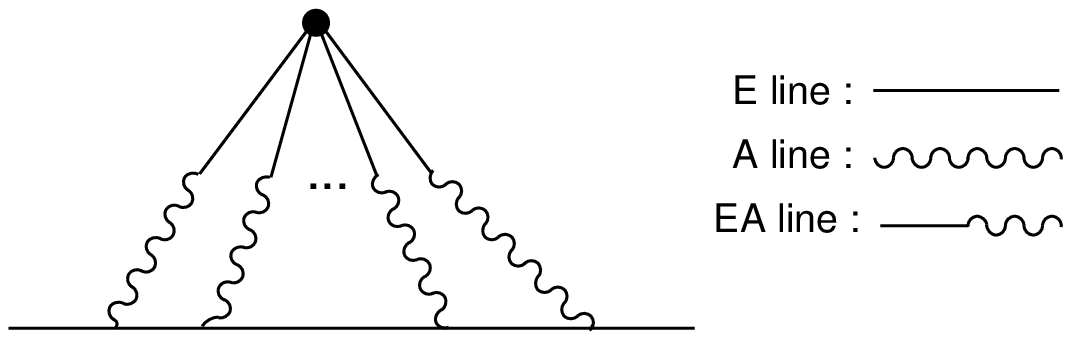}$$
\noindent{\eightpoint Fig. 6: The generalized vertex.}
\vskip.3truecm
\penalty-2000

\noindent Above $S(p)$ is given by (2.39$c$), and the self-energy generalizing
(2.22) is
$$
\Sigma_{SE}(p)=\sum(-i)^{n}nf_nI_n(p,p)\quad,\eqno(4.59)
$$
from which we find a two-particle irreducible kernel as given in Fig. 7,
$$
K(q, q'; p, p')=\sum_{n=2}^\infty (-i)^n f_n \sum_{l=1}^{n-1}I_l (q, q')
I_{n-l}(p'-q'; p-q)\quad .\eqno(4.60)
$$
\penalty-2000
$$\epsfxsize=3.5truecm\epsfbox{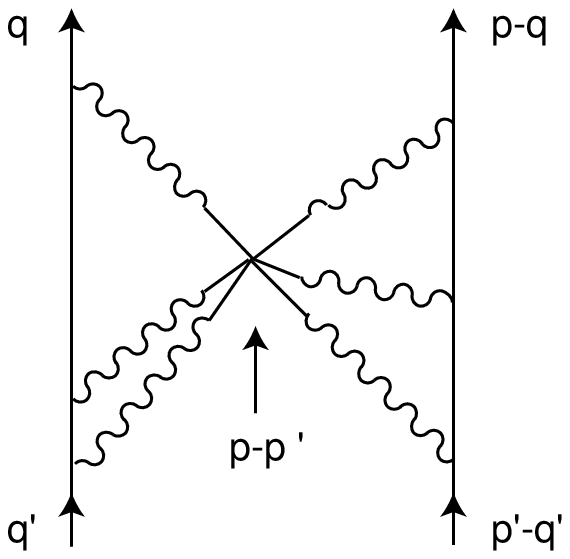}$$
\noindent{\eightpoint Fig. 7: Two-particle scattering with a generalized
vertex. }
\vskip.3truecm
\penalty-2000

The important technical point used in the deduction of (2.25) from (2.22) was
that it is possible to integrate over the  $(+)$ variables. This is also true
above, and one arrives at
$$
I_n(p,p')= \int_{\infty}^\infty {\rm d}k_1 \cdots {\rm d}k_{n-1}
{1\over p-k_1} \varepsilon (k_1){1\over k_1 - k_2} \varepsilon (k_2)  \cdots
\varepsilon (k_{n-1}) {1\over k_{n-1}-p'}\quad .\eqno(4.61)
$$

At this point it is necessary to introduce the infrared regulator. For $n=2$
one finds
$$
I_2(p,p')= {2\over p-p'}\ln \left\vert {p\over p'}\right\vert -\pi^2\varepsilon
(p)\delta(p-p')\quad .\eqno(4.62)
$$

In ref. [99] the authors introduced the generating functional
$$
u(p,p';z)= \sum_{n=0}^\infty z^{-n}I_n(p,p')\quad ,\eqno(4.63)
$$
upon defining $I_0(p,p')=\varepsilon (p)\delta(p-p')$ and $I_1(p,p')={P\over p
-p'}$. Such a function obeys a tractable integral equation. We multiply by $z$,
and separate the first term, which is just the $I_0(p,p')$ contribution, as can
be seen from eq. (4.61), that the remaining terms are of the form
$$
I_{n+1}(p,p')=\int{\rm d}k\,{P\over
p-k}\varepsilon(k)I_n(k,p')\quad.\eqno(4.64)
$$
Therefore one arrives at an integral equation for the generating functional, as
given by
$$
z\,u(p,p';z)=z\,\varepsilon(p)\delta(p-p')+\int{\rm d}k\,{P\over
p-k}\varepsilon
(k) u(k,p';z)\quad .\eqno(4.65)
$$

It is simple to find the result
$$
\int_{-\infty}^\infty {\rm d}k\,{P\over p-k}\varepsilon(k)\vert k\vert ^{2\nu}
=
\pi {\rm cot}\, (\pi\nu) \vert p\vert ^{2\nu} \quad .\eqno(4.66)
$$

We now use eqs. (4.61) and  (4.66) to find, after consecutive integrations
$$
\eqalignno{
z^{-n}\int I_n (p,k)\varepsilon (k)\vert k\vert^{2\nu}{\rm d}k=\,&z^{-n}\int
{\rm d}k_1 \cdots {\rm d}k_{n-1} {1\over p-k_1} \cdots {1\over k_{n-1}-k}
\varepsilon (k)\vert k\vert ^{2\nu}{\rm d}k\cr
= \, & z^{-n}(\pi {\rm cot}\,\pi\nu)^n\varepsilon (p)\vert p\vert^{2\nu}\quad .
&(4.67)\cr}
$$

Summing over $z$, we have
$$
\varepsilon (p) \int {\rm d}k\, u(p,k;z) \varepsilon (k) \vert k\vert ^{2\nu} =
{z\over z-\pi{\rm cot\, } \pi\nu}\varepsilon (p)\vert p\vert ^{2\nu}\quad .
\eqno(4.68)
$$

The solution to the integral equation (4.65) is unique, and can be found
by inspection, with the help of eq. (4.66). One finds\ref{99}
$$
u(p,p';z)={z\over z^2+\pi^2}\left[P{1\over p-p'}\left\vert {p\over p'}
\right\vert^{2\alpha(z)}+ z\varepsilon (p)\delta(p-p')\right]\quad ,\eqno(4.69)
$$
where $\alpha(z) = {1\over \pi}{\rm arc\, tan\,}{\pi\over z}$. The functions
$I_n(p,p')$ can now be found expanding (4.69) as a series in $z^{-1}$. We do
not need a systematic computation for them. We consider the steps analogous to
those used in (2.21) through (2.35). We can express the self-energy as
$$
\Sigma_{SE}(p)=\sum_n (-1)^nnf_nI_n=\oint{{\rm d}z\over 2\pi i}f'(z)u(p,p';iz)
\quad,\eqno(4.70)
$$
since only the term $ \sim z^{-1}$ survives in the right-hand side; therefore
one finds the right-hand side of the integral equation
$$
\eqalign{
\Biggl[ q_+-m^2&\left( {1\over p_-}+{1\over q_--p_-}\right)\Biggr]
\varphi(p_-;q) \cr
& =\left[ \Sigma_{SE} (p_-) + \Sigma_{SE} (q_--p_-)\right]\varphi(p_-;q) +
\int_0^{q_-}{\rm d}k_-K(p,k;q,q)\varphi(k_-,q) \quad ,\cr}
$$
which is the generalized counterpart of (2.34$a$). Its right-hand side reads
$$
\eqalignno{
2e^2\int {{\rm d}z\over 2\pi i} f'(z)&\Bigg\{ -\left[ u(q,q;iz) + u(p-q,p-q;iz)
\right] \phi(q) \cr
& + 2 \int _0^p {\rm d}k\, u(q,k;iz) u(p-k,p-q;iz)\phi(k)\Bigg\} \cr
=2e^2\int {{\rm d}z\over 2\pi i} & f'(z)\Bigg\{ - iz {z^2+\pi^2\over
(z^2-\pi^2)^2} 2\alpha(iz)\left({1\over q}+{1\over p-q}\right)\phi(q) \cr
& - {2z^2\over (z^2-\pi^2)^2} \int _0^p {\rm d}k\, P {1\over (q-k)^2}
\left[{q(p-k)\over (p-q)k}\right]^{2\alpha(iz)}\phi(k)\Bigg\} \quad
,&(4.71)\cr}
$$
from which one finally obtains the bound-state equation in the form
$$
\eqalignno{
\Bigl[\mu^2-\tau &\left( {1\over x} + {1\over 1-x}\right)\Bigr]\phi(x) \cr
 = \, & 2\pi \int {{\rm d}z\over 2\pi i}f'(z) \Big\{ - iz {z^2+\pi^2\over
(z^2-\pi^2)^2} 2\alpha(iz)\left({1\over x}+{1\over 1-x}\right)\phi(x) \cr
- & {2z^2\over (z^2-\pi^2)^2} \int _0^1 \!\!{\rm d}y\, P {1\over (x-y)^2}
\left[{x(1-y)\over (1-x)y}\right]^{2\alpha(iz)}\!\!\phi(y)\Big\} \,
.&(4.72)\cr}
$$

This result reveals that information can be obtained about the spectrum in
generalized QCD$_2$. There are corrections to 'tHooft's equation. In ref.
[99] such corrections have been exemplified for a quartic potential. In
particular, the authors showed that for an arbitrary potential $f(z)$ there is
always a massless eigenstate, arguing that it is a consequence of the chiral
$U(1)$ symmetry in the large-$N$ limit, where the $U(1)$ anomaly is suppressed
in lowest order of $1/N$.
\vskip 1cm
\penalty-3000
\noindent {\bf 5. Algebraic aspects of QCD$_2$ and integrability}
\vskip .5cm
\nobreak
\noindent We saw that two-dimensional QCD, although not exactly soluble, in
terms of free fields, is a theory from which some valuable results may be
obtained.
The $1/N$ expansion reveals a simple spectrum valid for weak coupling, while
the strong coupling offers the possibility of understanding the baryon as a
generalized sine-Gordon soliton. Moreover, the $1/N$ expansion of the
pure-gauge case may be performed, and the partition function is equivalent to
one of a string model described by a topological field theory, the Nambu--Goto
string action, and presumably terms preventing folds.

All such results point to a relatively simple structure, which could be
mirrored by an underlying symmetry algebra. In fact such algebraic structures
do exist. In the above-mentioned case of the large-$N$  expansion of pure
QCD$_2$, one finds a $W_\infty$-structure related to area-preserving
diffeomorphisms of the Nambu--Goto action. A $W_\infty$ structure for
gauge-invariant bilinears in the Fermi fields is constructed\ref{102} - (see
section 5.1). Such is an algebra which appears also in fermionic systems, and
in the description of the quantum Hall effect\ref{35}. Moreover, as shown
before, pure QCD$_2$ is equivalent to the $c=1$ matrix model,\ref{103} which
has also a representation in terms of non-relativistic fermions,\ref{104} and
contains a $W_\infty$ algebra\ref{36,105,106} as well. The problem is also
related to the Calogero--Sutherland models.\ref{107} The mass eigenstates build
a representation of the $W_\infty$ algebra as found in [102].

After bosonizing the theory, further algebraic functions of the fields turn out
to obey non-trivial conservation laws, as we will see. The theory can be
related to a product of several conformally invariant WZW sectors, a perturbed
WZW sector, all related by means of BRST constraints, which play a very
important role in gauge theories, as described in section 2. A dual
formulation exists and permits us to study the theory in two limits, both
strong and weak couplings. Finally, once displayed, the relation to Calogero
systems and further integrable models is also amenable to understanding in the
previous framework.
\vskip 1cm
\penalty-1000
\noindent {\bf 5.1 $W_\infty$ algebras for colourless bilinears}
\vskip .5cm
\nobreak
\noindent Let us consider the QCD$_2$ Lagrangian with massive fermions in the
fundamental representation, as in section 2. Using light-cone coordinates, the
Lagrangian is given by
$$
\eqalignno{
L = & -{1\over 4}\tr F_{\mu\nu}F^{\mu\nu}+\overline\psi(i\not\!\!D-m)\psi\cr
= \, & {1\over 8} \tr F^2_{+-} + \psi_-^\dagger(i\partial _+ + eA_+)\psi_- +
\psi_+^\dagger(i\partial _- + eA_-)\psi_+ - m(\psi_+^\dagger \psi_- +
\psi_-^\dagger\psi_+)\quad ,& (5.1)\cr}
$$
where we use the notation $\psi= \left( \matrix {\psi_-\cr \psi_+\cr}\right)$.

As usual, we get a considerable simplification while working at the
light-cone gauge, e.g. $A_+=0$, in which case $\psi_-$ decouples, up to the
mass term. Furthermore, we can also quantize the theory in the light
cone,\ref{143}
regarding $x^+$ as time and $x^-$ as space, in which case the equation of
motion of $\psi_+$ is actually a constraint equation, as we have done in
section 4 for different reasons. The light-cone Hamiltonian, which corresponds
to the $+$ component of the energy--momentum tensor, is given by the expression
$$
H \equiv P_+ = \int {\rm d}x^-\, \left[ {1\over 2}\tr E^2 + m (\psi_+^\dagger
\psi_- + \psi_-^\dagger\psi_+)\right] \quad ,\eqno(5.2)
$$
where $E= {1\over 2}\partial _+A_-$, the momentum canonically conjugated to
$A_-$; therefore the canonical quantization is achieved by the commutation rule
$$
\left[ A_- (x^-, x^+), E (y^-, x^+)\right] = i \delta (x^- - y^-)\quad .
\eqno(5.3)
$$

The equation of motion of $A_+$ is equivalent to the Gauss law
$$
\partial _-E^{ab} - ie [A_-, E]^{ab} + 2e \left({\psi^\dagger_-}^a\psi_-^b -
{1\over N} \delta^{ab} \psi_-^{\dagger c} \psi_-^c\right)=0\quad ,\eqno(5.4)
$$
where the last term drops out for the $U(N)$ case.

For the equation of motion of $\psi_+$, we obtain further constraints (no
light-cone time derivative), that is
$$
(i\partial _- + e A_-)\psi_+ - m \psi_- = 0 \quad , \eqno(5.5)
$$
and the corresponding complex conjugate equation. These constraints are
important in the computation of correlators involving $\psi_+$, otherwise we
do not need them. The Gauss law constraint can also be used to solve the
electric field in terms of the Fermi fields, once the full gauge arbitrariness
is fixed.

However we will follow another trend, defining the Wilson (open) operator by
means of the bilinear\ref{102}
$$
M_{\alpha \beta, i j}(x^-, y^-; x^+)=\psi_{i\alpha}(x^-, x^+)\,\E^{ie\int_{x^-}
^{y^-} A_-(z^-,x^+){\rm d}z^-}\psi^\dagger_{j\beta}(y^-,x^+)\quad ,\eqno(5.6)
$$
where $i,j = 1, \cdots ,F$ are indices of flavour. In the Hamiltonian
formalism one works at a given (light-cone) time $x^+$. Indeed, $H$ is
time-independent, and we can choose, for definiteness, a fixed time $x_0^+$. At
that given fixed point, one can always choose $A_-(x^-,x_0^+)=0$. The same
can be done in the case of the algebra obeyed by the bilinear. We are going to
compute the algebra at equal times, and in such a gauge one finds
$$
M_{\alpha\beta, ij} = \psi_{i\alpha }(x^-,x^+)\psi_{j\beta}^\dagger(y^-,x^+)
\quad , \eqno(5.7)
$$
for which it is simple to find the corresponding algebra, given the canonical
equal time anticommutator of the fermions:
$$\eqalignno{
\{\psi_-(x^-,x^+),\psi_-^\dagger (y^-,
x^+)\}=\,&\delta(x^--y^-)\quad,&(5.8a)\cr
[M_{ij} (x^-, y^-; x^+), M_{kl} ({x'}^-, {y'}^-; x^+)]  = \,  &
\delta_{jk}\delta(y^- - x^-)\,M_{il}(x^-, {y'}^-; x^+) -\cr
& - \delta_{il}\delta({y'}^- - x^-)\,M_{kj}({x'}^-, y^-; x^+) \quad
,&(5.8b)\cr}
$$
which is an infinite algebra of the type $W_\infty \otimes U(F)$. Similar
results have been found in ref. [144]. For the one-flavour case, it simplifies
to the usual $W_\infty$ algebra, of the type found in $c=1$ matrix models. In
fact, there are several types of $W$ algebras obeyed by bilinears constructed
out of representations of $U(F)$ groups. For off-critical perturbations of a
free-fermion system one finds a rather elaborate $W_{1+\infty}$ algebra as
obeyed by higher-spin currents.\ref{108} Here it is very interesting to notice
the rather simple structure obeyed by the above bilinear.

Still in the gauge $A_-=0$, achieved at a definite time, the electric field is
given by the expression
$$
E^{ab}(x^-,x^+)=-e\int{\rm d}y^-\,\varepsilon\left(x^--y^-\right)\,{\psi_-^
\dagger}^b(y^-, x^+)\psi_-^a(y^-, x^+)\quad ,\eqno(5.9)
$$
as deduced from the Gauss law, which after substitution upon the Hamiltonian
leads to the result
$$
\eqalignno{
H= {e^2\over 4N} \int {\rm d}x^-\, {\rm d}y^-\, \Big[ &
\psi^a_{i-}(x^-)\psi^{\dagger a}_{j-}(x^-)\,\vert x^--y^-\vert \, \psi^b_{j-}
(y^-) \psi^{\dagger b}_{i-}(y^-)  \cr
& - {1\over N} \psi^a_{i-}(x^-)\psi^{\dagger a}_{i-}(x^-)\,\vert x^--y^-\vert
\,\psi^b_{j-}(y^-) \psi^{\dagger b}_{j-}(y^-)  \cr
& - {im^2\over 4}\varepsilon\left(x^- - y^-\right)
\psi^a_{i-}(x^-)\psi^{\dagger a}_{i-}(y^-)\Big]\quad .&(5.10a) \cr}
$$

It can be written solely in terms of the previously defined bilinears as
$$
\eqalignno{
H= {N} \int {\rm d}x^-\, {\rm d}y^-\, \Big[ & {e^2\over 4} M_{--, ij}(x^-, y^-)
\vert x^--y^-\vert M_{--,ij}(x^-, y^-)  \cr
& - {e^2\over 4N} M_{--,ii}(x^-, x^-)\vert x^--y^-\vert M_{--,jj}(y^-, y^-)\cr
& - {im^2\over 4}\varepsilon\left(x^- - y^-\right) M_{--,ii}(x^- - y^-)\Big]
\quad.&(5.10b)
\cr}
$$

Finally, one can use the fermionic constraints at the given time to compute
$M_{++}$ and $M_{+-}$ in terms of $M_{--}$. The bilinears also obey quadratic
constraints, and can be seen to imply the bound-state structure obtained from
the large-$N$  expansion as proposed by 'tHooft.

The realization of the $W_\infty$ algebra in terms of the above fields $M(x,y)$
leads us to a close relation to string theory, and to 'tHooft's spectrum
derived in section 2.

It is natural, in view of the results of section 2, which are derived for large
$N$, to study the ``string" field $M(x,y)$ in such a limit, where it becomes
classical, being related to a (self-consistent Hartree-Fock) potential, where
the fermions move. It is, in fact, not difficult to obtain the solution to the
equation of motion obeyed by $M(x,y)$, fulfilling the constraint equation
($M^2(x,y)=M(x,y)$), obtained in ref. [102], to which we refer for details.
This constraint is a consequence of colour invariance $E^{ab}(x^-=\infty)=
E^{ab}(x^-=-\infty)$ (see eq. (5.9)). One can also define the baryon number
$B= \tr \, (1-M)$.

The classical solution of the constraint equations is given, in terms of the
Fourier transform, for a single flavour as
$$
M_0(k_-,k'_-;x^+) = \delta(k_--k'_-)\theta(k_-)\quad, \eqno(5.11)
$$
where the Fermi level was choosen for $B=0$. The fluctuations around (5.11) can
be computed for large $N$, as
$$
M= \E^{{i\over \sqrt N}w} \, M_0 \, \E^{-{i\over \sqrt N}w} \quad,\eqno(5.12)
$$
where $w$ represents a perturbation around the classical solution $M_0$. The
main result found in ref. [102] is that the $x^+$-Fourier transform of
$w^{-+}(k,k';x^+)=w(k,-k';x^+)$ for $k,k'>0$ is exactly 'tHooft's wave
function,
$$
w^{-+} (k,k',x^+) = \int {{\rm d}q^+\over 2\pi} \varphi(q_-=k_-+k'_-,q_+;x^+)
\E^{iq_+x^+} \quad ,\eqno(5.13)
$$
leading to (2.35) for $\varphi(q_-,q_+;x^+)$. This means that the mesons of
two-dimensional QCD form a representation of the $W$ algebra. As we mentioned,
in the bosonized version it has been argued\ref{142} that
the mesons obey an equation of motion presumably corresponding to  'tHooft's
equation, (2.35)  which would imply a rather explicit realization
of the given symmetry. For any number of
flavours one has a representation of $W_\infty\otimes U(N)$.\newfoot{$^{11)}$}
{Actually we have $W_{\infty +}\otimes W_{\infty -}$, see ref. [102] for
details.} The bilocals have been discussed in ref. [144] in the context of the
dynamics of hadrons in two dimensions, as well as in ref. [145].

\vskip 1cm
\penalty-1000
\noindent {\bf 5.2 Integrability and duality }
\vskip .5cm
\nobreak
\noindent We have seen in section 2 that after integrating out the fermions
and performing a set of field transformations we arrive at a product of
conformally invariant theories including a WZW theory with a non-local mass
term. We suppose that the BRST constraints define the physical states, and at
the Lagrangian level we consider the perturbed WZW action\ref{46,47}
$$
\eqalign{
S &= \Gamma[\beta] + {1\over 2} \mu^2 \tr \int {\rm d}^2z\, \left[\partial_+^
{-1}(\beta^{-1} \partial_+ \beta)\right] ^2\quad ,\cr
&= \Gamma[\beta] + {1\over 2} \mu^2 \Delta(\beta)\quad .\cr}\eqno(5.14)
$$

We will look for the Euler--Lagrange equations for $\beta$. It is not difficult
to find the variations:
$$
\eqalignno{
\delta \Gamma[\beta ]&= \left[{1\over 4\pi}\partial_-(\beta^{-1}\partial_+
\beta)\right]\beta^{-1}\delta\beta\quad ,&(5.15a)\cr
\delta \Delta(\beta) & = 2\Big( \partial_+^{-1} (\beta^{-1}\partial_+\beta)
 - \big[ \partial_+^{-2}
(\beta^{-1}\partial_+\beta),(\beta^{-1}\partial_+\beta)
\big] \Big)\beta^{-1}\delta \beta\quad .&(5.15b)\cr}
$$

Collecting the terms, we find it useful to define the current components
$$
\eqalign{
J_+^\beta&= \beta^{-1}\partial_+ \beta\quad ,\cr
J_-^\beta&= -4\pi\mu^2\partial_+^{-2}J_+^\beta=-4\pi\mu^{2}
\partial_+^{-2}(\beta^{-1}\partial_+\beta)\quad ,\cr}\eqno(5.16)
$$
which summarize the $\beta $ equation of motion as a zero-curvature condition
given by
$$
[{\cal L},{\cal L}]=[\partial_++J_+^\beta,\partial_- +J_-^\beta]=\partial_-
J_+^\beta-\partial_+J_-^\beta+[J_-^\beta, J_+^\beta ] =0 \quad .\eqno(5.17a)
$$
This is the integrability condition for the Lax pair\ref{110}
$$
{\cal L}_\mu M =0 \quad  , \quad {\rm with } \quad {\cal L}_\mu = \partial _
\mu - J^\beta_\mu \quad ,\eqno(5.17b)
$$
where $J_\pm^\beta=J^\beta_0\pm J^\beta_1$ and $M$ is the monodromy matrix.
This is not a Lax pair as in the usual non-linear $\sigma$-models,\ref{109}
where $J^\beta_\mu$ is a conserved current, and where we obtain a conserved
non-local charge from (5.17$a$), as well as higher local and non-local
conservation laws, derived from an extension of (5.17$a$) in terms of an
arbitrary spectral parameter.\ref{110} However, to a certain extent, the
situation is simpler in the present case, due to the rather unusual form of
the currents (5.16), which permits us to write the commutator appearing in
(5.17$a$) as a total derivative, in such a way that in terms of the current
$J_-^\beta$ we have
$$
\partial_+\left(4\pi\mu^2J_-^\beta+\partial_+\partial_-J_-^\beta+[J_-^\beta,
\partial_+J_-^\beta]\right)=0\quad .\eqno(5.18a)
$$

Therefore the quantity
$$
I_-^\beta (x^-) = 4\pi \mu^2 J_-^\beta (x^+,x^-) + \partial_+ \partial_-
J_-^\beta (x^+,x^-) + [J_-^\beta (x^+,x^-), \partial_+ J_-^\beta(x^+,x^-)]
 \quad \eqno(5.18b)
$$
does not depend on $x^+$, and it is a simple matter to derive an infinite
number
of conservation laws from the above. These are non-local conservation laws, as
is clear from (5.16).

This means that two-dimensional QCD is an integrable system!\ref{8,110,111}
Moreover, it corresponds to an off-critical perturbation of the WZW action. If
we write $\beta=\E^{i\phi}\sim 1+i\phi$,  we verify that the perturbing term
corresponds to a mass term for $\phi$. The next natural step is to obtain the
algebra obeyed by (5.18$b$), and its representation. However, there is a
difficulty presented by the non-locality of the perturbation. We now introduce
a further auxiliary field defining a dual action, local in all fields, and
representing the low-energy scales of the theory, and we later return to the
problem of finding the algebra obeyed by (5.18$b$).

Consider the $\Delta$-term of the action (5.14). We rewrite it introducing the
integral over a Gaussian field $C_-$ as
$$
\E^{{i\over 2}\mu^2\Delta}=\int{\cal D}C_-\,\E^{i\int{\rm d}^2x\,{1\over 2}\tr
(\partial_+C_-)^2-\mu\tr\int{\rm d}^2x\,C_-(\beta^{-1}\partial_+\beta)}\quad,
\eqno(5.19)
$$
where the left-hand side is readily obtained by completing the square in the
right-hand side.

Indeed, at this point we have two choices. We can proceed with the canonical
quantization of the action (5.14) with the non-local term substituted in terms
of the $C_-$ field-dependent expression obtained in the exponent of the
integrand of the right-hand side of eq. (5.19). Before that, motivated by
the presence of the auxiliary vector field $C_-$, we again make a change of
variables of the type
$$
\eqalignno{
C_- &= {i\over 4\pi \mu} W\partial_- W^{-1} \quad ,&(5.20a)\cr
{\cal D}C_- & = \E^{-ic_V \Gamma [W]} {\cal D} W \quad ,&(5.20b)\cr}
$$
together with the now very frequently used identity (2.78) in order to find a
dual action. We have for the $\beta$-partition function the expression
$$
{\cal Z} = \int {\cal D}\beta {\cal D}W \,\E^{i\Gamma[\beta] -i c_V \Gamma[W] +
{i\over 4\pi} \int {\rm d}^2x\, W \partial_-  W^{-1} \beta^{-1} \partial_+
\beta - i\int {\rm d}^2 x \,{1\over 2(4\pi \mu)^2} [\partial_+ (W\partial_-
W^{-1})]^2}\quad ,\eqno(5.21)
$$
from which we can separate the contribution $-\Gamma[\beta W] \equiv -
\Gamma[\widetilde \beta]$; after such man{\oe}uvre we are left with
$$
{\cal Z} = \int {\cal D} \widetilde \beta \,\E^{i\Gamma[\widetilde \beta]}\int
{\cal D} W\,\E^{-i(c_V+1)\Gamma[W]-{i\over 2(4\pi\mu)^2} \tr\int {\rm d}^2 z
\, [\partial_+(W\partial_- W^{-1})]^2}\quad .\eqno(5.22)
$$

The dual action now has a coupling constant corresponding to the inverse of the
initial charge. Therefore eq. (5.22) is appropriate to the study of a strongly
coupled limit. Notice that the procedure is, in a sense, similar to the one
used to obtain a dual action, where a non-dynamical field is introduced, and
one
eliminates the original fields by integration, leaving the so-called dual
formulation. (See refs. [112-114]  for further details on duality.)
We separate a further WZW-conformal piece, and we are left with a local massive
action for $W$. The drawback is the fact that now $W$ itself has an action
with a negative sign. Na\"{\i}vely it also describes massive excitations,
although a complete description of the spectrum can only be obtained after
disentangling the non-linear relations and imposing the BRST conditions.

For the sources, we replace $A_-$ (see eq. (2.108$c$)) by ${i\over e}
(U^{^{-1}}W\widetilde\beta^{^{-1}}\widetilde\Sigma)\partial_-(\widetilde
\Sigma^{^{-1}}\widetilde\beta W^{^{-1}}U)$. We also notice here that we have
dual descriptions of QCD$_2$. In the first, valid in the perturbative region,
for high energies, we find out a non-local perturbation of the WZW action. In
terms of $W$ the perturbation is local, but at the price of a negative sign in
the na\"{\i}ve kinetic term in the $W$ action, which is appropriate to
describing the low energy (strong coupling) regime of the theory. In spite of
such different complementary descriptions, both models are integrable. In the
weak coupling regime we found the conservation laws (5.14--17). In the
case of the $W$-theory, it is not difficult to find the equations of motion,
and again derive  similar relations for the quantity
$$
I_-^W (x^-)={1\over 4\pi}(c_V+1)J_-^W(x^+,x^-)+{1\over (4\pi\mu)^2}\partial
_+\partial_-J_-^W(x^+,x^-)+{1\over (4\pi\mu)^2}[J_-^W,\partial J_-
^W](x^+,x^-) ,\eqno(5.23)
$$
with $J_-^W=W\partial_-W^{-1}$ and $\partial_+I_-^W=0$, i.e. $I_-^W$ does not
depend on $x^+$. The conservation laws are local in this formulation.

Therefore, after finding isomorphic higher charges for both formulations, we
are motivated to find their corresponding algebras, and later quantize them.

To obtain the algebra obeyed by the previously found conserved charges, it is
easier to proceed with the canonical quantization,\ref{115} obtaining first
the Poisson algebra, and later the constraints and quantum commutators of the
model. In fact, from the computation of the fermion determinant, we have an
effective bosonic action that already takes into account some quantum
corrections, namely the fermionic loops have been summed up. Therefore, the
Poisson brackets already have quantum corrections arising from fermionic
loops. This fact minimizes the possibilities of anomalies in the full quantum
definition of the charges.\ref{116} As a matter of fact, we shall see that
quantum corrections are restricted to the introduction of renormalization
constants.

We thus have to deal with the action
$$
S = - (c_V+1) \Gamma[W] - {1\over 2(4\pi \mu)^2} \int {\rm d}^2 x \,
\left[ \partial _+ (W\partial _-W^{-1})\right]^2 \quad ,\eqno(5.24)
$$
with the WZW functional given by
$$
\Gamma[W] = {1\over 8\pi} \tr\int{\rm d}^2x \,\partial ^\mu W^{-1} \partial
_\mu W + {1\over 4\pi} \epsilon ^ {\mu\nu} \tr \int _0^1 {\rm d}r\int{\rm d}^2
x \,\hat W^{-1}\dot {\hat W} \hat W^{-1}\partial _\mu
\hat W \hat W^{-1}\partial _\nu \hat W \quad .\eqno(5.25)
$$
Due to the presence of higher derivatives in the above action, it is
convenient to introduce an auxiliary field and rewrite it in the equivalent
form
$$
S= -(c_V+1) \Gamma[W] + \tr {1\over 2}\int{\rm d}^2x\,\left(-B^2+{1\over 2\pi
 \mu } \partial _+ B \partial _-WW^{-1}\right)\quad ,\eqno(5.26)
$$
where (4.1) is obtained by completing the square in the $B$-term in (5.26). The
momentum canonically conjugated to the variable $W$ is
$$
\eqalign{
\Pi^W_{ij} &= {\partial S\over \partial \partial _0W_{ij}} = -{1\over 4\pi}
(c_V+1) \partial _0W^{-1}_{ji} - {1\over 4\pi} (c_V+1)A_{ji} + {1\over 4\pi
\mu}(W^{-1} \partial _+B)_{ji}\cr
& = \hat \Pi^W_{ij} - {1\over 4\pi} (c_V +1) A_{ji}\quad ,\cr}\eqno(5.27)
$$
where the first term is obtained from the principal $\sigma$-model term in the
WZW action, the second arises from the pure WZW term, and the third from the
interaction with the auxiliary field. It is convenient to separate the WZW
contribution $A_{ij}$ to the momentum, since the new variable $\widehat\Pi^W$
is local in the original fields. The treatment of the WZ term, leading to
$A_{ij}$, on the right-hand side above, follows closely ref. [115], see
also ref. [8]. An explicit form for $A_{ij}$ cannot be obtained in terms of
local fields, but we need only its derivatives, which are not difficult to
obtain, i.e.\ref{8,115}
$$
F_{ij;kl} = {\delta A_{ij}\over \delta W_{lk}} - {\delta A_{kl}\over
\delta W_{ji}} = \partial _1 W^{-1}_{il} W^{-1}_{kj} - W^{-1} _{il} \partial
_1 W^{-1}_{kj}\quad ,\eqno(5.28)
$$
in terms of which we have the Poisson-bracket relation
$$
\eqalignno{
\left\{ \hat \Pi^W_{ij}(x), \hat \Pi^W_{kl}(y)\right\} & = - {c_V +1\over
4\pi}\left( {\delta A_{lk} \over \delta W_{ij}} - {\delta A_{ji} \over \delta
W_{kl}}\right) \cr
&= {c_V+1 \over 4\pi} \left( \partial _1 W^{-1}_{jk} W^{-1}_{li} -
\partial _1 W^{-1} _{li} W^{-1}_{jk}\right)\delta(x^1-y^1)\quad .&(5.29)\cr}
$$
The momentum associated with the $B$ field is
$$
\Pi_{ij}^B = -{1\over 4\pi\mu} (W\partial _-W^{-1})_{ji}\quad .\eqno(5.30)
$$
We can now list the relevant field operators appearing in the definition of the
conservation law (5.23), that is
$$
\eqalign{
I^W_-& = {1\over 4\pi} (c_V +1) J^W_- + {1\over (4\pi\mu)^2} \partial_+
\partial_-J^W_- - {1\over (4\pi\mu)^2} [ J^W_-, \partial_+J^W_-]\quad ,\cr
\partial _+  I^W_- &= 0 \quad .\cr}\eqno(5.31)
$$
In terms of phase-space variables, they are
$$
\eqalign{
J^W_- = \, &W\partial _-W^{-1} = - 4\pi \mu \widetilde \Pi_B\quad ,\cr
\partial_+J^W_-=&-4\pi\mu\partial_+\widetilde \Pi_B=4\pi\mu B\quad ,\cr
\partial_+\partial_-J^W_-=\,& (4\pi\mu)^2\Bigl[W\widetilde{\hat \Pi}^W-(c_V+1)
\mu \widetilde \Pi_B\Bigr] \cr
& - (4\pi \mu)\mu (c_V+1) W'W^{-1}- 8 \pi\mu B'\quad ,\cr} \eqno(5.32)
$$
where the tilde means a transposition of the matrix indices. It is
straightforward to compute the Poisson algebra. We have
$$
\left\{I^W_{ij}(t,x), I^W_{kl}(t,y) \right\}  = \left[ I^W_{kj}
 \delta_{il} - I^W_{il}\delta_{kj}\right]\delta(x^1\!-\!y^1)-\alpha\delta^{il}
\delta^{kj}\delta'(x^1-y^1)\quad,\eqno(5.33)
$$
where $\alpha = {1\over 2\pi}(c_V+1)$. The affine algebra is thus realized
while acting on the current operator, since
$$
\eqalign{
\left\{ I^W_{ij}(t,x), J^W_{-kl}(t,y) \right\}=\,& (J^W_{-kj}\delta_{il} -
J^W_{-il}\delta_{kj})\delta(x^1-y^1)\cr
& + 2\delta_{il}\delta_{kj}\delta'(x^1-y^1)\quad ,\cr
\left\{ J^W_{ij}(t,x), J^W_{-kl}(t,y^1) \right\}=\,& 0\quad .\cr}\eqno(5.34)
$$
We thus obtain a current algebra for $I_-^W$, acting on $J^W_-$ with a central
extension. We shall return to this discussion later, after consideration of
the quantization of the charge.

The Hamiltonian density can also be computed, and we arrive at the phase-space
expression
$$
\eqalign{
H_W= \, & \widetilde {\hat \Pi}^W W' + 4\pi \mu\widetilde{\hat \Pi}^W\widetilde
\Pi ^B W - \widetilde \Pi^BB' - 4\pi \mu^2 (c_V +1)(\widetilde \Pi^B)^2\cr
& - 2(c_V+1)\mu\widetilde \Pi ^B W'W^{-1} + {1\over 4\pi} (c_V+1)(W'
W^{-1})^2 +{1\over 2}B^2\quad ,\cr}\eqno(5.35)
$$
where $B'=\partial _1B\, ,\, W'=\partial _1W$; the above Hamiltonian can be
rewritten in a quadratic form in terms of the currents, although in such a
case we also have velocities, due to the appearance of the time derivatives:
$$
H_W = \alpha \left( J_1^W\right)^{2} - {1\over (4\pi \mu)^2}\left[\partial
_+^2J^W_-J^W_+ +J^W_-\partial _-\partial _+J^W_- - (\partial _+J^W_-)^2\right]
\quad ,\eqno(5.36)
$$
where $J^W_1= {1\over 2}(J^W_+-J^W_-)$ and $J_+^W = W \partial _+W^{-1}$. At
this point we can compare the model with its $\beta$ formulation. In this case
we have the action
$$
S = \Gamma[\beta] + i\mu\tr \int {\rm d}^2 x\,C_- \beta ^{-1}\partial_+
\beta + {1\over 2} \tr \int {\rm d}^2x\, (\partial _+C_-)^2 \quad .\eqno(5.37)
$$

The canonical quantization proceeds straightforwardly, and the relevant
phase-space  expressions are obtained for $J_-^\beta $ in (5.16), which, due
to the $C_-$ equation of motion, read
$$
\eqalignno{
J^\beta_- &= -4\pi \mu^2 \partial _+^{-2} (\beta^{-1} \partial _+ \beta) =
4i\pi \mu C_-\quad ,&(5.38a)\cr
\Pi_- & = \partial _+ C_-\quad ,&(5.38b)\cr}
$$
while the $\beta$-momentum is given by
$$
\widetilde{\hat \Pi}^\beta_{ji}  = {1\over 4\pi} \partial _0 \beta^{-1}_{ji} +
\mu(C_-\beta^{-1})_{ji}\quad ,\eqno(5.39)
$$
where the hat above $\Pi^{{}^\beta}$ means that we have neglected the WZW
contribution as before,\ref{115} and as a consequence
$$\left\{ \widetilde{\hat \Pi}^\beta_{ji}(t,x), \widetilde{\hat \Pi}^\beta_{lk}
(t,y)\right\} =-{1\over 4\pi}\left( \partial_1\beta^{{}^{-1}}_{_{jk}}
\beta^{{}^{-1}}_{_{li}}-\partial_1\beta^{{}^{-1}}_{_{li}}
\beta^{{}^{-1}}_{_{jk}}\right)\delta(x-y)\quad .\eqno(5.40)
$$
{}From the definition of the canonical momentum associated with $C_-$ we have
$$
\partial _+ J^\beta_- = 4i\pi \mu\Pi _{-}\quad .\eqno(5.41)
$$
The conserved charge is
$$
\eqalign{
I^\beta_- &= 4\pi \mu^2J^\beta_- + \partial_+\partial_-J^\beta_-+[J^\beta_-
,\partial_+J^\beta_-]\quad,\cr
\partial _+I^\beta_-&=0\quad ;\cr} \eqno(5.42)
$$
therefore the situation is analogous to the one we found previously by
interchanging the $ (B, \Pi_B)$ phase-space  variables with $(\Pi_-, -C_-)$
(noticing the exchanged order).

At this point the Hamiltonian might be computed. However, we will postpone
that, since we will have to compute it in terms of more appropriate currents,
making the problem easier to formulate in terms of the constraints that are
hidden in the gauge-transformation properties.

We now come to the point where we should consider the quantization of the
symmetry current (5.42). Let us consider the problem in the $\beta$ language,
since the short-distance expansion depends on the high-energy behaviour of the
theory; since the only massive scale is the coupling constant, we have to
consider the weak coupling limit. This limit is better described by the
$\beta$ action. In such a case, we need the short-distance expansion of the
current $J^\beta _-=-4\pi\mu^2\partial_+^{-2}(\beta^{-1}\partial_+\beta)$ with
itself. Since the short-distance expansion is compatible with the weak
coupling limit, where the theory is conformally invariant, Wilson expansions
can be dealt with in the usual way.

Due to the renormalization of the higher charge, we cannot give an
interpretation of the field operator $I^\beta_{ij}$ by itself, but only to an
arbitrary linear combination involving the charge and the current. In any case,
since $I^\beta_{ij}$ is a right-moving field operator, it is natural to assume,
 in view of the Poisson algebra (5.33), that it obeys an algebra given
by\ref{117,118}
$$
I^\beta_{ij}(x^-) I^\beta_{kl}(y^-) = (I^\beta_{kj}\delta_{il}-I^\beta_{il}
\delta_{kj})(y^-){1\over x^--y^-} - \alpha {\delta^{il}\delta^{kj}\over(x^-
-y^-)^2}\quad .\eqno(5.43)
$$

For $J^\beta_{-ij}$ we are forced into a milder assumption. Indeed, the
equation $\partial_+J^\beta_{-ij}=0$ would be too simple to realize the whole
problem we are considering. In such a case we would be left with unequal time
commutators for the second of eqs. (5.34). But in any case, since $I^\beta_
{ij}$ is a right-moving field operator, the equal-time requirement in the
first of eqs. of (5.34) is also superfluous, and we get an operator-product
algebra of the type
$$
I^\beta_{-ij}(x^-)J^\beta_{-kl}(y^+,y^-)=(J^\beta_{-kj}\delta_{ij}-J^\beta_
{-il}\delta_{kj})(y^+,y^-){1\over x^--y^-}+2{\delta^{il}\delta^{kj}\over(x^-
-y^-)^2}\quad .\eqno(5.44)
$$
The second equation in (5.34) cannot be taken at arbitrary times, since $J^
\beta_-$ depends on both $x^+$ {\it and} $x^-$. Moreover, if $J^\beta_-$ were
purely right-moving, the second equation would imply, for unequal times, that
it is a trivial operator.

Some conclusions may be drawn for $J^\beta_-$. As we stressed above, $\partial
_+J^\beta_-$ cannot be zero\newfoot{$^{12)}$}{In the case where $J^\beta_-$ is
left-moving, we expect further modifications of the commutators.}, in the full
quantum theory; however, in view of (5.44), we conclude that left $(-)$
derivatives of this current are primary fields,\ref{118} since
$$
I^\beta_{ij}\partial_+^nJ^\beta_{-kl}={1\over x^--y^-}\left(\partial_+^nJ^\beta
_{-il}\delta_{kj}-\partial_+^nJ^\beta_{-kl}\delta_{il}\right)\quad .\eqno(5.45)
$$

Therefore, we expect an affine Lie algebra for $I^\beta(x^+)$, and
$\partial_+^n J_-^\beta$ should be primary fields depending on parameters
$x^-$.

Such an underlying structure is a rather unexpected result, since
it arose out of a non-linear relation obeyed by the current, which
can be traced back to an integrability condition of the model. Moreover, the
theory has an explicit mass term -- although free massive fermionic theories as
well as some off-critical perturbations of conformally invariant theories in
two
dimensions may contain affine Lie symmetry algebras.

On the current itself there is now a realization of such an algebra in the
right-moving sector.\ref{119}
\vskip .5cm
\penalty-1000
\noindent {\it Reobtaining the $U(1)$ case}
\vskip .5cm
\nobreak
\noindent Again, as in the case of the $\beta$-Lagrangian, we may study the
$U(1)$ limit by writing
$$
W = \E^{2i\sqrt \pi \Xi}\quad ,\eqno(5.46)
$$
to find
$$
L = - {1\over 2} \left(\partial _\mu \Xi\right)^2  + {\pi \over 2e^2} \left(
\partial^2\Xi\right)^2 \quad .\eqno(5.47)
$$

The $\Xi$ propagator is
$$
D_\Xi = {e^2\over \pi} {1\over p^2\left(p^2 - {e^2\over \pi}\right)} = -{1\over
p^2} + {1\over p^2-{e^2\over \pi}} \quad ,\eqno(5.48)
$$
which describes again a massive excitation corresponding to the previous
$\Sigma$ field (see eqs. (2.109--111)) and a massless negative metric
excitation.
\vskip 1cm
\penalty-1000
\noindent {\bf 5.3 Constraint structure of the theory}
\vskip .5cm
\nobreak
\noindent Consider the effective action
$$
S_{\rm eff} = \Gamma[\widetilde g] - (c_V+1)\Gamma[\Sigma] + \Gamma[\beta] -
{1\over 2}\mu^2 \int {\rm d}^2 x\, [\partial_+^{-1}(\beta^{-1}
\partial _+\beta)]^2 + S_{{\rm ghosts}}\quad .\eqno(5.49)
$$
Let us start by first coupling the fields  $(\widetilde g, \Sigma, {\rm
ghosts})
$ to external gauge fields
$$
A^{^{\rm ext}}_-={i\over e}V_{_{\rm ext}}\partial_-V_{_{\rm ext}}^{-1}\quad ,
\quad A^{^{\rm ext}}_+={i\over e}U_{_{\rm ext}}^{-1}\partial_+U_{_{\rm ext}}
\quad. \eqno(5.50)
$$

Such a coupling may be obtained by substituting each WZW functional as
prescribed in eq. (2.87). In the case of ghosts one has to perform a chiral
rotation, as in the discussion following eq. (2.143). Therefore, after such a
procedure and using again the invariance of the Haar measure to substitute $U_
{_{\rm ext}}(\widetilde g)V_{_{\rm ext}} \to (\widetilde g,\sigma)$, one finds
the effective action
$$
S_{\rm eff}(A)=\Gamma[\widetilde g]-(c_V+1)\Gamma[\Sigma]+S_{\rm ghosts}+[1-
(c_V+1) + c_V]\Gamma [U_{_{\rm ext}}V_{_{\rm ext}}]\quad .\eqno(5.51)
$$
Vanishing of the total central charge (i.e. the vanishing coefficient of the
last term above) tells us that the action does not depend on the external gauge
fields. Nevertheless, from minimal coupling the effective action can also be
written as
$$
\eqalignno{
S_{\rm eff}(A)=\,&S_{\rm eff}(0)-{1\over 4\pi}A^{^{\rm ext}}_+\left[ie
\widetilde g\partial_-\widetilde g^{-1}-ie(c_V+1)\Sigma\partial _-\Sigma^{-1}+
J_-({\rm ghost})\right]\cr
&-{1\over 4\pi} A^{^{\rm ext}}_-\left[ ie\widetilde g^{-1}\partial_+\widetilde
g - ie(c_V+1) \Sigma ^{-1} \partial _+\Sigma + J_+ ({\rm ghost}) \right] +
{\cal
O}(A^2)\, .&(5.52)\cr}
$$

Functionally differentiating the partition function once with
respect to $A^{^{\rm ext}}_+$ and separately with respect to
$A^{^{\rm ext}}_-$, and putting $A^{^{\rm ext}}_\pm =0$ we find the
constraints\ref{46,53}
$$
\eqalignno{
i\widetilde g\partial_-\widetilde g^{-1}-i(c_V+1)\Sigma\partial_-\Sigma^{-1}
+J_-({\rm ghosts})& \sim 0\quad ,&(5.53)\cr
i\widetilde g^{-1}\partial_+\widetilde g-i(c_V+1)\Sigma^{-1}\partial_+\Sigma
+J_+({\rm ghosts})& \sim 0\quad ,&(5.54)\cr}
$$
leading to two BRST charges $Q^{(\pm)}$ as discussed in ref. [53], which are
nilpotent. Therefore we find two first-class constraints.

The field $A^{^{\rm ext}}_+$ can also be coupled to the field $\beta$ instead
of
$\widetilde g$, since the system $(\beta, \Sigma, {\rm ghosts})$ has vanishing
central charge too. In such a case we have to disentangle the non-local
interaction considering instead of the third and fourth terms in (5.49), the
$\beta$ action
$$
S(\beta) = \Gamma[\beta]+ \int {\rm d}^2 x\, {1\over 2}(\partial _+C_-)^2 +
i\int {\rm d}^2 x \,\mu\, C_-\beta^{-1} \partial _+\beta\quad .\eqno(5.55)
$$

We make the minimal  substitution $\partial _+ \to\partial _+-ieA^{^{\rm ext}}
_+$, repeating the previous arguments for the $(\beta, \Sigma, {\rm ghosts})$
system, and we now arrive at the constraint (the minus gauging is not an
invariance if one includes the $\beta$ system):
$$
\beta \partial _-\beta ^{-1}+4i\pi\mu\beta C_-\beta^{-1} - i(c_V+1)\Sigma
\partial _-\Sigma^{-1} + J_-({\rm ghost}) \sim 0\quad .\eqno(5.56)
$$

One could na\"{\i}vely expect that, by repeating the previous arguments, one
obtains a system having a new set of first-class constraints. But if we
consider instead the equivalent system of constraints defined by the first set
(5.53), together with the difference of the $(-)$ currents, i.e. (5.54) and
(5.56) as given by
$$
\Omega_{ij}=(\beta \partial _-\beta^{-1})_{ij} + 4i\pi \mu (\beta C_-\beta
^{-1})_{ij} - (\widetilde g \partial _-\widetilde g^{-1})_{ij}\quad
,\eqno(5.57)
$$
one readily verifies that the latter cannot lead to a nilpotent
BRST charge due to the absence of ghosts. Therefore, it must be treated as a
second-class constraint.\ref{55} The Poisson algebra obeyed by $\Omega_{ij}$ is
$$
\eqalignno{
\left\{ \Omega_{ij}(t,x), \Omega_{kl}(t,y)\right\}&=(\widetilde \Omega_{il}
\delta_{kj} - \widetilde \Omega_{kj}\delta _{il})(t,x) \delta(x-y) + 2 \delta_
{il}\delta_{kj} \delta'(x-y)\quad ,&(5.58)\cr
{\widetilde \Omega} & = \widetilde g \partial _- \widetilde g^{-1}+
\beta\partial_-\beta^{-1}+4i\pi\mu\beta C_-\beta^{-1}\quad .&(5.59)\cr}
$$
(Notice the change of sign in $\widetilde \Omega$.) Using the above, we can
thus define the undetermined velocities, and no further constraint is
generated.

The fact that the theory possesses second-class constraints is very
annoying, since these cannot be realized by the usual cohomology construction.
Therefore, instead of building a convenient Hilbert space, one has to modify
the dynamics, since the usual relation between Poisson brackets and commutators
is replaced by the relation between Dirac brackets and commutators.

Nevertheless, as we will see, several nice structures unravelled so far remain
untouched after such a harsh mutilation.  Indeed, we shall see that there
is a rather deep separation between the ``right" currents, obeying equations
analogous to those written so far, and the ``left" currents, which will obey a
modified dynamics, due to the second-class constraints.

As a consequence of the definition of the canonical momenta, eq. (5.39), the
constraints have a simpler phase-space formulation, and are given by
$$
\Omega_{ij}=4\pi (\beta \widetilde {\hat \Pi}^\beta)_{ij} + \partial _1\beta
\beta^{-1} - 4\pi (\widetilde g \widetilde {\hat \Pi}^{\tilde g})_{ij}-\partial
_1\widetilde g \widetilde g^{-1}\quad ,\eqno(5.60)
$$
which has actually been used to compute (5.58). Notice that the structure of
the right-hand side of the phase-space expression is rather simple. Indeed, the
$C_-$ field just redefines the momentum associated with $\beta$, and the above
constraints are analogous to those appearing in the description of non-Abelian
chiral bosons,\ref{119} i.e. WZW theory with a constraint on a chiral current.
It follows that the Poisson algebra is very simple. Indeed, one
obtains\ref{119}
$$
\eqalign{
\left\{ \Omega_{ij}(x) , \Omega_{kl}(y)\right\}=\,& 16\pi\delta_{il}\delta_{kj}
\delta'(x^1\!-\!y^1) + 4\pi \Bigl[ (4\pi \beta \widetilde {\hat \Pi}^\beta +
\beta'\beta^{^{-1}} \!+\! 4\pi \widetilde g \widetilde {\hat \Pi} ^{\tilde g}\!
 +\!\widetilde g' g^{^{-1}})_{kj}\delta _{il}\cr
& -  (4\pi \beta \widetilde {\hat \Pi}^\beta +
\beta'\beta^{-1} + 4\pi \widetilde g \widetilde {\hat \Pi} ^{\tilde g} +
\widetilde g' g^{-1})_{il}\delta _{kj}\Bigr] \delta(x^1-y^1)\cr
=\,& 16\pi \delta_{il} \delta_{kj} \delta'(x^1-y^1) +  8\pi [ j_{-kj}
\delta_{il} - j_{-il} \delta_{kj} ] \delta (x^1-y^1)\quad ,\cr}\eqno(5.61)
$$
where $j_-= 4\pi \beta \widetilde {\hat \Pi}^\beta + \beta'\beta^{-1}$
satisfies the Poisson algebra
$$
\{ j_{-ij}, j_{-kl}\} = 8\pi \delta_{il} \delta_{kj} \delta'(x-y) + 4\pi
(j_{-kj} \delta_{il} - j_{-il}\delta_{kj})\delta(x-y)\quad .\eqno(5.62)
$$
The above expression also defines the $Q$ matrix
$$
Q_{ij;kl} = \left\{ \Omega_{ij}(x), \Omega_{kl}(y)\right\}\Big\vert_{{\rm equal
\, time}}\quad ,\eqno(5.63)
$$
which is not a combination of constraints, and therefore no further constraint
is generated by the Dirac algorithm. The inverse of the Dirac matrix is not
difficult to compute and we have the expression\ref{119}
$$
\eqalignno{
\left(Q^{-1}\right)_{ij;kl}=\,&{1\over 32\pi}\delta_{il}\delta_{kj}
\varepsilon(x)+\cr
& + {1\over 64\pi} (\delta_{il} j_{-jk} - \delta_{jk} j_{-li})\vert x \vert+\cr
& + {1\over 128\pi} (\delta_{ia} j_{-jb}-\delta_{jb} j_{-ai})
(\delta_{al} j_{-bk}-\delta_{bk}j_{-la}){1\over 2}x^2\varepsilon (x)+
&(5.64)\cr
& + {1\over 256\pi} (\delta_{ia}j_{-jb}-\delta_{jb}j_{-ai}) (\delta_{ca}j_{-bd}
-\delta_{bd} j_{-ac}) (\delta_{lc} j_{-dk}-\delta_{dk} j_{-cl}) {1\over 3} x^3
\varepsilon (x) + \cdots ,\cr}
$$
where $x$ is the space component of $x^\mu$.

The next step consists in replacing the Poisson brackets by Dirac
brackets. Thus we have to compute the Poisson brackets of the relevant
quantities with the constraints. We use
$$
\{A, B \}_{_{DB}} = \{ A, B \} _{_{PB}} - \{ A, \Omega_\alpha \} _{_{PB }}
Q_{\alpha\beta}^{-1} \{ \Omega_\beta, B \}_{_{PB}}\quad .\eqno(5.65)
$$

We will see that functions of
$$
J^\beta_+ = \beta^{-1}\partial _+\beta =- 4\pi \widetilde \Pi \beta +
\beta^{-1} \beta' + 4i\pi \mu C_-\quad \eqno(5.66)
$$
commute with $\Omega_\alpha$, and that their Dirac brackets coincide with
their Poisson brackets.

Canonical quantization through the Dirac formulation of the $\beta$ sector is
achieved by the formulae (5.38$a,b$), (5.39) and (5.41), from which we
obtain the phase space expression
$$
{1\over 4\pi} \beta^{-1}\partial _\pm\beta = - \widetilde {\hat \Pi}^\beta
\beta \pm {1\over 4\pi}\beta^{-1}\beta' + i\mu C_-\quad .\eqno(5.67)
$$

It is useful, in view of (5.58), to consider the combination
$$
{1\over 4\pi}\partial_-\beta \beta^{-1} = - \beta \widetilde {\hat\Pi}^\beta -
{1\over 4\pi} \beta' \beta^{-1} +i\mu\beta C_-\beta^{-1}\quad ;\eqno(5.68)
$$
or also, aiming at the expression of the constraint (5.58), which contains the
$C_-$ field, we have
$$
\beta \partial _-\beta^{-1} + 4i\pi \mu C_-\beta^{-1} = - 4\pi \beta
\widetilde {\hat \Pi}^\beta -  \beta' \beta^{-1}\quad .\eqno(5.69)
$$
Thus, in terms of phase-space variables the constraint is given by (5.60).
Using the above phase-space expressions we find that
$$
\eqalignno{
\left\{ J^\beta_-, \Omega \right\} &= 0\quad ,&(5.70)\cr
\left\{ [J^\beta_-, \partial _+J^\beta_-], \Omega \right\} &= \left\{ [C_-,
\Pi_-], \Omega \right\} = 0\quad .&(5.71)\cr}
$$
For $\left\{ \partial _+\partial _-J^\beta_-, \Omega \right\}$ we first have to
compute
$$
\eqalignno{
\partial _+\partial _-J^\beta_- & = \partial _+^2J^\beta_- - 2 (\partial _+
J^\beta_-)'\quad ,&(5.72)\cr
& =4\pi\mu^2 \beta^{-1}\partial_+ \beta - 2 (\Pi_-)'\quad .&(5.73)\cr}
$$
We use the fact that $\{\Pi'_-, \Omega\}=0$ and we are left with
$$
\beta ^{-1} \partial _+\beta =  -{4\pi} \widetilde {\hat \Pi}^\beta \beta +
 \beta ^{-1}\beta' + 4i\pi \mu C_-\quad .\eqno(5.74)
$$
Using now $\{ C_-, \Omega \} =0$,  we just have to consider
$$
j_{+ij} = \left( - 4\pi\widetilde {\hat \Pi}^\beta \beta +  \beta ^{-1}\beta'
\right)_{ij} \quad .\eqno(5.75)
$$
However, since $\{j_+,j_-\}=0$ we have $\{j_+,\Omega\}=0$! As a conclusion,
for the objects relevant to us, the Dirac algebra is the same as the Poisson
algebra! This is a non-trivial result, because it holds even though, due to
(5.60), the Dirac algebra obeyed by $\hat\Pi^\beta$ and $\beta$ changes
drastically, especially if we take into account the expression of the inverse
Dirac matrix (5.64), which is non-local and has an infinite number of terms!

In the duality transformation relating the $\beta$ and the $W$
fields, we also find interesting relations arising out of the constraint
structure of the theory. First let us perform a more detailed analysis of the
ghost structure. Going back to the transformations defined by (5.20) we have
the factor $(\det \partial _+ \, \det \partial _-)^{c_V}$ left out, which
contributes as
$$
{\cal Z}^{{\rm gh}'}=\int{\cal D}b'_{++}{\cal D}b'_{--}{\cal D}c'_+{\cal D}c'_-
\,\E^{-\tr\int{\rm d}^2 x\,(b'_{++}\partial_-c'_-+b'_{--}\partial _+c'_-)}
\quad .\eqno(5.76)
$$

The coupling of a subset of fields to an external gauge potential written in
the
form (5.50), can be made as in the usual way. If such a set has a vanishing
total central charge, the partition function does not depend on the gauge
potential, and we are led to constraints again. With the partition function
written in the $W$ language as in (5.26), and taking into account all
appropriate ghosts, we have various self-commuting constraints. Some of them,
such as
$$
\eqalignno{
J_{\tilde g} - (c_V+1)J_\Sigma + J_{{\rm ghost}} & \sim 0\quad ,&(5.77)\cr
J_{\tilde \beta} - (c_V+1)J_\Sigma + J_{{\rm ghost}} & \sim 0\quad,&(5.78)\cr}
$$
are the same as before, with the advantage that now $\widetilde \beta$ is a
pure WZW field, so that it can be simply identified with $\widetilde
g$, without further consequences. However, further constraints involving also
the $W$ field arise, such as
$$
J_+^{\tilde g} - (c_V+1)J_+^W + J_{+{\rm ghost}}  \sim 0\quad ,\eqno(5.79)
$$
so that  we have, as a consequence, the non-trivial second-class constraint
$$
J_+^\Sigma - J_+^W \sim 0 \quad ,\eqno(5.80)
$$
or, more explicitly,
$$
(c_V+1) \Sigma^{-1}\partial _+ \Sigma - (c_V + 1) W^{-1} \partial _+W + {1\over
\mu} W^{-1} \partial _+ B W =0 \quad . \eqno(5.81)
$$

We have proceeded  as in the $\beta$ formulation, but with the interaction of
the $A_-^{^{\rm ext}}$ field with the $W$, while in the (dual) $\beta$ case we
considered $A_+^{^{\rm ext}}$.

The phase-space expression is given in the formula
$$
\Omega ^{W,\Sigma} = - \widetilde {\hat \Pi}^WW + {1\over 4\pi} W^{-1} W' +
\widetilde {\hat \Pi} ^\Sigma\Sigma-{1\over 4\pi}\Sigma^{-1}\Sigma'\sim 0\quad
,\eqno(5.82)
$$
and resembles the $\beta$ formulation (see (5.60)). However, if we now
substitute the $B$ field from the constraint (5.80) back into the action we
find a non-local term. This means that while in the $\beta$ formulation, which
is non-local at the beginning, we end up with a local action after inserting
the constraint back, in the $W$ formulation, which is local at the beginning,
we end up with a non-local action; another feature of duality in both
formulations.

Keeping the Dirac algebra in mind, we substitute back the configuration-space
constraints into the action, maintaining  the phase-space  structure. In such a
case, using (5.57) and (2.78), we redefine $\beta g\equiv P\,,\,\beta=Pg^{-1}$,
 and find the effective action
$$
\eqalign{
S =\, &\Gamma[P] - {1\over 2\pi} g^{-1}\partial_+ g g^{-1}\partial_- g  -
{1\over 4\pi} P^{-1}\partial_+ P P^{-1}\partial_- P  + {1\over 4\pi} P^{-1}
\partial_+ P g^{-1}\partial_- g\cr
 & + {1\over 4\pi} P^{-1}\partial_- P g^{-1}
\partial_+ g + {1\over 4\pi} \partial_- g g^{-1} \partial_+ P P^{-1} -
{1\over 2\pi} \partial_- g g^{-1} P g^{-1} \partial_+ g g ^{-1} P\cr
& + {1\over 2(4\pi\mu)^2}\left[\partial _+\left\{ gP^{-1} g \partial _-
(g^{-1} P g^{-1})\right\}\right]^2 \quad .\cr}\eqno(5.83)
$$

The equation of motion/conservation law (5.18$a$) still holds, as previously
proved. From action (5.83) we can find the equations of motion. Notice that the
final action is a WZW theory off the critical point, a principal
$\sigma$-model, and current--current-type interactions between them.

For the dual formulation a further interesting structure arises. The constraint
is now
$$
\partial _+B = -\mu(c_V + 1) W \Sigma ^{-1} \partial _+(\Sigma W^{-1})
\quad .\eqno(5.84)
$$
As with to the above, we use (5.84) and (2.6) to introduce $S = W\Sigma$,
replacing the $W$ field. It is interesting enough to note that it is now the
dual formulation that is non-local due to the presence of the $B$ field. We
again arrive at the WZW theory for $S$, a principal $\sigma$-model term for
$\Sigma$, current--current-type interactions, and principal $\sigma$-model
terms for $S$. The latter are such that the (wrong) sign of the principal
$\sigma$ term in $\Gamma[S]$ changes, and we arrive at the WZW model with a
relative minus sign, or $\Gamma[S^{-1}]$!

However, the standard procedure for dealing with the constraints is to
substitute the phase-space expressions in the Hamiltonian. But in such a case,
the constraint (5.60) does not depend upon $C_-$, and leads just to a
connection
between the right-moving current of the $g$ sector, the left-moving current
being untouched by such a relation! Therefore, still in the present case, where
we have witnessed the appearance of second-class constraints, their main role
was to ensure the positive metric requirement, as we have seen by means of the
change of sign of the WZW action in the dual formulation. A rich algebraic
structure arises in both ($\beta$ and $W$) formulations of the theory\ref{47}.
\vskip 1cm
\penalty-1000
\noindent {\bf 5.4 Spectrum and comparison with the $1/N$ expansion}
\vskip .5cm
\nobreak
\noindent
Having recognized the role played by the $\beta$ action, we pass to a
discussion of the spectrum of the theory. The first step towards understanding
the model was taken by 'tHooft, proving that the bound states form a Regge
trajectory, valid for the weak coupling case (heavy quarks). Later,
Steinhardt\ref{22} studied the strong coupling case (light quarks) finding
the baryon as the soliton of a generalized sine-Gordon interaction.

Here we do not intend to provide a definite answer to such a complex question,
but some directions may be outlined from the computations performed. Indeed,
we have an appropriate formulation to deal separately with the two regimes: the
weak-coupling regime described by the $\beta$ action may be discussed
perturbatively. We will see that in the large-$N$  limit the relevant mass
parameter  is the one defined by 'tHooft, and we arrive at a possibility of
computing the exact mass spectrum, once the complicated constraint structure
is disentangled.

In order to understand the question concerning the spectrum, we first have to
know which is the mass of the simplest excitation, or the mass parameter
characterizing the theory. We thus consider the action
$$
S[\beta] = \Gamma[\beta] + {1\over 2} \mu^2 \int {\rm d}^2 x \left[
\partial _+^{-1} (\beta^{-1} \partial _+\beta)\right]^2\quad ,\eqno(5.85)
$$
and write a background quantum splitting for the $\beta$ field as
$$
\beta = \beta_0 \E^{i\xi}\quad ,\eqno(5.86)
$$
after which we have the background quantum splitting of the action up to
second order in the quantum field $\xi$. However, we have to be careful since,
in the large-$N$ limit, the second term is the zeroth-order Lagrangian, from
which we suppose that the $\xi$ field acquires a lowest mass $m_1$ to be
computed. The WZW term splits as
$$
\Gamma[\beta] = \Gamma[\beta_0] + {1\over 2} \int {\rm d}^2 x \,\,\beta_0^{-1}
\partial _\mu \beta_0 \, \xi \flex \partial _\nu \xi \,(g^{\mu\nu} + \epsilon
^{\mu\nu})\quad .\eqno(5.87)
$$

Using the fact that $\Gamma[\beta]$ is at the critical point, it is not
difficult to compute the $\beta_0^{-1}\partial _\mu\beta_0$ two-point function
at the one-loop order. We have the zeroth-order contribution from the second
term of (5.85), and the one-loop contribution, which leads to the result
$$
\beta^{-1}\partial_+\beta {\mu^2\over p_+^2} \beta^{-1}\partial_+\beta - N
{p_\mu p_\nu\over p^2}(g^{\mu\rho} + \epsilon^{\mu\rho})(g^{\nu\sigma} +
\epsilon^{\nu\sigma}) F(p)\beta^{-1}\partial_\rho\beta \beta^{-1} \partial_
\sigma \beta \quad ,\eqno(5.88)
$$
where
$$
F(p) = {1\over 2\pi}\sqrt{{p^2-4m_1^2\over p^2}}\ln {\sqrt{-p^2+4m_1^2} +
\sqrt{-p^2}\over\sqrt{-p^2+4m_1^2}-\sqrt{-p^2}}-{1\over\pi}\quad .\eqno(5.89)
$$
For $p^2=m_1^2$, we find that
$$
\beta_0^{-1}\partial_+\beta_0\, \beta_0^{-1}\partial_+\beta_0 {1\over p_+^2}
\left[ \mu^2 - 4 Nm_1^2 F(m_1^2)\right]\quad .\eqno(5.90)
$$

The zero of the two-point function contribution to the action is at
$$
m_1^2 = f e^2 N = f \alpha\quad ,\eqno(5.91)
$$
where $f$ is a numerical constant, in accordance with 'tHooft's
results.\ref{14}

That the second term of (5.85) has an extra factor $N$ arises from
the fact that the fermion loops are suppressed by a factor $1/N$. Since the
fermion loops contribute with a WZW functional, while the $\mu$ term stems
from the gauge-field self-interaction (see eqs. (2.108)) the factors of $N$
are correct. Moreover, it is exactly the given assignment that is compatible
with the planar expansion. Finally, we have to quote the fact that 'tHooft's
analysis for the bound state $\overline\psi\gamma_+\psi$ leads to a
Bethe--Salpeter equation compatible with the previous results, the methods
following closely his analysis.

More detailed information about the spectrum of the theory can be obtained
from the Hamiltonian formulation. From the action
$$
S= \Gamma[\beta] + \int {\rm d}^2 x {1\over 2} (\partial_+C_- )^2 + i\int
{\rm d}^2 x \mu C_-\beta^{-1}\partial _+\beta\quad ,\eqno(5.92)
$$
we obtain the canonical momenta
$$
\eqalignno{
\widetilde {\hat \Pi}^\beta & = {1\over 4\pi} \partial _0\beta^{-1} + i\mu
C_-\beta^{-1}\quad ,&(5.93)\cr
\Pi_-& = \partial _+C_-\quad ,&(5.94)\cr}
$$
and the Hamiltonian density
$$
\eqalignno{
H = \, &{1\over 2}\Pi_- (\Pi_--2C') - 2\pi (\widetilde{\hat \Pi}^\beta\beta)^2
-
{1\over 8\pi} (\beta^{-1}\beta')^2 \cr
& + 4\pi \mu \widetilde {\hat \Pi}^\beta\beta C_--2\pi\mu^2C_-^2-\mu\beta^{-1}
\beta'C_-\quad . &(5.95)\cr}
$$

The important currents are
$$
\eqalignno{
J^\beta_+&=\beta^{-1}\partial_+\beta =-4\pi\widetilde {\hat \Pi}^\beta \beta +
\beta^ {-1} \beta' + 4\pi \mu C_-\quad ,&(5.96)\cr
j^\beta_-&=\beta \partial _-\beta^{-1} + 4\pi \mu \beta C_-\beta^{-1} = 4\pi
\beta \widetilde {\hat \Pi}^\beta + \beta'\beta^{-1}\quad ,&(5.97)\cr}
$$
in terms of which the Hamiltonian density reads (notice that $j^\beta_-$ is
not related to $J_-^\beta$, eqs. (5.16) and (5.38a)):
$$
H_\beta = -{1\over 16\pi} \left[ \left( J_+^{\beta}\right)^2 + \left( j_-^
{\beta}\right)^2 \right] - {1\over 2} \mu J^\beta_+ C_- +
\pi \mu^2 C_-^2 + {1\over 2} \Pi\left(\Pi_- - 2C'\right)\quad .\eqno(5.98)
$$

{}From the previously discussed constraint structure (5.60), the current $j_-$
is
related to the free right-moving fermion current $j_-^g=g\partial_- g^{-1}$,
and
we will drop it in the discussion of the spectrum for $\beta$. Moreover, from
the Sugawara construction of the Virasoro algebra, in terms of the affine
algebra generators, we know that the Sugawara piece $H_{_+}=\!{-1\over
16\pi}(J_+^\beta)^2$ acquires a factor $ (c_V + 1)^{-1}\, {\mathop{=}\limits^
{SU(N)}} \,(N+1)^{-1}$ in the quantum theory. The $C_-^2$ terms are not known,
since the $C_-$ equation of motion is not easily solvable. Nevertheless, in
terms of $C_-$ and its conjugate momentum, the Hamiltonian is quadratic. If we
take it for granted that the zero-mode term is just the squared momentum,
moreover neglecting the $C_-J_+$ interaction, the Hamiltonian eigenstates have
masses $m_n$ obeying the Regge behaviour
$$
m_n^2 \sim n \, m_1^2 \quad .\eqno(5.99)
$$

Corrections to this equation can be obtained using a large-$N$  expansion for
the field $C_-$, a procedure that is at least possible upon considering the
large-$N$  limit of (5.92). There are in fact further eigenstates, and for a
choice of the normalization of the fields one finds asymptotically a low mass
eigenstate, compatible also with Steinhardt's baryon.\ref{22}
\vskip 1cm
\penalty-1000
\noindent {\bf 5.5 Integrability conditions and Calogero-type model}
\vskip .5cm
\nobreak
\noindent The issue of integrability has been discussed for a long time in the
literature. It started almost a century ago, with the observations of Korteweg
and de Vries\ref{120}, and today it is an area of research in itself,
including several applications.

In two-dimensional space-time the situation is simple. In fact, the
Coleman--Mandula theorem\ref{8,121} prevents the existence of higher
conservation laws in more than two dimensions. The theorem states that for
dimensions higher than two, the most general invariance group of a non-trivial
field theory is the product of the Poincar\'e group and an internal symmetry
group. Allowing anticommutators, we can at most have a supersymmetry algebra.
But in two dimensions there are integrable systems, with an infinite number of
conservation laws.\newfoot{$^{13)}$}{It is not yet established how one might
rephrase the property of integrability for real higher-dimensional systems. In
fact, due to the severe constraints imposed it should not be generally true as
higher conservations laws. However, several simplified models, important for
physical understanding, are integrable in a weak sense. In higher dimensions,
we can quote the self-dual Yang--Mills theory, as well as the $N=4$
supersymmetric Yang--Mills theory in four dimensions, and gravity in ten
dimensions\ref{122}. In such cases, however, the integrability properties,
although implying constraints, do not lead to a solution of the theory, or
imply the existence of higher conserved charges.}

Non-linear $\sigma$ models are two-dimensional counterparts of
four-dimensional  Yang--Mills theories, sharing several desired
properties.\ref{8} When such models are defined on a symmetric space they are
classically integrable --- as are their supersymmetric extensions. Such a fact,
when not spoiled by quantum anomalies, leads to a solution of the theory at the
$S$-matrix level.

It is natural to ask whether the two-dimensional counterpart of the Yang--Mills
theory is also integrable or not, and whether the $(3+1)$-dimensional
Yang--Mills theory itself displays, in some limit, such a property. The answer
to both questions seems to be positive, although most details should still be
worked out. In the case of two-dimensional Yang--Mills theory such an
information can be derived in two ways; first the pure-gauge model is
equivalent to a system of non-relativistic fermions, or to the $c=1$ matrix
model, a property which seems to hold true where there is fermionic matter to
begin with. Moreover, the bosonization of the model reveals the existence of a
Lax pair.

As far as the second question is concerned, namely concerning the
$(3+1)$-dimensional Yang--Mills theory, one has to consider the high-energy
scattering amplitudes. Such a problem was studied by several authors, and the
outcome is an effective two-dimensional theory describing the transverse
components of the momentum transfer. Verlinde and Verlinde\ref{123} arrived at
an effective two-dimensional model of two matrices, and also proved that the
effective interaction is described by the Lipatov vertex.\ref{124} Such a
vertex, on the other hand, leads to amplitudes obeying Hamiltonian equations
which seem to factorize into simple Hamiltonian systems,\ref{125-127} being
in principle integrable and solvable by the Bethe ansatz technique.\ref{128}

Although all such indications have to be further analyzed, and several details
must be understood, the possibilities of studying such models thus have a wide
avenue ahead, and high-energy scattering might prove to be equivalent to an
exactly soluble model, around which perturbative expansions can be performed.

We have seen that pure QCD$_2$ in the Coulomb gauge $(A_0=0)$, on a cylinder
whose cross section has length $L$, has a Hamiltonian given by eq. (3.102),
and that the Gauss law, corresponding to the $A_0$ equation of motion, was
written in eq. (3.105). We also defined the Wilson line in eq. (3.104), and
the ``dressed"  electric field in eq. (3.103). Due to the Gauss law, the
dressed electric field $V(x)$ is a constant, ($\partial_1V(x)=0$), therefore
$V(x)=V(0)$. We defined also the Wilson loop around the cylinder as
$W_{\rm cyl}= W[0,L]$ (now we specify $W_{\rm cyl}$), and we immediately
verified, writing $V(0)= V(L)$, that
$$
[W_{\rm cyl}, \dot A_1(0)]=0 \quad.\eqno(5.100)
$$

The time derivative of the cylindric Wilson loop $\dot W_{\rm cyl}$ can also be
computed from the known formula for the derivative of the exponential of an
algebra-valued object, that is
$$
\dot W_{\rm cyl}= ie \int {\rm d}x\, W[0,x] \dot A_1(x) W[x,L] = ie \int
{\rm d}x\,V(x)= ie LV(0)\quad,\eqno(5.101)
$$
or in terms of $W_{\rm cyl}$ and $\dot A_1$:
$$
\dot W_{\rm cyl} = ie L W_{\rm cyl}\dot A_1(0)=ieL\dot A_1(0)W_{\rm cyl}\quad,
\eqno(5.102)
$$
which together with (5.100) implies $[W_{\rm cyl}, \dot W_{\rm cyl}] =0$.

Rewriting $\dot A_1$ in terms of the cylindric Wilson loop, one obtains the
Hamiltonian (3.108), that is a one-dimensional unitary matrix model.

Moreover, we notice also that as matrices, $W_{\rm cyl}$ and $\dot W_{\rm cyl}$
commute, therefore the space of states only contains singlets! This means that
we have to deal only with the eigenvalues of the cylindric Wilson variable
$W_{\rm cyl}$.

On the other hand, if we consider the Hamiltonian\ref{88}
$$
H = \tr \left[ -{1\over 2}{\delta^2\over \delta \phi_{ij}^2} + U(\phi)\right]
\quad ,\eqno(5.103)
$$
where $U(\phi)$ depends only on the eigenvalues, that is it does not depend on
angular variables, we obtain eigenfunctions of the type
$$
\psi = {\chi\over \Delta(\lambda)} \quad ,\eqno(5.104)
$$
where $ \Delta(\lambda) = \prod\limits_{i<j}(\lambda_i<\lambda_j)$ is the
Vandermonde determinant. In such a case the expectation of the Hamiltonian is
$$
\int\prod_i{\rm d}\lambda_i\,{\cal D}\Omega\sum\left({\partial\psi\over\partial
\phi_{ij}}\right)\Delta^2(\lambda_i)={\rm volume}\int\prod_i{\rm
d}\lambda_i\sum
\left({\partial\psi\over\partial\lambda_i}\right)^2\Delta^2(\lambda_i)\quad ,
\eqno(5.105)
$$
where we used ${\cal D}\phi = {\cal D}\Omega{\cal D}\lambda\Delta^2(\lambda)$.
Writing $\psi$ in terms of $\chi$ we find $\int\prod{\cal D}\lambda_i\sum
\left( {\partial \chi\over \partial \lambda_i}\right)^2$ and therefore may use
as Hamiltonian
$$
H = \sum_i \left[ - {1\over 2} {\partial^2\over \partial \lambda_i^2} +
U(\lambda_i)\right] \quad ,\eqno(5.106)
$$
for wave functions $\chi$, or else for $\psi$
$$
H= - {1\over 2\Delta(\lambda)} \sum {\partial^2\over \partial\lambda_i^2}
\Delta(\lambda) + U(\lambda_i) + \sum {\pi_{ij}^2 + \overline \pi_{ij}^2 \over
(\lambda_i -\lambda_j)^2} \quad ,\eqno(5.107)
$$
where $\pi_{ij}$ and $\overline \pi_{ij}$ are the generators of (left and
right)
rotations. Due to the presence of the Vandermonde determinant, which is
antisymmetric under the exchange of two ``particles", the problem boils down
to the theory of $N$ fermions in the non-relativistic limit.

The introduction of a Wilson line along the time at $x=0$ is rather
instructive.\ref{88} It can be seen as a static source at $x=0$, and in that
case we consider the Euclidian action
$$
S_E = {1\over 4}\int {\rm d}^2x\, \tr F_{\mu\nu}F_{\mu\nu}+\int {\rm d}t\,
\overline \psi (i\partial _t-eA_0^a(x=0)T^a + M)\psi \quad .\eqno(5.108)
$$

The partition function as computed in ref. [88] in a series of $\E^{-TM}$ is
obtained by noticing that there are $({\rm dim \, }R)$ independent fermions
with
energy $M-{i\over T}w_n$, where $\E^{i w_n}$ are the eigenvalues of the Wilson
loops in the representation $R$. We are interested in the Hamiltonian, which
in the gauge $A_0=0$ reads
$$
H = {1\over 2}\int _0^L {\rm d}x\, \tr \dot A_1^2 + M \overline \psi \psi
\quad ,\eqno(5.109)
$$
and the Gauss law has a source term, that is
$$
\nabla_1 F_{10} = \partial _1\dot A_1 + ie[A_1, \dot A_1] = e \overline \psi
T^a
\psi \tau^a \delta(x-L+\epsilon)\quad ;\eqno(5.110)
$$
we again find it useful to introduce the field
$$
V(x)= W[0,x] \dot A_1(x) W[x,L] \quad ,\eqno(5.111)
$$
whose derivative is concentrated at $x=L-\epsilon$,
$$
\partial _1V= e W[0,{L-\epsilon}]\overline \psi T^a\psi \tau^a
W[{L-\epsilon},L]
\delta(x-L+\epsilon)\quad ,\eqno(5.112)
$$
which implies that $V(x)$ is almost always constant, with a singularity at
$x=L-\epsilon$, such that
$$
V(L)= V(0)+ e W(\overline \psi \psi)\quad .\eqno(5.113)
$$

The last term implies that the commutator of the Wilson loop $W$ with the
electric field no longer vanishes,
$$
[W, \dot A_1(0)] = eW\overline \psi\psi \quad .\eqno(5.114)
$$

The time derivative of the Wilson line is easily obtained from the derivative
of
the ordered exponential, and reads
$$
\dot W = i e \int _0^L{\rm d}x\, V(x)\quad ,\eqno(5.115)
$$
which can be integrated due to the constancy of $V(x)$ (up to the
$\delta$-function in eq. (5.112)). Upon use of (5.111) we find
$$
\dot W = ieL\dot A_1(0)W\quad ,\eqno(5.116)
$$
which implies, together with (5.114), that
$$
[\dot W, W^{-1}] = ie^2L\overline \psi \psi\quad .\eqno(5.117)
$$

Using again the diagonalized form of the matrix $W$
$$
W = U\Lambda U^\dagger\quad ,\eqno(5.118)
$$
the constraint leads to the equation
$$
2U^\dagger\dot U - \Lambda U^\dagger \dot U \Lambda^\dagger - \Lambda^\dagger
U^\dagger \dot U \Lambda = - i e^2 L J\quad ,\eqno(5.119)
$$
where $ J=U^\dagger \overline \psi\psi U$. Defining $\Omega= U^\dagger \dot U$,
one rewrites (5.119) as
$$
\Omega_{ij}\left( 2- \E^{i(\theta_i-\theta_j)} - \E^{-i(\theta_i-\theta_j)}
\right) = -ie^2LJ_{ij}\quad ,\eqno(5.120)
$$
which enables us to compute $\Omega_{ij}$; moreover the Hamiltonian
$$
H=-{1\over 2e^2L}\tr\left(W^{-1}\dot W\right)^2+M\overline\psi\psi\eqno(5.121)
$$
turns into
$$
H = {1\over 2e^2L} \sum \dot \theta_i^2  + {1\over 2}e^2L \sum _{i\ne j}{J_{ij}
J_{ji}\over 4 {\rm sin}^2 (\theta_i - \theta_j)/2}\quad ,\eqno(5.122)
$$
which is a theory of non-relativistic fermions interacting via a two-body
potential analogous to that of the Calogero--Sutherland model.\ref{129}

Such results are in close analogy with those obtained in ref. [111] relating
Yang--Mills theory with sources and the Calogero--Sutherland model. The main
issue is the fact that the Yang--Mills Hamiltonian is quadratic in momentum,
generating the motion of a free system on the cotangent bundle of the Lie
algebra; after diagonalization, such a matrix model describes fermions
interacting with a well-defined Sutherland-type potential. The study of ref.
[104] points also to the same direction, relating QCD$_2$ on a cylinder to a
one-dimensional matrix model of the type introduced by Kazakov and
Migdal.\ref{130}
\vskip 1cm
\penalty-1000
\noindent {\bf 5.6 QCD at high energies and two-dimensional field theory }
\vskip .5cm
\nobreak
\noindent The fact that high-energy scattering amplitudes have a corresponding
description in terms of two-dimensional field theory is remarkable, and
deserves further study. A deeper insight into such a problem was obtained in
the work of Verlinde and Verlinde,\ref{123} who starting from
$(3+1)$-dimensional Yang--Mills theory, studied the limit where the incoming
energy squared $s$ is much larger than the exchanged momentum squared $t$.
This will be related to the so-called leading logarithmic approximation
(LLA).\ref{124-126, 131-133} By means of a scaling argument, they found a
two-dimensional theory and the corresponding correction.

The problem is understood by considering the splitting of the
$(3+1)$-dimensional coordinates into ``fast coordinates" $x^\alpha=(x^+, x^-)=
(x+t,x-t)$, whose Fourier counterparts are large, and ``slow coordinates"
$z^i=(y,z)$, which describe large-distance physics.

The crucial observation made by the authors is that due to the Lorentz
contraction in the direction of the motion of the fast particles, the field
strength will be of the form of a shock wave, non-vanishing only on a
hyper-plane passing through the trajectory of the particle; in this case, the
wave function of a test particle is submitted to a gauge rotation, which
depends on the transverse distance ($z^i$ above), explaining the
two-dimensional nature of the effective interaction, and providing a
Lagrangian mechanism first discovered in terms of the Feynman diagrammatic
expansion.\ref{132}

One first redefines the ``fast coordinates" as
$$
x^\alpha \to \lambda x^\alpha \quad ,\eqno(5.123)
$$
which scales the incoming energy as
$$
s' = \lambda^2 s \quad . \eqno(5.124)
$$
In such a case, one chooses $\lambda \sim 1/\sqrt s$, which tends to zero in
the
desired limit (short distance in the fast coordinates) and keeps $s'$ fixed. In
such a case, the four-dimensional Yang--Mills action
$$
S = - {1\over 4}\int {\rm d}^4x\, \tr F_{\mu\nu}F^{\mu\nu}\eqno(5.125)
$$
transforms into
$$
\eqalignno{
S'= \, & \tr \int {\rm d}^4x'\, \left[ -{1\over 4\lambda^2}F^2_{\alpha\beta} -
{1\over 2}\left( F_{\alpha i}\right)^2 - {\lambda^2\over 4}\left( F_{ij}
\right)^2\right]\cr
= \, & \int{\rm d}^4x'\, \tr \left[ {1\over 2} E^{\alpha \beta}F_{\alpha \beta
}
+ {1\over 2} \left( F_{\alpha i}\right)^2 + {\lambda^2\over 4} \left( E_{\alpha
\beta}\right)^2 + {\lambda^2\over 4}\left(F_{ij}\right)^2\right]\quad,&(5.126)
\cr}
$$
where one uses the fact that the gauge potential transforms as
$$
A_i \to A_i \quad , \quad A_\alpha \to \lambda^{-1}A_\alpha\quad .\eqno(5.127)
$$
An auxiliary antisymmetric field $E_{\alpha \beta}$ has been introduced above.

The $\lambda\to 0$ limit may be singular and terms having $\lambda$ as a
coefficient may be necessary at a later stage. For the time being, we just
consider the zeroth-order approximation, namely neglecting the terms containing
positive powers of $\lambda$.

One may also include matter fields, writing the action
$$
S = \int {\rm d}^4x\, \left[ -{1\over 2} \tr \left( E^{\alpha \beta} F_{\alpha
\beta} + F^{\alpha i}F_{\alpha i}\right) + \overline \psi \gamma^\alpha \left(
\partial _\alpha + ie A_\alpha\right)\psi\right]\quad ,\eqno(5.128)
$$
where the quark-mass term also disappears in such a limit due to the factor
$\lambda^2$ from the measure (notice that $\psi\to\lambda^{-1/2}\psi$). Due to
the absence of the mass term, the fermion action factorizes into two chiral
terms. In the above action, the auxiliary field $(E_{\alpha\beta})$ acts as a
Lagrange multiplier for the constraint $F_{\alpha \beta}=0$, or in light-cone
components $F_{+-}=0$, which is equivalent to the previous observation that
the main contribution follows from transverse configurations in the gauge
field.

The chirality conservation of the action (5.128) implies that the current
satisfies
$$
\nabla_+ j_- = \nabla_- j_+ =0 \quad .\eqno(5.129)
$$

The gauge-field equation of motion can be written perturbatively in terms of
the parameter $\lambda$ and from its perturbative solution the Lipatov vertex
may be obtained. Indeed, keeping $\lambda$ as a parameter, one develops the
gauge field and current in a coupling perturbative expansion as
$$
\eqalignno{
A_{i,\alpha}= & \sum _{n\ge 0} A_{i,\alpha}^{(n)}e^n\quad ,&(5.130a)\cr
j_\alpha = & \sum _{n\ge 0} j_\alpha ^{(n)} e^n \quad ,&(5.130b)\cr}
$$
where the current $j_\alpha$ is chirally conserved in the high-energy limit.
Plugging the above eqs. (5.130) back into the equation of motion
$$
\nabla_\mu F^{\mu\nu}= j^\nu\quad ,\eqno(5.131)
$$
one obtains for the first few perturbative equations, in the Lorentz gauge
$$
\eqalignno{
A_i^{(0)} =\, & 0\quad ,&(5.132a)\cr
\partial _i^2A_\alpha^{(0)} = & -j_\alpha ^{(0)}\quad ,&(5.132b)\cr
\partial ^2A_i^{(1)} =\, & [A_\alpha^{(0)}, \partial _iA_\alpha^{(0)}]\quad ,
&(5.132c)\cr
\partial ^2A_\alpha^{(1)} =\, & j_\alpha^{(1)} + \lambda^{-2} [A_\beta^{(0)},
\partial _\beta A_\alpha^{(0)}]\quad ,&(5.132d)\cr}
$$
where $\partial^2=\lambda^{-2}\partial^\alpha\partial_\alpha-\partial_i^2$.
Recall that in the Lorentz gauge $\partial_+A_-^{(0)}=0=\partial_-A_+^{(0)}$.
One can use the chiral conservation law of the currents in order to eliminate
them order by order in perturbation theory. From the first-order result,
$$
\partial^2A_-^{(1)}= - \left[ {1\over \partial_+}A_+^{(0)}, \partial _i^2
A_-^{(0)}\right] + \lambda^{-2} \left[ A_+^{(0)}, \partial
_-A_-^{(0)}\right]\quad ,\eqno(5.133)
$$
we write for the diagram of Fig. 8 the Lipatov vertex\ref{124}
\penalty-2000
$$\epsfxsize=5truecm\epsfbox{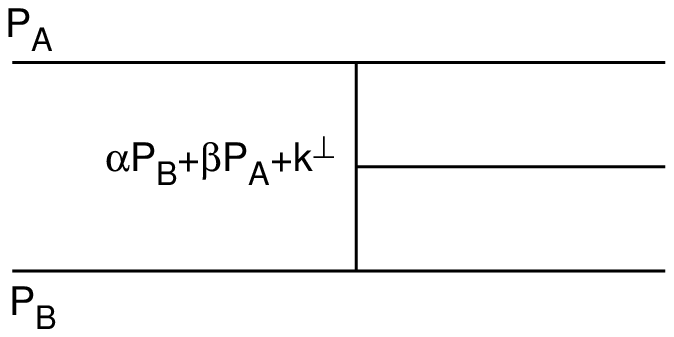}$$
\noindent{\eightpoint Fig. 8: The Lipatov vertex}
\vskip.3truecm
\penalty-2000

$$
C_\mu = - k^\perp_{i_\mu} - {k'}^\perp_{i_\mu} + P_{A_\mu}\left({2k_i^2\over
\alpha_i s} + \beta_i\right) - P_{B_\mu}\left({2{k'_i}^2\over \beta_i s} +
\alpha_i\right)\quad ,\eqno(5.134)
$$
where $\alpha$ and $\beta$ correspond to the Sudakov decomposition of the
momenta of the gluons
$$
k_i= \alpha_i P_A + \beta_i P_B + k^\perp_i\quad ,\eqno(5.135)
$$
so that $k^\perp_i\cdot P_A=0=k_i^\perp P_B$, that is $k_i^\perp$ is the
momentum in the impact parameter space. In such a case, the curvature tensor
is proportional to $\lambda^2$:
$$
\eqalignno{
F_{+-}^{(1)}=\, &\partial _+A_-^{(1)} - \partial _-A_+^{(1)}-ie\left[
A_+^{(0)},
A_-^{(0)}\right] \cr
=\, & 2\lambda^2 {1\over \partial ^\alpha \partial _\alpha} \left[ \partial_i
A_+^{(0)}, \partial_i A_-^{(0)}\right]\quad .&(5.136)\cr}
$$

In the effective action (5.126), the term $\lambda^{-2}F^2_{+-}$ is replaced
by the lagrange multiplier, that is $E^{(1)}=\lambda^{-2}F_{+-}$. Therefore,
the perturbation theory result is equivalent to (5.126) at $\lambda=0$. This
permits a good simplification of the theory, since $A_\alpha$ is now a pure
gauge field. The theory is now described by two Wilson lines summarized by
$$
W_\pm \vert z\vert = \tr P \, {\rm exp\,}\left( ie\int _{-\infty}^\infty {\rm
d}x^\pm \, A_\pm (z)\right)\quad .\eqno(5.137)
$$
The two-point function of two right-moving quarks is given by
$$
\eqalignno{
&\langle T \overline\psi(x_\alpha, z_i)\psi(x'_\alpha, z'_i)\rangle \cr
& = -\delta^{(2)}(z\!-\!z')\!\left(\!\delta(x^-\!-\!{x'}^-)\theta (x^+\!-\!
{x'}^+)+{1\over x^-\!-\!{x'}^-\!+\!i\epsilon }\right)\!P \E^{ie
\int_x^{x'}\!{\rm d}x^+ A_+(z)}\, . &(5.138)\cr}
$$

The fact that the gauge field $A_\alpha(z)$ has pure gauge content, permits an
effective simplification of the theory. Indeed, we have
$$
A_\alpha = {i\over e}\partial_\alpha  UU^{-1}\quad ,\eqno(5.139)
$$
and the action reads
$$
S[U, A_i] = {1\over 2e^2}\int {\rm d}^4x\, \tr [\partial _\alpha (U^{-1}
D_iU)]^2\quad ,\eqno(5.140)
$$
where $D_i=\partial _i - ieA_i$. If one redefines the field $A_i \to \widetilde
A_i = {i\over e}U^{-1}D_iU$, every local interaction disappears, and only the
Wilson-line description survives. One can use for them the definition
$$
P \E^{ie \int {\rm d}x^+\, A_+(z)} = g_2(z)g_1^{-1}(z)\quad ,\eqno(5.141)
$$
where the variables $g_A(z)$ (resp. $h_A(z)$ for $A_-$), $A=1,2$ are dynamical
variables.

Due to the $U$ equations of motion, the integration over ${\rm d}x^+$ and
${\rm d}x^-$ can be performed in the action for classical configurations,
leaving only the value of the fields at boundaries, corresponding to the
above-defined (effectively two-dimensional) fields $g_A, h_A$. Indeed
$$
\eqalignno{
S[U, A] = & {1\over 2e^2}\int {\rm d}^2z\, \tr
\int {\rm d}x^+ \,\partial_+\left( U^{-1} D_i U \right)
\int {\rm d}x^- \,\partial_-\left( U^{-1} D_- U \right)\cr
= & {1\over 2e^2}\int {\rm d}^2z\, M^{AB}\tr \left( g_A^{-1} D_i^+
g_A\,h_B^{-1}
D_i^- h_B \right)\quad ,&(5.142)\cr}
$$
where $M^{11}=M^{22}=1=-M^{12}=-M^{21}$, and $D_i^\pm $ contains the boundary
fields $a_i^\pm (z)$.

High-energy scattering thus has a very simplified description as compared to
the full difficulty of general scattering. The simplification comes from a
general issue connected with an effective dimensional reduction occurring in
such a limit, as exemplified in the simple case of one-loop high-energy
fermionic scattering in QED, as described\ref{132,133} in Fig. 9
\penalty-2000
$$\epsfxsize=10truecm\epsfbox{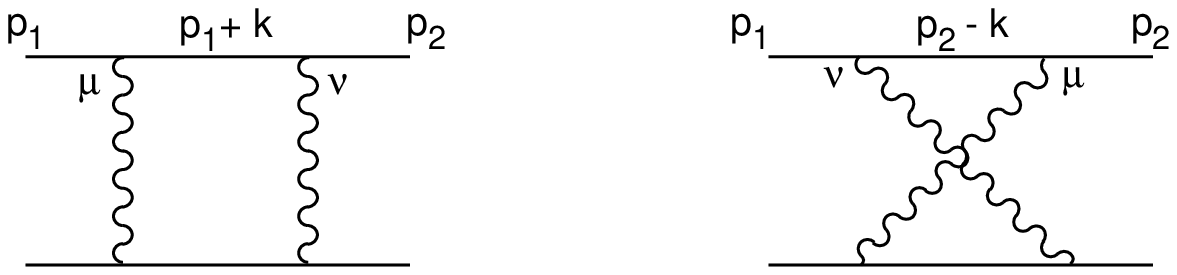}$$
\noindent{\eightpoint Fig. 9: Fermionic high-energy scattering in
four-dimensional QED.}
\vskip.3truecm
\penalty-2000

The upper line of the diagram is given respectively by
$$
{\gamma_\nu(\not \!p_1+\not \! k + m)\gamma_\mu\over (p_1+k)^2 -m^2 +i\epsilon}
\quad \quad {\rm and} \quad
{\gamma_\mu(\not \!p_2-\not \! k + m)\gamma_\nu\over (p_2+k)^2 -m^2 +i\epsilon}
\quad .\eqno(5.143)
$$

In the high-energy limit, we use the external fermions equation of motion
$\overline u_2\gamma_\mu u_1={p_{1_\mu}\over m}\chi_2^\dagger\chi_1$, where
$\chi$ are bispinors. Moreover, we work on shell, and define $k_\pm=k_0
\pm k_3$, where the $z$-axis is chosen in the centre-of-mass system as the
direction of $\vec p_1$. Adding the two contributions in (5.143), with all
above ingredients we arrive at
$$
{p_{1_\mu}p_{1_\nu}\over m w}\left({1\over q_-+i\epsilon} + {1\over -q_-+i
\epsilon}\right)= -2\pi i {p_{1_\mu}p_{1_\nu}\over m w}\delta (q_-) \quad
,\eqno(5.144)
$$
where $w$ is the centre-of-mass energy. Notice already the presence of the
first $\delta$-function in one of the non-transverse components of the
momentum. From the remaining part of the diagram, one obtains a term
proportional to ${1\over -q_++ i\epsilon}$, whose principal part vanishes due
to symmetry, and one remains with a second delta term ($\delta(q_+)$).
Therefore the scattering is described, in the loop variables, by an effective
two-dimensional theory.

Such an idea as applied to the scattering of high-energy hadrons in QCD leads
to
the possibility of summing an infinite series of diagrams, and leads to the
so-called Reggeization of the gluons, which up to an effective Regge form
factor given by
$$
s^{\alpha(t)-1}\eqno(5.145)
$$
($s$ and $t$ are the Mandelstam variables, to be appropriately defined below),
are described by the Born approximation of the perturbative expansion. This is
the leading logarithmic approximation (LLA)\ref{131}. Such a study of
high-energy scattering has been performed by several authors, both in
QED\ref{133} and QCD.\ref{124,131,132}

There is a surprising regularity in the QCD results,\ref{131} where in higher
orders the only new issue, in the LLA, namely $\alpha_s\ln {s\over M^2}\approx
1$, are the logarithmic factors $\left(\alpha_s\ln {s\over M^2}\right)^J$.

Taken separately, the diagrams  violate the Froissart bound, leading to
non-unitary amplitudes. A previous knowledge of Reggeon field theory leads
to a solution of the problem. Indeed, higher orders are necessary, and as a
matter of fact, the above-mentioned regularity of higher-order computations in
the LLA can be summarized by the Reggeon factor, given by eq. (5.145). A
non-perturbative solution should obey unitarity, which relates elastic and
non-elastic scattering through the equation
$$
{1\over s}{\rm Im}\, T_{2\to 2} = \sum_{n=0}^\infty\int {\rm d}\Omega_n
\, T_{2\to 2+n}(s+i\epsilon)T_{2\to 2+n}(s-i\epsilon)\quad,\eqno(5.146)
$$
where ``$n$" is the number of new particles produced and $\{{\rm d}\Omega_n\}$
is the integration over the phase space of the intermediate states, to be
defined in detail in eq. (5.162). This means that we have to supplement the LLA
with an infinite subclass of diagrams, in such a way that unitarity holds.

The amplitude resulting from the LLA, and matching the lowest-order
computations in perturbation theory, leads to the Reggeized amplitude
$$
T_{2\to 2} = - {\E^{-i\pi \alpha(t)}-1\over t-M^2}s^{\alpha(t)} g t^a_{bc}\,
gt^{a'}_{cb'}\quad ,\eqno(5.147)
$$
where $\alpha(t)= 1 + (t-M^2)\beta(t)$ and
$$
\beta(p^2)= e^2N\int {{\rm d}^2k\over (2\pi)^3} {1\over (k_{\perp}^2+M^2)
[(k-p)^2_{\perp}+M^2]}\quad ,\eqno(5.148)
$$
where $p_{\perp}$ is the component of momentum perpendicular to the incoming
direction, or equivalently the impact parameter space. Also $ s= (P_A+P_B)^2$
and $t=(P_A-P'_A)^2$.

In the high-energy limit the transferred momentum has mostly transverse
components, $q=(P_A-P'_A)\simeq (0,0,q_{\perp})$ and $s >>s_1\sim s_2\sim
\cdots
\sim s_{n+1}>>q^2_{1_{\perp}}q^2_{n+1_{\perp}} \sim M^2$, where we define,
according to Lipatov\ref{124}
$$
\eqalignno{
s &   = (P_A+P_B)^2     \simeq 2P_AP_B \quad ,\cr
s_i & = (k_i+k_{i+1})^2 \simeq 2k_ik_{i+1} \quad ,\cr
k_i & = q_i - q_{i+1} \quad ,\cr
k_0 & \equiv P'_A = P_A-q_0\, ,\, k_{n+1} \equiv P'_B = P_B + q_{n+1}\quad ,&
(5.149)\cr}
$$
and momentum conservation requires $P_A+P_B = \sum _{i=0}^{n+1}k_i$.

As we have already mentioned, order-by-order computations, in the high-energy
limit have been performed by several authors, and the common feature in
non-Abelian gauge theories is the fact that corrections exponentiate, leading
to the so-called Reggeization of the gluon. This means that in the LLA,
previously
defined, one can just compute the Born amplitude, substituting the vertex by
Lipatov's vertex, which includes the energy-dependent correction $s^{\alpha(t)
}$. Thus we write for the $2\to 2+n$ amplitude, in the LLA approximation, the
result
$$
\eqalignno{
A_{2\to  2+n}&(q_i)\cr
 =\,  & 2se {(s_{01}/M^2)^{w(q_1)}\over q_1^2} e t^{a_1}_{i_1,
i_2}(C(q_1,q_2)\cdot \varepsilon(k_1)) {(s_{12}/M^2)^{w(q_2)}\over q_2^2} e
t^{a_2}_{i_2, i_3}(C(q_2,q_3)\cdot \varepsilon(k_2))\cr
& \times \cdots  e t^{a_n}_{i_n, i_{n+1}}(C(q_n,q_{n+1})\cdot \varepsilon(k_n))
{(s_{n,n+1}/M^2)^{w(q_{n+1})}\over q_{n+1}^2}\quad ,&(5.150)\cr}
$$
where $\varepsilon_\mu(k)$ is the polariazation vetor of the Reggeon. The
trajectory of the Reggeized gluon is
$$
w(q)=-{e^2N\over 2(2\pi)^3}\int{{\rm d}^2k\,q^2\over k^2(q-k)^2}\quad,
\eqno(5.151)
$$
and the result (5.150) can be summarized by
$$
A^{{\rm LLA}}=A^{{\rm tree}}s_1^{w(q_1)}\cdots s_{n+1}^{w(q_{n+1})}\quad,
\eqno(5.152)
$$
which corresponds to the leading logarithmic approximation (LLA). The form of
the vertex for effective gluon production as computed by Lipatov\ref{124}
$$
C^\mu (q_i,q_{i+1}) = - {q^\mu_i}^\perp - {q^\mu_{i+1}}^\perp + P_A^\mu
\left({2q_i^2\over \alpha_i s} + \beta_i\right)- P_B^\mu
\left({2q_{i+1}^2\over \beta_i s }+ \alpha_i\right)\quad ,\eqno(5.153)
$$
has been discussed previously in the comparison with the Lagrangian methods.

The fact that unitarity has to be satisfied in the $s$-channel for all
subenergy
variables has been emphasized by Bartels.\ref{131} This is the moment where
it is crucial to use unitarity equations to compute the discontinuity in the
elastic-scattering amplitude (5.146). In order to use the unitarity equation,
we have to rewrite the kinematical variables in a suitable form.

Since details are generally not available in the literature, we present the
whole derivation of the proof of integrability of the high energy scattering
amplitude.

We have to use again Sudakov's decomposition (5.135) for the outgoing gluons
on shell in the energy limit $(k_i^2=s\alpha_i\beta_i-(k_i^\perp)^2=0)$, we
find
$$
\alpha_i\beta_i = {(k_i^\perp)^2\over s}\quad , \eqno(5.154)
$$
and
$$
s_{i,i+1} = (\alpha_{i+1}\beta_i + \alpha_i\beta_{i+1})s - 2k_i^\perp\cdot
k_{i+1}^\perp\quad,\eqno(5.155)
$$
where, in general, the first term dominates, and $\alpha_i\beta_{i+1}+\alpha_
{i+1}\beta_i\sim 1$. Moreover, substituting $\alpha_i\beta_i={(k_i^\perp)^2
\over s}$, we find that in the multi-Regge limit $\left({\alpha_{i+1}\over
\alpha_i}+{\beta_{i+1}\over\beta_i}\right)$ must be large, and since $\alpha_0
\sim 1$, and $\beta_{n+1}\sim 1$, we find the order
$$
1\sim \alpha_0 >> \alpha_1 >> \cdots >>\alpha_{n+1} \sim q^2_{n+1}/s\quad ,
\eqno(5.156)
$$
and
$$
q^2_0/s \sim \beta_0 << \beta_1 \cdots << \beta_{n+1}\sim 1\quad .\eqno(5.157)
$$
Thus we obtain for the partial energy variables the results
$$
\eqalignno{
s_{i,i+1} & = \alpha_i \beta_{i+1}s\quad ,\cr
\prod_{i=0}^ns_{i,i+1} & = \alpha_0\beta_{n+1}s \prod_{i=1}^n (k_i^\perp)^2
\quad .&(5.158)\cr}
$$

The phase-space integral is obtained from
$$
\eqalignno{
{{\rm d}^4k\over (2\pi)^3}\delta (k^2)&= s{\rm d}\alpha{\rm d}\beta{{\rm d}^2
k^\perp \over 2(2\pi)^3} \delta(s\alpha \beta - k_\perp^2)\quad ,\cr
& = {{\rm d}\alpha\over \alpha}{{\rm d}^2k^\perp\over 2(2\pi)^3 }  \quad ,&
(5.159)\cr}
$$
leading to the physical momentum volume
$$
\prod^{n+1}_{i=0} {{\rm d}^4k_i\over (2\pi)^3}\delta (k_i^2) = \prod_{i=0}^n
{{\rm d}s_{i,i+1}\over s_{i,i+1}} \prod_{i=0}^{n+1}{{\rm d}^2k_i^\perp\over
2(2\pi)^3} {{\rm d}\alpha_0\over \alpha_0}\quad .\eqno(5.160)
$$
{}From this expression we find the total phase-space volume which reads
$$
\eqalignno{
{\rm d}\Omega_n & = (2\pi)^4 \delta^4\left(P_A+P_B-\sum_{i=0}^{n+1}k_i\right)
\prod_{i=0}^{n+1}{{\rm d}^4k_i\over (2\pi)^3}\delta(k_i^2)\cr
& = {\pi\over s} \prod_{i=0}^{n+1}{{\rm d}\alpha_i\over \alpha_i}\prod_{i=1}^
{n+1}{{\rm d}^2k_i^\perp\over (2\pi)^3}\delta(1-\alpha_0)\delta(1-\beta_{n+1})
\quad .&(5.161)\cr}
$$

We can now integrate over $s_{i,i+1}$ instead of $\alpha_i$ and use eq.
(5.158) as well as integrate over $k_0^\perp$ to arrive at
$$
{\rm d}\Omega_n= {\pi\over s}\left[\prod_{i=0}^n\left({\rm d}s_{i,i+1}
{{\rm d}^2k_{i+1}^\perp\over (2\pi)^3}\right)\right]\delta\left(\prod s_{i,i+1}
 - s\prod(k_i^\perp)^2\right)\quad .\eqno(5.162)
$$

The Mellin transformation is now easily performed since the $s$ variable enters
in a simple way as an overall factor, as well as in the argument of the delta
function. Thus we have
$$
\eqalignno{
\int \int _0^\infty {{\rm d}s\over M^2} \left({s\over M^2}\right)^{-w-1} s
{\cal F}(s){\rm d}\Omega_n = \,&\pi \int {{\rm d}s_{0,1}\over M^2}\left(\prod_
{i=1}^n{{\rm d}s_{i,i+1}\over (k_i^\perp)^2}\right)\left({s^{\rm tot}\over M^2}
\right)^{-w-1}\cr
& \times \prod_{i=1}^{n+1}{{\rm d}^2k_i^\perp\over (2\pi)^3}{\cal F}
(s^{\rm tot})\quad ,&(5.163)\cr}
$$
where $s^{\rm tot}= s_{0,1}\prod_{i=0}^n{s_{i,i+1}\over (k_i^\perp)^2}$.

Notice now that for $i\le n$, $q_i\sim f(k_i^\perp, \alpha_i; q_{i+1})\sim
h(k_i^\perp, s_{i-1,i}; q_i)$. Thus all $s_{i,i+1}$ integrations, but for
$i=m$, are non-trivial.

The denominators $q_i^2 \simeq - (\vec q_i)^2$ are factored to the integration
over the perpendicular momenta. The partial energy integration (over
$s_{i,i+1}$) is not too complicated now. From $A_{2\to 2+n}$ and its conjugate
we find terms of the type
$$
\left[{s_{i,i+1}\over (k_i^\perp)^2}\right]^{-w-1}\left[s_{i,i+1}\over M^2
\right]^{w(q_i) + w(q-q_i)} \quad .\eqno(5.164)
$$

{}From the LLA $s_{i,i+1}= s\alpha_i\beta_{i+1}>>s \alpha_i\beta_i
= (k_i^\perp)^2$; therefore, using the latter as a lower bound for integration,
we find for the contribution (5.164) for the elastic scattering
$$
\int_{(k_i^\perp)^2}^\infty {{\rm d}s_{i,i+1}\over (k_i^\perp)^2} \left[
{s_{i,i+1}\over (k_i^\perp)^2}\right]^{-w-1}\left[s_{i,i+1}\over M^2
\right]^{w(q_i) + w(q-q_i)}= {1\over w-w(q_i)- w(q-q_i)}\quad .\eqno(5.165)
$$

The sum over polarizations may be performed as usual, and we find
$$
\eqalignno{
\sum_a C^\mu (q_i,q_{i+1}) C^\nu (q-q_i,q-q_{i+1})\epsilon_\mu^{(a)} (k_i)
\epsilon_\nu^{(a)} (k_i) & = C_\mu (q_i,q_{i+1}) C^\mu (q-q_i,q-q_{i+1})\cr
& = K(q_i, q_{i+1}\vert q)\quad ,&(5.166)\cr}
$$
defining a kernel that will be used in the derivation of the Bethe--Salpeter
equation, and which can be computed from the expression of the Lipatov vertex
(5.153). One first defines the function $F(q,q')$ by
$$
A(w,t)= \int {\rm d}^2q'F(q,q')\phi(q,q')\prod_{i=2}^{n+1}\int {\rm d}^2q_i
{K(q_{i-1},q_i\vert q)\over q_i^2(q-q_i)^2 [w-w(q_i)-w(q-q_i)]}\quad ,
\eqno(5.167)
$$
where
$$
F(q,q_1) = \sum_{n=1}^\infty \left({e^2N\over (2\pi)^3}\right)^n {1\over
q_1^2(q-q_1)^2[w-w(q_1)-w(q-q_1)]}\quad .\eqno(5.168)
$$

The vertex function $\phi(q,q')$ has to be used in order to define the hadronic
state. The Bethe--Salpeter equation obeyed by $F(q,q_1)$ can be simply read
from the above series representation:
$$
q_1^2(q-q_1)^2[w-w(q_1)-w(q-q_1)]F(qq_1)= {e^2N\over (2\pi)^3} \left[ 1+ \int
{\rm d}^2q'_1 K(q'_1,q-1\vert q)F(q,q'_1)\right] \quad .\eqno(5.169)
$$
The resulting amplitude is essentially two-dimensional\ref{134-136}, and
it is useful to define the Fourier transformation in the impact parameter space
as
$$
\eqalignno{
\delta^2(q-q')f_w(k,k';q)=\, & {1\over (2\pi)^8}\int \prod_1^2 {\rm d}^2\rho_r
\prod_1^2 {\rm d}^2\rho_{r'} f_w(\rho_1,\rho_2,\rho_{1'},\rho_{2'})\cr
& \times  \E^{ik\rho_1+ i(q-k)\rho_2-ik'\rho_{1'}-i(q'-k')\rho_{2'}}\quad .
& (5.170)\cr}
$$

Lipatov rewrote the Bethe--Salpeter equation in such a way that conformal
invariance in two-dimensional space can be verified, that is
$$
\eqalignno{
w&{\partial^2\over \partial \rho_1^2}{\partial^2\over \partial
\rho_2^2}f_w(\rho_1,\rho_2) \cr
 = \,& (2\pi)^4\delta^2(\rho_1\!-\!\rho_{1'})\delta^2(\rho_2\!-\!\rho_{2'})\!
+\!{e^2N\over (2\pi)^3}\Bigg\{ (2\pi)^2 \delta^2(\rho_1-\rho_2)
\left({\partial\over \partial \rho_1} + {\partial\over \partial
\rho_2}\right)^2f_w(\rho_1,\rho_2)\cr
+ &{\partial^2\over \partial \rho_1^2}\int {{\rm d}^2\rho_0\over
\vert \rho_{01}\vert^2}\left[{\partial^2\over \partial \rho_2^2}
f_w(\rho_0,\rho_2) - {\vert \rho_{12}\vert^2\over \vert \rho_{01}\vert
^2 + \vert \rho_{02}\vert ^2}{\partial^2\over \partial \rho_2^2}
f_w(\rho_1,\rho_2)  \right]\cr
+ &{\partial\over \partial \rho_2^2}\int {{\rm d}^2\rho_0\over
\vert \rho_{02}\vert^2}\left[{\partial^2\over \partial \rho_1^2}
f_w(\rho_1,\rho_0) - {\vert \rho_{12}\vert^2\over \vert \rho_{01}\vert
^2 + \vert \rho_{02}\vert ^2}{\partial^2\over \partial \rho_1^2}
f_w(\rho_1,\rho_2)  \right]\Bigg\}\quad, &(5.171)\cr}
$$
where $\rho_{ij}=\rho_i-\rho_j$. In terms of $f_w(\rho_1,\rho_2,\rho_{1'},
\rho_{2'})$, the Bethe--Salpeter
equation is an integro-differential equation, with quartic derivatives with
respect to $\rho_i$. The problem, although appearing far from the solubility
from
the point of view of such a difficult differential equation, is facilitated
from the fact that it is conformally invariant in the impact parameter space.
In two dimensions, conformal invariance means factorization in terms of
holomorphic and anti-holomorphic functions. Neglecting the contact terms, the
remaining integrals are not difficult to compute. Consider
$$
\int {{\rm d}^2\rho_0\over \vert \rho_{01}\vert^2}{\partial^2\over\partial
\rho_2^2} f(\rho_0,\rho_2)=\int {{\rm d}^2\rho_0\over \vert \rho_{01}
\vert^2} \E^{-iP\rho_{01}}{\partial^2\over \partial \rho_2^2}
f(\rho_1,\rho_2)\quad ,\eqno(5.172)
$$
where $P$ is the generator of translations. The above integral is in general
infrared-divergent, but as proved by Lipatov, considering all contributions the
infrared divergence will cancel (see also next term in (5.171)). Neglecting
this divergence, the integral corresponds to the massless boson
propagator in two dimensions with the phase-space variables interchanged, and
one obtains\ref{8} the result $2\pi \, \ln P^2$. For the remaining term we have
$$
\eqalignno{
\int {{\rm d}^2\rho_0\over\vert\rho_{01}\vert^2}{\vert\rho_{12}\vert^2\over
\vert\rho_{01}\vert^2 + \vert\rho_{02}\vert^2}{\partial^2\over \partial
\rho_2^2}f(\rho_1,\rho_2)& = \int {\rm d}^2 \rho_0 \left[ {1\over\vert
\rho_{01}\vert^2} - {2\over \vert\rho_{01}\vert^2 + \vert\rho_{02}\vert^2 }
\right]{\partial^2\over \partial \rho_2^2}f(\rho_1,\rho_2)\cr
&=2\pi\ln\vert\rho_{12}\vert^2{\partial^2\over \partial \rho_2^2}
f(\rho_1,\rho_2)\quad.&(5.173)\cr}
$$

Writing the derivative terms in momentum space, and separating the holomorphic
and anti-holomorphic pieces we obtain the pair Hamiltonian
$$
H_{ik} = P_i^{-1}\ln \rho_{ik}P_i + P_k^{-1}\ln \rho_{ik}P_k + \ln P_iP_k
\quad .\eqno(5.174)
$$

The final situation is outlined in ref. [125], where Lipatov states that the
Bethe--Salpeter equation, for $n$ reggeized gluons for a large number $(N)$ of
colours, is described in terms of a wave function
$f_w(\vec\rho_1, \cdots, \vec\rho_n; \vec\rho_0)$ satisfying factorization
$$
f_w(\vec\rho_1, \cdots, \vec\rho_n; \vec\rho_0)= \sum f^2(\rho_{10}, \cdots,
\rho_{n0}) \overline f^2(\overline \rho_{10}, \cdots ,\overline \rho_{n0})\quad
,\eqno(5.175)
$$
where $\rho_{ij}= \rho_i-\rho_j$ and $\overline \rho_{ij}= \overline\rho_i -
\overline \rho_j$ are the coordinates of the gluons, and these functions
satisfy independent Schr\"odinger equations
$$
\eqalignno{
Hf^2 = \,& \epsilon f^2 \quad ,&(5.176a)\cr
\overline H\,\overline f^2=\,&\overline\epsilon\overline f^2\quad.&(5.176b)\cr}
$$

{}From the previous analysis, an extremely interesting physics describing the
high-energy scattering of QCD arises. Thus, recapitulating, gluons interact in
such a way that they build up collective states, the Reggeons, interacting
almost as free particles, but conveniently dressed. Unitarity imposes strong
conditions. In particular it is possible to compute the partial-wave
scattering amplitude via unitarity condition, since the plain LLA leads to
amplitudes that do not satisfy the Froissart bound. The unitary partial-wave
scattering amplitude can be proved to satisfy a Bethe--Salpeter equation. The
solutions to such an equation describe the singlet composite states built out
of Reggeons, the first state with vacuum quantum numbers being the
Pomeron.\ref{137} The Bethe--Salpeter equation is a Hamiltonian equation
describing an effective interaction for pairs of gluons, being of the form
$$
{\cal H} = -{e^2\over 2\pi}\sum H_{ij}\, t^a_it^a_j \quad ,\eqno(5.177)
$$
where $(i,j)$ are the pairs of Reggeons. Such two-particle Hamiltonians
${\cal H}_{ij}$, as presented in (5.174), are effectively two-dimensional, and
may be interpreted as describing a lattice model with pair interactions, the
size of the lattice being the number of external Reggeons.

A big simplification arises in the large-$N$ (number of colours) limit, where
the theory is planar, and thus the external Reggeons interact only with their
nearest neighbours. In such a case one substitutes in eq. (5.177), $t^a_it^a_j
\to-{N\over 2}\delta_{i,j+1}$. Faddeev and Korchemsky\ref{127} proved that the
Hamiltonian thus obtained corresponds to the XXX Heisenberg chain with spin
zero.

The form of such an identification arises from the computation of the
Heisenberg Hamiltonian in terms of the solution of the Yang--Baxter equation.
The solution is given by\ref{138}
$$
R_{ij}(\lambda) = {\Gamma(i\lambda-2s) \Gamma(i\lambda+2s+1)\over
\Gamma(i\lambda -J_{ij}) \Gamma(i\lambda +J_{ij}+1)}\quad ,\eqno(5.178)
$$
where $s$ is the spin and the operator $J_{ij}$ acts on the quantum space
$h_i\otimes h_j$, $h_i$ corresponding to the $i^{{\rm th}}$ site of the
Heisenberg lattice, and satisfies
$$
J_{ij}(J_{ij}+1)=(\vec s_i+\vec s_j)^2=2\vec s_i\cdot\vec s_j+2s(s+1)\quad .
\eqno(5.179)
$$

The form of the Hamiltonian is
$$
\eqalignno{
H & = \sum H_{j j+1}\quad ,&(5.180)\cr
H_{j j+1} & = - i{{\rm d}\over {\rm d}\lambda} \ln R_{j j+1}(\lambda)\Bigg\vert
_{\lambda =0}\quad .&(5.181)\cr}
$$

Astonishingly enough, Faddeev and Korchemsky\ref{127} computed this
Hamiltonian and verified that for $s=0$ it is given by eq. (5.174)! Therefore
one can use the Bethe ansatz to obtain the solution of high-energy
QCD.\ref{128}
\vskip 1cm
\penalty-3000
\noindent {\bf 6. Conclusions}
\vskip .5cm
\nobreak
\noindent After twenty years of development, QCD$_2$ stays in an outstanding
position in the way towards non-perturbative comprehension of strong
interactions. The large-$N$  limit of the theory revealed a desirable structure
for the mesonic spectrum, whose higher levels display a Regge behaviour. These
properties were later generalized for fermions in the adjoint representation,
an important step towards understanding the theory in higher dimensions,
since in such a case adjoint matter substitutes the lack of the transverse
degrees of freedom of the gauge field in two dimensions. Further properties of
the perturbative theory are also in accordance with expectations for strong
interactions, and it proves therefore to have certain advantages over the
usual non-linear $\sigma$-models in the description of strong interactions by
means of simplified models.

The central issue of the computation of the non-Abelian fermionic determinant
is the key to understanding the theory, bypassing the severe question of
confinement since it provides an effective theory for the description of the
mesonic bound states, opening the possibility of understanding baryons as
solitons of the effective interactions.  The full QCD problem can be dealt
with using these methods. There are paralels with the QED developments, but
even
well known facts as the role played by negative metric states\ref{140} are
further complicated in the present case, where the coset construction needs to
be advocated.

The string interpretation of pure Yang--Mills theory, as well as its
Landau--Ginzburg-type generalizations connected the previous picture to that of
non-critical string theory. The relevance of these developments is found in the
basis they form for a deeper understanding of the role of non-critical string
theory in the realm of strong interactions. Although far from being realized,
such is aparently the correct way to understand strong interactions at
intermediate energies. A very general formulation of QCD$_2$ has been recently
studied including many features discussed here.\ref{141}

Finally, the question of high energy scattering in strong interactions is
linked with integrable models, a property shared in full by two-dimensional
QCD. Thus one sees the important role of higher symmetries algebras, spectrum
generating algebras, and integrability conditions, which might give a clue to
the full solution of the theory.

\vskip 1cm
\nobreak

\noindent {\bf Acknowledgements:}

\noindent The authors wish to thank E. Kiritsis for the reading of the
manuscript and sugestions, L. Lipatov and D. Kutasov for several clarifying
letters, and G. Korchemsky for discussions and a detailed file containing his
computations. Discussions and suggestions of A. Dhar, T.T, Wu, and A. Zadra
are also thankfully acknowledged. This work was partially supported by CAPES
(E.A.), Brazil, under
contract No.1526/93-4, and partially by the World Laboratory (M.C.B.A.).

\vskip 1cm
\penalty-1000
\noindent {\bf Appendix A}
\vskip .5cm
\nobreak
\noindent In this appendix we summarize our notation and conventions. In most
of the review we work in Minkowski two-dimensional space, but we give here the
necessary dictionary for translating results to Euclidian space. For the metric
and $\epsilon $ tensor we use
$$
g^{00} = - g^{11} = 1 \quad , \quad \epsilon _{01} = - \epsilon ^{01} = 1 \quad
;\eqno(A.1)
$$
the gamma matrices are
$$
\gamma^0 = \pmatrix {0&1\cr 1& 0\cr} \quad ,\quad
\gamma^1 = \pmatrix {0&-1\cr 1& 0\cr} \quad ,\quad
\gamma_5 = \pmatrix {1&0\cr 0& -1\cr} \quad , \quad
\gamma_\mu \gamma_5 = \epsilon _{\mu\nu} \gamma^\nu\quad ,\eqno(A.2)
$$
in such a way that $\gamma_\pm = \gamma_0 \pm \gamma_1$, $\gamma_+\gamma_- =2$
and we use also
$$
\epsilon^{\mu\nu}\epsilon_{\rho\sigma} = \delta^\mu_\sigma\delta^\nu_\rho -
\delta ^\mu_\rho \delta^\nu_\sigma\quad .\eqno(A.3)
$$
The definition of the light-cone variables
$$
J_+=J_0+J_1\quad ,\quad A_+=A_0+A_1\quad  \eqno(A.4)
$$
leads to extra factors of 2, as for example in eq. (2.62). The tilde is
generically reserved for the definitions of pseudo-vectors
$$
\widetilde A_\mu =\epsilon_{\mu\nu}A^\nu \quad ,\quad {\rm or}\quad
\widetilde D_\mu =\epsilon_{\mu\nu}D^\nu \quad .\eqno(A.5)
$$
Notice that $\partial_\mu\widetilde\partial_\nu -\partial_\nu\widetilde
\partial_\mu = \epsilon_{\mu\nu}\dal$, where $\dal $ is the d'Alembertian. The
massless propagator obeying $\dal D(x)=\delta (x)$ is $D(x)=-{i\over 4\pi}
\ln \, (-x^2+i\epsilon)$. In Minkowski space,
$$
x^\mu = (x^0, x^1)\quad ,\quad  \partial _\pm = \partial _0 \pm \partial _1
\quad ,\quad x^\pm = x^0 \pm x^1\quad .\eqno(A.6)
$$
On the other hand, in Euclidian space:
$$
x_\mu = (x_1, x_2)\, ,\,  \overline \partial = \partial _1 - i \partial_2
\equiv \partial _-^E \, ,\, \partial = \partial _1 + i \partial_2 \equiv
\partial _+^E \, ,\, z = x_1- ix_2 \, ,\, \overline z = x_1 + ix_2 \, .
\eqno(A.7)
$$
In order to translate from one space to the other, we have $x_2=ix_0$, implying
(notice the important $(-)$ sign!)
$$
\overline \partial \longleftrightarrow - \partial _-\quad ,\quad \partial
\longleftrightarrow \partial _+\quad .\eqno(A.8)
$$
Notice also that
$$
{\partial \over \partial \overline z} = {1\over 2}\overline \partial
\quad ,\quad  {\partial \over \partial z} = {1\over 2}\partial\quad .\eqno(A.9)
$$
With these conventions,
$$
F_{\mu\nu}F_{\mu\nu}  = 2F_{12}^2 = -{1\over 2}F_{z\bar z}^2\quad {\rm and}
\quad
F_{z \bar z}  = - i F_{12} + i F_{21} = - 2i F_{12}\quad . \eqno(A.10)
$$
Path integrals are always performed in Euclidian space, while in the canonical
quantization we use the Minkowski version:
$$
{\partial\over \partial x^\pm} = {1\over 2}\partial _\pm\quad ,\quad {\rm d}
x^+\, {\rm d}x^- = \Bigg\vert\Bigg\vert \matrix {1&1\cr1&-1\cr}
\Bigg\vert\Bigg\vert \, {\rm d}^2x\, = 2 {\rm d}^2x \quad .\eqno(A.11)
$$
The gauge field and the covariant derivative, in the direct and adjoint
representations, and the $\tau $ matrices are
$$
\eqalignno{
A_\mu & = \sum_a \tau^a A_\mu^a \equiv \tau^a A_\mu^a\quad ,& (A.12a)\cr
D_\mu & = \partial _\mu - ie A_\mu \quad ,&(A12b)\cr
\nabla_\mu^{ab} & =\partial_\mu\delta^{ab}+ef^{acb}A^c_\mu\quad,&(A.12c)\cr
[\tau^a,\tau^b] & = i f^{abc}\tau^c \quad ,\quad \tr \tau^a\tau^b = \delta^{ab}
\quad .&(A.12d)\cr}
$$
The gauge field strength is
$$
\eqalignno{
F_{\mu\nu} = & \partial _\mu A_\nu - \partial _\nu A_\mu - ie[A_\mu,A_\nu]\quad
,&(A.13a)\cr
F^a_{\mu\nu} = & \partial _\mu A^a_\nu - \partial _\nu A^a_\mu + e f^{abc}
A^b_\mu A^c_\nu\quad .&(A.13b)\cr}
$$

The covariant derivative on fermions in the fundamental representation,
$\psi^i$, are defined by (A.12$a$)
$$
(D_\mu \psi)_i = \partial _\mu \psi_i - ie (A_\mu\psi)_i  \quad ,\eqno(A.14)
$$
while for fermions in the adjoint representation, $\psi^a$, we have
$$
(\nabla_\mu\psi)^a = \partial _\mu\psi^a + e f^{acb}A_\mu^{\, c}\psi^b=
\partial
_\mu \psi^a - ie [A_\mu, \psi]^a\quad .\eqno(A.15)
$$
In two dimensions, it is true that
$$
F^{\mu\nu}=-{1\over 2}\epsilon^{\mu\nu}\epsilon_{\rho\sigma} F^{\rho\sigma}
\quad .\eqno(A.16)
$$
\vskip 1cm
\penalty-1000
\noindent {\bf Appendix B}
\vskip .5cm
\nobreak
\noindent Considering maps of a closed orientable two-dimensional surface
${\cal M}_g$ of genus $g$, onto another two-dimensional surface ${\cal M}_G$
of genus $G$, a relation between the genera $G, g$ and the winding number
$(n)$ of the mapping can be obtained. In the case of smooth maps, this
relation is given by Kneser's formula\ref{139}
$$
2(g-1) \ge 2n (G-1)\quad .\eqno(B.1)
$$

To understand this bound one uses the idea of covering maps, which in case of
smooth maps do not have branch points or collapsed handles. It can be proved
that such smooth maps can be continuously deformed into the covering maps. To
construct them, first consider a closed orientable two-dimensional surface
${\cal M}_G$ of genus $G>1$, as in Fig. 10a. Then cut this surface along a
cycle, as in Fig. 10b, leading to a surface of genus $(G-1)$. This surface is
topologically equivalent to the surface of Fig. 10c. Now add $n$ copies of the
same topologically equivalent surface, as in Fig. 10d, obtaining a surface of
genus $n(G-1)$. Finally close the covering gluing, the circles $a$ and $b$
forming a surface with an additional handle, ending up with an $n$-fold
covering
of ${\cal M}_G$ by ${\cal M}_g $, where the genera and the winding number are
related by
$$
g= n(G-1) + 1\quad .\eqno(B.2)
$$
\penalty-2000
$$\epsfysize=4.5truecm\epsfbox{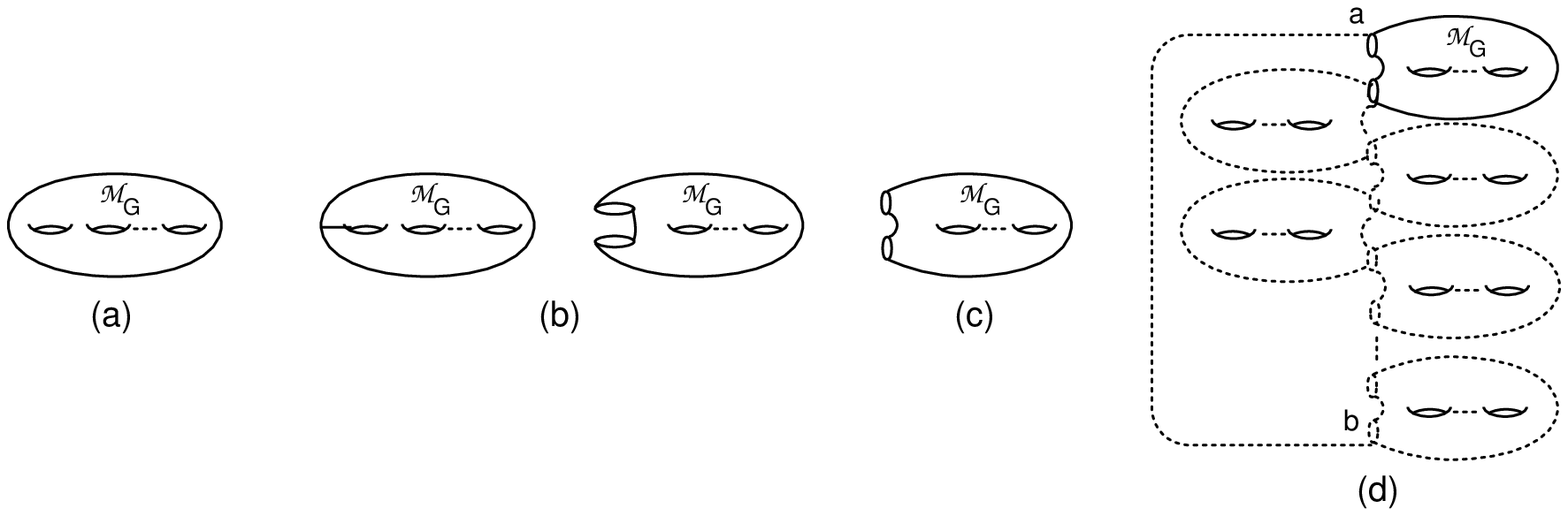}$$
\noindent{\eightpoint Fig. 10: Covering space to obtain the Kneser formula.}
\vskip.3truecm
\penalty-2000

This is not the whole history, because there are no smooth maps of a genus
$g>1$
surface that wind around it more than once. This means that we have to include
branch points or collapsed tubes, modifying the above relation to
$$
2(g-1) = 2n (G-1) + B \quad ,\eqno(B.3)
$$
where $B$ is the total branching number. The above relation is known as the
Riemann--Hurwitz theorem. (See ref. [83] for more details.)

In fact the relation (B.2) pictures only the surfaces that saturate the bound
given by Kneser's formula; adding extra handles to ${\cal M}_g $, the
Euler characteristic can only increase.
\vskip 1cm
\penalty-1000
\noindent {\bf Appendix C }
\vskip .5cm
\nobreak
\noindent A character is the trace of the matrix in a given representation;
that is, given a representation $R$
$$
R \colon \, g \to T(g)\quad ,\eqno(C.1)
$$
the character is
$$
\chi[g] = \tr \, T(g)\quad .\eqno(C.2)
$$

Its utility lies upon the fact that it is invariant under a unitarity
transformation, and the representation is fully decomposed in terms of
uniquely characterized irreducible representations.

In a given representation, characters may be computed in terms of the
eigenvalues of the given element. In the case of $U(N)$, the eigenvalues are
phases $\epsilon_j = \E^{i\varphi_j}$, and the irreducible representations are
labelled by $N$ integers $n_i$, ordered as $n_1\ge n_2 \ge \cdots \ge n_N$, and
it has been proved that the character of $U$ is a ratio of the determinants
(see refs. [76, 77] for details)
$$
\chi_{\{\lambda_i\}}[U]={\det\,\epsilon_i^{n_j+N-j}\over\det\,\epsilon_i^{N-j}}
={\Delta_f (\epsilon)\over \Delta_0 (\epsilon)}\quad .\eqno(C.3)
$$

The dimension of the representation is simply obtained as the character
computed on the unit element, that is
$$
{\rm dim \, }R = \tr_R \, 1 = \chi_R [1]\quad .\eqno(C.4)
$$

Consider a manifold $M$ on which the group acts. It is clear that tensors also
provide representations for the group action. The irreducible representations
are classified by means of the symmetry of the tensor with respect to indices
exchange, in a Young tableau as in Fig. 11; there, the
i$^{\rm th}$ row has $n_i$ elements, $n_i \ge n_{i+1}$, and the tensor is
constructed in such a way as to be symmetric in the indices in a same row, and
later antisymmetrized with respect to indices in the same column:
\penalty-2000
$$\epsfysize=2truecm\epsfbox{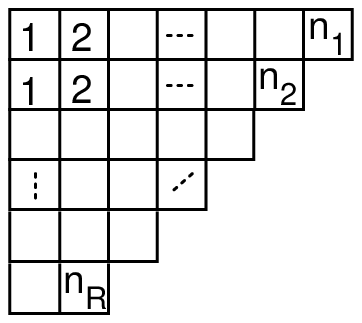}$$
\nobreak
\noindent{\eightpoint Fig. 11: A general Young tableau.}
\vskip.3truecm
\penalty-2000

The Young frame is then given in terms of the properties under permutations.
The length of the rows, $n_1\ge n_2\ge \cdots \ge n_k > 0$ defines the
representation. They are the analogue of the total angular momentum. A
standard  tableau is defined as the one where the integers $1,2,\cdots ,N\,$,
are distributed in non-decreasing order from the left in every row and
in increasing order from the top in every column. The $n_i$'s above represent
the highest-weight vectors.

The best property of characters, however, is the fact that they obey
orthogonality and completeness relations, once an invariant metric is defined.
This means that functions of the group can be  decomposed in a ``Fourier
series", where the ``Fourier elements" are the characters.

Let us denote by ${\cal D}_{\alpha \beta}^R(g)$ a $d_R$-dimensional
representation in terms of a matrix. The integral
$$
\int {\cal D}g\, {\cal D}_{\alpha \beta}^R(g){\cal D}_{\gamma \delta}^S(g^{-1})
\quad ,\eqno(C.5)
$$
where ${\cal D}g$ is the Haar measure, cannot be non-vanishing unless $R=S$,
otherwise one would be able to construct a homomorphism from $R$ to $S$. On
the other hand, when they are equal, the result must be proportional to
$\delta_{\alpha \delta}\delta_{\beta\gamma}$, and one finds
$$
\int {\cal D}g\, {\cal D}_{\alpha \beta}^R(g){\cal D}_{\delta\gamma}^S(g^{-1})
= {1\over {\rm dim \,}R }\,
\delta_{RS}\delta_{\alpha\gamma}\delta_{\beta\delta}
\quad .\eqno(C.6)
$$

Along the same lines one also finds  (for proofs, we refer to the mathematical
literature; see ref. [76])
$$
\sum_{R,\alpha,\beta}({\rm dim\, }R)\,{\cal D}_{\alpha \beta}^R(g)
{\cal D}_{\beta \alpha}^R(h^{-1}) = \delta (g,h)\quad .\eqno(C.7)
$$

These relations imply, for characters, the results
$$
\eqalignno{
\int {\cal D}g\, \chi_R[g] \, \chi_S[g] & = \delta_{RS}\quad ,&(C.8a)\cr
\sum_R ({\rm dim \, }R) \, \chi_R[gh^{-1}] & = \delta(g,h)\quad .&(C.8b)\cr}
$$

Moreover
$$
\int {\cal D}g\, \chi_R[g] \, \chi_S[g^{-1}h] = {1\over {\rm dim \, }R }
\delta_{RS} \chi_R[h] \quad ,\eqno(C.9)
$$
which upon use of Haar measure invariance implies
$$
\int {\cal D}g\, \chi_R[g h] \, \chi_S[g^{-1} m] = {1\over {\rm dim \, }R }
\delta_{RS} \chi_R[h m] \quad .\eqno(C.10)
$$

Using (C.6), we can also show that
$$
\int {\cal D}g\, \chi_R[g h g^{-1} m]  = {1\over {\rm dim \, }R } \chi_R[h]
\chi_R [m]\quad ,\eqno(C.11)
$$

In view of the orthogonality relation of the characters:
$$
\int {\rm d}V\, \chi_{R_1}[U_{p_1}V]\, \chi_{R_2}[V^+ U_{p_2}] = {1\over
{\rm dim}\, R}\, \delta_{R_1, R_2}\, \chi[U_1 U_2] \quad,\eqno(C.12)
$$
and finally we write down the combination formula for characters
$$
\sum_{\sigma \in S_n}d_R^{-1}\chi_R[\sigma]\chi_R[\varphi] = \sum_\sigma
\chi_R[\sigma\varphi]\quad,\eqno(C.13)
$$

\vskip .5cm
\penalty-330
\noindent {\it Casimir and dimensions}
\vskip .5cm
\nobreak
\noindent Besides the characters, important concepts in the theory use of
representations are the Casimir operators, and the dimension. The latter can be
simply obtained from the character computed at the unit element, namely
$$
{\rm dim \, }R = \chi_R [1]\quad .\eqno(C.13)
$$

The Casimirs are invariants under the group transformation, and the
or\-der-\-``$p$" Ca\-si\-mir is given by the trace of the product
of all possible
$p$ elements of the group, that is
$$
C_p = \sum _{\{ i_i\}} A_{i_1 i_2} A_{i_2 i_3} \cdots A_{i_p i_1}\quad .
\eqno(C.14)
$$

We will need here the quadratic Casimir
$$
C_2 = \tr \, A^2\quad ,\eqno(C.15)
$$

Given the Young tableau, which is equivalent to giving their representation,
the
computation of Casimirs and dimensions is a well-defined problem in group
theory, and the result is known. The actual procedure is however too long to be
discussed in full generality. For $U(N)$ or $SU(N)$, the results are rather
simple. The quadratic Casimir can be computed acting with
$$
X_2 = E_{-\alpha} E_{\alpha} + H^2\quad ,\eqno(C.16)
$$
on a highest-weight vector. One finds, for the eigenvalue:
$$
C_2(R)=m^2+\sum_\alpha\vec\alpha\cdot\vec m=m^2+2\vec m\cdot\vec r\quad.
\eqno(C.17)
$$

The values of $r_i$ can be found in ref. [76], $2r_i = N+1-2i$, and one finds
$$
C_2(R) = \sum _{i=1}^N n_i (n_i + 1 - 2i) + Nn\quad ,\eqno(C.18)
$$
for $U(N)$, with $\sum n_i=n$. For $SU(N)$ the last term must be dropped, and
one has to use $ \tilde n_i=n_i-{1\over N}\sum n_i$, and after some simple
algebra one finds
$$
C_2^{SU(N)}(R) = \sum n_i (n_i + 1 - 2i) + Nn - {n^2\over N} \quad .\eqno(C.19)
$$

The dimension is given by Weyl's formula and we have, using $h_i=n_i+N-i$,
$$
{\rm dim \, } R = {\prod_{i<j} (h_i - h_j)\over \prod _{i<j} (n_i - n_j +j-i)}
\quad .\eqno(C.20)
$$

\vskip 1cm
\penalty-300
\centerline {\bf References}
\vskip 1cm
\nobreak
\refer[[1]/C.G. Callan, Phys. Rev. {\bf D 2} (1970) 1541]

\refer[/K. Symanzik, Commun. Math. Phys. {\bf 18} (1970) 227;  Springer Tracts
in Mod. Phys. {\bf 57} (1971) 222]

\refer[/D.J. Gross and F.A. Wilczek, Phys. Rev. Lett. {\bf 30} (1973) 1346]

\refer[/H.D. Politzer, Phys. Rep. {\bf 14} (1974) 129]

\refer[[2]/G. K\"allen, {\it Quantum electrodynamics} (Springer Verlag, Berlin,
1972)]

\refer[[3]/W. Marciano and H. Pagels, Phys. Rep. {\bf C 36} (1978) 137]

\refer[[4]/S. Weinberg, Phys. Rev. Lett. {\bf 19} (1967) 1264]

\refer[/A. Salam, in {\it Elementary particle theory}, ed. N. Svartholm
(Almquist and Wiksell, Stockholm, 1968)]

\refer[[5]/P. van Nieuwenhuizen, Phys. Rep. {\bf 68} (1981) 189]

\refer[/E. Witten, Phys. Lett. {\bf B 105} (1981) 267]

\refer[[6]/J.H. Schwarz, ed. {\it Superstrings, The first fifteen years of
superstrings theory} (World Scientific, Singapore, 1985)]

\refer[[7]/M.B. Green, J.H. Schwarz and E. Witten, {\it Superstrings theory}
(Cambridge University Press, 1987)]

\refer[[8]/E. Abdalla, M.C.B. Abdalla and K. Rothe, {\it Non-perturbative
methods in two-\-dimen\-sional quantum field theory} (World Scientific,
Singapore, 1991)]

\refer[[9]/J. Schwinger, Phys. Rev. {\bf 128} (1962) 2425; Theor. Phys. (IAEA)
Vienna (1963) 88]

\refer[[10]/J. Lowenstein and J.A. Swieca, Ann. Phys. {\bf 68} (1971) 172]

\refer[[11]/K. Fredenhagen, {\it Criteria for quark confinement}, Proceedings
Math. Phys. 499, Marseille 1986]

\refer[/K. Fredenhagen and M. Marcu, Phys. Rev. Lett. {\bf 56} (1986) 223]

\refer[[12]/J.A. Swieca, Fortsch. Phys. {\bf 25} (1977) 303--326]

\refer[[13]/G. 'tHooft, Nucl. Phys. {\bf B 72} (1974) 461]

\refer[[14]/G. 'tHooft, Nucl. Phys. {\bf B 75} (1974) 461]

\refer[[15]/A.M. Polyakov and P.B. Wiegmann, Phys. Lett. {\bf B 131} (1983)
121; {\bf B 141} (1984) 223]

\refer[[16]/J. Wess and B. Zumino. Phys. Lett. {\bf B 37} (1971) 95]

\refer[[17]/E. Witten, Commun. Math. Phys. {\bf 92} (1984) 455]

\refer[[18]/P. di Vecchia, B. Durhuus and J.L. Petersen, Phys. Lett.
{\bf B 144} (1984) 245]

\refer[/J.L. Petersen, Acta Phys. Polonica {\bf B 16} (1985) 271]

\refer[/Y. Frishman and J. Sonnenschein, Nucl. Phys. {\bf B 294} (1987) 801]

\refer[[19]/D. Gonzales and A.N. Redlich, Phys. Lett. {\bf B 147} (1984) 150;
Nucl. Phys. {\bf B 256} (1985) 621]

\refer[/E. Abdalla and M.C.B. Abdalla, Nucl. Phys. {\bf B 255} (1985) 392]

\refer[[20]/V. Baluni, Phys. Lett. {\bf B 90} (1980) 407]

\refer[[21]/P. Mitra and P. Roy, Phys. Lett. {\bf B 79} (1978) 469]

\refer[/A. Patrascioiu, Phys. Rev. {\bf D 15} (1977) 3592]

\refer[[22]/P.J. Steinhardt, Nucl. Phys. {\bf B 176} (1980) 100]

\refer[[23]/D. Gepner, Nucl. Phys. {\bf B 252} (1985) 481]

\refer[[24]/Y. Frishman and J. Sonnenschein, Phys. Rep. {223} (1993) 309]

\refer[[25]/T.T. Wu, Phys. Lett. {\bf B 71} (1977) 142]

\refer[[26]/Y. Frishman, C.T. Sachrajda, H. Abarbanel and R. Blankenbecler,
Phys. Rev. {\bf D 15} (1977) 2275]

\refer[[27]/C.G. Callan, N. Coote and D.J. Gross, Phys. Rev. {\bf D 13} (1976)
1649]

\refer[[28]/S.J. Brodsky and G.R. Farrar, Phys. Rev. Lett. {\bf 31} (1973)
1153; Phys. Rev. {\bf D 11} (1975) 1309]

\refer[[29]/E.D. Bloom and F.J. Gilman, Phys. Rev. Lett. {\bf 25} (1970) 1140;
Phys. Rev. {\bf D 4} (1971) 2901]

\refer[[30]/E. Abdalla, M.C.B. Abdalla, D. Dalmazi and A. Zadra,  Lecture
Notes in Physics, vol. M20 (Springer Verlag, Heidelberg, 1994)]

\refer[[31]/A. D'Adda, P. di Vecchia and M. L\"uscher, Nucl. Phys.
{\bf B 146} (1978) 63]

\refer[[32]/D.J. Gross and A. Neveu, Phys. Rev. {\bf D 10} (1974) 3235]

\refer[[33]/E. Abdalla, M.C.B. Abdalla and M. Gomes, Phys. Rev. {\bf D 23}
(1981) 1800; {\bf D 25} (1982) 452; {\bf D 27} (1983) 825]

\refer[[34]/E. Abdalla, B. Berg and P. Weisz, Nucl. Phys. {\bf B 157} (1979)
387]

\refer[[35]/A. Capelli, C. Trugenberger and G. Zemba, Nucl. Phys. {\bf B 396}
(1993) 465; Phys. Rev. Lett. {\bf 72} (1994) 72]

\refer[/S. Iso, D. Karabali and B. Sakita, Nucl. Phys. {\bf B 388} (1992) 700]

\refer[/D. Karabali, Nucl. Phys. {\bf B 428} (1994) 531]

\refer[[36]/A. Dhar, G. Mandal and S. Wadia, Mod. Phys. Lett. {\bf A 8} (1993)
3557]

\refer[/D. Das, A. Dhar, G. Mandal and S. Wadia, Mod. Phys. Lett. {\bf A 7}
(1992) 71]

\refer[[37]/K. Fujikawa, Phys. Rev. Lett. {\bf 42} (1979) 1195; {\bf 44}
(1980) 1733;]

\refer[/K. Fujikawa, Phys. Rev. {\bf D 21} (1980) 2848; {\bf D 22} (1980)
1499 (E); {\bf D 23} (1981) 2262]

\refer[[38]/N.K. Nielsen, K.D. Rothe and B. Schroer, Nucl. Phys. {\bf B 160}
(1979) 330]

\refer[[39]/H. Partovi, Phys. Lett. {\bf B 80} (1978) 377]

\refer[/C. Sorensen and G.H. Thomas, Phys. Rev. {\bf D 21} (1980) 1625]

\refer[[40]/M. Einhorn, Phys. Rev. {\bf D 14} (1976) 3451]

\refer[[41]/M. Einhorn, S. Nussimov and E. Rabinovici, Phys. Rev. {\bf D 15}
(1977) 2282]

\refer[[42]/I.O. Stamatescu and T.T. Wu, Nucl. Phys. {\bf B 143} (1978)
503--520]

\refer[[43]/R. Brout, F. Englert and J-M. Fr\`ere, Nucl. Phys. {\bf B 134}
(1978) 327]

\refer[/P. Mitra, Phys. Lett. {\bf B 92} (1980) 324]

\refer[[44]/S.P. Novikov, Usp. Mat. Nauk {\bf 37} (1982) 3]

\refer[[45]/V. Knizhnik and A.B. Zamolodchikov, Nucl. Phys. {\bf B 247} (1984)
83]

\refer[[46]/E. Abdalla and M.C.B. Abdalla, to appear in Int. J. Mod. Phys.
{\bf A}]

\refer[[47]/E. Abdalla and M.B.C. Abdalla, Phys. Lett. {\bf  B 337} (1994)
347--353]

\refer[[48]/M.B. Halpern, Phys. Rev. {\bf D13} (1976) 337; {\bf D 12} (1975)
1684]

\refer[[49]/A.G. Izergin, V.E. Korepin, M.A. Semenov-Tian-Shanskii and L.D.
Faddeev, Theor. J. Math. Phys. {\bf 38} (1979) 1]

\refer[[50]/G.D. Date, Y. Frishman and J. Sonnenschein, Nucl. Phys.
{\bf B 283} (1987) 365]

\refer[[51]/P. Goddard, A. Kent and D. Olive, Phys. Lett. {\bf B 152} (1985)
88; Commun. Math. Phys. {\bf 103} (1986) 105.]

\refer[/K. Bardacki, E. Rabinovici and B. S\"aring, Nucl. Phys. {\bf B 299}
(1988) 151]

\refer[[52]/K. Bardacki and M. Halpern, Phys. Rev. {\bf D 3} (1971) 2493]

\refer[/K. Gawedzki and A. Kupiainen, Phys. Lett. {\bf B 215} (1988) 119]

\refer[[53]/D. Karabali and H.J. Schnitzer, Nucl. Phys. {\bf B 329} (1990)
649]

\refer[[54]/P. Goddard and D. Olive, Int. J. Mod. Phys. {\bf A 1} (1986) 303]

\refer[[55]/P.A.M. Dirac, {\it Lectures on quantum mechanics} (Yeshiva Univ.
Press, New York, 1964); Can. J. Math. {\bf 2} (1950) 129]

\refer[[56]/S. Adler, Phys. Rev. {\bf 177} (1969) 2426; In {\it Lectures on
elementary particle physics and quantum field theory}, eds. S. Deser et al.,
Vol. I (MIT Press, Boston, 1970);]

\refer[/S.B. Treiman, R. Jackiw, B. Zumino and E. Witten, {\it Current algebra
and anomalies} (World Scientific, Singapore, 1985)]

\refer[[57]/K. Harada and I. Tsutsui, Phys. Lett. {\bf B 183} (1987) 311]

\refer[[58]/O. Babelon, F.A. Schaposnik and C.M. Viallet, Phys. Lett.
{\bf B 177} (1986) 385]

\refer[[59]/H.O. Girotti and K.D. Rothe, Int. J. Mod. Phys. {\bf A 4} (1989)
3041;]

\refer[/K. Harada and H. Mukaido, Z. Phys. {\bf C 48} (1990) 151--158]

\refer[[60]/R. Jackiw and R. Rajaraman, Phys. Rev. Lett. {\bf 54} (1985) 1219]

\refer[[61]/H. Lehmann, K. Symanzik and W. Zimmermann, Nuovo Cimento {\bf 1}
(1955) 205]

\refer[[62]/A.S. Wightman, Phys. Rev. {\bf 101} (1956) 860]

\refer[[63]/C. Detar, J. Finkelstein and C.I. Tan (eds.), {\it A passion for
physics}, Essays in Honor of Geoffrey Chew (World Scientific, Singapore, 1985)]

\refer[[64]/M. Jacob (ed.) {\it Dual theory}, Physics Reports reprint vol. 1
(North Holland, Amsterdam, 1974)]

\refer[[65]/G. Veneziano, Nuovo Cimento {\bf A 53} (1968) 190]

\refer[[66]/S. Elitzur, Phys. Rev. {\bf D 12} (1975) 3978]

\refer[[67]/K. Wilson, Phys. Rev. {\bf 179} (1969) 1499]

\refer[[68]/C. Itzykson and J.M. Drouffe, {\it Statistical field theory}, vol.
1 (Cambridge Univ. Press, 1989)]

\refer[[69]/A. Migdal, Sov. Phys. JETP {\bf 42} (1976) 413]

\refer[[70]/D. Gross, Nucl. Phys. {\bf B 400} (1993) 161]

\refer[[71]/D. Gross and W. Taylor, Nucl. Phys. {\bf B 400} (1993) 181]

\refer[[72]/D. Gross and W. Taylor, Nucl. Phys. {\bf B 403} (1993) 395]

\refer[[73]/S. Cordes, G. Moore and S. Ramgoolam, Rutgers preprint, RU-94-20,
hep-th/9402107; Lectures on 2D Yang--Mills theory, equivariant cohomology and
topological string theory, Trieste Spring School, April 1994]

\refer[[74]/S. K\"oster, Durham preprint, DTP 94-33, hep-th/9408140]

\refer[[75]/S. Okubo, J. Math. Phys. {\bf 18} (1977) 2382]

\refer[[76]/A.O. Barut and R. Raczka, {\it Theory of groups representations
and applications} (Polish Sci. Publ., 1977)]

\refer[[77]/H. Weyl, {\it The classical groups} (Princeton Univ. Press, 1946)]

\refer[[78]/D. Gross and E. Witten, Phys. Rev. {\bf D 21} (1980) 446]

\refer[[79]/V.A. Kazakov and I.K. Kostov, Nucl. Phys. {\bf B 176} (1980) 199]

\refer[[80]/V.A. Kazakov, Nucl. Phys. {\bf B 179} (1981) 283; Phys. Lett.
{\bf B 128} (1983) 316]

\refer[[81]/Y. Makeenko and A. Migdal, Phys. Lett. {\bf B 88} (1979) 135]

\refer[[82]/M. Douglas and V.A. Kazakov, Phys. Lett. {\bf B 319} (1993) 219]

\refer[[83]/H. Farkas and I. Kra, {\it Riemann surfaces} (Springer Verlag,
Heidelberg, 1980)]

\refer[[84]/D.J. Gross, Two-dimensional Quantum Gravity, Jerusalem Winter
School for Theoretical Physics, eds. D.J. Gross, T. Piran and S. Weinberg
(World
Scientific, Singapore, 1991)]

\refer[[85]/A. Jevicki and B. Sakita, Nucl. Phys. {\bf B 165} (1980) 511;
{\bf B 185} (1981) 89]

\refer[[86]/S.R. Das and A. Jevicki, Mod. Phys. Lett. {\bf A 5} (1990)
1639--1650]

\refer[[87]/J.A. Minahan and A.P. Polychronakos, Phys. Lett. {\bf B 312}
(1993) 155]

\refer[[88]/J.A. Minahan and A.P. Polychronakos, Phys. Lett. {\bf B 326} (1994)
288]

\refer[[89]/M.R. Douglas, Rutgers preprint RU-93-13, hep-th/9311130]

\refer[[90]/E. Brezin, C. Itzykson, G. Parisi and J.B. Zuber, Commun. Math.
Phys. {\bf 59} (1978) 35]

\refer[[91]/I. Bars and F. Green, Phys. Rev. {\bf D 20} (1979) 3311]

\refer[[92]/S. Dalley and I. Klebanov, Phys. Rev. {\bf D 47} (1993) 2517]

\refer[[93]/D. Kutasov, Nucl. Phys. {\bf B 414} (1994) 33]

\refer[[94]/G. Bhanot, K. Demeterfi and I.R. Klebanov, Phys. Rev. {\bf D 48}
(1993) 4980; Nucl. Phys. {\bf B 418} (1994) 15]

\refer[[95]/J. Boorstein and D. Kutasov, Univ. of Chicago preprint EFI-94-42,
hep-th/9409128]

\refer[[96]/J. Polchinski and A. Strominger, Phys. Rev. Lett. {\bf 67}
(1991) 1681]

\refer[[97]/J. Polchinski, Phys. Rev. Lett. {\bf 68} (1992) 1267 ]

\refer[[98]/J.J. Atick and E. Witten, Nucl. Phys. {\bf B 310} (1988) 291]

\refer[[99]/M.R. Douglas, K. Li and M. Staudacher, Nucl. Phys. {\bf B 420}
(1994) 118--140]

\refer[[100]/O. Ganor, J. Sonnenschein and S. Yankielowicz, Nucl. Phys.
{\bf B 434} (1995) 139]

\refer[[101]/B. Rusakov and S. Yankielowicz, Phys. Lett. {\bf B 339} (1994)
258--262]

\refer[[102]/A. Dhar, G. Mandal and S.R. Wadia, Phys. Lett. {\bf B 329} (1994)
15--26]

\refer[[103]/D.J. Gross and I. Klebanov, Nucl. Phys. {\bf B 352} (1990) 671]

\refer[[104]/M. Caselle, A. D'Adda, L. Magnea and S. Panzeri, Nucl. Phys.
{\bf B 416} (1994) 751]

\refer[[105]/J. Avan and A. Jevicki, Phys. Lett. {\bf B 226} (1991) 35;
{\bf B 272} (1991) 17]

\refer[/A. Jevicki, Nucl. Phys. {\bf B 376} (1992) 75]

\refer[[106]/D. Fairlie and C. Zachos, Phys. Lett. {\bf B 224} (1989) 101]

\refer[/D. Fairlie, P. Fletcher and C. Zachos, J. Math. Phys. {\bf 31} (1990)
1088]

\refer[/I. Bakas, Commun. Math. Phys. {\bf 134} (1990) 487]

\refer[/C.N. Pope, L.J. Romans and X. Shen, Nucl. Phys. {\bf B 339} (1990) 191]

\refer[/E. Kiritsis and I. Bakas, Nucl. Phys. {\bf B 343} (1990) 185,E
{\bf B 350} (1991) 512; Int. J. Mod. Phys. {\bf A 6} (1991) 2871; Mod. Phys.
Lett. {\bf A 5} (1990) 2039]

\refer[[107]/J.A. Minahan and A.P. Polychronakos, Phys. Rev. {\bf B 50} (1994)
4236]

\refer[[108]/E. Abdalla, M.C.B. Abdalla, G. Sotkov and  M. Stanishkov, to
appear in Int. J. Mod. Phys. {\bf A}]

\refer[/G. Mandal, A. Sengupta and S.R. Wadia, Mod. Phys. Lett. {\bf A 6}
(1991)
1465]

\refer[[109]/K. Pohlmeyer, Commun. Math. Phys. {\bf 46} (1976) 207]

\refer[/M. L\"uscher and K. Pohlmeyer, Nucl. Phys. {\bf B 137} (1978) 46]

\refer[/H. Eichenherr and M. Forger, Nucl. Phys. {\bf B 164} (1980) 528;
{\bf B 282} (1987) 745]

\refer[[110]/L.D. Faddeev, {\it Lectures on quantum inverse scattering method},
in Nankai Lectures on Mathematical Physics and Integrable Systems, ed. X.C.
Song, (World Scientific, Singapore, 1990)]

\refer[[111]/A. Gorsky and N. Nekrasov, Nucl. Phys. {\bf B 414} (1994) 213]

\refer[[112]/C.P. Burgess and F. Quevedo, Phys. Lett. {\bf B 329} (1994) 457]

\refer[[113]/L. Alvarez-Gaum\'e, in 1993 Summer School in High Energy Physics,
eds. E. Gava, A. Masiero, K.S. Narain, S. Randjbar-Daemi and Q. Shafi, p. 256,
(World Scientific, Singapore)]

\refer[/E. Alvarez, L. Alvarez-Gaum\'e and Y. Lozano, Nucl. Phys. {\bf B 424}
(1994) 155]

\refer[[114]/K. Kikkawa and M. Yamasaki, Phys. Lett. {\bf B 149} (1984) 357]

\refer[/T.H. Buscher, Phys. Lett. {\bf B 194} (1987) 51; {\bf B 201} (1988)
466]

\refer[/E. Alvarez and M.A.R. Osorio, Phys. Rev. {\bf D 40} (1989) 1150]

\refer[/E. Kiritsis, Mod. Phys. Lett. {\bf A 6} (1991) 2871]

\refer[/M. Rocek and E. Verlinde, Nucl. Phys. {\bf B 373} (1992) 630]

\refer[/E. Kiritsis, Nucl. Phys. {\bf B 405} (1993) 109]

\refer[/A. Giveon and E. Kiritsis, Nucl. Phys. {\bf B 411} (1994) 487]

\refer[[115]/E. Abdalla and K. Rothe, Phys. Rev. {\bf D 361} (1987) 3190]

\refer[[116]/E. Abdalla, M. Forger and M. Gomes, Nucl. Phys. {\bf B 210} (1982)
181]

\refer[[117]/V.G. Knizhnik and A.B. Zamolodchikov, Nucl. Phys. {\bf B 247}
(1984) 83]

\refer[[118]/A.A. Belavin, A.M. Polyakov and A.B. Zamolodchikov, Nucl. Phys.
{\bf B 241} (1994) 333]

\refer[[119]/E. Abdalla and M.C.B. Abdalla, Phys. Rev. {\bf D 40} (1989) 491]

\refer[[120]/A.C. Scott, F.Y.F. Chu and D.W. Mc Laughlin, Proceedings IEEE
{\bf 61} (1973) 1443]

\refer[[121]/S. Coleman and J. Mandula, Phys. Rev. {\bf 159} (1967) 1251]

\refer[[122]/M.K. Prasad, A. Sinha and L.L. Chau-Wang, Phys. Rev. Lett.
{\bf 43} (1979) 750]

\refer[/I.Ya Aref'eva and I.V. Volovich, Phys. Lett. {\bf B 149} (1984) 131]

\refer[/L.L. Chau, M.K. Prasad and A. Sinha, Phys. Rev. {\bf D 23} (1981) 2321;
{\bf D 24} (1981) 1574]

\refer[/E. Witten, Phys. Lett. {\bf B 77} (1978) 394; Nucl. Phys. {\bf B 266}
(1986) 245]

\refer[/E. Abdalla, M Forger and M. Jacques, Nucl. Phys. {\bf B 307} (1988)
198]

\refer[/J. Avan, H.J. de Vega and J.M. Maillet, Phys. Lett. {\bf B 171} (1986)
255]

\refer[/J. Harnad, J. Hurtubise, M. L\'egar\'e and S. Shnider, Nucl. Phys.
{\bf B 256} (1985) 609]

\refer[[123]/E. Verlinde and H. Verlinde, Princeton preprint PUPT-1319,
hep-th/9302022]

\refer[[124]/L.N. Lipatov, Nucl. Phys. {\bf B 365} (1991) 614]

\refer[[125]/L.N. Lipatov, JETP Lett. {\bf 59} (1994) 596]

\refer[[126]/L.N. Lipatov, Phys. Lett. {\bf B 251} (1990) 284; {\bf 309} (1993)
394; Sov. Phys. JETP {\bf 63} (1986) 904]

\refer[/Ya. Balitsky and L.N. Lipatov, Sov. J. Nucl. Phys. {\bf  28} (1978)
822]

\refer[[127]/L.D. Faddeev and G.P. Korchemsky, Phys. Lett. {\bf B 342} (1995)
311]

\refer[[128]/G.P. Korchemsky, Stony Brook preprint ITP-SB-94-62,
hep-ph/9501232]

\refer[[129]/F. Calogero, J. Math. Phys. {\bf 10} (1969) 2191, 2197; {\bf 12}
(1971) 419]

\refer[/B. Sutherland, Phys. Rev. {\bf A 4} (1971) 2019; {\bf A 5} (1972) 1372;
Phys. Rev. Lett. {\bf 34} (1975) 1083]

\refer[[130]/V.A. Kazakov and A.A. Migdal, Nucl. Phys. {\bf B 397} (1993) 214]

\refer[[131]/J. Bartels, Nucl. Phys. {\bf B 151} (1979) 293, {\bf B 175}
(1980) 365]

\refer[[132]/H. Cheng and T.T. Wu, {\it Expanding protons: Scattering at high
energies} (MIT Press, Boston, 1987)]

\refer[[133]/H. Cheng and T.T. Wu, Phys. Rev. {\bf 182} (1969) 1852, 1868,
1873, 1899]

\refer[[134]/E.A. Kuraev, L.N. Lipatov and V.S. Fadin, Sov. Phys. JETP
{\bf 44} (1976) 443]

\refer[[135]/E.A. Kuraev, L.N. Lipatov and V.S. Fadin, Sov. Phys. JETP
{\bf 45} (1977) 199]

\refer[[136]/L.N. Lipatov, Sov. Phys. JETP {\bf 63} (1986) 904]

\refer[[137]/B.H. Gribov, JETP {\bf 53} (1967) 654]

\refer[/H.D.I. Abarbanel and J.B. Bronzan, Phys. Rev. {\bf D 9} (1974) 2397]

\refer[[138]/V.O. Tarosov, L.A. Takhtajan and L.D. Faddeev, J. Theor. Math.
Phys. {\bf 57} (1983) 163]

\refer[/P.P. Kulish, N.Yu. Reshetikin and E.K. Sklyanin, Lett. Math. Phys.
{\bf 5} (1981) 393]

\refer[[139]/H. Zieschang, E. Vogt and H.-D. Coldewey, Lecture Notes in
Mathematics, vol. 122 (Springer Verlag, Heidelberg, 1970)]

\refer[[140]/Th. Jolic{\oe}ur and J.C. Le Guillou, Int. J. Mod. Phys. {\bf A 8}
(1993) 1923]

\refer[[141]/D. Kutasov and A. Schwimmer, Univ. of Chicago preprint EFI-94-67,
hep-th/9501024]

\refer[[142]/M. Bochicchio, Rome Univ. preprint Rome-1089-1995, hep-th/9503013]

\refer[[143]/K. Hornbostel, S.J. Brodsky and H.C. Pauli, Phys. Rev. {\bf D 41}
(1990) 3814]

\refer[/P. Gaete, J. Gamboa and I. Schmidt, Phys. Rev. {\bf D 49} (1994) 5621]

\refer[[144]/S.G. Rajeev, Int. J. Mod. Phys. {\bf A 9} (1994) 5583--5624]

\refer[[145]/I. Pesando, Mod. Phys. Lett. {\bf A 9} (1994) 2927]

\refer[/M. Cavicchi, Int. J. Mod. Phys. {\bf A 10} (1995) 167--198]
\end